\documentclass[12pt]{article}

\usepackage{graphicx}
\usepackage{amsmath,amssymb}
\usepackage{bm,bbm}
\usepackage{graphicx, color}
\usepackage{wrapfig}

\begin{document}
\title{
\begin{flushright}
\ \\*[-80pt] 
\begin{minipage}{0.2\linewidth}
\normalsize
KUNS-2260 \\*[50pt]
\end{minipage}
\end{flushright}
{\Large \bf 
Non-Abelian Discrete Symmetries \\ in 
Particle Physics
\\*[20pt]}}

\author{
\centerline{
Hajime~Ishimori$^{1}$, \
Tatsuo~Kobayashi$^{2}$, \ 
Hiroshi~Ohki$^{2}$,}  \\
\centerline{
Hiroshi Okada$^{3}$, \ 
Yusuke~Shimizu$^{1}$ \
and \  Morimitsu~Tanimoto$^{4}$ }
\\*[20pt]
\centerline{
\begin{minipage}{\linewidth}
\begin{center}
$^1${\it \normalsize
Graduate~School~of~Science~and~Technology,~Niigata~University, \\ 
Niigata~950-2181,~Japan } \\
$^2${\it \normalsize 
Department of Physics, Kyoto University, 
Kyoto 606-8502, Japan} \\
$^3${\it \normalsize 
Centre for Theoretical Physics, The British University in Egypt, 
El-Sherouk City, 11837,  Egypt} \\
$^4${\it \normalsize
Department of Physics, Niigata University,~Niigata,  950-2181, Japan } 
\end{center}
\end{minipage}}
\\*[50pt]}
\vskip 2 cm
\date{\small
\centerline{ \bf Abstract}
\begin{minipage}{0.9\linewidth}
\medskip 
We  review  pedagogically non-Abelian discrete groups,
which  play  an important role in the particle physics.
We  show  group-theoretical aspects for many concrete groups, 
such as representations, their tensor products.
We  explain   how to derive, conjugacy classes, 
characters, representations, and tensor products for 
these groups (with a finite number).
We discussed  them explicitly for $S_N$, $A_N$, 
$T'$, $D_N$, $Q_N$, $\Sigma(2N^2)$, $\Delta(3N^2)$, $T_7$, 
$\Sigma(3N^3)$ and $\Delta(6N^2)$,
 which have been  applied for  model building in the particle physics.
We also present   typical flavor models by using 
 $A_4$, $S_4$, and $\Delta (54)$ groups.
Breaking patterns of discrete groups and 
decompositions of multiplets are important 
for applications of the  non-Abelian discrete symmetry.
We  discuss   these  breaking patterns of 
the non-Abelian discrete group, 
which are  a powerful tool for model buildings.
We also review briefly  about anomalies of non-Abelian discrete symmetries
by using the path integral approach.
\end{minipage}
}

\begin{titlepage}
\maketitle
\thispagestyle{empty}
\clearpage
\thispagestyle{empty}
\tableofcontents
\thispagestyle{empty}
\end{titlepage}


\def\2tvec#1#2{
\left(
\begin{array}{c}
#1  \\
#2  \\   
\end{array}
\right)}

\def\mat2#1#2#3#4{
\left(
\begin{array}{cc}
#1 & #2 \\
#3 & #4 \\
\end{array}
\right)
}

\def\Mat3#1#2#3#4#5#6#7#8#9{
\left(
\begin{array}{ccc}
#1 & #2 & #3 \\
#4 & #5 & #6 \\
#7 & #8 & #9 \\
\end{array}
\right)
}

\def\3tvec#1#2#3{
\left(
\begin{array}{c}
#1  \\
#2  \\   
#3  \\
\end{array}
\right)}

\def\4tvec#1#2#3#4{
\left(
\begin{array}{c}
#1  \\
#2  \\   
#3  \\
#4  \\
\end{array}
\right)}

\def\5tvec#1#2#3#4#5{
\left(
\begin{array}{c}
#1  \\
#2  \\
#3  \\
#4  \\
#5  \\
\end{array}
\right)}

\def\L{\left}
\def\R{\right}

\def\pl{\partial}

\def\lra{\leftrightarrow}



\section{Introduction}

Symmetries play an important role in particle physics.
Continuous (and local)  symmetries such as 
Lorentz, Poincare and gauge symmetries are essential 
to understand several phenomena, which happen in 
particle physics  like  
strong, weak and electromagnetic interactions 
among particles.
Discrete symmetries such as $C$, $P$ and $T$ are also important.

Furthermore, Abelian discrete symmetries, $Z_N$, are also often 
imposed in order to control allowed couplings 
in model building for particle physics, 
in particular model building beyond the standard model.
For example, $R$-parity and matter parities are assumed 
in supersymmetric standard models to forbid the fast proton decay.
Such parities are also important from the viewpoint of 
dark matter.
In addition to Abelian discrete symmetries, 
non-Abelian discrete symmetries were applied for model building 
of particle physics recently, in particular to understand 
the flavor physics.

There are many free parameters in the standard model 
including its extension with neutrino mass terms 
and most of them are originated from the flavor sector, 
i.e. Yukawa couplings of quarks and leptons.
 The flavor symmetries are introduced to control 
  Yukawa couplings in the three generations
although the origin of the generations is unknown.
The quark masses and mixing angles have been discussed
in the standpoint of the flavor symmetries.
The discovery of neutrino masses and the  neutrino mixing
 \cite{Pontecorvo:1957cp,Maki:1962mu}
 has stimulated  the work of the flavor symmetries.
Recent experiments of the neutrino oscillation 
go into a  new  phase  of precise  determination of
 mixing angles and mass squared  differences  
\cite{Schwetz:2008er,Fogli:2008jx,Fogli:2009zza,GonzalezGarcia:2010er}, 
which indicate the tri-bimaximal mixing  for three flavors 
 in the lepton sector  
\cite{Harrison:2002er,Harrison:2002kp,Harrison:2003aw,Harrison:2004uh}. 
These large  mixing angles are completely  
 different from the quark mixing ones.
Therefore, it is very important
to find a natural model that leads to these mixing patterns
 of quarks and leptons with good accuracy.
Non-Abelian discrete symmetries 
are studied to apply for flavor physics, 
that is, model building to derive 
experimental values of quark/lepton masses 
and mixing angles by assuming 
non-Abelian discrete flavor symmetries 
of quarks and leptons.
Especially, the lepton mixing  has been intensively discussed
in non-Abelian discrete flavor symmetries 
as seen, e.g. 
in the review by Altarelli and Feruglio \cite{Altarelli:2010gt}.

The flavor symmetry may be  a remnant of the higher 
dimensional space-time symmetry, 
after it is broken down to the 4-dimensional Poincare symmetry 
through compactification, e.g. via orbifolding.
Actually, it  was shown how the flavor symmetry $A_4$ (or $S_4$) can arise 
if the three fermion generations are taken to live on the fixed points of a 
specific 2-dimensional orbifold \cite{Altarelli:2006kg}. 
Further non-Abelian discrete symmetries can arise in a similar setup
 \cite{Adulpravitchai:2009id}.

Superstring theory is a promising candidate for unified theory 
including gravity.
Certain string modes correspond to gauge bosons, 
quarks, leptons, Higgs bosons and gravitons as well as 
their superpartners.
Superstring theory predicts six extra dimensions.
Certain classes of discrete symmetries can be derived 
from superstring theories.
A combination among geometrical symmetries of a compact space 
and stringy selection rules for couplings enhances 
discrete flavor symmetries.  
For example, $D_4$ and $\Delta(54)$ flavor symmetries can be obtained
in heterotic orbifold models  
\cite{Kobayashi:2004ya,Kobayashi:2006wq,Ko:2007dz}.
In addition to these flavor symmetries, the $\Delta(27)$ flavor symmetry 
can be derived from magnetized/intersecting D-brane models
\cite{Abe:2009vi,Abe:2009uz,Abe:2010ii}.

There is another possibility that 
non-Abelian discrete groups are  originated from the breaking of  continuous 
(gauge) flavor symmetries
\cite{deMedeirosVarzielas:2005qg,Adulpravitchai:2009kd,Frampton:2009pr}.

Thus, the non-Abelian discrete  symmetry can arise from the underlying theory,
e.g. the string theory or compactification via orbifolding.
Also, the non-Abelian discrete-symmetries are interesting tools 
for controlling the flavor structure in model building from 
the bottom-up approach.
Hence, the non-Abelian flavor symmetries could become a bridge 
between the low-energy physics and the underlying theory. 
Therefore, it is quite important to study  
the properties of non-Abelian groups.

Non-Abelian continuous groups are well-known 
and of course there are several good reviews and books.
On the other hand, non-Abelian discrete symmetries 
may not be familiar to all of particle physicists 
compared with non-Abelian continuous symmetries.
However, non-Abelian discrete symmetries have become important 
tools for model building, in particular 
for the flavor physics. 
Our purpose of this article is to 
review pedagogically non-Abelian discrete groups with minding 
particle phenomenology and 
show group-theoretical aspects for many concrete groups explicitly, 
such as representations and their tensor products 
\cite{miller}-\cite{Ludl:2009ft}.
We show these aspects in detail for 
$S_N$ \cite{Pakvasa:1977in}-\cite{s4modelsoon}, 
$A_N$  \cite{Ma:2001dn}-\cite{Everett:2008et}, 
$T'$ \cite{Frampton:1994rk},\cite{Chen:2007afa}-\cite{Frampton:2009fw}, 
$D_N$  \cite{Bergshoeff:1995cg}-\cite{Blum:2009nh}, 
$Q_N$  \cite{Babu:2004tn}-\cite{Frigerio:2007nn},
$\Sigma(2N^2)$ \cite{Ma:2007ia},
$\Delta(3N^2)$  \cite{Branco:1983tn}-\cite{King:2009ap}, 
$T_7$\cite{Hagedorn:2008bc},
$\Sigma(3N^3)$  \cite{Hagedorn:2008bc},
and  
$\Delta(6N^2)$ groups
 \cite{King:2009ap},\cite{Escobar:2008vc}-\cite{Ishimori:2009ew}.
We explain pedagogically how to derive conjugacy classes, 
characters, representations and tensor products for 
these groups (with a finite number) when algebraic relations are given.
Thus, the readers could apply for other groups.

In applications for particle physics, 
the breaking patterns of discrete groups and decompositions 
of multiplets are also important.
Such aspects are studied in this paper.

Symmetries at the tree level are not always 
symmetries in quantum theory.
If symmetries are anomalous, breaking terms 
are induced by quantum effects.
Such anomalies are important in applications for particle 
physics.
Here, we study such anomalies for discrete symmetries 
\cite{Frampton:1994rk},\cite{Krauss:1988zc}-\cite{Luhn:2008xh} 
and show anomaly-free conditions explicitly for 
the above concrete groups.
If flavor symmetries are stringy symmetries, 
these anomalies may also be controlled by 
string dynamics, i.e. anomaly cancellation.

This article is organized as follows.
In section 2, we summarize basic group-theoretical 
aspects, which are necessary in 
the rest of sections.
The readers, which are familiar to group theory, 
can skip section 2.
In sections 3 to 12, we present non-Abelian discrete groups,
$S_N$, $A_N$, $T'$, $D_N$, $Q_N$, $\Sigma(2N^2)$,  $\Delta(3N^2)$,
$T_7$, $\Sigma(3N^3)$,  and  
$\Delta(6N^2)$, respectively.
In section 13, the breaking patterns of the non-Abelian discrete groups
are discussed.
In section 14, we review the anomaly  of non-Abelian flavor symmetries,
 which  is a   topic 
in the particle physics, 
and show the anomaly-free conditions explicitly for the above 
concrete groups.
In section 15, typical flavor models 
with the non-Abelian discrete symmetries are presented.
Section 16 is devoted to summary.
In appendix A, useful theorems on finite group theory are presented.
In appendices B and C, we show representation bases of $S_4$ and 
$A_4$, respectively, which are different from 
those in sections 3 and 4.

\clearpage

\section{Finite groups}

In this section, we summarize basic aspects on 
group theory, which are necessary in the following sections.
We use several theorems without their proofs, 
in order for the readers to read easily.
However, proofs of useful theorems are given in 
Appendix A.
(See also e.g. Refs.~\cite{miller,Hamermesh,Georgi:1982jb,Ludl:2009ft}.)
 
A group, $G$,  is a set, where multiplication is defined 
such that the following properties are satisfied:
\begin{enumerate}
\item {\bf Closure}

If $a$ and $b$ are elements of the group $G$, 
$c=ab$ is also its element.

\item{\bf Associativity}

$(ab)c=a(bc)$ for $a,b,c \in G$.

\item{\bf Identity}

The group $G$ includes an identity element $e$, which 
satisfies $ae = ea =a$ for any element $a \in G$.
 
\item{\bf Inverse}

The group $G$ includes an inverse element $a^{-1}$ for 
any element $a \in G$ such that $aa^{-1}=a^{-1}a=e$.

\end{enumerate}

The {\bf order} is the number of elements in $G$.
The order of a finite group is finite.
The group $G$ is called {\bf Abelian} if 
all of their elements are commutable each other, i.e. 
$ab = ba$.
If all of elements do not satisfy the commutativity, 
the group is called {\bf non-Abelian}.
One of simple finite groups is the cyclic group $Z_N$, which consists of 
\begin{eqnarray}
\{ e, a, a^2, \cdots, a^{N-1} \},
\end{eqnarray}
where  $a^N = e$.
The $Z_N$ group can be represented as 
discrete rotations, whose generator $a$ corresponds to 
$2\pi/N$ rotation.
The $Z_N$ group is Abelian.
We focus on non-Abelian discrete symmetries 
in the following sections.

If a subset $H$ of the group $G$ is also a group, 
$H$ is called the {\bf subgroup} of $G$.
The order of the subgroup $H$ must be a divisor of the 
order of $G$.
That is {\bf Lagrange's theorem}.
(See Appendix A.) 
If a subgroup $N$ of $G$ satisfies $g^{-1}Ng=N$ 
for any element $g \in G$, the subgroup $N$ is called 
a {\bf normal subgroup} or an {\bf invariant subgroup}.
The subgroup $H$ and normal subgroup $N$ of $G$ 
satisfy $HN=NH$ and it is a subgroup of $G$, where 
$HN$  denotes
\begin{equation}
\{h_i n_j | h_i \in H, n_j \in N    \},
\end{equation}
and $NH$ denotes a similar meaning.

When $a^h=e$ for an element $a \in G$, 
the number $h$ is called the {\bf order} of $a$.
The elements, $\{ e, a, a^2, \cdots, a^{h-1} \}$, form 
a subgroup, which is the Abelian $Z_h$ group with the order $h$.

The elements $g^{-1}ag$ for $g \in G$ are called elements 
conjugate to the element $a$.
The set including all elements to conjugate to an element $a$ of $G$, 
$\{ g^{-1}ag, \ \forall g \in G \}$, 
is called a {\bf conjugacy class}.
All of elements in a conjugacy class have the same order 
since 
\begin{eqnarray}
(gag^{-1})^h=ga(g^{-1}g)a(g^{-1}g)\cdot\cdot\cdot ag^{-1}=ga^{h}g^{-1}
=geg^{-1}=e .
\end{eqnarray}
The conjugacy class including the identity $e$ consists of 
the single element $e$.

We consider two groups, $G$ and $G'$, and a map $f$ of  
$G$ on  $G'$.
This map is {\bf homomorphic} only if 
the map preserves the multiplication structure, that is,
\begin{equation}
f(a)f(b) = f(ab),
\end{equation}
for $a,b \in G$.
Furthermore, the map is {\bf isomorphic} when 
the map is one-to-one correspondence.

A {\bf representation} of $G$ is a homomorphic map of elements of $G$ onto 
matrices, $D(g)$ for $g \in G$.
The representation matrices should satisfy 
$D(a)D(b)=D(c)$ if $ab=c$ for $a,b,c \in G$.
The vector space $v_j$, on which representation matrices act, 
is called a {\bf representation space} such as 
$D(g)_{ij} v_j$ $(j = 1,\cdots,n)$.
The dimension $n$ of the vector space $v_j$ $(j = 1,\cdots,n)$
is called as a {\bf dimension} of the representation.
A subspace in the representation space is called 
{\bf invariant subspace} if $D(g)_{ij}v_j$ for any vector $v_j$ 
in the subspace and any element $g \in G$ also corresponds 
to a vector in the same subspace.
If a representation has an invariant subspace, 
such a representation is called {\bf reducible}.
A representation is {\bf irreducible} if it has no invariant subspace.
In particular, a representation is called {\bf completely reducible} 
if $D(g)$ for $g \in G$ are written as the following 
block diagonal form,
\begin{eqnarray}
\left(
\begin{array}{cccc}
D_1(g) & 0 &  &    \\
0 & D_2(g) &  &     \\
  &    & \ddots &   \\
  &    &      & D_r(g)  \\
\end{array}
\right),
\end{eqnarray}
where each $D_\alpha(g)$ for $\alpha=1,\cdots, r$ is irreducible.
This implies that a reducible representation $D(g)$ is 
the direct sum of $D_\alpha(g)$,
\begin{eqnarray}
 \sum_{\alpha =1}^r \oplus D_\alpha(g).
\end{eqnarray}
Every (reducible) representation of a fine group 
is completely reducible.
Furthermore, every representation  of a fine group 
is equivalent to a unitary representation.(See Appendix A.)
The simplest (irreducible) representation is 
found that $D(g)=1$ for all elements $g$, that is, a trivial singlet.
The matrix representations satisfy the following orthogonality 
relation,
\begin{eqnarray}
\sum_{g \in G}  D_\alpha(g)_{i \ell} D_\beta(g^{-1})_{mj} = 
\frac{N_G}{d_\alpha} \delta_{\alpha \beta}\delta_{ij} \delta_{\ell m},
\end{eqnarray}
where $N_G$ is the order of $G$ and $d_\alpha$ is the dimension of 
the $D_\alpha(g)$.(See Appendix A.)

The {\bf character} $\chi_D(g)$ of a representation $D(g)$
is the trace of the representation matrix,
\begin{eqnarray}
\chi_D(g) = {\rm tr}~D(g) = \sum_{i=1}^{d_\alpha} D(g)_{ii}.
\end{eqnarray}
The element conjugate to $a$ has the same character 
because of the property of the trace,
\begin{eqnarray}
{\rm tr}~D(g^{-1}a g) = {\rm tr}~\left( D(g^{-1})D(a) D(g) \right)
= {\rm tr}~D(a),
\end{eqnarray}
that is, the characters are constant in a conjugacy class.
The characters satisfy the following orthogonality relation,
\begin{eqnarray}\label{eq:character-1}
\sum_{g \in G} \chi_{D_\alpha}(g)^* \chi_{D_\beta}(g) 
= N_G \delta_{\alpha \beta},
\end{eqnarray}
where $N_G$ denotes the order of a group $G$.
(See Appendix A.)
That is, the characters of different irreducible representations 
are orthogonal and different from each others.
{\it Furthermore it is found that the number of irreducible representations 
must be equal to the number of conjugacy classes.}
(See Appendix A.)
In addition, they satisfy the following orthogonality relation,
\begin{eqnarray}\label{eq:character-2}
\sum_{\alpha} \chi_{D_\alpha}(g_i)^* \chi_{D_\alpha}(g_j) 
= \frac{N_G}{n_i} \delta_{C_i C_j},
\end{eqnarray}
 where $C_i$ denotes the conjugacy class of $g_i$ 
and $n_i$ denotes the number of elements in the conjugacy class $C_i$.
(See Appendix A.)
That is, the above equation means that the right hand side 
is equal to $\frac{N_G}{n_i}$ if $g_i$ and $g_j$ belong to 
the same conjugacy class, and that otherwise it must vanish.
A trivial singlet, $D(g)=1$ for any $g\in G$, 
must always be included.
Thus, the corresponding character satisfies $\chi_1(g)=1$ 
for any $g\in G$.

Suppose that there are $m_n$ $n$-dimensional irreducible 
representations, that is, $D(g)$ are represented by $(n \times n)$ matrices.
The identity $e$ is always represented by the 
$(n \times n)$ identity matrix.
Obviously, the character $\chi_{D_\alpha}(C_1)$ for 
the conjugacy class $C_1=\{ e\}$ is found that 
$\chi_{D_\alpha}(C_1) = n$ for the $n$-dimensional representation.
Then, the orthogonality relation (\ref{eq:character-2}) requires 
\begin{eqnarray}\label{eq:character-2-e}
&&\sum_\alpha[\chi_\alpha(C_1)]^2=\sum_nm_nn^2=m_1+4m_2+9m_3+\cdots=N_G,
\end{eqnarray}
where $m_n \geq 0$.
Furthermore, $m_n$ must satisfy 
\begin{eqnarray}\label{eq:sum-dim}
&&\sum_nm_n = 
{\rm the~number~of~conjugacy~classes},
\end{eqnarray}
because the number of irreducible representations 
is equal to the number of conjugacy classes.
Eqs.~(\ref{eq:character-2-e}) and (\ref{eq:sum-dim}) 
as well as Eqs.~(\ref{eq:character-1}) and (\ref{eq:character-2})
are often used in the following sections to determine 
characters.

We can construct a lager group from more than two groups, 
$G_i$, by a certain product.
A rather simple one is the {\bf direct product}.
We consider e.g. two groups $G_1$ and $G_2$.
Their direct product is denoted as $G_1 \times G_2$, 
and its multiplication rule is defined as 
\begin{equation}
(a_1,a_2)(b_1,b_2)=(a_1b_1,a_2b_2),
\end{equation}
for $a_1,b_1 \in G_1$ and $a_2,b_2 \in G_2$.

The {\bf semi-direct product} is more non-trivial product 
between two groups $G_1$ and $G_2$, and it is defined such as 
\begin{equation}
(a_1,a_2)(b_1,b_2)=(a_1f_{a_2}(b_1),a_2b_2),
\end{equation}
for $a_1,b_1 \in G_1$ and $a_2,b_2 \in G_2$, 
where $f_{a_2}(b_1)$ denotes a homomorphic map from $G_2$ to $G_1$.
This semi-direct product is denoted as $G_1 \rtimes _f G_2$.
We consider the group G, its subgroup $H$, and normal subgroup $N$, 
whose elements are $h_i$ and $n_j$, respectively.
When $G=NH=HN$ and $N \cap H = \{ e \}$, the semi-direct product 
$N \rtimes_f H$ is isomorphic to $G$, $G \simeq  N \rtimes_f H$, 
where we use the map $f$ as 
\begin{equation}
f_{h_i}(n_j) = h_i n_j (h_i)^{-1}.
\end{equation}
For the notation of the semi-direct product, 
we will often omit $f$ and denote it as $N \rtimes H$.

\clearpage



\section{$S_N$}
\label{sec:SN}

All possible permutations among $N$ objects $x_i$ with 
$i=1,\cdots, N$, form a group,
\begin{eqnarray}
(x_1, \cdots, x_n) \rightarrow (x_{i_1},\cdots,x_{i_N}).
\end{eqnarray} 
This group is the so-called $S_N$ with the order $N !$, 
and $S_N$ is often called as the symmetric group.
In the following we show concrete aspects on 
$S_N$ for smaller $N$.
The simplest one of $S_N$ except the trivial $S_1$ 
is $S_2$, which consists of 
two permutations,
\begin{eqnarray}
(x_1, x_2) \rightarrow (x_1, x_2), 
\qquad (x_1, x_2) \rightarrow (x_2, x_1).
\end{eqnarray} 
This is nothing but $Z_2$, that is Abelian.
Thus, we start with $S_3$.

\subsection{$S_3$}
\label{subsec:S3}

\begin{figure}[t]
\unitlength=1mm
\begin{picture}(170,50)
\hspace{5cm}
\includegraphics[width=7cm]{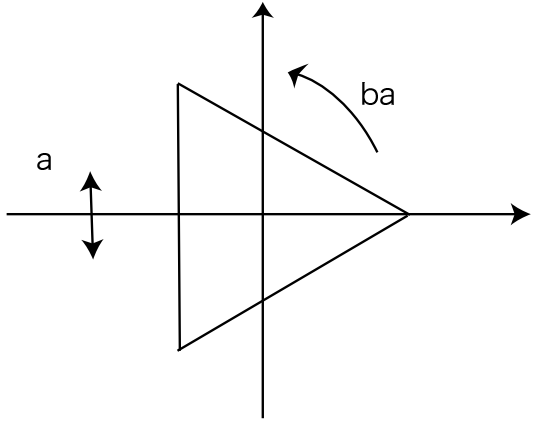}
\end{picture}
\vspace{-0.5cm}
\caption{The $S_3$ symmetry of an equilateral triangle}
\label{fig:S3}
\end{figure}

$S_3$ consists of all permutations among 
three objects, $(x_1,x_2,x_3)$ and its order 
is equal to $3 ! = 6$.
All of six elements correspond to the following 
transformations,
\begin{eqnarray}\label{eq:s3-permutation}
e   &:& (x_1,x_2,x_3)\to (x_1,x_2,x_3), \nonumber \\
a_1 &:& (x_1,x_2,x_3)\to (x_2,x_1,x_3),\nonumber \\
a_2 &:& (x_1,x_2,x_3)\to (x_3,x_2,x_1),\nonumber \\
a_3 &:& (x_1,x_2,x_3)\to (x_1,x_3,x_2),\\
a_4 &:& (x_1,x_2,x_3)\to (x_3,x_1,x_2),\nonumber \\
a_5 &:& (x_1,x_2,x_3)\to (x_2,x_3,x_1).\nonumber 
\end{eqnarray}
Their multiplication forms a closed algebra, e.g.
\begin{eqnarray}
a_1a_2 &:&(x_1,x_2,x_3)\to (x_2,x_3,x_1), \nonumber\\
a_2a_1 &:&(x_1,x_2,x_3)\to (x_3,x_1,x_2),\\
a_4a_2 &:&(x_1,x_2,x_3)\to (x_1,x_3,x_2), \nonumber
\end{eqnarray}
i.e.
\begin{eqnarray}
a_1a_2=a_5,\quad a_2a_1=a_4,\quad a_4a_2=a_2a_1a_2=a_3 .
\end{eqnarray}
Thus, by defining $a_1=a,a_2=b$, 
all of elements are written as 
\begin{eqnarray}
\{e,a,b,ab,ba,bab\}.
\end{eqnarray}
Note that $aba=bab$.
The $S_3$ group is a symmetry of an equilateral triangle 
as shown in Figure \ref{fig:S3}.
The elements $a$ and $ab$ correspond to 
a reflection and the $2\pi/3$ rotation, respectively.

\vskip .5cm
{$\bullet$ \bf Conjugacy classes}

These elements are classified to three conjugacy classes,
\begin{eqnarray}
C_1:\{e\},\quad C_2:\{ab,ba\},\quad C_3:\{a,b,bab\}.
\end{eqnarray}
Here, the subscript of $C_n$, $n$, denotes the number of elements 
in the conjugacy class $C_n$.
Their orders are found as 
\begin{eqnarray}
(ab)^3=(ba)^3=e, \qquad a^2=b^2=(bab)^2=e.
\end{eqnarray}
The elements $\{e,ab,ba\}$ correspond to even permutations, 
while the elements $\{a,b,bab\}$ are odd permutations.

\vskip .5cm
{$\bullet$ \bf Characters and representations}

Let us study irreducible representations of $S_3$.
The number of irreducible representations must be equal to 
three, because there are three conjugacy classes.
We assume that there are $m_n$ $n$-dimensional representations, 
that is, $D(g)$ are represented by $(n \times n)$ matrices.
Here, $m_n$ must satisfy $\sum_n m_n =3$.
Furthermore, the orthogonality relation (\ref{eq:character-2-e}) requires 
\begin{eqnarray}\label{eq:character-2-S3}
&&\sum_\alpha[\chi_\alpha(C_1)]^2=\sum_nm_nn^2=m_1+4m_2+9m_3+\cdots=6,
\end{eqnarray}
where $m_n \geq 0$.
This equation has only two possible solutions, 
$(m_1,m_2)=(2,1)$ and $(6,0)$, but 
only the former $(m_1,m_2)=(2,1)$ satisfies $m_1+m_2=3$.
Thus, irreducible representations of $S_3$ include 
two singlets ${\bf 1}$ and ${\bf 1'}$ , and a doublet ${\bf 2}$.
We denote their characters by $\chi_1(g), \chi_{1'}(g)$ 
and $\chi_2(g)$, respectively.
Obviously, it is found that $\chi_1(C_1)=\chi_{1'}(C_1)=1$ 
and $\chi_2(C_1)=2$.
Furthermore, one of singlet representations corresponds to 
a trivial singlet, that is, $\chi_1(C_2)=\chi_1(C_3)=1$.
The characters, which are not fixed at this stage, are 
$\chi_{1'}(C_2)$, $\chi_{1'}(C_3)$, $\chi_{2}(C_2)$ 
and $\chi_{2}(C_3)$.
Now let us determine them.
For a non-trivial singlet ${\bf 1'}$, representation 
matrices are nothing but characters, $\chi_{1'}(C_2)$ and 
$\chi_{1'}(C_3)$.
They must satisfy 
\begin{eqnarray}
\left( \chi_{1'}(C_2)\right)^3 =1, \qquad \left( \chi_{1'}(C_3) 
\right)^2=1.
\end{eqnarray}
Thus, $\chi_{1'}(C_2)$ is one of $1$, $\omega$ and $\omega^2$, where 
$\omega = \exp[2\pi i/3]$, and 
$\chi_{1'}(C_3) $ is $1$ or $-1$.
On top of that, the orthogonality relation (\ref{eq:character-1}) requires 
\begin{eqnarray}
\sum_g \chi_{1}(g) \chi_{1'}(g)= 1 + 2\chi_{1'}(C_2)
+3\chi_{1'}(C_3) =0.
\end{eqnarray}
Its unique solution is obtained by $\chi_{1'}(C_2)=1$ and 
$\chi_{1'}(C_3)=-1$.
Furthermore, the orthogonality relations (\ref{eq:character-1}) and 
(\ref{eq:character-2})  require 
\begin{eqnarray}
\sum_g \chi_{1}(g) \chi_{2}(g) &=& 2 + 2\chi_{2}(C_2)
+3\chi_{2}(C_3) =0, \\
\sum_\alpha \chi_{\alpha}(C_1)^* \chi_{\alpha}(C_2) 
&=& 1 + \chi_{1'}(C_2)+2\chi_{2}(C_2) =0.
\end{eqnarray}
Their solution is written by $\chi_2(C_2)=-1$ and 
$\chi_2(C_3)=0$.
These results are shown in Table~\ref{tab:S3-character}.

\begin{table}[t]
\begin{center}
\begin{tabular}{|c|c|c|c|c|}
\hline
     &$h$&$\chi_1$&$\chi_{1'}$&$\chi_2$ \\ \hline
$C_1$&$1$&  $1$   &   $1$     &   $2$   \\ \hline
$C_2$&$3$&  $1$   &   $1$     &   $-1$  \\ \hline
$C_3$&$2$&  $1$   &   $-1$    &   $0$   \\ 
\hline
\end{tabular}
\end{center}
\caption{Characters of $S_3$ representations}
\label{tab:S3-character}
\end{table}

Next, let us figure out representation matrices $D(g)$ of 
$S_3$ by using the character in Table~\ref{tab:S3-character}.
For singlets, their characters are nothing but representation 
matrices.
Thus, let us consider representation matrices $D(g)$ 
for the doublet, where $D(g)$ are $(2 \times 2)$ unitary matrices.
Obviously, $D_2(e)$ is the  $(2 \times 2)$ identity matrices.
Because of $\chi_2(C_3)=0$, one can diagonalize one element of 
the conjugacy class $C_3$.
Here we choose e.g. $a$ in $C_3$ as the diagonal element,
\begin{eqnarray}
a = \left(
\begin{array}{cc}  
1 & 0 \\
0 & -1 \\
\end{array}
\right).
\end{eqnarray}
The other elements in $C_3$  as well as $C_2$ are 
non-diagonal matrices.
Recalling $b^2=e$, we can write 
\begin{eqnarray}
b=\mat2{\cos\theta}{\sin\theta}{\sin\theta}{-\cos\theta}, \quad
bab=\mat2{\cos2\theta}{\sin2\theta}{\sin2\theta}{-\cos2\theta} .
\end{eqnarray}
Then, we can write elements in $C_2$ as 
\begin{eqnarray}
ab=\mat2{\cos\theta}{\sin\theta}{-\sin\theta}{\cos\theta}, \quad
ba=\mat2{\cos\theta}{-\sin\theta}{\sin\theta}{\cos\theta} .
\end{eqnarray}
Recall that the trace of elements in $C_2$ is equal to $-1$.
Then, it is found that 
$\cos\theta=-1/2 $, 
that is, 
$\theta={2\pi}/{3},{4\pi}/{3}.$
When we choose $\theta={4\pi}/{3}$, 
we obtain the matrix representation of $S_3$ as 
\begin{eqnarray}\label{eq:s3-2-rep}
&&e=\mat2{1}{0}{0}{1} ,\quad a=\mat2{1}{0}{0}{-1}, \quad
 b=\mat2{-\frac12}{-\frac{\sqrt{3}}{2}}{-\frac{\sqrt{3}}{2}}{\frac12} ,
 \nonumber \\
&&ab=\mat2{-\frac12}{-\frac{\sqrt{3}}{2}}{\frac{\sqrt{3}}{2}}{-\frac12},\quad
ba=\mat2{-\frac12}{\frac{\sqrt{3}}{2}}{-\frac{\sqrt{3}}{2}}{-\frac12},\quad
bab=\mat2{-\frac12}{\frac{\sqrt{3}}{2}}{\frac{\sqrt{3}}{2}}{\frac12} .
\end{eqnarray}

\vskip .5cm
{$\bullet$ \bf Tensor products}

Finally, we consider tensor products of irreducible 
representations.
Let us start with the tensor products of 
two doublets, $(x_1,x_2)$ and $(y_1,y_2)$.
For example, each element $x_i y_j$ is transformed under $b$ as 
\begin{eqnarray}
x_1y_1&\to&\frac{x_1y_1+3x_2y_2+\sqrt{3}(x_1y_2+x_2y_1)}{4}, 
\nonumber \\
x_1y_2&\to&\frac{\sqrt{3}x_1y_1-\sqrt{3}x_2y_2-x_1y_2+3x_2y_1}{4},
\nonumber  \\
x_2y_1&\to&\frac{\sqrt{3}x_1y_1-\sqrt{3}x_2y_2-x_2y_1+3x_1y_2}{4}, 
\\
x_2y_2&\to&\frac{3x_1y_1+x_2y_2-\sqrt{3}(x_1y_2+x_2y_1)}{4}. 
\nonumber
\end{eqnarray}
Thus, it is found that
\begin{eqnarray}
b(x_1y_1+x_2y_2) = (x_1y_1+x_2y_2),\quad
b(x_1y_2-x_2y_1) = - (x_1y_2-x_2y_1) .
\end{eqnarray}
That implies these linear combinations 
correspond to the singlets,
\begin{eqnarray}
{\bf 1}:x_1y_1+x_2y_2,\quad {\bf 1'}:x_1y_2-x_2y_1  .
\end{eqnarray}
Furthermore, it is found that 
\begin{eqnarray}
b \2tvec{x_2y_2-x_1y_1}{x_1y_2+x_2y_1} = 
\mat2{-\frac12}{-\frac{\sqrt{3}}{2}}{-\frac{\sqrt{3}}{2}}{\frac12}
\2tvec{x_2y_2-x_1y_1}{x_1y_2+x_2y_1} .
\end{eqnarray}
Hence, $(x_2y_2-x_2y_2,x_1y_2+x_2y_1)$ corresponds to 
the doublet, i.e.
\begin{eqnarray}
{\bf 2}=\2tvec{x_2y_2-x_1y_1}{x_1y_2+x_2y_1} .
\end{eqnarray}

Similarly, we can study the tensor product of 
the doublet $(x_1,x_2)$ and the singlet ${\bf 1'}$ $y'$.
Their products $x_iy'$ transform under $b$ as 
\begin{eqnarray}
x_1y' &\to&\frac12x_1y'+\frac{\sqrt{3}}{2}x_2y' , 
\nonumber \\
x_2y' &\to&\frac{\sqrt{3}}{2}x_1y'-\frac12x_2y' .
\end{eqnarray}
That implies those form a doublet, 
\begin{eqnarray}
{\bf 2}:\2tvec{-x_2y'}{x_1y'} .
\end{eqnarray}

These results are summarized as follows,
\begin{eqnarray}
&&\2tvec{x_1}{x_2}_{\bf 2}\otimes\2tvec{y_1}{y_2}_{\bf 2}
=(x_1y_1+x_2y_2)_{\bf 1}+(x_1y_2-x_2y_1)_{{\bf 1}'}
+\2tvec{x_1y_2+x_2y_1}{x_1y_1-x_2y_2}_{\bf 2}, \nonumber \\
&&\2tvec{x_1}{x_2}_{\bf 2}\otimes (y')_{{\bf
    1}'}=\2tvec{-x_2y'}{x_1y'}_{\bf 2}, \\
&&(x')_{{\bf 1}'}\otimes(y')_{{\bf 1}'}=(x'y')_{\bf 1}  .\nonumber
\end{eqnarray}
In addition, obviously the tensor product of two trivial 
singlets corresponds to a trivial singlet.

Tensor products are important to applications for 
particle phenomenology.
Matter and Higgs fields may be assigned to have certain representations of 
discrete symmetries.
The Lagrangian must be invariant under discrete symmetries.
That implies that n-point couplings corresponding to a 
trivial singlet can appear in Lagrangian.

In addition to the above (real) representation of $S_3$, 
another representation, i.e. the complex representation, 
is often used in the literature.
Here, we mention about changing representation bases.
All permutations of $S_3$ in Eq.~(\ref{eq:s3-permutation}) are 
represented on the reducible triplet $(x_1,x_2,x_3)$ as 
\begin{eqnarray}\label{eq:s3-3-rep}
 & & \Mat3{1}{0}{0}{0}{1}{0}{0}{0}{1},~\Mat3{1}{0}{0} {0}{0}{1} {0}{1}{0},~
\Mat3{0}{1}{0} {1}{0}{0} {0}{0}{1}, \nonumber \\
 & & \Mat3{0}{1}{0} {0}{0}{1} {1}{0}{0},~
\Mat3{0}{0}{1} {0}{1}{0}{1}{0}{0},~\Mat3{0}{0}{1} {1}{0}{0} {0}{1}{0}.
\end{eqnarray}
We change the representation through the unitary transformation,  
$U^{\dagger}gU$, e.g. by using the unitary matrix $U_{\rm tribi}$, 
\begin{eqnarray}
U_{\rm tribi}=\Mat3{\sqrt{2/3}}{1/\sqrt3}{0}
{-1/\sqrt6}{1/\sqrt3}{-1/\sqrt2} 
{-1/\sqrt6}{1/\sqrt3}{1/\sqrt2}.
\end{eqnarray}
Then, the six elements of $S_3$ are written as 
\begin{eqnarray}
& & \Mat3{1}{0}{0}{0}{1}{0}{0}{0}{1},~\Mat3{1}{0}{0}{0}{1}{0} {0}{0}{-1},~
\Mat3{1}{0}{0}{0}{-\frac12}{-\frac{\sqrt3}2} {0}{-\frac{\sqrt3}2}{\frac12},
\nonumber \\
& & \Mat3{1}{0}{0}{0}{-\frac12}{-\frac{\sqrt3}2}
{0}{\frac{\sqrt3}2}{-\frac12},
~\Mat3{1}{0}{0}{0}{-\frac12}{\frac{\sqrt3}2} {0}{\frac{\sqrt3}2}{\frac12},
~\Mat3{1}{0}{0}{0}{-\frac12}{\frac{\sqrt3}2} {0}{-\frac{\sqrt3}2}{-\frac12}.
\end{eqnarray}
Note that this form is completely reducible and that 
the (right-bottom) $(2 \times 2)$ submatrices are 
exactly the same as those for the doublet representation (\ref{eq:s3-2-rep}).
The unitary matrix $U_{\rm tribi}$ is called the tri-bimaximal matrix 
and plays a role in the neutrino physics as studied in section 15.

We can use another unitary matrix $U$ in order to obtain 
a completely reducible form from the reducible representation 
matrices (\ref{eq:s3-3-rep}).
For example, let us use the following matrix,
\begin{eqnarray}
U_{w}=\frac1{\sqrt3}\Mat3{1}{1}{1} {1}{w}{w^2} {1}{w^2}{w}.
\end{eqnarray}
Then,  the six elements of $S_3$ are written as 
\begin{eqnarray}
 & & \Mat3{1}{0}{0} {0}{1}{0} {0}{0}{1},~\Mat3{1}{0}{0} {0}{0}{1} {0}{1}{0},~
\Mat3{1}{0}{0} {0}{0}{w^2} {0}{w}{0}, \nonumber \\
 & & \Mat3{1}{0}{0} {0}{w}{0} {0}{0}{w^2},~
\Mat3{1}{0}{0} {0}{0}{w} {0}{w^2}{0},
~\Mat3{1}{0}{0} {0}{w^2}{0} {0}{0}{w}.
\end{eqnarray}
The (right-bottom) $(2 \times 2)$ submatrices correspond to 
the doublet representation in the different basis, that is, 
the complex representation.
This unitary matrix is called the magic matrix.
In different bases, the multiplication rule does not change.
For example, we obtain ${\bf 2} \times {\bf 2} = {\bf 1}+{\bf 1}'+{\bf
  2}$ in both the real and complex bases.
However, elements of doublets in the left hand side are 
written in a different way.

\subsection{$S_4$}
\label{subsec:S4}

\begin{figure}[t]
\unitlength=1mm
\begin{picture}(170,70)
\hspace{4.5cm}
\includegraphics[width=7cm]{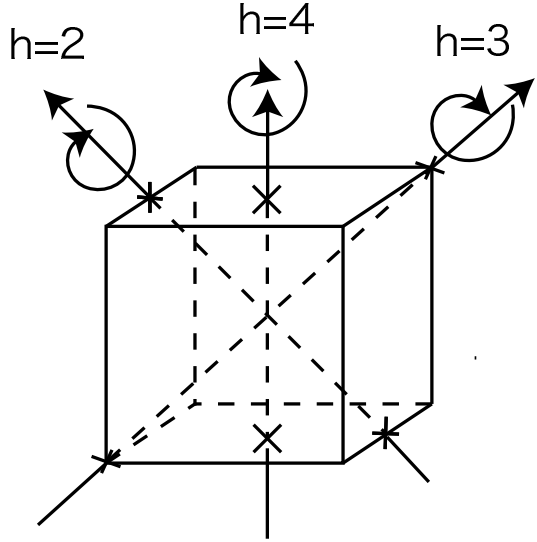}
\end{picture}
\vspace{-0.5cm}
\caption{The $S_4$ symmetry of a cube.
This figure shows the transformations corresponding 
to the $S_4$ elements with $h=2,3$ and 4.
Note that the group can be also considered 
as the regular octahedron in a way similar to a cube. }
\label{fig:S4}
\end{figure}

$S_4$ consists of all permutations among 
four objects, $(x_1,x_2,x_3,x_4)$,
\begin{eqnarray}
 (x_1,x_3,x_2,x_4),\quad \to\quad (x_i,x_j,x_k,x_l), 
\end{eqnarray}
and the order of $S_4$ 
is equal to $4 ! = 24$.
We denote all of $S_4$ elements as 
\begin{eqnarray}
&&a_1:(x_1,x_3,x_2,x_4),~a_2:(x_2,x_1,x_4,x_3),
~a_3:(x_3,x_4,x_1,x_2),~a_4:(x_4,x_3,x_2,x_1), 
\nonumber \\
&&b_1:(x_1,x_4,x_2,x_3),~b_2:(x_4,x_1,x_3,x_2),
~b_3:(x_2,x_3,x_1,x_4),~b_4:(x_3,x_2,x_4,x_1), 
\nonumber \\
&&c_1:(x_1,x_3,x_4,x_2),~c_2:(x_3,x_1,x_2,x_4),
~c_3:(x_4,x_2,x_1,x_3),~c_4:(x_2,x_4,x_3,x_1), 
\nonumber \\
&&d_1:(x_1,x_2,x_4,x_3),~d_2:(x_2,x_1,x_3,x_4),
~d_3:(x_4,x_3,x_1,x_2),~d_4:(x_3,x_4,x_2,x_1),\\
&&e_1:(x_1,x_3,x_2,x_4),~e_2:(x_3,x_1,x_4,x_2),
~e_3:(x_2,x_4,x_1,x_3),~e_4:(x_4,x_2,x_3,x_1), \nonumber \\
&&f_1:(x_1,x_4,x_3,x_2),~f_2:(x_4,x_1,x_2,x_3),
~f_3:(x_3,x_2,x_1,x_4),~f_4:(x_2,x_3,x_4,x_1), \nonumber 
\end{eqnarray}
where we have shown the ordering of four objects 
after permutations.
The $S_4$ is a symmetry of a cube as shown in 
Figure \ref{fig:S4}.

It is obvious that $x_1+x_2+x_3+x_4$ is invariant under 
any permutation of $S_4$, that is, a trivial singlet.
Thus, we use the vector space, which is orthogonal to 
this singlet direction, 
\begin{eqnarray}\label{eq:S4-3}
{\bf 3}:\3tvec{A_x}{A_y}{A_z}=
\3tvec{x_1+x_2-x_3-x_4}{x_1-x_2+x_3-x_4}{x_1-x_2-x_3+x_4} ,
\end{eqnarray}
in order to construct matrix representations of $S_4$, 
that is, a triplet representation.
In this triplet vector space, all of $S_4$ elements are 
represented by the following matrices,
 \begin{eqnarray}\label{eq:S4-3-2}
&&
a_1=\Mat3{1}{0}{0} {0}{1}{0} {0}{0}{1},\quad
a_2=\Mat3{1}{0}{0} {0}{-1}{0}{0}{0}{-1},  \nonumber \\
&&
a_3=\Mat3{-1}{0}{0} {0}{1}{0} {0}{0}{-1},\quad
a_4=\Mat3{-1}{0}{0} {0}{-1}{0} {0}{0}{1},\quad  \nonumber \\
&&
b_1=\Mat3{0}{0}{1} {1}{0}{0} {0}{1}{0},\quad
b_2=\Mat3{0}{0}{1} {-1}{0}{0} {0}{-1}{0},  \nonumber  \\
&&
b_3=\Mat3{0}{0}{-1} {1}{0}{0} {0}{-1}{0},\quad
b_4=\Mat3{0}{0}{-1} {-1}{0}{0} {0}{1}{0},\quad  \nonumber \\
&&
c_1=\Mat3{0}{1}{0} {0}{0}{1} {1}{0}{0},\quad
c_2=\Mat3{0}{1}{0} {0}{0}{-1} {-1}{0}{0},  \nonumber \\
&&
c_3=\Mat3{0}{-1}{0} {0}{0}{1} {-1}{0}{0},\quad
c_4=\Mat3{0}{-1}{0} {0}{0}{-1} {1}{0}{0},\quad \nonumber \\
&&
d_1=\Mat3{1}{0}{0} {0}{0}{1} {0}{1}{0},\quad
d_2=\Mat3{1}{0}{0} {0}{0}{-1} {0}{-1}{0},   \\
&&
d_3=\Mat3{-1}{0}{0} {0}{0}{1} {0}{-1}{0},\quad
d_4=\Mat3{-1}{0}{0} {0}{0}{-1} {0}{1}{0},\quad  \nonumber \\
&&
e_1=\Mat3{0}{1}{0} {1}{0}{0} {0}{0}{1},\quad
e_2\Mat3{0}{1}{0} {-1}{0}{0} {0}{0}{-1},    \nonumber \\
&&
e_3=\Mat3{0}{-1}{0} {1}{0}{0} {0}{0}{-1},\quad
e_4=\Mat3{0}{-1}{0} {-1}{0}{0} {0}{0}{1},\quad  \nonumber \\
&&
f_1=\Mat3{0}{0}{1} {0}{1}{0} {1}{0}{0},\quad
f_2=\Mat3{0}{0}{1} {0}{-1}{0} {-1}{0}{0},  \nonumber    \\
&&
f_3=\Mat3{0}{0}{-1} {0}{1}{0} {-1}{0}{0},\quad
f_4=\Mat3{0}{0}{-1} {0}{-1}{0} {1}{0}{0}. \nonumber 
\end{eqnarray}

\vskip .5cm
{$\bullet$ \bf Conjugacy classes}

The $S_4$ elements can be classified by the order $h$ of each element, 
where $a^h=e$, as 
\begin{eqnarray}
\begin{array}{ccc}
h=1\quad:\quad & \{a_1\}, \\
h=2\quad:\quad &\{a_2,a_3,a_4,d_1,d_2,e_1,e_4,f_1,f_3 \},  \\
h=3\quad:\quad &\{b_1,b_2,b_3,b_4,c_1,c_2,c_3,c_4\}, \\
h=4\quad:\quad &\{d_3,d_4,e_2,e_3,f_2,f_4\} . 
\end{array}
\end{eqnarray}

Moreover, they are classified by the conjugacy classes as 
\begin{eqnarray}
\begin{array}{ccc}
 C_1~: &\{a_1\} , & h=1,  \\
 C_3~: &~\{a_2,a_3,a_4\} , & h=2, \\
 C_6~: &~\{d_1,d_2,e_1,e_4,f_1,f_3\}, & h=2, \\
 C_8~: &~\{b_1,b_2,b_3,b_4,c_1,c_2,c_3,c_4\}, & h=3,  \\
 C_{6'}~: &~\{d_3,d_4,e_2,e_3,f_2,f_4\}, & h=4. 
\end{array}
\end{eqnarray}

\vskip .5cm
{$\bullet$ \bf Characters and representations}

Thus, $S_4$ includes five conjugacy classes, 
that is, there are five irreducible representations.
For example, all of elements are written as multiplications of 
$b_1$ in $C_8$ and $d_4$ in $C_{6'}$, which satisfy 
\begin{eqnarray}
& & (b_1)^3=e, \quad (d_4)^4=e, \quad 
 d_4 (b_1)^2d_4 = b_1, \quad d_4 b_1 d_4 = b_1 (d_4)^2b_1.
\end{eqnarray} 
The orthogonality relation (\ref{eq:character-2-e}) requires 
\begin{eqnarray}\label{eq:chracter-2-S4}
&&\sum_\alpha[\chi_\alpha(C_1)]^2=\sum_nm_nn^2=m_1+4m_2+9m_3+\cdots=24,
\end{eqnarray}
like Eq.~(\ref{eq:character-2-S3}), and 
$m_n$ also satisfy $m_1+m_2+m_3+\cdots= 5$, because 
there must be five irreducible representations.
Then, their unique solution is obtained as 
$(m_1,m_2,m_3)=(2,1,2)$.
That is, irreducible representations of 
$S_4$ include two singlets ${\bf 1}$ and ${\bf 1}'$, 
one doublet ${\bf 2}$, and two triplets ${\bf 3}$ and 
${\bf 3}'$, where ${\bf 1}$ corresponds to a trivial singlet 
and ${\bf 3}$ corresponds to (\ref{eq:S4-3}) and (\ref{eq:S4-3-2}).
We can compute the character for each representation 
by an analysis similar to $S_3$ in the previous subsection.
The results are shown in Table \ref{tab:S4-character}.

\begin{table}[t]
\begin{center}
\begin{tabular}{|c|c|c|c|c|c|c|}
\hline
        &$h$&$\chi_1$&$\chi_{1'}$&$\chi_2$&$\chi_3$&$\chi_{3'}$ \\ \hline
$C_1$   &$1$&  $1$   &   $1$     &  $2$   &   $3$  &   $3$    \\ \hline
$C_3$   &$2$&  $1$   &   $1$     &  $2$   &   $-1$ &  $-1$    \\ \hline
$C_6$   &$2$&  $1$   &   $-1$    &  $0$   &   $1$  &  $-1$    \\ \hline 
$C_{6'}$&$4$&  $1$   &   $-1$    &  $0$   &   $-1$ &  $1$     \\ \hline
$C_8$   &$3$&  $1$   &   $1$     & $-1$   &   $0$  &   $0$    \\
\hline
\end{tabular}
\end{center}
\caption{Characters of $S_4$ representations}
\label{tab:S4-character}
\end{table}

For ${\bf 2}$, the representation matrices are written as 
e.g. 
\begin{eqnarray}\label{eq:S4-2}
&&
a_2({\bf2})=\mat2{1}{0}{0}{1},~b_1({\bf2})=\mat2{\omega}{0}{0}{\omega^2},
\nonumber \\
&&
d_1({\bf2})=d_3({\bf2})=d_4({\bf2})=\mat2{0}{1}{1}{0}.
\end{eqnarray}
For ${\bf 3}'$, the representation matrices are written as 
e.g. 
\begin{eqnarray}
&& 
a_2({\bf3'})=\Mat3{1}{0}{0} {0}{-1}{0} {0}{0}{-1},~
b_1({\bf3'})=\Mat3{0}{0}{1} {1}{0}{0} {0}{1}{0}, \\
&&
d_1({\bf3'})=\Mat3{-1}{0}{0} {0}{0}{-1} {0}{-1}{0},~ 
d_3({\bf3'})=\Mat3{1}{0}{0} {0}{0}{-1} {0}{1}{0},~
d_4({\bf3'})=\Mat3{1}{0}{0} {0}{0}{1} {0}{-1}{0}.\nonumber 
\end{eqnarray}
Note that $a_2({\bf 3'}) = a_2({\bf 3})$ and 
$b_1({\bf 3'}) = b_1({\bf 3})$, but $d_1({\bf3'}) = - d_1({\bf3}) $, 
$d_3({\bf3'})= - d_3({\bf3})$ and $d_4({\bf3'})= - d_4({\bf3})$.
This aspect would be obvious from the above character table.

\vskip .5cm
{$\bullet$ \bf Tensor products}

Finally, we show the tensor products.
The tensor products of ${\bf 3} \times {\bf 3}$ can be 
decomposed as 
\begin{eqnarray}
({\bf A})_{\bf3}\times({\bf B})_{\bf3}=({\bf A}\cdot{\bf B})_{\bf1}
+\2tvec{ {\bf A}\cdot\Sigma\cdot{\bf B} } { {\bf A}\cdot\Sigma^*\cdot{\bf B} }_{\bf2}
+\3tvec{\{A_yB_z\}} {\{A_zB_x\}} {\{A_xB_y\}}_{\bf3}
+\3tvec{\L[A_yB_z\R]} {\L[A_zB_x\R]} {\L[A_xB_y\R]}_{\bf3'} ,
\end{eqnarray}
where 
\begin{eqnarray}
{\bf A}\cdot{\bf B}&=&A_xB_x+A_yB_y+A_zB_z, \nonumber  \\
\{A_iB_j\}&=&A_iB_j+A_jB_i, \nonumber \\
\L[A_yB_z\R]&=&A_iB_j-A_jB_i, \\
{\bf A}\cdot\Sigma\cdot{\bf B} &=&A_xB_x+\omega A_yB_y+\omega^2A_zB_z,
\nonumber \\
{\bf A}\cdot\Sigma^*\cdot{\bf B} &=&A_xB_x+\omega^2 A_yB_y+\omega
A_zB_z . \nonumber 
\end{eqnarray}

The tensor products of other representations are also 
decomposed as e.g.
\begin{eqnarray}
({\bf A})_{\bf3'}\times({\bf B})_{\bf3'}=({\bf A}\cdot{\bf B})_{\bf1}
+\2tvec{ {\bf A}\cdot\Sigma\cdot{\bf B} } { {\bf A}\cdot\Sigma^*\cdot{\bf B} }_{\bf2}
+\3tvec{\{A_yB_z\}} {\{A_zB_x\}} {\{A_xB_y\}}_{\bf3}
+\3tvec{\L[A_yB_z\R]} {\L[A_zB_x\R]} {\L[A_xB_y\R]}_{\bf3'}, \\
({\bf A})_{\bf3}\times({\bf B})_{\bf3'}=({\bf A}\cdot{\bf B})_{\bf1'}
+\2tvec{ {\bf A}\cdot\Sigma\cdot{\bf B} } {-{\bf A}\cdot\Sigma^*\cdot{\bf B} }_{\bf2}
+\3tvec{\{A_yB_z\}} {\{A_zB_x\}} {\{A_xB_y\}}_{\bf3'}
+\3tvec{\L[A_yB_z\R]} {\L[A_zB_x\R]} {\L[A_xB_y\R]}_{\bf3},
\end{eqnarray}
and 
\begin{eqnarray}
({\bf A})_{\bf2}\times({\bf B})_{\bf2}=
\{A_xB_y\}_{\bf1}+\L[A_xB_y\R]_{\bf1'}
+\2tvec{A_yB_y} {A_xB_x}_{\bf2} ,
\end{eqnarray}

\begin{eqnarray}
\2tvec{A_x}{A_y}_{\bf 2}\times \3tvec{B_x}{B_y}{B_z}_{\bf 3}
&=&\3tvec{(A_x+A_y)B_x}{(\omega^2A_x+\omega A_y)B_y}{(\omega
  A_x+\omega^2 A_y)B_z}_{\bf 3}+ 
\3tvec{(A_x-A_y)B_x}{(\omega^2A_x-\omega A_y)B_y}{(\omega A_x-\omega^2
  A_y)B_z}_{{\bf 3}'}, \\
\2tvec{A_x}{A_y}_{\bf 2}\times \3tvec{B_x}{B_y}{B_z}_{{\bf 3}'}
&=&\3tvec{(A_x+A_y)B_x}{(\omega^2A_x+\omega A_y)B_y}{(\omega
  A_x+\omega^2 A_y)B_z}_{{\bf 3}'}+ 
\3tvec{(A_x-A_y)B_x}{(\omega^2A_x-\omega A_y)B_y}{(\omega A_x-\omega^2
  A_y)B_z}_{\bf 3} .
\end{eqnarray}
Furthermore, we have ${\bf 3}\times {\bf 1'}={\bf 3}'$ and 
${\bf 3}'\times {\bf 1'}={\bf 3}$ and  
${\bf 2}\times {\bf 1'}={\bf 2}$.

In the literature, several bases are used for $S_4$.
The decomposition of tensor products, 
${\bf r} \times {\bf r}' = \sum_m {\bf r}_m$, 
does not depend on the basis.
For example, we obtain ${\bf 3} \times {\bf  3}' = {\bf 1}' + {\bf 2} 
+ {\bf 3} +{\bf  3}'$ in any basis.
However, the multiplication rules written by components 
depend on the basis, which we use.
We have used the basis (\ref{eq:S4-2}).
In appendix B, we show the relations between 
several bases and give explicitly 
the multiplication rules in terms of components.


Similarly, we can study the $S_N$ group with $N>4$.
Here we give a brief comment on such groups.
The $S_N$ group with $N>4$ has only one invariant subgroup, 
that is, the alternating group, $A_N$.
The $S_N$ group has two one-dimensional representations: 
one is trivial singlet, that is, invariant 
under all the elements (symmetric representation), 
the other is pseudo singlet, that is, symmetric 
under the even permutation-elements 
but antisymmetric under the odd permutation-elements.  
Group-theoretical aspects of $S_5$ are derived from those 
of $S_4$ by applying a theorem of Frobenuis(Frobenuis formula), 
graphical method(Young tableaux), recursion formulas 
for characters (branching laws). 
The details are given in, e.g., the text book of \cite{Hamermesh}.
Such analysis would be extended recursively from $S_N$ to 
$S_{N+1}$.

\clearpage



\section{$A_N$}

All even permutations among $S_N$ form a group, which is 
called $A_N$ with the order $(N !)/2$.
It is often called the alternating group.
For example, among $S_3$ in subsection \ref{subsec:S3} 
the even permutations include
\begin{eqnarray}
e   &:& (x_1,x_2,x_3)\to (x_1,x_2,x_3), \nonumber\\
a_4 &:& (x_1,x_2,x_3)\to (x_3,x_1,x_2),\\
a_5 &:& (x_1,x_2,x_3)\to (x_2,x_3,x_1), \nonumber
\end{eqnarray}
while the odd permutations include
\begin{eqnarray}
a_1 &:& (x_1,x_2,x_3)\to (x_2,x_1,x_3), \nonumber\\
a_2 &:& (x_1,x_2,x_3)\to (x_3,x_2,x_1),\\
a_3 &:& (x_1,x_2,x_3)\to (x_1,x_3,x_2). \nonumber
\end{eqnarray}
The three elements, $\{e, a_4, a_5 \}$ form the 
group $A_3$.
Since $(a_4)^2=a_5$ and $(a_4)^3=e$, 
the group $A_3$ is nothing but $Z_3$.
Thus, the smallest non-Abelian group is $A_4$.

\subsection{$A_4$}
\label{sec:A4}

\begin{figure}[t]
\unitlength=1mm
\begin{picture}(170,70)
\hspace{4.5cm}
\includegraphics[width=7cm]{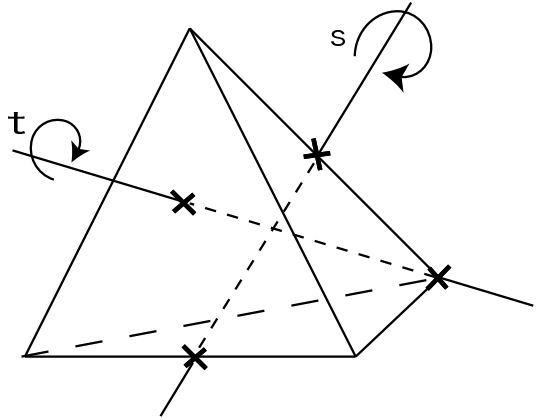}
\end{picture}
\vspace{-0.5cm}
\caption{The $A_4$ symmetry of tetrahedron.}
\label{fig:A4}
\end{figure}

All even permutations of $S_4$ form $A_4$, whose 
order is equal to $(4 !)/2=12$.
The $A_4$ group is the symmetry of a tetrahedron 
as shown in Figure \ref{fig:A4}.
Thus, the $A_4$ group is often denoted as $T$.
Using the notation in subsection \ref{subsec:S4}, 
all of 12 elements are denoted as 
 \begin{eqnarray}\label{eq:A4-ABC}
&&
a_1=\Mat3{1}{0}{0} {0}{1}{0} {0}{0}{1},\quad
a_2=\Mat3{1}{0}{0} {0}{-1}{0}{0}{0}{-1}, \nonumber \\
&&
a_3=\Mat3{-1}{0}{0} {0}{1}{0} {0}{0}{-1},\quad
a_4=\Mat3{-1}{0}{0} {0}{-1}{0} {0}{0}{1}, \nonumber \\
&&
b_1=\Mat3{0}{0}{1} {1}{0}{0} {0}{1}{0},\quad
b_2=\Mat3{0}{0}{1} {-1}{0}{0} {0}{-1}{0}, \nonumber \\
&&
b_3=\Mat3{0}{0}{-1} {1}{0}{0} {0}{-1}{0},\quad
b_4=\Mat3{0}{0}{-1} {-1}{0}{0} {0}{1}{0},\\
&&
c_1=\Mat3{0}{1}{0} {0}{0}{1} {1}{0}{0},\quad
c_2=\Mat3{0}{1}{0} {0}{0}{-1} {-1}{0}{0}, \nonumber \\
&&
c_3=\Mat3{0}{-1}{0} {0}{0}{1} {-1}{0}{0},\quad
c_4=\Mat3{0}{-1}{0} {0}{0}{-1} {1}{0}{0}. \nonumber 
\end{eqnarray}
{}From  these forms, it is found obviously that $A_4$ is isomorphic to 
$\Delta(12) \simeq (Z_2 \times Z_2) \rtimes Z_3$, 
which is explained in section \ref{sec:Delta-3n}.

They are classified by the conjugacy classes as 
\begin{eqnarray}
\begin{array}{ccc}
C_1~: & \{a_1\}, & h=1,  \\
C_3~: & \{a_2,a_3,a_4\}, & h=2,  \\
 C_4~:& \{b_1,b_2,b_3,b_4,\}, & h=3, \\
 C_{4'}~:& \{c_1,c_2,c_3,c_4,\}, & h=3,
\end{array}
\end{eqnarray}
where we have also shown the orders of each element in 
the conjugacy class by $h$.
There are four conjugacy classes and 
there must be four irreducible representations, 
i.e. $m_1+m_2+m_3 + \cdots = 4$.

The orthogonality relation (\ref{eq:character-2}) requires 
\begin{eqnarray}\label{eq:chracter-2-A4}
&&\sum_\alpha[\chi_\alpha(C_1)]^2=\sum_nm_nn^2=m_1+4m_2+9m_3+\cdots=12,
\end{eqnarray}
for $m_i$, which satisfy $m_1+m_2+m_3 + \cdots = 4$.
The solution is obtained as $(m_1,m_2,m_3)=(3,0,1)$.
That is, the $A_4$ group has three singlets, 
${\bf 1}$, ${\bf 1}'$, and ${\bf 1}''$, and a single triplet 
${\bf 3}$, where the triplet corresponds to (\ref{eq:A4-ABC}).

Another algebraic definition of 
$A_4$ is often used in the literature.
We denote $a_1 = e$, $a_2=s$ and $b_1=t$.
They satisfy the following algebraic relations,
\begin{eqnarray}\label{eq:T-st}
s^2=t^3=(st)^3=e.
\end{eqnarray}
The closed algebra of these elements, $s$ and $t$, is 
defined as the $A_4$.
It is straightforward to write all of $a_i, b_i$ and $c_i$ 
elements by $s$ and $t$.
Then, the conjugacy classes are rewritten as 
\begin{eqnarray}
\begin{array}{ccc}
C_1:&\{e\}, & h=1, \\
C_3:&\{s,tst^2,t^2st\},& h=2, \\
C_4:&\{t,ts,st,sts\}, & h=3,\\
C_{4'}:&\{t^2,st^2,t^2s,tst\}, & h=3. 
\end{array}
\end{eqnarray}
Using them, we study characters.
First, we consider characters of three singlets.
Because $s^2=e$, the characters of $C_3$ have two possibilities, 
$\chi_\alpha(C_3) = \pm 1$.
However, the two elements, $t$ and $ts$, belong to 
the same conjugacy class $C_4$.
That means that $\chi_\alpha(C_3)$ should have the unique value,
 $\chi_\alpha(C_3) =  1$.
Similarly, because of $t^3=e$, the characters $\chi_\alpha(t)$
can correspond to three values, i.e. $\chi_\alpha(t)=\omega^n$,  
$n=0,1,2$, and 
all of these three values are consistent with the above 
structure of conjugacy classes.
Thus, all of three singlets, ${\bf 1}$, ${\bf 1'}$ and 
${\bf 1''}$  are classified by 
these three values of  $\chi_\alpha(t)=1, \omega$ and $\omega^2$, 
respectively.
Obviously, it is found that $\chi_\alpha(C_{4'}) =
(\chi_\alpha(C_{4}))^2$.
Thus, the generators such as $s=a_2,t=b_1,t^2=c_1$ are represented 
on the non-trivial singlets ${\bf 1'}$ and ${\bf 1''}$ 
as 
\begin{eqnarray}
&&
s({\bf 1'})={a_2}({\bf 1'})=1,
\quad t({\bf 1'})={b_1}({\bf  1'})=\omega,
\quad t^2({\bf 1'})={c_1}({\bf 1'})=\omega^2,
\nonumber \\ &&
s({\bf 1''})={a_2}({\bf 1''})=1,
\quad t({\bf 1''})={b_1}({\bf  1''})=\omega^2,
\quad t^2({\bf 1''})={c_1}({\bf 1''})=\omega .
\end{eqnarray}
These characters are shown in Table~\ref{tab:A4-character}.
Next, we consider the characters for the triplet representation.
Obviously, the matrices in Eq.~(\ref{eq:A4-ABC}) correspond to 
the triplet representation.
Thus, we can obtain their characters.
Its result is also shown in Table~\ref{tab:A4-character}.

The tensor product of ${\bf 3} \times {\bf 3}$ can 
be decomposed as 
\begin{eqnarray}
({\bf A})_{\bf3}\times({\bf B})_{\bf3}&=&({\bf A}\cdot{\bf B})_{\bf1}
+({\bf A}\cdot\Sigma\cdot{\bf B})_{\bf1'}
+({\bf A}\cdot\Sigma^*\cdot{\bf B})_{\bf1''}  \nonumber \\
& & +\3tvec{\{A_yB_z\}} {\{A_zB_x\}} {\{A_xB_y\}}_{\bf3}
+\3tvec{\L[A_yB_z\R]} {\L[A_zB_x\R]} {\L[A_xB_y\R]}_{\bf3}.
\end{eqnarray}

\begin{table}[t]
\begin{center}
\begin{tabular}{|c|c|c|c|c|c|}
\hline
        &$h$&$\chi_1$&$\chi_{1'}$&$\chi_{1''}$&$\chi_3$ \\ \hline
$C_1$   &$1$&   $1$  &    $1$    &    $1$     &$3$   \\ \hline
$C_3$   &$2$&   $1$  &    $1$    &    $1$     &$-1$  \\ \hline
$C_4$   &$3$&   $1$  & $\omega$  & $\omega^2$ &$0$   \\ \hline
$C_{4'}$&$3$&   $1$  & $\omega^2$&  $\omega$  &$0$   \\
\hline
\end{tabular}
\end{center}
\caption{Characters of $A_4$ representations}
\label{tab:A4-character}
\end{table}

\subsection{$A_5$}
\label{sec:A5}

\begin{figure}[t]
\unitlength=1mm
\begin{picture}(120,50)
\hspace{4.5cm}
\includegraphics[width=7cm]{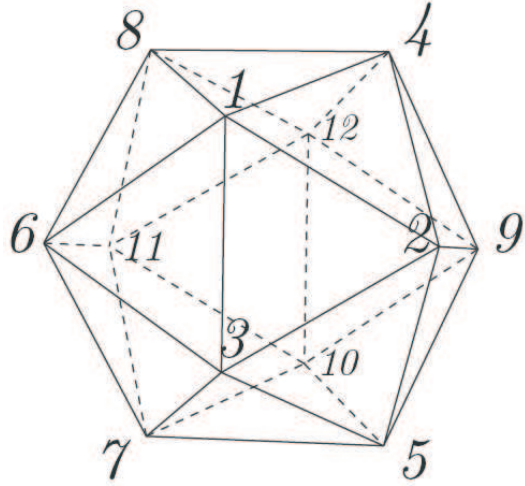}
\end{picture}
\vspace{-0.5cm}
\caption{The regular icosahedron.}
\label{fig:num-ico}
\end{figure}

Here, we mention briefly about the $A_5$ group.
The $A_5$ group is isomorphic to the symmetry of a regular icosahedron.
Thus, it is pedagogical to explain group-theoretical aspects 
of $A_5$ as the symmetry of a regular icosahedron~\cite{shirai}.
As shown in Figure \ref{fig:num-ico}, 
a regular icosahedron consists of 20 identical equilateral 
triangular faces, 30 edges and 12 vertices. 
The icosahedron is dual to a dodecahedron, whose 
symmetry is also isomorphic to $A_5$.
The $A_5$ elements correspond to all the proper rotations 
of the icosahedron.
Such rotations are classified into five types, 
that is, the $0$ rotation (identity), 
$\pi$ rotations about the midpoint of each edge, 
rotations by $2\pi/3$ about axes through the center of each face, 
and rotations by $2\pi/5$ and $4\pi/5$ about an axis through each
vertex.
Following ~\cite{shirai}, we label the vertex by number 
$n=1,\cdots,12$ in Figure \ref{fig:num-ico}.
Here, we define two elements $a$ and $b$ such that  
$a$ corresponds to the rotation by $\pi$
about the midpoint of the edge between vertices 1 and 2 
while $b$ corresponds the rotation by $2\pi/3$ 
about the axis through the center of the triangular face 
10-11-12. 
That is, these two elements correspond to the transformations acting on the 
12 vertices as follows,
\begin{eqnarray}
a : (1,2,3,4,5,6,7,8,9,10,11,12) \to (2,1,4,3,8,9,12,5,6,11,10,7), 
\nonumber  \\
b : (1,2,3,4,5,6,7,8,9,10,11,12) \to (2,3,1,5,6,4,8,9,7,11,12,10). 
\nonumber  
\end{eqnarray}
Then the product $ab$ is given by the following transformation as
\begin{eqnarray}
ab : (1,2,3,4,5,6,7,8,9,10,11,12) \to (3,2,5,1,9,7,10,6,4,12,11,8), 
\nonumber  
\end{eqnarray}
which is the rotation by $2\pi/5$ about the axis through the vertex 2.
All of the $A_5$ elements are written by products of these 
elements, which satisfy 
\begin{eqnarray}
a^2=b^3=(ab)^5=e .
\end{eqnarray}

The order of $A_5$ is equal to $(5 !)/2=60$.
All of the $A_5$ elements, i.e. all the rotations of 
the icosahedron,  are classified into 
five conjugacy classes as follows,
\begin{eqnarray}
C_1   &:& \{ e \}, \nonumber \\
C_{15}&:& \{ 
a(12),a(13),a(14),a(16),a(18),
a(23),a(24),a(25),a(29),a(35) \}, \nonumber\\
&&\ \ 
a(36),a(37),a(48),a(49),a(59),   \}, \nonumber \\
C_{20} &:& \{ 
b(123),b(124),b(126),b(136),b(168),b(235),b(249), \\
&&\ \ 
b(259),b(357),b(367), \text{ and  their inverse elements} \}, \nonumber \\
C_{12}  &:& \{  
c(1),c(2),c(3),c(4),c(5),c(6), \text{ and their inverse elements}
\}, \nonumber \\
C_{12}' &:& \{ 
c^2(1),c^2(2),c^2(3),c^2(4),c^2(5),c^2(6), 
\text{ and their inverse elements}
\}, \nonumber 
\end{eqnarray}
where $a(km)$, $b(kmn)$ and $c(k)$ 
denote the rotation by $\pi$ about the 
midpoint of the edge $k-m$, 
the rotation by $2\pi/3$ about the axis through 
the center of the face $k-m-n$ and the 
rotation by $2\pi/5$ about the axis through the vertex $k$. 
The conjugacy classes, $C_1$, $ C_{15}$, $C_{20}$, 
$C_{12}$ and $C_{12}'$, include 1, 15, 20, 12 and 12 elements, 
respectively.
Since obviously 
$\left( a(km)\right)^2=\left( b(kmn) \right)^3=\left( c(k) \right)^5=e$, 
we find $h=2$ in $C_{15}$, $h=3$ in $C_{20}$, 
$h=5$ in $C_{12}$ and $h=5$ in  $C_{12}'$, 
where $h$ denotes the order of each element in the 
conjugacy class, i.e. $g^h=e$.
The orthogonality relations (\ref{eq:character-2-e}) and (\ref{eq:sum-dim})
for $A_5$ 
lead to
\begin{eqnarray}
&&m_1+4m_2+9m_3+16m_4+25m_5 + \cdots =60,\\
&&m_1+m_2+m_3+m_4+m_5 + \cdots =5 .
\end{eqnarray}
The solution is found as $(m_1,m_2,m_3,m_4,m_5)=(1,0,2,1,1)$.
Therefore the $A_5$ group has one trivial singlet, ${\bf 1}$, 
two triplets, ${\bf 3}$ and ${\bf 3}'$, one quartet, ${\bf 4}$, and 
one quintet, ${\bf 5}$.
The characters are shown in Table \ref{A5-character}.
Instead of $a$ and $b$, we use the generators, 
$s=a$ and $t=bab$,  
which satisfy 
\begin{equation}
s^2=t^5=(t^2st^3st^{-1}stst^{-1})^2=e.
\end{equation}
The generators, $s$ and $t$, are represented as~\cite{shirai},
 \begin{eqnarray}
&&
 s=\frac{1}{2}\Mat3{-1}{\phi}{\frac{1}{\phi}} {\phi}{\frac{1}{\phi}}{1} {\frac{1}{\phi}}{1}{-\phi},
 \quad
 t=\frac{1}{2}\Mat3{1}{\phi}{\frac{1}{\phi}}
 {-\phi}{\frac{1}{\phi}}{1} {\frac{1}{\phi}}{-1}{\phi}, 
\qquad {\rm~~on~~{\bf 3}}, 
 \\&&
  s =\frac{1}{2}\Mat3{-\phi}{\frac{1}{\phi}}{1} {\frac{1}{\phi}}{-1} {\phi} {1}{\phi}{\frac{1}{\phi}},
 \quad
t =\frac{1}{2}\Mat3{-\phi}{-\frac{1}{\phi}}{1}
{\frac{1}{\phi}}{1}{\phi} {-1}{\phi}{-\frac{1}{\phi}}, 
\qquad {\rm~~on~~{\bf 3}'}, 
 \end{eqnarray}
 
 \begin{eqnarray}
&& s=\frac{1}{4}
\left(
\begin{array}{cccc}
-1 & -1 & -3 & -\sqrt5\\
-1 & 3 & 1 & -\sqrt5 \\
-3 & 1 & -1 & \sqrt5 \\
-\sqrt5 & -\sqrt5 & \sqrt5 & -1\\
\end{array}
\right),
 \nonumber \\
&& t =\frac{1}{4}
\left(
\begin{array}{cccc}
-1 & 1 & -3 & \sqrt5\\
-1 & -3 & 1 & \sqrt5 \\
3 & 1 & 1 & \sqrt5 \\
\sqrt5 & -\sqrt5 & -\sqrt5 & -1\\
\end{array}
\right),  \qquad {\rm~~on~~{\bf 4}}, 
 \end{eqnarray}

\begin{eqnarray}
&& s =\frac{1}{2}
\left(
\begin{array}{ccccc}
\frac{1-3\phi}{4} & \frac{\phi^2}{2} & -\frac{1}{2\phi^2} & \frac{\sqrt5}{2} & \frac{\sqrt3}{4\phi}\\
\frac{\phi^2}{2} &1 & 1 & 0 & \frac{\sqrt3}{2\phi}\\
-\frac{1}{2\phi^2} &1 & 0 & -1 & -\frac{\sqrt3\phi}{2}\\
\frac{\sqrt5}{2} &0 & -1 & 1 & -\frac{\sqrt3}{2}\\
\frac{\sqrt3}{4\phi} & \frac{\sqrt3}{2\phi} & -\frac{\sqrt3\phi}{2} & -\frac{\sqrt3}{2} & \frac{3\phi-1}{4}\\
\end{array}
\right),  \nonumber
\\
&& t =\frac{1}{2}
\left(
\begin{array}{ccccc}
\frac{1-3\phi}{4} & -\frac{\phi^2}{2} & -\frac{1}{2\phi^2} & -\frac{\sqrt5}{2} & \frac{\sqrt3}{4\phi}\\
\frac{\phi^2}{2} &-1 & 1 & 0 & \frac{\sqrt3}{2\phi}\\
\frac{1}{2\phi^2} &1 & 0 & -1 & \frac{\sqrt3\phi}{2}\\
-\frac{\sqrt5}{2} &0 & 1 & 1 & \frac{\sqrt3}{2}\\
\frac{\sqrt3}{4\phi} & -\frac{\sqrt3}{2\phi} & -\frac{\sqrt3\phi}{2} & \frac{\sqrt3}{2} & \frac{3\phi-1}{4}\\
\end{array}
\right), \qquad {\rm~~on~~{\bf 5}}, 
 \end{eqnarray}
where $\phi=\frac{1+\sqrt5}{2}$. 
Furthermore, the multiplication rules are also shown 
in Table \ref{A5-muliplication-rule}.

\begin{table}[t]
\begin{center}
\begin{tabular}{|c|c|c|c|c|c|c|}
\hline
        &$h$&$1$&$3$&$3'$&$4$&$5$ \\ \hline
$C_1$   &$1$&   $1$  &    $3$    &    $3$     &$4$&$5$   \\ \hline
$C_{15}$   &$2$&   $1$  &    $-1$    &    $-1$     &$0$&$1$  \\ \hline
$C_{20}$   &$3$&   $1$  & $0$  & $0$ &$1$&$-1$   \\ \hline
$C_{12}$   &$5$&   $1$  & $\phi$  & $1-\phi$ &$-1$ &$0$  \\ \hline
$C_{12'}$&$5$&   $1$  & $1-\phi$&  $\phi$  &$-1$ &$0$  \\
\hline
\end{tabular}
\end{center}
\caption{Characters of $A_5$ representations, where $\phi=\frac{1+\sqrt5}{2}$.}
\label{A5-character}
\end{table}

\begin{table}[t]
\begin{center}
\begin{tabular}{|c|}
\hline
 ${\bf 3}\otimes{\bf 3}={\bf1}\oplus{\bf3}\oplus{\bf5}$ \\ \hline
${\bf 3'}\otimes{\bf 3'}={\bf1}\oplus{\bf3'}\oplus{\bf5}$   \\ \hline
${\bf 3}\otimes{\bf 3'}={\bf4}\oplus{\bf5}$  \\ \hline
${\bf 3}\otimes{\bf 4}={\bf 3'}\oplus{\bf4}\oplus{\bf5}$   \\ \hline
${\bf 3'}\otimes{\bf 4}={\bf 3}\oplus{\bf4}\oplus{\bf5}$  \\ \hline
${\bf 3}\otimes{\bf 5}={\bf 3}\oplus{\bf 3'}\oplus{\bf4}\oplus{\bf5}$  \\ \hline
${\bf 3'}\otimes{\bf 5}={\bf 3}\oplus{\bf 3'}\oplus{\bf4}\oplus{\bf5}$  \\
\hline
${\bf 4}\otimes{\bf 4}={\bf 1}\oplus{\bf 3}\oplus{\bf 3'}\oplus{\bf4}\oplus{\bf5}$  \\
\hline
${\bf 4}\otimes{\bf 5}={\bf 3}\oplus{\bf 3'}\oplus{\bf4}\oplus{\bf5}\oplus{\bf5}$  \\
\hline
${\bf 5}\otimes{\bf 5}={\bf 1}\oplus{\bf 3}\oplus{\bf 3'}\oplus{\bf4}\oplus{\bf4}\oplus{\bf5}\oplus{\bf5}$  \\
\hline
\end{tabular}
\end{center}
\caption{Multiplication rules for the $A_5$ group.}
\label{A5-muliplication-rule}
\end{table}

Here we show some parts of tensor products \cite{Everett:2008et},
\begin{eqnarray}
{\3tvec {x_1}  {x_2} {x_3}}_{{\bf 3}} 
\otimes 
{\3tvec {y_1}  {y_2} {y_3}}_{{\bf 3}}
&=&
(x_1y_1+x_2y_2+x_3y_3)_{{\bf 1}}
 \oplus
{ \3tvec {x_3y_2 - x_2y_3 }  {x_1y_3-x_3y_1 }  {x_2y_1-x_1y_2}}_{{\bf 3}}
\nonumber\\
& \oplus &
{ \5tvec {x_2y_2 - x_1y_1 }  {x_2y_1+x_1y_2 }  {x_3y_2+x_2y_3}  {x_1y_3+x_3y_1}  
 {-\frac{1}{\sqrt3}(x_1y_1+x_2y_2-2x_3y_3)} }_{{\bf 5}},
\end{eqnarray}

\begin{eqnarray}
{\3tvec {x_1}  {x_2} {x_3}}_{{\bf 3'}} 
\otimes 
{\3tvec {y_1}  {y_2} {y_3}}_{{\bf 3'}}
&=&
(x_1y_1+x_2y_2+x_3y_3)_{{\bf 1}}
 \oplus
{ \3tvec {x_3y_2 - x_2y_3 }  {x_1y_3-x_3y_1 }  {x_2y_1-x_1y_2}}_{{\bf 3'}}
\nonumber\\
& \oplus &
{ \5tvec {\frac{1}{2}(-\frac{1}{\phi}x_1y_1 -\phi x_2y_2+\sqrt5 x_3y_3) } 
 {x_2y_1+x_1y_2 }  
 {-(x_3y_1+x_1y_3)}  
 {x_2y_3+x_3y_2}  
 {\frac{1}{2\sqrt3}((1-3\phi)x_1y_1+(3\phi-2)x_2y_2+x_3y_3)} }_{{\bf 5}},
\end{eqnarray}

\begin{equation}
{\3tvec {x_1}  {x_2} {x_3}}_{{\bf 3}} 
\otimes 
{\3tvec {y_1}  {y_2} {y_3}}_{{\bf 3'}}
=
{ \4tvec {\frac{1}{\phi}x_3y_2-\phi x_1y_3 } {\phi x_3y_1+\frac{1}{\phi}x_2y_3 }  
{-\frac{1}{\phi}x_1y_1+\phi x_2y_2}  {x_2y_1-x_1y_2+x_3y_3}}_{{\bf 4}}
\oplus
{ \5tvec {\frac{1}{2}(\phi^2x_2y_1 +\frac{1}{\phi^2} x_1y_2-\sqrt5 x_3y_3) } 
 {-(\phi x_1y_1+\frac{1}{\phi}x_2y_2 )}  
 {\frac{1}{\phi}x_3y_1-\phi x_2y_3}  
 {\phi x_3y_2+\frac{1}{\phi}x_1y_3}  
 {\frac{\sqrt3}{2}(\frac{1}{\phi}x_2y_1+\phi x_1y_2+x_3y_3)} }_{{\bf 5}}.
\end{equation}

For ${\bf 4}\otimes {\bf4}$ and  ${\bf 5}\otimes {\bf5}$, they include 
the simple form of trivial singlets as
\begin{eqnarray}
&&
{\4tvec {x_1}  {x_2} {x_3} {x_4}}_{{\bf 4}} 
\otimes 
{\4tvec {x_1}  {y_2} {y_3} {y_4}}_{{\bf 4}}
\rightarrow
(x_1y_1+x_2y_2+x_3y_3+x_4y_4)_{{\bf 1}} \ ,
\\&&
{\5tvec {x_1}  {x_2} {x_3} {x_4} {x_5}}_{{\bf 5}} 
\otimes 
{\5tvec {y_1}  {y_2} {y_3} {y_4} {y_5}}_{{\bf 5}}
\rightarrow
(x_1y_1+x_2y_2+x_3y_3+x_4y_4+x_5y_5)_{{\bf 1}} \ .
\end{eqnarray}

\clearpage



\section{$T'$}
\label{sec:T'}

The $T'$ group is the double covering group of $A_4 =T$.
Instead of Eq.~(\ref{eq:T-st}), we consider the following 
algebraic relations,
\begin{eqnarray}\label{eq:T'-st}
s^2=r, \qquad r^2=t^3=(st)^3=e, \qquad rt=tr.
\end{eqnarray}
The closed algebra including $r, s$ and $t$ is the $T'$ group.
It consists of 24 elements.

\vskip .5cm
{$\bullet$ \bf Conjugacy classes}

All of 24 elements in $T'$ are classified by their orders as
\begin{eqnarray}
\begin{array}{ccc}
h=1 \quad :\quad &\{e\},  \\
h=2 \quad :\quad &\{r\},  \\
h=3 \quad :\quad &\{t,t^2,ts,st,rst^2,rt^2s,rtst,rsts\}, \\
h=4 \quad :\quad &\{s,rs,tst^2,t^2st,rtst^2,rt^2st\},  \\
h=6 \quad :\quad &\{rt,rst,rts,rt^2,sts,st^2,t^2s,tst\}. 
\end{array}
\end{eqnarray}
Furthermore, they are classified into seven conjugacy 
classes as 
\begin{eqnarray}
\begin{array}{ccc}
C_1   :&\{e\}, &h=1,   \\
C_{1'}  :&\{r\}, & h=2,  \\
C_4    :&\{t,rsts,st,ts\}, & h=3,  \\
C_{4'}  :&\{t^2,rtst,rt^2s,rst^2\}, & h=3,\\
C_6    :&\{s,rs,tst^2,t^2st,rtst^2,rt^2st\}, & h=4,   \\
C_{4''} :&\{rt,sts,rst,rts\}, & h=6,  \\
C_{4'''}:&\{rt^2,tst,t^2s,st^2\} ,  & h=6 .
\end{array}
\end{eqnarray}

\vskip .5cm
{$\bullet$ \bf Characters and representations}

The orthogonality relations (\ref{eq:character-2-e}) and (\ref{eq:sum-dim})
for $T'$ 
lead to
\begin{eqnarray}
&&m_1+2^2m_2+3^2m_3+ \cdots=24,\\
&&m_1+m_2+m_3 + \cdots =7 .
\end{eqnarray}
The solution is found as $(m_1,m_2,m_3)=(3,3,1)$.
That is, there are three singlets, three doublets and a 
triplet.

Now, let us study characters.
The analysis on $T'$ is quite similar to one on 
$A_4=T$.
First, we start with singlets.
Because of $s^4=r^2=e$, there are four possibilities 
for $\chi_\alpha(s)=(i)^n$ ($n=0,1,2,3$).
However, since $t$ and $ts$ belong to the same conjugacy class, 
$C_4$, the character consistent with the structure of 
conjugacy classes is obtained as 
$\chi_\alpha(s)= 1$ for singlets.
That also means $\chi_\alpha(r)= 1$.
Then, similarly to $A_4=T$, three singlets are classified by 
three possible values of  
$\chi_\alpha(t)=\omega^n$.
That is, three singlets, ${\bf 1}$, ${\bf 1'}$, and 
${\bf 1''}$  are classified by 
these three values of  $\chi_\alpha(t)=1, \omega$ and $\omega^2$, 
respectively.
These are shown in Table \ref{tab:T'-character}.

Next, let us study three doublet representations 
${\bf 2}$, ${\bf 2}'$ and ${\bf 2}''$, and a triplet 
representation ${\bf 3}$ for $r$.
The element $r$ commutes with all of elements.
That implies by the Shur's lemma that $r$ can 
be represented by 
\begin{eqnarray}
\lambda_{2,2',2''} 
\begin{pmatrix}
1 & 0 \\
0 & 1
\end{pmatrix},
\end{eqnarray}
on  ${\bf 2}$, ${\bf 2}'$ and ${\bf 2}''$ 
and 
\begin{eqnarray}
\lambda_{3} 
\begin{pmatrix}
1 & 0 & 0\\
0 & 1 & 0\\
0 & 0 & 1
\end{pmatrix},
\end{eqnarray}
on ${\bf 3}$.
In addition, possible values of $\lambda_{2,2',2''} $ and 
$\lambda_{3} $ must be equal to $\lambda_{2,2',2''}=\pm 1$ 
and $\lambda_{3} =\pm 1$ because of $r^2=e$.
That is, possible values of characters are obtained as 
$\chi_2(r),\chi_{2'}(r), \chi_{2''}(r)=\pm 2$ and 
$\chi_3(r)=\pm 3$.
Furthermore, the second orthogonality relation between 
$e$ and $r$ leads to 
\begin{eqnarray}
\sum_{\alpha} \chi_{D_\alpha}(e)^* \chi_{D_\alpha}(r)
= 3 +  2\chi_2(r) + 2\chi_{2'}(r) + 2\chi_{2''}(r) +  
3\chi_3(r)= 0.
\end{eqnarray}
Here we have used $\chi_{1,1',1''}(r)=1$.
Its solution is obtained as 
$\chi_2(r)=\chi_{2'}(r)=\chi_{2''}(r)=- 2$ and 
$\chi_3(r)= 3$.
These are shown in Table \ref{tab:T'-character}.
That is, the $r$ element is represented by 
\begin{eqnarray}\label{eq:T'-r-2}
r=-
\begin{pmatrix}
1 & 0 \\
0 & 1
\end{pmatrix},
\end{eqnarray}
on  ${\bf 2}$, ${\bf 2}'$ and ${\bf 2}''$, 
and 
\begin{eqnarray}
r= 
\begin{pmatrix}
1 & 0 & 0\\
0 & 1 & 0\\
0 & 0 & 1
\end{pmatrix},
\end{eqnarray}
on ${\bf 3}$.

Now, let us study doublet representation of $t$.
We use the basis diagonalizing $t$.
Because of $t^3=e$, the element $t$ could be written as 
\begin{eqnarray}
\begin{pmatrix}
\omega^k & 0 \\
0 & \omega^\ell
\end{pmatrix},
\end{eqnarray}
with $k,\ell = 0,1,2$.
However, if $k=\ell$, the above matrix would become proportional to 
the $(2 \times 2)$ identity matrix, 
that is, the element $t$ would also commute with all of elements.
That is nothing but a singlet representation.
Then, we should have the condition $k \neq \ell$.
As a result, there are three possible values for the trace of 
the above values as $\omega^k+\omega^\ell=-\omega^n$ with 
$k,\ell,n=0,1,2$ and $k \neq \ell, \ell \neq n, n \neq k$.
That is, the characters of $t$ for three doublets, 
${\bf 2}$, ${\bf 2}'$ and ${\bf 2}''$ are 
classified as $\chi_2(t) = -1, \chi_{2'}(t) = -\omega$ 
and $\chi_{2''}(t) = -\omega^2$.
These are shown in Table \ref{tab:T'-character}.
Then, the element $t$ is represented by 
 \begin{eqnarray}\label{eq:T'-t-2}
t = 
\begin{pmatrix}
\omega^2 & 0 \\
0 & \omega
\end{pmatrix}, \qquad {\rm on~~} {\bf 2},
\end{eqnarray}
 \begin{eqnarray}
t = 
\begin{pmatrix}
1 & 0 \\
0 & \omega^2
\end{pmatrix}, \qquad {\rm on~~} {\bf 2}',
\end{eqnarray}
 \begin{eqnarray}
t = 
\begin{pmatrix}
\omega & 0 \\
0 & 1
\end{pmatrix}, \qquad {\rm on~~} {\bf 2}''.
\end{eqnarray}
Since we have found the explicit $(2 \times 2)$ matrices 
for $r$ and $t$ on all of three doublets, 
it is straightforward to calculate the explicit forms 
of $rt$ and $rt^2$, which belong to the conjugacy classes, 
$C_{4}'$ and $C_4''$ , respectively.
Then, it is also straightforward to compute 
the characters of $C_{4}'$ and $C_4''$ for doublets 
by such explicit forms of $(2 \times 2)$ matrices for $rt$ and $rt^2$.
They are shown in Table \ref{tab:T'-character}.

In order to determine the character of $t$ for 
the triplet, $\chi_{3}(t)$, 
we use the second orthogonality relation between $e$ and $t$, 
\begin{eqnarray}\label{eq:T'-e-t}
\sum_{\alpha} \chi_{D_\alpha}(e)^* \chi_{D_\alpha}(t)
= 0.
\end{eqnarray}
Note that all of the characters $\chi_\alpha(t)$ 
except $\chi_{3}(t)$ have 
been derived in the above.
Then, the above orthogonality relation (\ref{eq:T'-e-t}) 
requires $\chi_{3}(t)=0$, that is, $\chi_{3}(C_4)=0$.
Similarly, it is found that 
$\chi_{3}(C_4')=\chi_{3}(C_4'')=\chi_{3}(C_4''')=0$ 
as shown in Table \ref{tab:T'-character}.
Now, we study the explicit form of the $(3 \times 3)$ 
matrix for $t$ on the triplet.
We use the basis to diagonalize $t$.
Since $t^3=e$ and $\chi_{3}(t)=0$, we can obtain 
 \begin{eqnarray}
t = 
\begin{pmatrix}
 1 & 0 & 0 \\
0 & \omega & 0 \\
0 & 0 & \omega^2
\end{pmatrix},\qquad {\rm on~~} {\bf 3}.
\end{eqnarray}

Finally, we study the characters of $C_6$ including $s$ 
for the doublets and the triplet.
Here, we use the first orthogonality relation between the trivial 
singlet representation and the doublet representation ${\bf 2}$ 
\begin{eqnarray}\label{eq:T'-1-2}
\sum_{g \in G} \chi_{1}(g)^* \chi_{2}(g) 
= 0.
\end{eqnarray}
Recall that all of characters except $\chi_{2}(C_6)$ 
have been already given.
This orthogonality relation (\ref{eq:T'-1-2}) requires 
$\chi_{2}(C_6) =0$.
Similarly, we find that  $\chi_{2'}(C_6) =\chi_{2''}(C_6) =0$.
Furthermore, the character of $C_6$ for the triplet 
$\chi_{3}(C_6)$ is also determined by using the 
orthogonality relation $\sum_{g \in G} \chi_{1}(g)^* \chi_{2}(g) 
= 0$ with the other known characters.
As a result, we obtain $\chi_{3}(C_6) = -1$.
Now, we have completed all of characters in the $T'$ group, which are 
summarized in Table \ref{tab:T'-character}.
Then, we study the explicit form of $s$ on the doublets and triplet.
On the doublets, the element must be the 
$(2 \times 2)$ unitary matrix, which satisfies 
$\text{tr}(s)=0$ and $s^2=r$.
Recall that the doublet representation for $r$ is already 
obtained in Eq.~(\ref{eq:T'-r-2}).
Thus, the element $s$ could be represented as 
\begin{eqnarray}\label{eq:T'-s-2}
s=-\frac{1}{\sqrt3}
\begin{pmatrix}
i & \sqrt2 p \\
\sqrt2 \bar p & -i
\end{pmatrix},\quad p=i\phi,
\end{eqnarray}
on the doublet representations. 
For example, for ${\bf 2}$ this representation of $s$ satisfies 
\begin{equation}
\text{tr}(st)=-\frac{i}{\sqrt3}(\omega^2-\omega)=-1,
\end{equation}
so the ambiguity of $p$ cannot be removed.

Here, we summarize the doublet and triplet representations,
\begin{eqnarray}
t=\mat2{\omega^2}{0}{0}{\omega},
\quad
r=\mat2{-1}{0}{0}{-1},
\quad
s=-\frac{1}{\sqrt{3}}\mat2{i}{\sqrt{2}p}{-\sqrt{2}\bar p}{-i} 
~~~{\rm  on}~~{\bf 2} ,
\end{eqnarray}

\begin{eqnarray}
t=\mat2{1}{0}{0}{\omega^2},\quad r=\mat2{-1}{0}{0}{-1},\quad
s=-\frac{1}{\sqrt{3}}\mat2{i}{\sqrt{2}p}{-\sqrt{2}\bar p}{-i}
~~~{\rm  on}~~{\bf 2}' ,
\end{eqnarray}

\begin{eqnarray}
t=\mat2{\omega}{0}{0}{1},\quad r=\mat2{-1}{0}{0}{-1},\quad
s=-\frac{1}{\sqrt{3}}\mat2{i}{\sqrt{2}p}{-\sqrt{2}\bar p}{-i}
~~~{\rm  on}~~{\bf 2}'' ,
\end{eqnarray}

\begin{eqnarray}
t=\Mat3{1}{0}{0}{0}{\omega}{0}{0}{0}{\omega^2},\quad
r=\Mat3{1}{0}{0}{0}{1}{0}{0}{0}{1},\ \ 
s=\frac13\Mat3{-1}{2p_1}{2p_1p_2} {2\bar p_1}{-1}{2p_2} {2\bar p_1\bar p_2}{2\bar p_2}{-1}
~{\rm  on}~~{\bf 3} ,
\end{eqnarray}
where $p_1=e^{i\phi_1}$ and $p_2=e^{i\phi_2}$.

\vskip .5cm
{$\bullet$ \bf Tensor products}

From the above relations, complete tensor products can be determined.
First, we study the tensor product of  ${\bf 2}$ and ${\bf 2}$,
i.e.
\begin{eqnarray}
\2tvec{x_1}{x_2}_{\bf2}\otimes\2tvec{y_1}{y_2}_{\bf2}.
\end{eqnarray}
Then, we investigate the transformation property of 
elements $x_iy_j$ for $i,j=1,2$ under $t$, $r$ and $s$.
Then, it is found that 
\begin{eqnarray}
\2tvec{x_1}{x_2}_{\bf 2(2')}\otimes\2tvec{y_1}{y_2}_{\bf 2(2'')}
=
\left(\frac{x_1y_2-x_2y_1}{\sqrt{2}}\right)_{\bf 1}
\oplus\3tvec{\frac{i}{\sqrt{2}}p_1p_2\bar p(x_1y_2+x_2y_1)}{p_2\bar p^2x_1y_1}{x_2y_2}_{\bf3}.
\end{eqnarray}
Similarly, we can obtain 
\begin{eqnarray}
\2tvec{x_1}{x_2}_{\bf2'(2)}\otimes\2tvec{y_1}{y_2}_{\bf2'(2'')}
&=\left(\frac{x_1y_2-x_2y_1}{\sqrt{2}}\right)_{\bf1''}
\oplus
\3tvec{p_1\bar p^2x_1y_1}{x_2y_2}{\frac{i}{\sqrt{2}}\bar p\bar
  p_2(x_1y_2+x_2y_1)}_{\bf3}, 
\\
\2tvec{x_1}{x_2}_{\bf2''(2)}\otimes\2tvec{y_1}{y_2}_{\bf2''(2')}
&=\left(\frac{x_1y_2-x_2y_1}{\sqrt{2}}\right)_{\bf1'}
\oplus
\3tvec{x_2y_2}{\frac{i}{\sqrt{2}}\bar p\bar p_1(x_1y_2+x_2y_1)}{\bar p^2\bar p_1\bar p_2x_1y_1}_{\bf3}.
\end{eqnarray}
Furthermore, we can compute other products such as 
$\bf2\times2'$, $\bf2\times2''$ and $\bf2'\times2''$.
Then, it is found that 
\begin{eqnarray}
\bf2\times 2'=2''\times 2'', \qquad 2\times2''=2'\times 2', \qquad
2'\times 2''=2\times 2 .
\end{eqnarray}
Moreover, a similar analysis leads to 
\begin{eqnarray}
\3tvec{x_1}{x_2}{x_3}_{\bf3}\otimes\3tvec{y_1}{y_2}{y_3}_{\bf3}
&=&
[x_1y_1+p^2_1p_2(x_2y_3+x_3y_2)]_{\bf1}\nonumber
\\&\oplus&
[x_3y_3+\bar p_1\bar p^2_2(x_1y_2+x_2y_1)]_{\bf1'}
\oplus
[(x_2y_2+\bar p_1p_2(x_1y_3+x_3y_1)]_{\bf1''}\nonumber
\\&\oplus&
\3tvec{2x_1y_1-p^2_1p_2(x_2y_3+x_3y_3)}
{2p_1p^2_2x_3y_3-x_1y_2-x_2y_1}{2p_1\bar p_2x_2y_2-x_1y_3-x_3y_1}_{\bf3}
\nonumber
\\&\oplus&
\3tvec{x_2y_3-x_3y_2}{\bar p^2_1\bar p_2(x_1y_2-x_2y_1)}
{\bar p^2_1\bar p_2(x_3y_1-x_1y_3)}_{\bf3},
\end{eqnarray}
\begin{eqnarray}
\2tvec{x_1}{x_2}_{\bf2,2',2''}\otimes\3tvec{y_1}{y_2}{y_3}_{\bf3}
&=&
  \left(\begin{array}{c}
 -i\sqrt2pp_1x_2y_2+x_1y_1
\\ 
i\sqrt2\bar pp_1p_2x_1y_3-x_2y_1
\end{array}\right)_{\bf2,2',2''}
\nonumber\\&\oplus &
 \left(\begin{array}{c}
-i\sqrt2pp_2x_2y_3+x_1y_2
\\
i \sqrt2\bar p\bar p_1x_1y_1-x_2y_2
\end{array}\right)_{\bf2',2'',2}
\nonumber\\&\oplus&
 \left(\begin{array}{c}
 -i\sqrt2p\bar p_1\bar p_2 x_2y_1+x_1y_3
 \\
i\sqrt2\bar p\bar p_2x_1y_2-x_2y_3
\end{array}\right)_{\bf2'',2,2'},
\end{eqnarray}
\begin{eqnarray}
&&(x)_{\bf1'(1'')}\otimes\2tvec{y_1}{y_2}_{\bf2,2',2''}
=
\2tvec{xy_1}{xy_2}_{\bf2'(2''),2''(2),2(2')},\\
&&(x)_{\bf1'}\otimes\3tvec{y_1}{y_2}{y_3}_{\bf3}
=
\3tvec{xy_3}{\bar p^2_1\bar p_2xy_1}{\bar p_1\bar p^2_2xy_2}_{\bf3},\quad
(x)_{\bf1''}\otimes\3tvec{y_1}{y_2}{y_3}_{\bf3}
=
\3tvec{xy_2}{\bar p_1 p_2xy_3}{\bar p^2_1\bar p_2xy_1}_{\bf3}.
\end{eqnarray}
The representations for $p'$ can be in general obtained 
by transforming $p$ as follows,
\begin{eqnarray}
&&
\Phi_2(p')= \left(\begin{array}{cc}
 1 & 0 
 \\
0 & e^{-i\gamma}
\end{array}\right)
\Phi_2(p), \quad p'=pe^{i\gamma},
\\
&& \Phi_3(p')= \left(\begin{array}{ccc}
 1 & 0 & 0
 \\
0 & e^{-i\gamma} & 0
\\
0 & 0 & e^{-i(\alpha+\beta)}
\end{array}\right)
\Phi_3(p), \quad p'_1=p_1e^{i\alpha},\quad p'_2=p_2e^{-i\beta}.
\end{eqnarray}

If one takes the parameters $p=i$ and $p_1=p_2=1$, 
then the generator $s$ is simplified as
\begin{eqnarray}
& & s=
-\frac{i}{\sqrt3}
 \left(\begin{array}{cc}
 1 & \sqrt2 
 \\
\sqrt2  & -1
\end{array}\right),\quad {\rm on~~} {\bf 2}, \\
 & & \nonumber \\
& & s=
 \left(\begin{array}{ccc}
 -1 & 2 & 2
 \\
2 & -1 & 2
\\
2 & 2 & -1
\end{array}\right),\quad {\rm on~~} {\bf 3}. \nonumber
\end{eqnarray}
\begin{table}[t]
\begin{center}
\begin{tabular}{|c|c|c|c|c|c|c|c|c|}
\hline
          &$h$&$\chi_1$ &$\chi_{1'}$&$\chi_{1''}$
              &$\chi_2$ &$\chi_{2'}$&$\chi_{2''}$&$\chi_3$ \\ \hline
$C_1$     &$1$&$1$      &$1$        &$1$          
              &$2$      &$2$        &$2$         &$3$\\ \hline
$C_{1'}$  &$2$&$1$      &$1$        &$1$        
              &$-2$     &$-2$       &$-2$        &$3$\\ \hline
$C_4$     &$3$&$1$      &$\omega$   &$\omega^2$        
              &$-1$     &$-\omega$  &$-\omega^2$ &$0$\\ \hline
$C_4'$    &$3$&$1$      &$\omega^2$ &$\omega$        
              &$-1$     &$-\omega^2$&$-\omega$   &$0$\\ \hline
$C_{4''}$ &$6$&$1$      &$\omega$   &$\omega^2$        
              &$1$      &$\omega$   &$\omega^2$  &$0$\\ \hline
$C_{4'''}$&$6$&$1$      &$\omega^2$ &$\omega$        
              &$1$      &$\omega^2$ &$\omega$    &$0$\\ \hline
$C_6$     &$4$&$1$      &$1$        &$1$        
              &$0$      &$0$        &$0$         &$-1$\\ \hline
\end{tabular}
\end{center}
\caption{Characters of ${T'}$ representations}
\label{tab:T'-character}
\end{table}

These tensor products can be also simplified as 

\begin{eqnarray}
&&\2tvec{x_1}{x_2}_{\bf2(2')}\otimes\2tvec{y_1}{y_2}_{\bf2(2'')}
=\left(\frac{x_1y_2-x_2y_1}{\sqrt{2}}\right)_{\bf1}
\oplus \3tvec{\frac{x_1y_2+x_2y_1}{\sqrt{2}}}{-x_1y_1}{x_2y_2}_{\bf3},
\\&&
\2tvec{x_1}{x_2}_{\bf2'(2)}\otimes\2tvec{y_1}{y_2}_{\bf2'(2'')}
=\left(\frac{x_1y_2-x_2y_1}{\sqrt{2}}\right)_{\bf1''}
\oplus \3tvec{-x_1y_1}{x_2y_2}{\frac{x_1y_2+x_2y_1}{\sqrt{2}}}_{\bf3}, 
\\&&
\2tvec{x_1}{x_2}_{\bf2''(2)}\otimes\2tvec{y_1}{y_2}_{\bf2''(2')}
=\left(\frac{x_1y_2-x_2y_1}{\sqrt{2}}\right)_{\bf1'}
\oplus \3tvec{x_2y_2}{\frac{x_1y_2+x_2y_1}{\sqrt{2}}}{-x_1y_1}_{\bf3},
\end{eqnarray}
\begin{eqnarray}
\3tvec{x_1}{x_2}{x_3}_{\bf3}\otimes\3tvec{y_1}{y_2}{y_3}_{\bf3}
&=&
[x_1y_1+x_2y_3+x_3y_2]_{\bf1}\nonumber\\
&\oplus&[x_3y_3+x_1y_2+x_2y_1]_{\bf1'}
\oplus(x_2y_2+x_1y_3+x_3y_1]_{\bf1''}\nonumber
\\&\oplus&
\3tvec{2x_1y_1-x_2y_3-x_3y_3}
{2x_3y_3-x_1y_2-x_2y_1}{2x_2y_2-x_1y_3-x_3y_1}_{\bf3}
\nonumber
\\&\oplus&
\3tvec{x_2y_3-x_3y_2}{x_1y_2-x_2y_1}
{x_3y_1-x_1y_3}_{\bf3},
\end{eqnarray}
\begin{eqnarray}
\2tvec{x_1}{x_2}_{\bf2,2',2''}\otimes\3tvec{y_1}{y_2}{y_3}_{\bf3}
&=&
  \left(\begin{array}{c}
 \sqrt2x_2y_2+x_1y_1
 \\
\sqrt2x_1y_3-x_2y_1
\end{array}\right)_{\bf2,2',2''}
\oplus
 \left(\begin{array}{c}
\sqrt2x_2y_3+x_1y_2
\\
\sqrt2x_1y_1-x_2y_2
\end{array}\right)_{\bf2',2'',2}
\nonumber\\&\oplus&
 \left(\begin{array}{c}
 \sqrt2 x_2y_1+x_1y_3
 \\
\sqrt2x_1y_2-x_2y_3
\end{array}\right)_{\bf2'',2,2'},
\end{eqnarray}
\begin{eqnarray}
&&(x)_{\bf1'(1'')}\otimes\2tvec{y_1}{y_2}_{\bf2,2',2''}
=
\2tvec{xy_1}{xy_2}_{\bf2'(2''),2''(2),2(2')},\\
&&(x)_{\bf1'}\otimes\3tvec{y_1}{y_2}{y_3}_{\bf3}
=
\3tvec{xy_3}{xy_1}{xy_2}_{\bf3},\quad
(x)_{\bf1''}\otimes\3tvec{y_1}{y_2}{y_3}_{\bf3}
=
\3tvec{xy_2}{xy_3}{xy_1}_{\bf3}.
\end{eqnarray}

When $p=e^{i\pi/12}$ and $p_1=p_2=\omega$, 
then the representation and their tensor products are given in the Ref. \cite{Feruglio:2007uu}.

\clearpage


\section{$D_N$}
\label{sec:DN}

\subsection{Generic aspects}

$D_N$ is a symmetry of a regular polygon with $N$ sides 
and it is often called as the dihedral group.
It is isomorphic to $Z_N \rtimes Z_2$ and 
it is also denoted by $\Delta(2N)$.
It consists of cyclic rotation, $Z_N$ and 
reflection.
That is, it is generated by two generators $a$ and $b$, 
which act on $N$ edges $x_i$ ($i=1,\cdots,N$) of $N$-polygon as 
\begin{eqnarray}\label{eq:Dn-n}
a&:&(x_1,x_2 \cdots, x_N) \rightarrow (x_{N},x_{1}\cdots,x_{N-1}),\\
b&:&(x_1,x_2 \cdots, x_N) \rightarrow (x_1,x_N \cdots, x_2).
\end{eqnarray} 
These two generators satisfy 
\begin{eqnarray}\label{eq:Dn-AB}
a^N=e, \qquad b^2=e, \qquad bab=a^{-1} ,
\end{eqnarray}
where the third equation is equivalent to $aba=b$.
The order of $D_N$ is equal to $2N$, 
and all of $2N$ elements are written as $a^mb^k$ 
with $m=0,\cdots, N-1$ and $k=0,1$.
The third equation in (\ref{eq:Dn-AB}) 
implies that the $Z_N$ subgroup 
including $a^m$ is a normal subgroup of $D_N$.
Thus, $D_N$ corresponds to a semi-direct product 
between $Z_N$ including $a^m$ and $Z_2$ including $b^k$, i.e. 
$Z_N \rtimes Z_2$.
Eq.~(\ref{eq:Dn-n}) corresponds to the (reducible) $N$-dimensional 
representation.
The simple doublet representation is written as
 \begin{eqnarray}\label{eq:Dn-2}
a=\left(
\begin{array}{cc}
\cos 2\pi /N & -\sin 2 \pi /N \\
 \sin 2 \pi /N & \cos 2\pi /N 
\end{array}
\right), \qquad 
b = \left(
 \begin{array}{cc}
 1 & 0 \\
0 & -1
\end{array}
\right).
\end{eqnarray} 

\vskip .5cm
{$\bullet$ \bf Conjugacy classes}

Because of the algebraic relations (\ref{eq:Dn-AB}), 
it is found that $a^m$ and $a^{N-m}$ belong to 
the same conjugacy class and also $b$ and $a^{2m}b$
belong to the same conjugacy class.
When $N$ is even, $D_N$ has the following $3+N/2$ conjugacy 
classes,
\begin{eqnarray}
\begin{array}{ccc}
 C_1:&\{ e \}, &  h=1,\\
 C_2^{(1)}:&\{ a, a^{N-1} \}, &  h=N,\\
 \vdots  & \vdots &  \vdots , \\
 C_2^{(N/2-1)}:&\{ a^{N/2-1}, a^{N/2+1} \},&   h=N/gcd(N,N/2-1),\\
 C_1': &\{ a^{N/2} \}, & h=2,\\
 C_{N/2}: &\{ b, a^2b, \cdots, a^{N-2}b \}, &  h=2,\\
 C_{N/2}': &\{ ab, a^3b, \cdots, a^{N-1}b \},  & h=2,
\end{array}
\end{eqnarray}
where we have also shown the orders of each element in 
the conjugacy class by $h$.
That implies that there are $3+N/2$ irreducible representations.
Furthermore, 
the orthogonality relation (\ref{eq:character-2-e}) requires 
\begin{eqnarray}\label{eq:chracter-2-Dn-even}
&&\sum_\alpha[\chi_\alpha(C_1)]^2=\sum_nm_nn^2=m_1+4m_2+9m_3+\cdots=2N,
\end{eqnarray}
for $m_i$, which satisfies $m_1+m_2+m_3+\cdots= 3+N/2$.
The solution is found as $(m_1,m_2) = (4,N/2-1)$.
That is, there are four singlets and $(N/2-1)$ doublets.

On the other hand, when $N$ is odd, $D_N$ has the following $2+(N-1)/2$ conjugacy 
classes,
\begin{eqnarray}
\begin{array}{ccc}
 C_1:& \{ e \}, & h=1,  \\
 C_2^{(1)}:& \{ a, a^{N-1} \}, & h=N,\\
\vdots &  \vdots & \vdots,\\
 C_2^{(N-1)/2}:& \{ a^{(N-1)/2}, a^{(N+1)/2} \}, & h=N/gcd(N,(N-1)/2),\\
 C_{N}: & \{ b, ab, \cdots, a^{N-1}b \}, & h=2.
\end{array}
\end{eqnarray}
That is, there are $2+(N-1)/2$ irreducible representations.
Furthermore, 
the orthogonality relation (\ref{eq:character-2-e}) requires the same equation 
as (\ref{eq:chracter-2-Dn-even})
for $m_i$, which satisfies $m_1+m_2+m_3+\cdots= 2+(N-1)/2$.
The solution is found as $(m_1,m_2) = (2,(N-1)/2)$.
That is, there are two singlets and $(N-1)/2$ doublets.

\vskip .5cm
{$\bullet$ \bf Characters and representations}

First of all, we study on singlets.
When $N$ is even, there are four singlets.
Because of $b^2=e$ in $C_{N/2}$ and $(ab)^2=e$ in $C_{N/2}'$, 
the characters $\chi_\alpha(g)$ for four singlets 
should satisfy $\chi_\alpha(C_{N/2})=\pm 1$ and 
 $\chi_\alpha(C_{N/2}')=\pm 1$.
Thus, we have four possible combinations of $\chi_\alpha(C_{N/2})=\pm 1$ and 
 $\chi_\alpha(C_{N/2}')=\pm 1$ and they correspond to 
four singlets, ${\bf 1}_{\pm \pm}$, which are shown in 
Table \ref{tab:Dn-even-character}.

Similarly, we can study $D_N$ with $N=$ odd, which has two singlets.
Because of $b^2=e$ in $C_{N}$, 
the characters $\chi_\alpha(g)$ for two singlets 
should satisfy $\chi_\alpha(C_{N})=\pm 1$.
Since both $b$ and $ab$ belong to the same conjugacy class $C_{N}$, 
the characters $\chi_\alpha(a)$ for two singlets must 
always satisfy $\chi_\alpha(a)=1$.
Thus, there are two singlets, ${\bf 1}_+$ and  ${\bf 1}_-$.
Their characters are determined by whether the conjugacy class 
includes $b$ or not as shown in Table  \ref{tab:Dn-odd-character}.

Next, we study doublet representations, that is, 
$(2 \times 2)$ matrix representations.
Indeed, Eq.~(\ref{eq:Dn-2}) corresponds to one of doublet 
representations.
Similarly, $(2 \times 2)$ matrix representations for 
generic doublet ${\bf 2}_k$ are obtained by replacing 
\begin{eqnarray}
a \rightarrow a^k.
\end{eqnarray}
That is, $a$ and $b$ are 
represented for the doublet   ${\bf 2}_k$  as 
 \begin{eqnarray}\label{eq:Dn-2k}
a=\left(
\begin{array}{cc}
\cos 2\pi k/N & -\sin 2 \pi k/N \\
 \sin 2 \pi k/N & \cos 2\pi k/N 
\end{array}
\right), \qquad 
b = \left(
 \begin{array}{cc}
 1 & 0 \\
0 & -1
\end{array}
\right),
\end{eqnarray} 
where $k=1,\cdots,N/2-1$ for $N=$ even and 
$k=1,\cdots,(N-1)/2$ for $N=$ odd.
When we write the doublet ${\bf 2}_k$ as 
\begin{eqnarray}
{\bf 2}_k = \left(
\begin{array}{c}
x_k \\
y_{k}
\end{array}
\right),
\end{eqnarray}
the generator $a$ is the $Z_N$ rotation on 
the two-dimensional real coordinates $(x_k,y_k)$ and 
the generator $b$ is the reflection along $y_k$, i.e. 
$y_k \rightarrow -y_k$.
These transformations can be represented on 
the complex coordinate $z_k$ and its conjugate $\bar z_{-k}$.
These bases are transformed as 
\begin{eqnarray}\label{eq:r-c-trans}
\left(
\begin{array}{c}
z_k \\
\bar z_{-k}
\end{array}
\right) = 
U \left(
\begin{array}{c}
x_k \\
y_{k}
\end{array}
\right) , \qquad 
U=\left(
\begin{array}{cc}
1 & i \\
1 & -i
\end{array}
\right).
\end{eqnarray}
Then, in the complex basis, the generators, $a$ and $b$, 
can be obtained as $\tilde a = UaU^{-1}$ and $\tilde b = UbU^{-1}$,
 \begin{eqnarray}\label{eq:Dn-2k-2}
\tilde a=\left(
\begin{array}{cc}
\exp 2\pi ik/N & 0 \\
 0 & \exp -2\pi ik/N
\end{array}
\right), \qquad 
\tilde b = \left(
 \begin{array}{cc}
 0 & 1 \\
1 & 0
\end{array}
\right).
\end{eqnarray}
This complex basis may be useful.
For instance,  the generator $\tilde a$ is the diagonal matrix.
That implies that in the doublet ${\bf 2}_k$, which is denoted by 
\begin{eqnarray}
{\bf 2}_k = \left(
\begin{array}{c}
z_k \\
\bar z_{-k}
\end{array}
\right),
\end{eqnarray}
each of up and down components, $z_k$ and $\bar z_{-k}$, has 
the definite $Z_N$ charge.
That is, $Z_N$ charges of $z_k$ and $\bar z_{-k}$ are equal to 
$k$ and $-k$, respectively. 
The characters of these matrices for the doublets ${\bf 2}_k$ 
are obtained and those are shown in Tables \ref{tab:Dn-even-character}
and \ref{tab:Dn-odd-character}.
These characters satisfy the orthogonality relations (\ref{eq:character-1}) and 
 (\ref{eq:character-2}).

\begin{table}[t]
\begin{center}
\begin{tabular}{|c|c|c|c|c|c|c|}
\hline
   & $\!\!h\!\!$ & $\!\!\chi_{1_{++}}\!\!$ & $\!\!\chi_{1_{+-}}\!\!$ 
   & $\!\!\chi_{1_{-+}}\!\!$  & $\!\!\chi_{1_{--}}\!\!$ & $\!\!\chi_{2k}\!\!$  \\ \hline
$\!\!C_1\!\!$ & $\!\!1\!\!$ &  $\!\!1\!\!$ & $\!\!1\!\!$ & $\!\!1\!\!$ & $\!\!1\!\!$ & $\!\!2\!\!$ \\ \hline
$\!\!C_2^{1}\!\!$ & $\!\!N\!\!$ & $\!\!1\!\!$ & $\!\!-1\!\!$ & $\!\!-1\!\!$ & $\!\!1\!\!$ & $\!\!2\cos(2\pi k/N)\!\!$ \\ \hline
$\!\!\vdots \!\!$ & &   &      &      &   &    \\ \hline
$\!\!C_2^{N/2-1}\!\!$ & $\!\!N/gcd(N,N/2-1)\!\!$ & $\!\!1\!\!$ & $\!\!(-1)^{(N/2-1)}\!\!$ & $\!\!(-1)^{(N/2-1)}\!\!$ 
& $\!\!1\!\!$ & $\!\!2\cos(2\pi k(N/2-1)/N)\!\!$ \\ \hline
$\!\!C_1'\!\!$ & $\!\!2\!\!$ &  $\!\!1\!\!$ & $\!\!(-1)^{N/2}\!\!$  & $\!\!(-1)^{N/2}\!\!$  & $\!\!1\!\!$ & $\!\!-2\!\!$ \\ \hline
$\!\!C_{N/2}\!\!$ & $\!\!2\!\!$ & $\!\!1\!\!$ & $\!\!1\!\!$ & $\!\!-1\!\!$ & $\!\!-1\!\!$ & $\!\!0\!\!$ \\ \hline
$\!\!C_{N/2}'\!\!$ & $\!\!2\!\!$ & $\!\!1\!\!$ & $\!\!-1\!\!$ & $\!\!1\!\!$ & $\!\!-1\!\!$ & $\!\!0\!\!$ \\ \hline
\end{tabular}
\end{center}
\caption{Characters of $D_{N={\rm even}}$ representations}
\label{tab:Dn-even-character}
\end{table}

\begin{table}[t]
\begin{center}
\begin{tabular}{|c|c|c|c|c|}
\hline
   & $h$ & $\chi_{1_{+}}$ & $\chi_{1_{-}}$ &  $\chi_{2k}$  \\ \hline
$C_1$ & 1 &  1 & 1  & 2 \\ \hline
$C_2^{1}$ & $N$ & 1 & $1$ & $2\cos(2\pi k/N)$  \\ \hline
$\vdots $ & &   &    &    \\ \hline
$C_2^{(N-1)/2}$ & $N/gcd(N,(N-1)/2)$ & 1  & 1 & $2\cos (2\pi k(N-1)/2N)$  \\ \hline
$C_{N}$ & 2 & 1 & -1  & 0 \\ \hline
\end{tabular}
\end{center}
\caption{Characters of $D_{N={\rm odd}}$ representations}
\label{tab:Dn-odd-character}
\end{table}

\vskip .5cm
{$\bullet$ \bf Tensor products}

Now, we study the tensor products.
First, we consider the $D_N$ group with $N=$ even.
Let us start with ${\bf 2}_k \times {\bf 2}_{k'}$, i.e.
\begin{eqnarray}
\left(
\begin{array}{c}
z_k \\
\bar z_{-k}
\end{array}
\right)_{{\bf 2}_k} \otimes 
\left(
\begin{array}{c}
z_{k'} \\
\bar z_{-k'}
\end{array}
\right)_{{\bf 2}_{k'}},
\end{eqnarray}
where $k,k'=1,\cdots,N/2-1$.
Note that $z_kz_{k'}$,  $z_k\bar z_{-k'}$,  $\bar z_{-k}z_{k'}$ 
and $\bar z_{-k}\bar z_{-k'}$ have define $Z_N$ changes, i.e. 
$k+k'$, $k-k'$,  $-k+k'$ and  $-k-k'$, respectively.
For the case with $k+k'\neq N/2$ and $k-k'\neq 0$, 
they decompose two doublets as 
\begin{eqnarray}
\left(
\begin{array}{c}
z_k \\
\bar z_{-k}
\end{array}
\right)_{{\bf 2}_k} \otimes 
\left(
\begin{array}{c}
z_{k'} \\
\bar z_{-k'}
\end{array}
\right)_{{\bf 2}_{k'}} = 
\left(
\begin{array}{c}
z_kz_{k'} \\
\bar z_{-k}\bar z_{-k'}
\end{array}
\right)_{{\bf 2}_{k+k'}}
\oplus
\left(
\begin{array}{c}
z_k\bar z_{-k'} \\
\bar z_{-k}z_{k'}
\end{array}
\right)_{{\bf 2}_{k-k'}}.
\end{eqnarray}
When $k+k'=N/2$, the matrix $a$ is 
represented on the above (reducible) doublet $(z_kz_{k'},\bar z_{-k}\bar z_{-k'})$
as 
\begin{eqnarray}
a\left(
\begin{array}{c}
z_kz_{k'} \\
\bar z_{-k}\bar z_{-k'}
\end{array}
\right) = \left(
\begin{array}{cc}
  -1 & 0 \\
 0   & -1 
\end{array}
\right) \left(
\begin{array}{c}
z_kz_{k'} \\
\bar z_{-k}\bar z_{-k'}
\end{array}
\right).
\end{eqnarray}
Since $a$ is proportional to the $(2\times 2)$ identity matrix
for $(z_kz_{k'},\bar z_{-k}\bar z_{-k'})$ with $k+k'=N/2$, we can diagonalize 
another matrix $b$ in this vector space $(z_kz_{k'},\bar z_{-k}\bar z_{-k'})$.
Such a basis is obtained as 
$(z_kz_{k'}+ \bar z_{-k}\bar z_{-k'},z_kz_{k'}- \bar z_{-k}\bar z_{-k'})$ and 
their eigenvalues of $b$ are obtained as 
\begin{eqnarray}
b\left(
\begin{array}{c}
z_kz_{k'} + \bar z_{-k}\bar z_{-k'}\\
z_kz_{k'} - \bar z_{-k}\bar z_{-k'}
\end{array}
\right) = \left(
\begin{array}{cc}
  1 & 0 \\
 0   & -1 
\end{array}
\right) \left(
\begin{array}{c}
z_kz_{k'} + \bar z_{-k}\bar z_{-k'}\\
z_kz_{k'} - \bar z_{-k}\bar z_{-k'}
\end{array}
\right).
\end{eqnarray}
Thus, $z_kz_{k'}+ \bar z_{-k}\bar z_{-k'}$ and $z_kz_{k'}- \bar z_{-k}\bar z_{-k'}$
correspond to ${\bf 1}_{+-}$ and ${\bf 1}_{-+}$, respectively.

In the case of  $k-k'=0$, a similar decomposition happens 
for the (reducible) doublet $(z_k\bar z_{-k'},\bar z_{-k}z_{k'})$.
The matrix $a$ is the $(2\times 2)$ identity matrix 
on the vector space $(z_k\bar z_{-k'},\bar z_{-k}z_{k'})$ 
with $k-k'=0$.
Then, we take the basis 
$(z_k\bar z_{-k'}+ \bar z_{-k}z_{k'},z_k\bar z_{-k'}- \bar z_{-k}z_{k'})$, 
where $b$ is diagonalized.
That is, $z_k\bar z_{-k'}+ \bar z_{-k}z_{k'}$ and $z_k\bar z_{-k'}- \bar z_{-k}z_{k'}$ 
correspond to ${\bf 1}_{++}$ and ${\bf 1}_{--}$, respectively.

Next, we study the tensor products of 
the doublets ${\bf 2}_k$  and singlets, e.g. 
${\bf 1}_{--} \times {\bf 2}_k$.
Here we denote the vector space for the singlet  ${\bf 1}_{--}$ 
by $w$, where $aw=w$ and $bw=-w$.
Then, it is found that $(wz_k,-w\bar z_k)$ is nothing but 
the doublet ${\bf 2}_k$, that is, 
${\bf 1}_{--} \times {\bf 2}_k = {\bf 2}_k$.
Similar results are obtained for other singlets.
Furthermore, it is straightforward to study 
the tensor products among singlets.
Hence, the tensor products of $D_N$ irreducible representations 
with $N=$even are summarized as 
\begin{eqnarray}
\left(
\begin{array}{c}
z_k \\
\bar z_{-k}
\end{array}
\right)_{{\bf 2}_k} \otimes 
\left(
\begin{array}{c}
z_{k'} \\
\bar z_{-k'}
\end{array}
\right)_{{\bf 2}_{k'}} = 
\left(
\begin{array}{c}
z_kz_{k'} \\
\bar z_{-k}\bar z_{-k'}
\end{array}
\right)_{{\bf 2}_{k+k'}}
\oplus
\left(
\begin{array}{c}
z_k\bar z_{-k'} \\
\bar z_{-k}z_{k'}
\end{array}
\right)_{{\bf 2}_{k-k'}},
\end{eqnarray}
for $k+k'\neq N/2$ and $k-k'\neq 0$, 
\begin{eqnarray}
\!\left(
\begin{array}{c}
z_k \\
\bar z_{-k}
\end{array}
\right)_{{\bf 2}_k}\!
\otimes 
\!\left(
\begin{array}{c}
z_{k'} \\
\bar z_{-k'}
\end{array}
\right)_{{\bf 2}_{k'}}\! &=& 
\!\left(z_kz_{k'} + \bar z_{-k}\bar z_{-k'} \right)_{{\bf 1}_{+-}}\!
\oplus 
\!\left(z_kz_{k'} - \bar z_{-k}\bar z_{-k'} \right)_{{\bf 1}_{-+}}\!
\nonumber \\ & & \oplus
\!\left(
\begin{array}{c}
z_k\bar z_{-k'} \\
\bar z_{-k}z_{k'}
\end{array}
\right)_{{\bf 2}_{k-k'}}\!,
\end{eqnarray}
for $k+k'= N/2$ and $k-k'\neq 0$, 
\begin{eqnarray}
\!\left(
\begin{array}{c}
z_k \\
\bar z_{-k}
\end{array}
\right)_{{\bf 2}_k}\! 
\otimes 
\!\left(
\begin{array}{c}
z_{k'} \\
\bar z_{-k'}
\end{array}
\right)_{{\bf 2}_{k'}}\! &=& 
\!\left(z_k \bar z_{-k'} + \bar z_{-k}z_{k'} \right)_{{\bf 1}_{++}}\!
\oplus 
\!\left(z_k\bar z_{-k'} - \bar z_{-k}z_{k'} \right)_{{\bf 1}_{--}}\!
\nonumber \\ & & \oplus 
\!\left(
\begin{array}{c}
z_kz_{k'} \\
\bar z_{-k}\bar z_{-k'}
\end{array}
\right)_{{\bf 2}_{k+k'}}\!,
\end{eqnarray}
for $k+k'\neq N/2$ and $k-k'= 0$, 
\begin{eqnarray}
\left(
\begin{array}{c}
z_k \\
\bar z_{-k}
\end{array}
\right)_{{\bf 2}_k} \otimes 
\left(
\begin{array}{c}
z_{k'} \\
\bar z_{-k'}
\end{array}
\right)_{{\bf 2}_{k'}} & =&  
\left(z_k \bar z_{-k'} + \bar z_{-k}z_{k'} \right)_{{\bf 1}_{++}}
\oplus \left(z_k\bar z_{-k'} - \bar z_{-k}z_{k'} \right)_{{\bf
    1}_{--}} 
\nonumber \\
& & \oplus \left(z_kz_{k'} + \bar z_{-k}\bar z_{-k'} \right)_{{\bf 1}_{+-}}
\oplus \left(z_kz_{k'} - \bar z_{-k}\bar z_{-k'} \right)_{{\bf 1}_{-+}} ,
\end{eqnarray}
for $k+k'= N/2$ and $k-k'= 0$, 
\begin{eqnarray}
 & & \left( w\right)_{{\bf 1}_{++}} \otimes 
\left(
\begin{array}{c}
z_k \\
\bar z_{-k}
\end{array}
\right)_{{\bf 2}_k}= 
\left(
\begin{array}{c}
wz_k \\
w\bar z_{-k}
\end{array}
\right)_{{\bf 2}_k}, \quad 
\left( w\right)_{{\bf 1}_{--}} \otimes 
\left(
\begin{array}{c}
z_k \\
\bar z_{-k}
\end{array}
\right)_{{\bf 2}_k}= 
\left(
\begin{array}{c}
wz_k \\
-w \bar z_{-k}
\end{array}
\right)_{{\bf 2}_k},  \nonumber\\
 & & \left( w\right)_{{\bf 1}_{+-}} \otimes 
\left(
\begin{array}{c}
z_k \\
\bar z_{-k}
\end{array}
\right)_{{\bf 2}_k}= 
\left(
\begin{array}{c}
w\bar z_{-k} \\
wz_{k}
\end{array}
\right)_{{\bf 2}_k}, \quad 
\left( w\right)_{{\bf 1}_{-+}} \otimes 
\left(
\begin{array}{c}
z_{k} \\
\bar z_{-k} 
\end{array}
\right)_{{\bf 2}_k}= 
\left(
\begin{array}{c}
w\bar z_{-k} \\
- wz_{k}
\end{array}
\right)_{{\bf 2}_k},
\end{eqnarray}
\begin{eqnarray}
{\bf 1}_{s_1s_2} \otimes {\bf 1}_{s'_1s'_2}
= {\bf 1}_{s''_1s''_2},
\end{eqnarray}
where $s''_1=s_1s'_1$ and $s''_2=s_2s'_2$.

Similarly, we can analyze the tensor products of 
$D_N$ irreducible representations with $N=$ odd.
Its results are summarized as follows,
\begin{eqnarray}
\left(
\begin{array}{c}
z_k \\
\bar z_{-k}
\end{array}
\right)_{{\bf 2}_k} \otimes 
\left(
\begin{array}{c}
z_{k'} \\
\bar z_{-k'}
\end{array}
\right)_{{\bf 2}_{k'}} = 
\left(
\begin{array}{c}
z_kz_{k'} \\
\bar z_{-k}\bar z_{-k'}
\end{array}
\right)_{{\bf 2}_{k+k'}}
\oplus
\left(
\begin{array}{c}
z_k\bar z_{-k'} \\
\bar z_{-k}z_{k'}
\end{array}
\right)_{{\bf 2}_{k-k'}},
\end{eqnarray}
for $k-k'\neq 0$, where $k,k'=1,\cdots, N/2-1$, 
\begin{eqnarray}
\!\left(
\begin{array}{c}
z_k \\
\bar z_{-k}
\end{array}
\right)_{{\bf 2}_k}\! \otimes 
\!\left(
\begin{array}{c}
z_{k'} \\
\bar z_{-k'}
\end{array}
\right)_{{\bf 2}_{k'}}\! &=& 
\!\left(z_k \bar z_{-k'} + \bar z_{-k}z_{k'} \right)_{{\bf 1}_{+}}\!
\oplus 
\!\left(z_k\bar z_{-k'} - \bar z_{-k}z_{k'} \right)_{{\bf 1}_{-}}\!
\nonumber \\ & & \oplus 
\!\left(
\begin{array}{c}
z_kz_{k'} \\
\bar z_{-k}\bar z_{-k'}
\end{array}
\right)_{{\bf 2}_{k+k'}},\!
\end{eqnarray}
for $k-k'= 0$, 
\begin{equation}
\left( w\right)_{{\bf 1}_{++}} \otimes 
\left(
\begin{array}{c}
z_k \\
\bar z_{-k}
\end{array}
\right)_{{\bf 2}_k}= 
\left(
\begin{array}{c}
wz_k \\
w\bar z_{-k}
\end{array}
\right)_{{\bf 2}_k}, \qquad 
\left( w\right)_{{\bf 1}_{-}} \otimes 
\left(
\begin{array}{c}
z_k \\
\bar z_{-k}
\end{array}
\right)_{{\bf 2}_k}= 
\left(
\begin{array}{c}
wz_k \\
-w \bar z_{-k}
\end{array}
\right)_{{\bf 2}_k}, 
\end{equation}
\begin{eqnarray}
{\bf 1}_{s} \otimes {\bf 1}_{s'}
= {\bf 1}_{s''},
\end{eqnarray}
where $s''=ss'$.

Note that the above multiplication rules are the same 
between the complex basis and the real basis.
For example, we obtain that 
in both bases 
${\bf 2}_k \otimes {\bf 2}_{k'} = {\bf 2}_{k+k}' + {\bf 2}_{k-k'}$ 
for 
$k+k' \neq N/2$ and $k-k' \neq 0$.
However, elements of doublets are written in a different way, 
although those transform as (\ref{eq:r-c-trans}).

\subsection{$D_4$}

Here, we give simple examples of $D_N$.
The smallest non-Abelian group in $D_N$ is $D_3$.
However, $D_3$ corresponds to a group of all possible 
permutations of three objects, that is, $S_3$.
Thus, we show  $D_4$ and $D_5$ as simple examples.

\begin{figure}[t]
\unitlength=1mm
\begin{picture}(170,50)
\hspace{5cm}
\includegraphics[width=6cm]{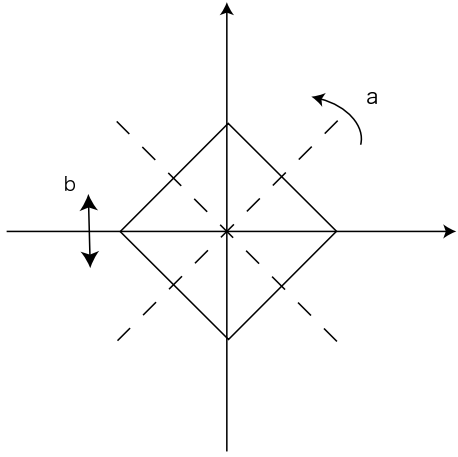}
\end{picture}
\vspace{-0.5cm}
\caption{The $D_4$ symmetry of a square}
\label{fig:D4}
\end{figure}

The $D_4$ is the symmetry of a square, 
which is generated by the $\pi/2$ rotation $a$ and 
the reflection $b$, where they satisfy 
$a^4=e$, $b^2=e$ and $bab=a^{-1}$.
(See Figure \ref{fig:D4}.)
Indeed, the $D_4$ consists of the eight elements, 
$a^m b^k$ with $m=0,1,2,3$ and $k=0,1$.
The $D_4$ has the following five conjugacy classes, 
\begin{eqnarray}
\begin{array}{ccc}
 C_1:&\{ e \}, &  h=1,\\
 C_2:&\{ a, a^{3} \}, &  h=4,\\
 C_1': &\{ a^{2} \}, & h=2,\\
 C_{2}': &\{ b, a^2b \}, &  h=2,\\
 C_{2}'': &\{ ab, a^3b \},  & h=2,
\end{array}
\end{eqnarray}
where we have also shown the orders of each element in 
the conjugacy class by $h$.

The $D_4$ has four singlets, ${\bf 1}_{++}$, 
${\bf 1}_{+-}$, ${\bf 1}_{-+}$ and ${\bf 1}_{--}$, 
and one doublet ${\bf 2}$.
The characters are shown in Table \ref{tab:D4-character}.
The tensor products are obtained as 
\begin{eqnarray}
\left(
\begin{array}{c}
z \\
\bar z
\end{array}
\right)_{{\bf 2}} \otimes 
\left(
\begin{array}{c}
z' \\
\bar z'
\end{array}
\right)_{{\bf 2}} & =&  
\left(z \bar z' + \bar z z' \right)_{{\bf 1}_{++}}
\oplus \left(z \bar z' - \bar z z' \right)_{{\bf 1}_{--}} \nonumber\\
& & \oplus \left(z z' + \bar z \bar z' \right)_{{\bf 1}_{+-}}
\oplus \left(z z' - \bar z \bar z' \right)_{{\bf 1}_{-+}} ,
\end{eqnarray}
\begin{eqnarray}
 & & \left( w\right)_{{\bf 1}_{++}} \otimes 
\left(
\begin{array}{c}
z \\
\bar z
\end{array}
\right)_{{\bf 2}}= 
\left(
\begin{array}{c}
wz \\
w\bar z
\end{array}
\right)_{{\bf 2}}, \qquad 
\left( w\right)_{{\bf 1}_{--}} \otimes 
\left(
\begin{array}{c}
z \\
\bar z
\end{array}
\right)_{{\bf 2}}= 
\left(
\begin{array}{c}
w z \\
-w \bar z
\end{array}
\right)_{{\bf 2}},  \nonumber\\
 & & \left( w\right)_{{\bf 1}_{+-}} \otimes 
\left(
\begin{array}{c}
z \\
\bar z
\end{array}
\right)_{{\bf 2}}= 
\left(
\begin{array}{c}
w\bar z \\
wz
\end{array}
\right)_{{\bf 2}}, \qquad 
\left( w\right)_{{\bf 1}_{-+}} \otimes 
\left(
\begin{array}{c}
z \\
\bar z
\end{array}
\right)_{{\bf 2}}= 
\left(
\begin{array}{c}
w\bar z \\
- wz
\end{array}
\right)_{{\bf 2}},  
\end{eqnarray}
\begin{eqnarray}
{\bf 1}_{s_1s_2} \otimes {\bf 1}_{s'_1s'_2}
= {\bf 1}_{s''_1s''_2},
\end{eqnarray}
where $s''_1=s_1s'_1$ and $s''_2=s_2s'_2$.

\begin{table}[t]
\begin{center}
\begin{tabular}{|c|c|c|c|c|c|c|}
\hline
   & $h$ & $\chi_{1_{++}}$ & $\chi_{1_{+-}}$ 
   & $\chi_{1_{-+}}$  & $\chi_{1_{--}}$ & $\chi_{2}$  \\ \hline
$C_1$ & 1 &  1 & 1 & 1 & 1 & 2 \\ \hline
$C_2$ & $4$ & 1 & $-1$ & $-1$ & $1$ & $0$ \\ \hline
$C_1'$       & 2 &  1 & $1$  & $1$  & 1 & $-2$ \\ \hline
$C_2'$ & 2 & 1 & 1 & $-1$ & $-1$ & 0 \\ \hline
$C_2''$ & 2 & 1 & -1 & $1$ & $-1$ & 0 \\ \hline
\end{tabular}
\end{center}
\caption{Characters of $D_{4}$ representations}
\label{tab:D4-character}
\end{table}

\subsection{$D_5$}

\begin{figure}[t]
\unitlength=1mm
\begin{picture}(170,50)
\hspace{5cm}
\includegraphics[width=5cm]{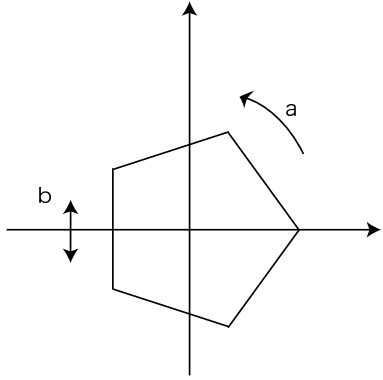}
\end{picture}
\vspace{-0.5cm}
\caption{The $D_5$ symmetry of a regular pentagon}
\label{fig:D5}
\end{figure}

The $D_5$ is the symmetry of a regular pentagon, 
which is generated by the $2\pi/5$ rotation $a$ and 
the reflection $b$.
See Figure \ref{fig:D5}.
They satisfy that $a^5=e$, $b^2=e$ and $bab=a^{-1}$.
The $D_5$ includes the 10 elements, $a^mb^k$ with $m=0,1,2,3,4$ 
and $k=0,1$.
They are classified into the following four conjugacy 
classes,
\begin{eqnarray}
\begin{array}{ccc}
 C_1:& \{ e \}, & h=1,  \\
 C_2^{(1)}:& \{ a, a^{4} \}, & h=5,\\
 C_2^{(2)}:& \{ a^{2}, a^{3} \}, & h=5,\\
 C_{5}: & \{ b, ab, a^2b, a^3b, a^4b\}, & h=2.
\end{array}
\end{eqnarray}

The $D_5$ has two singlets, ${\bf 1}_+$ and ${\bf 1}_-$, 
and two doublets, ${\bf 2}_1$ and ${\bf 2}_2$. 
Their characters are shown in Table \ref{tab:D5-character}.

\begin{table}[t]
\begin{center}
\begin{tabular}{|c|c|c|c|c|c|}
\hline
   & $h$ & $\chi_{1_{+}}$ & $\chi_{1_{-}}$ &  $\chi_{2_1}$ & $\chi_{2_2}$   \\ \hline
$C_1$ & 1 &  1 & 1  & 2 & 2 \\ \hline
$C_2^{1}$ & $5$ & 1 & $1$ & $2\cos(2\pi /5)$ & $2\cos(4\pi /5)$  \\ \hline
$C_2^{2}$ & $5$ & 1  & 1 & $2\cos (4\pi /5)$ & $2\cos (8\pi /5)$  \\ \hline
$C_{5}$ & 2 & 1 & -1  & 0 & 0 \\ \hline
\end{tabular}
\end{center}
\caption{Characters of $D_{5}$ representations}
\label{tab:D5-character}
\end{table}

The tensor products are obtained as 
\begin{equation}
\left(
\begin{array}{c}
z \\
\bar z
\end{array}
\right)_{{\bf 2}_2} \otimes 
\left(
\begin{array}{c}
z' \\
\bar z'
\end{array}
\right)_{{\bf 2}_{1}} = 
\left(
\begin{array}{c}
z z' \\
\bar z \bar z '
\end{array}
\right)_{{\bf 2}_{2}}
\oplus
\left(
\begin{array}{c}
z \bar z' \\
\bar z z'
\end{array}
\right)_{{\bf 2}_{1}},
\end{equation}
\begin{equation}
\left(
\begin{array}{c}
z_k \\
\bar z_{-k}
\end{array}
\right)_{{\bf 2}_k} \otimes 
\left(
\begin{array}{c}
z'_{k} \\
\bar z'_{-k}
\end{array}
\right)_{{\bf 2}_{k}} = 
\left(z_k \bar z'_{-k} + \bar z_{-k}z'_{k} \right)_{{\bf 1}_{+}}
\oplus \left(z_k\bar z'_{-k} - \bar z_{-k}z'_{k} \right)_{{\bf 1}_{-}}
\oplus \left(
\begin{array}{c}
z_kz'_{k} \\
\bar z_{-k}\bar z'_{-k}
\end{array}
\right)_{{\bf 2}_{2k}},
\end{equation}
\begin{equation}
 \left( w\right)_{{\bf 1}_{+}} \otimes 
\left(
\begin{array}{c}
z_k \\
\bar z_{-k}
\end{array}
\right)_{{\bf 2}_k}= 
\left(
\begin{array}{c}
wz_k \\
w\bar z_{-k}
\end{array}
\right)_{{\bf 2}_k}, \qquad 
\left( w\right)_{{\bf 1}_{-}} \otimes 
\left(
\begin{array}{c}
z_k \\
\bar z_{-k}
\end{array}
\right)_{{\bf 2}_k}= 
\left(
\begin{array}{c}
wz_k \\
-w \bar z_{-k}
\end{array}
\right)_{{\bf 2}_k}, 
\end{equation}
\begin{equation}
{\bf 1}_{s} \otimes {\bf 1}_{s'}
= {\bf 1}_{s''},
\end{equation}
where $s''=ss'$.

\clearpage


\section{$Q_N$}
\label{sec:QN}

\subsection{Generic aspects}

The binary dihedral group $Q_N$ with $N=$ even consists of 
the elements, $a^mb^k$ 
with $m=0,\cdots, N-1$ and $k=0,1$, 
where the generators $a$ and $b$ satisfy
 \begin{eqnarray}\label{eq:Qn-AB}
a^N=e, \qquad b^2=(a^{N/2}), \qquad b^{-1}ab=a^{-1} .
\end{eqnarray}
The order of $Q_N$ is equal to $2N$.
The generator $a$ can be represented by the same 
$( 2 \times 2)$ matrices as $D_N$, i.e. 
 \begin{eqnarray}
a=\left(
\begin{array}{cc}
\exp 2\pi ik/N & 0 \\
 0 & \exp -2\pi ik/N
\end{array}
\right).
\end{eqnarray}
Note that $a^{N/2}=e$ for $k=$ even and 
$a^{N/2}=-e$ for $k=$ odd.
That leads to that $b^2=e$ for $k=$ even 
and $b^2 =-e$ for $k=$ odd.
Thus, the generators  $a$ and $b$ are represented 
by $(2 \times 2)$ matrices, e.g. as  
 \begin{eqnarray}\label{eq:Qn-2k-2}
a=\left(
\begin{array}{cc}
\exp 2\pi ik/N & 0 \\
 0 & \exp -2\pi ik/N
\end{array}
\right), \qquad 
b = \left(
 \begin{array}{cc}
 0 & i \\
 i & 0
\end{array}
\right),
\end{eqnarray}
for $k=$ odd, 
 \begin{eqnarray}\label{eq:Qn-2k-2-2}
a=\left(
\begin{array}{cc}
\exp 2\pi ik/N & 0 \\
 0 & \exp -2\pi ik/N
\end{array}
\right), \qquad 
b = \left(
 \begin{array}{cc}
 0 & 1 \\
 1 & 0
\end{array}
\right),
\end{eqnarray}
 for $k=$ even.

\vskip .5cm
{$\bullet$ \bf Conjugacy classes}

By use of the algebraic relations (\ref{eq:Qn-AB}), 
the elements are classified into the $(3+N/2)$ conjugacy classes as
\begin{eqnarray}
\begin{array}{ccc}
 C_1:&\{ e \}, &  h=1,\\
 C_2^{(1)}:&\{ a, a^{N-1} \}, &  h=N,\\
\vdots & \vdots & \vdots , \\
 C_2^{(N/2-1)}:&\{ a^{N/2-1}, a^{N/2+1} \},&   h=N/gcd(N,N/2-1),\\
 C_1': &\{ a^{N/2} \}, & h=2,\\
 C_{N/2}: &\{ b, a^2b, \cdots, a^{N-2}B \}, &  h=4,\\
 C_{N/2}': &\{ ab, a^3b, \cdots, a^{N-1}B \},  & h=4,
\end{array}
\end{eqnarray}
where we have also shown the orders of each element in 
the conjugacy class by $h$.
These are almost the same as the conjugacy classes of 
$D_N$ with $N=$ even.
There must be the $(3+N/2)$  irreducible representations, 
and similarly to $D_{N={\rm~even}}$ there are four singlets 
and $(N/2-1)$ doublets.

\vskip .5cm
{$\bullet$ \bf Characters and representations}

The characters of $Q_N$ for doublets are 
the same as those of $D_{N={\rm~even}}$,
and are shown in Tables \ref{tab:Qn-4n-character} and 
\ref{tab:Qn-4n+2-character}. 
The characters of $Q_N$ for singlets depend on 
the value of $N$.
First, we consider the case with $N=4n$, where 
we have the relation,
\begin{eqnarray}\label{eq:Qn-4n-AB}
b^2=a^{2n}.
\end{eqnarray}
Because of $b^4=e$ in $C_{N/2}$, the characters 
$\chi_\alpha(b)$ for four singlets must satisfy 
$\chi_\alpha(b)= e^{\pi i n/2}$ with $n=0,1,2,3$.
In addition, the element $ba^2$ belongs to the same 
conjugacy class as $b$.
That implies $\chi_\alpha(a^2)= 1$ for four singlets.
Then, by using  Eq.~(\ref{eq:Qn-4n-AB}), we find 
$\chi_\alpha(b^2)= 1$, that is, $\chi_\alpha(b)= \pm 1$.
Thus, the characters of $Q_N$ with 
$N=4n$ for singlets are the same as those of 
$D_{N={\rm~even}}$ and are shown in Table \ref{tab:Qn-4n-character}.

Next, we consider four singlets of $Q_N$ for $N=4n+2$ 
and in this case we have the relation,
\begin{eqnarray}\label{eq:Qn-4n+2-AB}
b^2=a^{2n+1}.
\end{eqnarray}
Since $b$ and $a^2b$ are included in the same 
conjugacy class, the characters 
$\chi_\alpha(a^2)$ for four singlets must satisfy $\chi_\alpha(a^2)=1$, 
that is, there are two possibilities, $\chi_\alpha(a)= \pm 1$.
When $\chi_\alpha(a)= 1$, the relation \eqref{eq:Qn-4n+2-AB} leads to the two 
possibilities $\chi_\alpha(b)= \pm 1$.
On the other hand, when $\chi_\alpha(a)= -1$, 
the relation \eqref{eq:Qn-4n+2-AB} leads to the two 
possibilities $\chi_\alpha(b)= \pm i$.
Then, totally there are four possibilities corresponding 
to the four singlets.
Note that $\chi_\alpha(a) = \chi_\alpha(b^2)$ for all of 
singlets.

\begin{table}[t]
\begin{center}
\begin{tabular}{|c|c|c|c|c|c|c|}
\hline
   & $\!\!h\!\!$ & $\!\!\chi_{1_{++}}\!\!$ & $\!\!\chi_{1_{+-}}\!\!$ 
   & $\!\!\chi_{1_{-+}}\!\!$  & $\!\!\chi_{1_{--}}\!\!$ & $\!\!\chi_{2k}\!\!$  \\ \hline
$\!\!C_1\!\!$ & $\!\!1\!\!$ & $\!\!1\!\!$ & $\!\!1\!\!$ & $\!\!1\!\!$ & $\!\!1\!\!$ & $\!\!2\!\!$ \\ \hline
$\!\!C_2^{1}\!\!$ & $\!\!N\!\!$ & $\!\!1\!\!$ & $\!\!-1\!\!$ & $\!\!-1\!\!$ & $\!\!1\!\!$ & $\!\!2\cos(2\pi k/N)\!\!$ \\ \hline
$\!\!\vdots\!\! $ & &   &      &      &   &    \\ \hline
$\!\!C_2^{N/2-1}\!\!$ & $\!\!N/gcd(N,N/2-1)\!\!$ & $\!\!1\!\!$ & $\!\!(-1)^{(N/2-1)}\!\!$ & $\!\!(-1)^{(N/2-1)}\!\!$ & $\!\!1\!\!$ & $\!\!2\cos(2\pi k(N/2-1)/N)\!\!$ \\ \hline
$\!\!C_1'\!\!$ & $\!\!2\!\!$ & $\!\!1\!\!$ & $\!\!(-1)^{N/2}\!\!$ & $\!\!(-1)^{N/2}\!\!$  & $\!\!1\!\!$ & $\!\!-2\!\!$ \\ \hline
$\!\!C_{N/2}\!\!$ & $\!\!4\!\!$ & $\!\!1\!\!$ & $\!\!1\!\!$ & $\!\!-1\!\!$ & $\!\!-1\!\!$ & $\!\!0\!\!$ \\ \hline
$\!\!C_{N/2}'\!\!$ & $\!\!4\!\!$ & $\!\!1\!\!$ & $\!\!-1\!\!$ & $\!\!1\!\!$ & $\!\!-1\!\!$ & $\!\!0\!\!$ \\ \hline
\end{tabular}
\end{center}
\caption{Characters of $Q_{N}$ representations for $N=4n$}
\label{tab:Qn-4n-character}
\end{table}

\begin{table}[t]
\begin{center}
\begin{tabular}{|c|c|c|c|c|c|c|}
\hline
   & $\!\!h\!\!$ & $\!\!\chi_{1_{++}}\!\!$ & $\!\!\chi_{1_{+-}}\!\!$ 
   & $\!\!\chi_{1_{-+}}\!\!$  & $\!\!\chi_{1_{--}}\!\!$ & $\!\!\chi_{2k}\!\!$  \\ \hline
$\!\!C_1\!\!$ & $\!\!1\!\!$ &  $\!\!1\!\!$ & $\!\!1\!\!$ & $\!\!1\!\!$ & $\!\!1\!\!$ & $\!\!2\!\!$ \\ \hline
$\!\!C_2^{1}\!\!$ & $\!\!N\!\!$ & $\!\!1\!\!$ & $\!\!-1\!\!$ & $\!\!-1\!\!$ & $\!\!1\!\!$ & $\!\!2\cos(2\pi k/N)\!\!$ \\ \hline
$\!\!\vdots \!\!$ & &   &      &      &   &    \\ \hline
$\!\!C_2^{N/2-1}\!\!$ & $\!\!N/gcd(N,N/2-1)\!\!$ & $\!\!1\!\!$ & $\!\!(-1)^{(N/2-1)}\!\!$ & $\!\!(-1)^{(N/2-1)}\!\!$ 
& $\!\!1\!\!$ & $\!\!2\cos(2\pi k(N/2-1)/N)\!\!$ \\ \hline
$\!\!C_1'\!\!$ & $\!\!2\!\!$ & $\!\!1\!\!$ & $\!\!(-1)^{N/2}\!\!$  & $\!\!(-1)^{N/2}\!\!$  & $\!\!1\!\!$ & $\!\!-2\!\!$ \\ \hline
$\!\!C_{N/2}\!\!$ & $\!\!4\!\!$ & $\!\!1\!\!$ & $\!\!i\!\!$ & $\!\!-i\!\!$ & $\!\!-1\!\!$ & $\!\!0\!\!$ \\ \hline
$\!\!C_{N/2}'\!\!$ & $\!\!4\!\!$ & $\!\!1\!\!$ & $\!\!-i\!\!$ & $\!\!i\!\!$ & $\!\!-1\!\!$ & $\!\!0\!\!$ \\ \hline
\end{tabular}
\end{center}
\caption{Characters of $Q_{N}$ representations for $N=4n+2$}
\label{tab:Qn-4n+2-character}
\end{table}

\vskip .5cm
{$\bullet$ \bf Tensor products}

Furthermore, similarly to $D_N$ with $N=$ even, 
the tensor products of $Q_N$ irreducible representations 
can be analyzed.
The results for $Q_N$ with $N=4n$ 
are obtained as 
\begin{eqnarray}
\left(
\begin{array}{c}
z_k \\
\bar z_{-k}
\end{array}
\right)_{{\bf 2}_k} \otimes 
\left(
\begin{array}{c}
z_{k'} \\
\bar z_{-k'}
\end{array}
\right)_{{\bf 2}_{k'}} = 
\left(
\begin{array}{c}
z_kz_{k'} \\
 (-1)^{kk'} \bar z_{-k}\bar z_{-k'}
\end{array}
\right)_{{\bf 2}_{k+k'}}
\oplus
\left(
\begin{array}{c}
z_k\bar z_{-k'} \\
 (-1)^{kk'}\bar z_{-k}z_{k'}
\end{array}
\right)_{{\bf 2}_{k-k'}},
\end{eqnarray}
for $k+k'\neq N/2$ and $k-k'\neq 0$, 
\begin{eqnarray}
\left(
\begin{array}{c}
z_k \\
\bar z_{-k}
\end{array}
\right)_{{\bf 2}_k} \otimes 
\left(
\begin{array}{c}
z_{k'} \\
\bar z_{-k'}
\end{array}
\right)_{{\bf 2}_{k'}} &=& 
\left(z_kz_{k'} + (-1)^{kk'} \bar z_{-k}\bar z_{-k'} \right)_{{\bf 1}_{+-}}
\oplus \left(z_kz_{k'} - (-1)^{kk'} \bar z_{-k}\bar z_{-k'}
\right)_{{\bf 1}_{-+}} \nonumber \\
& & \oplus
\left(
\begin{array}{c}
z_k\bar z_{-k'} \\
 (-1)^{kk'}\bar z_{-k}z_{k'}
\end{array}
\right)_{{\bf 2}_{k-k'}},
\end{eqnarray}
for $k+k'= N/2$ and $k-k'\neq 0$, 
\begin{eqnarray}
\left(
\begin{array}{c}
z_k \\
\bar z_{-k}
\end{array}
\right)_{{\bf 2}_k} \otimes 
\left(
\begin{array}{c}
z_{k'} \\
\bar z_{-k'}
\end{array}
\right)_{{\bf 2}_{k'}} &=& 
\left(z_k \bar z_{-k'} +  (-1)^{kk'} \bar z_{-k}z_{k'} \right)_{{\bf 1}_{++}}
\oplus \left(z_k\bar z_{-k'} - (-1)^{kk'} \bar z_{-k}z_{k'} \right)_{{\bf 1}_{--}}
 \nonumber \\ & & \oplus \left(
\begin{array}{c}
z_kz_{k'} \\
 (-1)^{kk'} \bar z_{-k}\bar z_{-k'}
\end{array}
\right)_{{\bf 2}_{k+k'}},
\end{eqnarray}
for $k+k'\neq N/2$ and $k-k'= 0$, 
\begin{eqnarray}
\left(
\begin{array}{c}
z_k \\
\bar z_{-k}
\end{array}
\right)_{{\bf 2}_k} \otimes 
& \left(
\begin{array}{c}
 z_{k'} \\
\bar z_{-k'}
\end{array}
\right)_{{\bf 2}_{k'}}  = 
\left(z_k \bar z_{-k'} + (-1)^{kk'} \bar z_{-k}z_{k'} \right)_{{\bf 1}_{++}}
\oplus  \left(z_k\bar z_{-k'} -   (-1)^{kk'}\bar z_{-k}z_{k'}
\right)_{{\bf 1}_{--}} \nonumber \\
& \oplus \left(z_kz_{k'} +  (-1)^{kk'}\bar z_{-k}\bar z_{-k'} \right)_{{\bf 1}_{+-}}
\oplus \left(z_kz_{k'} -  (-1)^{kk'} \bar z_{-k}\bar z_{-k'} \right)_{{\bf 1}_{-+}} ,
\end{eqnarray}
for $k+k'= N/2$ and $k-k'= 0$, 
\begin{eqnarray}
 & & \left( w\right)_{{\bf 1}_{++}} \otimes 
\left(
\begin{array}{c}
z_k \\
\bar z_{-k}
\end{array}
\right)_{{\bf 2}_k}= 
\left(
\begin{array}{c}
wz_k \\
w\bar z_{-k}
\end{array}
\right)_{{\bf 2}_k}, \quad 
\left( w\right)_{{\bf 1}_{--}} \otimes 
\left(
\begin{array}{c}
z_k \\
\bar z_{-k}
\end{array}
\right)_{{\bf 2}_k}= 
\left(
\begin{array}{c}
wz_k \\
-w \bar z_{-k}
\end{array}
\right)_{{\bf 2}_k},   \nonumber\\
 & & \left( w\right)_{{\bf 1}_{+-}} \otimes 
\left(
\begin{array}{c}
z_k \\
\bar z_{-k}
\end{array}
\right)_{{\bf 2}_k}= 
\left(
\begin{array}{c}
w\bar z_{-k} \\
wz_{k}
\end{array}
\right)_{{\bf 2}_k}, \quad 
\left( w\right)_{{\bf 1}_{-+}} \otimes 
\left(
\begin{array}{c}
z_{k} \\
\bar z_{-k}
\end{array}
\right)_{{\bf 2}_k}= 
\left(
\begin{array}{c}
w\bar z_{-k} \\
- wz_{k}
\end{array}
\right)_{{\bf 2}_k},
\end{eqnarray}
\begin{eqnarray}
{\bf 1}_{s_1s_2} \otimes {\bf 1}_{s'_1s'_2}
= {\bf 1}_{s''_1s''_2},
\end{eqnarray}
where $s''_1=s_1s'_1$ and $s''_2=s_2s'_2$.
Note that some minus signs are different from 
the tensor products of $D_N$.

Similarly, the tensor products of $Q_N$ with 
$N=4n+2$ are obtained as 
\begin{eqnarray}
\left(
\begin{array}{c}
z_k \\
\bar z_{-k}
\end{array}
\right)_{{\bf 2}_k} \otimes 
\left(
\begin{array}{c}
z_{k'} \\
\bar z_{-k'}
\end{array}
\right)_{{\bf 2}_{k'}} = 
\left(
\begin{array}{c}
z_kz_{k'} \\
  (-1)^{kk'} \bar z_{-k}\bar z_{-k'}
\end{array}
\right)_{{\bf 2}_{k+k'}}
\oplus
\left(
\begin{array}{c}
z_k\bar z_{-k'} \\
 (-1)^{kk'}\bar z_{-k}z_{k'}
\end{array}
\right)_{{\bf 2}_{k-k'}},
\end{eqnarray}
for $k+k'\neq N/2$ and $k-k'\neq 0$, 
\begin{eqnarray}
\left(
\begin{array}{c}
z_k \\
\bar z_{-k}
\end{array}
\right)_{{\bf 2}_k} \otimes 
\left(
\begin{array}{c}
z_{k'} \\
\bar z_{-k'}
\end{array}
\right)_{{\bf 2}_{k'}} &=& 
\left(z_kz_{k'} + \bar z_{-k}\bar z_{-k'} \right)_{{\bf 1}_{+-}}
\oplus \left(z_kz_{k'} - \bar z_{-k}\bar z_{-k'} \right)_{{\bf 1}_{-+}}
\nonumber \\ 
& & \oplus 
\left(
\begin{array}{c}
z_k\bar z_{-k'} \\
(-1)^{kk'}\bar z_{-k}z_{k'}
\end{array}
\right)_{{\bf 2}_{k-k'}},
\end{eqnarray}
for $k+k'= N/2$ and $k-k'\neq 0$, 
\begin{eqnarray}
\left(
\begin{array}{c}
z_k \\
\bar z_{-k}
\end{array}
\right)_{{\bf 2}_k} \otimes 
\left(
\begin{array}{c}
z_{k'} \\
\bar z_{-k'}
\end{array}
\right)_{{\bf 2}_{k'}} &=& 
\left(z_k \bar z_{-k'} +(-1)^{kk'} \bar z_{-k}z_{k'} \right)_{{\bf 1}_{++}}
\oplus \left(z_k\bar z_{-k'} -(-1)^{kk'} \bar z_{-k}z_{k'} \right)_{{\bf 1}_{--}}
\nonumber \\ & & \oplus \left(
\begin{array}{c}
z_kz_{k'} \\
(-1)^{kk'} \bar z_{-k}\bar z_{-k'}
\end{array}
\right)_{{\bf 2}_{k+k'}},
\end{eqnarray}
for $k+k'\neq N/2$ and $k-k'= 0$, 
\begin{eqnarray}
\left(
\begin{array}{c}
z_k \\
\bar z_{-k}
\end{array}
\right)_{{\bf 2}_k} \otimes 
\left(
\begin{array}{c}
z_{k'} \\
\bar z_{-k'}
\end{array}
\right)_{{\bf 2}_{k'}} & =&  
\left(z_k \bar z_{-k'} +(-1)^{kk'} \bar z_{-k}z_{k'} \right)_{{\bf 1}_{++}}
\oplus \left(z_k\bar z_{-k'} -(-1)^{kk'} \bar z_{-k}z_{k'}
\right)_{{\bf 1}_{--}} \nonumber \\
& & \oplus\left(z_kz_{k'} + \bar z_{-k}\bar z_{-k'} \right)_{{\bf 1}_{+-}}
\oplus \left(z_kz_{k'} - \bar z_{-k}\bar z_{-k'} \right)_{{\bf 1}_{-+}} ,
\end{eqnarray}
for $k+k'= N/2$ and $k-k'= 0$, 
\begin{eqnarray}
 & & \left( w\right)_{{\bf 1}_{++}} \otimes 
\left(
\begin{array}{c}
z_k \\
\bar z_{-k}
\end{array}
\right)_{{\bf 2}_k}= 
\left(
\begin{array}{c}
wz_k \\
w\bar z_{-k}
\end{array}
\right)_{{\bf 2}_k}, \quad 
\left( w\right)_{{\bf 1}_{--}} \otimes 
\left(
\begin{array}{c}
z_k \\
\bar z_{-k}
\end{array}
\right)_{{\bf 2}_k}= 
\left(
\begin{array}{c}
wz_k \\
-w \bar z_{-k}
\end{array}
\right)_{{\bf 2}_k},  \nonumber\\
 & & \left( w\right)_{{\bf 1}_{+-}} \otimes 
\left(
\begin{array}{c}
z_k \\
\bar z_{-k}
\end{array}
\right)_{{\bf 2}_k}= 
\left(
\begin{array}{c}
w\bar z_{-k} \\
wz_{k}
\end{array}
\right)_{{\bf 2}_k}, \quad 
\left( w\right)_{{\bf 1}_{-+}} \otimes 
\left(
\begin{array}{c}
z_{k} \\
\bar z_{-k}
\end{array}
\right)_{{\bf 2}_k}= 
\left(
\begin{array}{c}
w\bar z_{-k} \\
- wz_{k}
\end{array}
\right)_{{\bf 2}_k},
\end{eqnarray}
\begin{eqnarray}
{\bf 1}_{s_1s_2} \otimes {\bf 1}_{s'_1s'_2}
= {\bf 1}_{s''_1s''_2},
\end{eqnarray}
where $s''_1=s_1s'_1$ and $s''_2=s_2s'_2$.

\subsection{$Q_4$}

Here we give simple examples.
In this subsection, we show the results on $Q_4$ and 
in the next subsection we show $Q_6$.

The $Q_4$ has the eight elements, $a^mb^k$, for $m=0,1,2,3$ and  $k=0,1$, 
where $a$ and $b$ satisfy $a^4=e$, $b^2=a^2$ and $b^{-1}ab=a^{-1}$.
These elements are classified into 
the five conjugacy classes,
\begin{eqnarray}
\begin{array}{ccc}
 C_1:&\{ e \}, &  h=1,\\
 C_2:&\{ a, a^{3} \}, &  h=4,\\
 C_1': &\{ a^{2} \}, & h=2,\\
 C_{2}': &\{ b, a^2b  \}, &  h=4,\\
 C_{2}'': &\{ ab, a^3b \},  & h=4,
\end{array}
\end{eqnarray}
where we have also shown the orders of each element in 
the conjugacy class by $h$.

The $Q_4$ has four singlets, ${\bf 1}_{++}$, 
${\bf 1}_{+-}$, ${\bf 1}_{-+}$ and ${\bf 1}_{--}$, 
and one doublet ${\bf 2}$.
The characters are shown in Table \ref{tab:Q4-character}.
The tensor products are obtained as 
\begin{eqnarray}
\left(
\begin{array}{c}
z \\
\bar z
\end{array}
\right)_{{\bf 2}} \otimes 
\left(
\begin{array}{c}
z' \\
\bar z'
\end{array}
\right)_{{\bf 2}} & = & 
\left(z \bar z' - \bar z z' \right)_{{\bf 1}_{++}}
\oplus \left(z \bar z' + \bar z z' \right)_{{\bf 1}_{--}} \nonumber\\
& & \oplus \left(z z' - \bar z \bar z' \right)_{{\bf 1}_{+-}}
\oplus \left(z z' + \bar z \bar z' \right)_{{\bf 1}_{-+}} ,
\end{eqnarray}
\begin{eqnarray}
 & & \left( w\right)_{{\bf 1}_{++}} \otimes 
\left(
\begin{array}{c}
z \\
\bar z
\end{array}
\right)_{{\bf 2}}= 
\left(
\begin{array}{c}
wz \\
w\bar z
\end{array}
\right)_{{\bf 2}}, \quad 
\left( w\right)_{{\bf 1}_{--}} \otimes 
\left(
\begin{array}{c}
z \\
\bar z
\end{array}
\right)_{{\bf 2}}= 
\left(
\begin{array}{c}
wz \\
-w \bar z
\end{array}
\right)_{{\bf 2}},  \nonumber\\
 & & \left( w\right)_{{\bf 1}_{+-}} \otimes 
\left(
\begin{array}{c}
z \\
\bar z
\end{array}
\right)_{{\bf 2}}= 
\left(
\begin{array}{c}
w\bar z \\
wz
\end{array}
\right)_{{\bf 2}}, \quad 
\left( w\right)_{{\bf 1}_{-+}} \otimes 
\left(
\begin{array}{c}
z \\
\bar z
\end{array}
\right)_{{\bf 2}}= 
\left(
\begin{array}{c}
w\bar z \\
- wz
\end{array}
\right)_{{\bf 2}},
\end{eqnarray}
\begin{eqnarray}
{\bf 1}_{s_1s_2} \otimes {\bf 1}_{s'_1s'_2}
= {\bf 1}_{s''_1s''_2},
\end{eqnarray}
where $s''_1=s_1s'_1$ and $s''_2=s_2s'_2$.
Note that some minus signs are different from 
the tensor products of $D_4$.

\begin{table}[t]
\begin{center}
\begin{tabular}{|c|c|c|c|c|c|c|}
\hline
   & $h$ & $\chi_{1_{++}}$ & $\chi_{1_{+-}}$ 
   & $\chi_{1_{-+}}$  & $\chi_{1_{--}}$ & $\chi_{2}$  \\ \hline
$C_1$ & 1 &  1 & 1 & 1 & 1 & 2 \\ \hline
$C_2$ & $4$ & 1 & $-1$ & $-1$ & $1$ & $2\cos(\pi /2)$ \\ \hline
$C_1'$       & 2 &  1 & $1$  & $1$  & 1 & $-2$ \\ \hline
$C_{2}'$ & 4 & 1 & 1 & $-1$ & $-1$ & 0  \\ \hline
$C_{2}''$ & 4 & 1 & -1 & $1$ & $-1$ & 0 \\ \hline
\end{tabular}
\end{center}
\caption{Characters of $Q_{4}$ representations }
\label{tab:Q4-character}
\end{table}

\subsection{$Q_6$}

The $Q_6$ has the 12 elements, $a^mb^k$, for $m=0,1,2,3,4,5$ and  $k=0,1$, 
where $a$ and $b$ satisfy $a^6=e$, $b^2=a^3$ and $b^{-1}ab=a^{-1}$.
These elements are classified into 
the six conjugacy classes,
\begin{eqnarray}
\begin{array}{ccc}
 C_1:&\{ e \}, &  h=1,\\
 C_2^{(1)}:&\{ a, a^{5} \}, &  h=6,\\
 C_2^{(2)}:&\{ a^{2}, a^{4} \},&   h=3,\\
 C_1': &\{ a^{3} \}, & h=2,\\
 C_{3}: &\{ b, a^2b, a^{4}b \}, &  h=4,\\
 C_{3}': &\{ ab, a^3b, a^{5}b \},  & h=4.
\end{array}
\end{eqnarray}

The $Q_6$ has four singlets, ${\bf 1}_{++}$, 
${\bf 1}_{+-}$, ${\bf 1}_{-+}$, and ${\bf 1}_{--}$, 
and two doublets, ${\bf 2}_1$ and ${\bf 2}_2$.
The characters are shown in Table \ref{tab:Q6-character}.
The tensor products are obtained as 
\begin{eqnarray}
\left(
\begin{array}{c}
z \\
\bar z
\end{array}
\right)_{{\bf 2}_2} \otimes 
\left(
\begin{array}{c}
z' \\
\bar z'
\end{array}
\right)_{{\bf 2}_{1}} = 
\left(z z' - \bar z \bar z' \right)_{{\bf 1}_{+-}}
\oplus \left(z z' + \bar z \bar z' \right)_{{\bf 1}_{-+}}
\oplus 
\left(
\begin{array}{c}
z \bar z' \\
\bar z z'
\end{array}
\right)_{{\bf 2}_{1}},
\end{eqnarray}
\begin{eqnarray}
\left(
\begin{array}{c}
z \\
\bar z
\end{array}
\right)_{{\bf 2}_k} \otimes 
\left(
\begin{array}{c}
z' \\
\bar z'
\end{array}
\right)_{{\bf 2}_{k}} = 
\left(z \bar z' - \bar z z' \right)_{{\bf 1}_{++}}
\oplus \left(z \bar z' + \bar z z' \right)_{{\bf 1}_{--}}
\oplus \left(
\begin{array}{c}
z z' \\
- \bar z \bar z'
\end{array}
\right)_{{\bf 2}_{k'}},
\end{eqnarray}
for $k,k'=1,2$ and $k'\neq k$, 
\begin{eqnarray}
\hspace{-3mm} & & \left( w\right)_{{\bf 1}_{++}} \otimes 
\left(
\begin{array}{c}
z_k \\
\bar z_{-k}
\end{array}
\right)_{{\bf 2}_k}= 
\left(
\begin{array}{c}
wz_k \\
w\bar z_{-k}
\end{array}
\right)_{{\bf 2}_k}, \quad 
\left( w\right)_{{\bf 1}_{--}} \otimes 
\left(
\begin{array}{c}
z_k \\
\bar z_{-k}
\end{array}
\right)_{{\bf 2}_k}= 
\left(
\begin{array}{c}
wz_k \\
-w \bar z_{-k}
\end{array}
\right)_{{\bf 2}_k},  \nonumber\\
\hspace{-3mm} & & \left( w\right)_{{\bf 1}_{+-}} \otimes 
\left(
\begin{array}{c}
z_k \\
\bar z_{-k}
\end{array}
\right)_{{\bf 2}_k}= 
\left(
\begin{array}{c}
w\bar z_{-k} \\
wz_{k}
\end{array}
\right)_{{\bf 2}_k}, \quad 
\left( w\right)_{{\bf 1}_{-+}} \otimes 
\left(
\begin{array}{c}
z_{k} \\
\bar z_{-k}
\end{array}
\right)_{{\bf 2}_k}= 
\left(
\begin{array}{c}
w\bar z_{-k} \\
- wz_{k}
\end{array}
\right)_{{\bf 2}_k},
\end{eqnarray}
\begin{eqnarray}
{\bf 1}_{s_1s_2} \otimes {\bf 1}_{s'_1s'_2}
= {\bf 1}_{s''_1s''_2},
\end{eqnarray}
where $s''_1=s_1s'_1$ and $s''_2=s_2s'_2$.

\begin{table}[t]
\begin{center}
\begin{tabular}{|c|c|c|c|c|c|c|c|}
\hline
   & $h$ & $\chi_{1_{++}}$ & $\chi_{1_{+-}}$ 
   & $\chi_{1_{-+}}$  & $\chi_{1_{--}}$ & $\chi_{2_1}$ & $\chi_{2_2}$  \\ \hline
$C_1$ & 1 &  1 & 1 & 1 & 1 & 2 & 2 \\ \hline
$C_2^{1}$ & $6$ & 1 & $-1$ & $-1$ & $1$ & $2\cos(2\pi /6)$ &$2\cos(4\pi /6)$ \\ \hline
$C_2^{2}$ & $3$ & 1 & $1$ & $1$ & 1 
& $2\cos(4\pi /6)$ & $2\cos(8\pi /6)$\\ \hline
$C_1'$  & $2$ &  $1$ & $1$  & $1$  & $1$ & $-2$ & $2$  \\ \hline
$C_{3}$ & $4$ & $1$ & $i$ & $-i$ & $-1$ & $0$ & $0$ \\ \hline
$C_{3}'$ & $4$ & $1$ & $-i$ & $i$ & $-1$ & $0$ & $0$  \\ \hline
\end{tabular}
\end{center}
\caption{Characters of $Q_{6}$ representations}
\label{tab:Q6-character}
\end{table}

\clearpage


\section{$\Sigma (2N^2)$}
\label{sec:sigma-2n}

\subsection{Generic aspects}

The discrete group $\Sigma (2N^2)$ is isomorphic 
to $(Z_N \times Z_N')\rtimes Z_2$.
We denote the generators of $Z_N$ and $Z_N'$ by $a$ and $a'$, 
respectively, and the $Z_2$ generator is written by $b$.
They satisfy
\begin{eqnarray}
& & a^N = {a'}^N = b^2 = e, \nonumber \\
& &  aa' = a'a, \qquad bab=a'.\label{daisu-kankei}
\end{eqnarray}
Using them, all of $\Sigma (2N^2)$ elements are written as 
\begin{eqnarray}
 & & g=b^{k}a^{m}a'^{n} ,
\end{eqnarray}
for $k=0,1$ and $m,n=0,1,...,N-1$.

These generators, $a$, $a'$ and $b$, are represented, e.g. as 
\begin{eqnarray}\label{eq:sigma-2-1}
a=\mat2{1}{0} {0}{\rho},
\qquad a'=\mat2{\rho}{0} {0}{1}, \qquad b=\mat2{0}{1}{1}{0},
\end{eqnarray}
where $\rho=e^{2\pi i/N}$.
Then, all of $\Sigma (2N^2)$ elements are written as 
\begin{eqnarray}
\mat2{\rho^m}{0}{0}{\rho^n}, \qquad \mat2{0}{\rho^m}{\rho^n}{0}.
\end{eqnarray}

\vskip .5cm
{$\bullet$ \bf Conjugacy classes}

Now, let us study the conjugacy classes of $\Sigma (2N^2)$.
We obtain the algebraic relations,
\begin{eqnarray}
& & b(a^la'^m)b^{-1}= a^ma'^l, \qquad b(ba^la'^m)b^{-1}=ba^ma'^l,
\nonumber \\
& & a^k(ba^la'^m)a^{-k}=ba^{l-k}a'^{m+k}, \qquad 
a'^k(ba^la'^m)a'^{-k}=ba^{l+k}a'^{m-k}.
\end{eqnarray}
Then, it is found that the $\Sigma (2N^2)$  group has 
the following conjugacy classes,
\begin{eqnarray}
\begin{array}{ccc}
 C_1:&\{ e \}, &  h=1,\\
 C_1^{(1)}:&\{ aa'\}, &  h=N,\\
\vdots & \vdots & \vdots \\
 C_1^{(k)}:&\{ a^ka'^k\}, &  h=N/gcd(N,k),\\
 \vdots  & \vdots &\vdots \\
 C_1^{(N-1)}:&\{ a^{N-1}a'^{N-1} \},&   h=N/gcd(N,N-1),\\
 C_{N}'^{(k)}:&\{
 ba^{k},ba^{k-1}a',\cdots,ba'^k,\cdots,ba^{k+1}a'^{N-1}\},
&   h=2N/gcd(N,k),\\
 C_2^{(l,m)}:&\{ a^{l}a'^{m}~,~a'^{l}a^{m} \},&   h=N/gcd(N,l,m),\\
\end{array}
\end{eqnarray}
where $l > m$ for $l,m = 0, \cdots,N-1$.
The number of conjugacy classes $C_2^{(l,m)}$ is equal to 
$N(N-1)/2$.
The total number of conjugacy classes of $\Sigma (2N^2)$  
is equal to $N(N-1)/2+N+N=(N^2+3N)/2$.

\vskip .5cm
{$\bullet$ \bf Characters and representations}

The orthogonality relations (\ref{eq:character-2-e}) and (\ref{eq:sum-dim})
for $\Sigma (2N^2)$ 
lead to
\begin{eqnarray}
&&m_1+2^2m_2+ \cdots =2N^2,\\
&&m_1+m_2+ \cdots =(N^2+3N)/2 .
\end{eqnarray}
The solution is found as $(m_1,m_2)=(2N,N(N-1)/2)$.
That is, there are $2N$ singlets and $N(N-1)/2$ doublets.

First of all, we study on singlets.
Since $a$ and $a'$ belong to the same conjugacy class $C_2^{(1,0)}$, 
the characters $\chi_\alpha (g)$ for singlets should satisfy 
$\chi_\alpha (a) = \chi_\alpha (a') $.
Furthermore, because of $b^2=e$ and $a^N=e$, 
possible values of $\chi_\alpha (g)$ for singlets are 
found as $\chi_\alpha (a) = \rho^n$ and $\chi_\alpha (b) = \pm 1$.
Then, totally we have $2N$ combinations, which correspond to 
$2N$ singlets, ${\bf 1}_{\pm n}$ for $n=0,1,\cdots, N-1$.
These characters are summarized in Table \ref{tab:Sigma-character}.

Next, we study doublet representations.
Indeed, the matrices (\ref{eq:sigma-2-1}) correspond to a 
doublet representation.
Similarly, $(2 \times 2)$ matrix representations for 
generic doublets ${\bf 2}_{p,q}$ are written by replacing 
\begin{eqnarray}
a \to a^pa'^q ~~~{\rm and }~~~ a' \to a^qa'^p.
\end{eqnarray}
That is, for doublets ${\bf 2}_{p,q}$, the generators 
$a$ and $a'$ as well as $b$ are represented as 
\begin{eqnarray}\label{eq:sigma-2-gene}
a=\mat2{\rho^q}{0} {0}{\rho^p},
\qquad a'=\mat2{\rho^p}{0} {0}{\rho^q}, \qquad b=\mat2{0}{1}{1}{0}.
\end{eqnarray}
We denote the doublet ${\bf 2}_{p,q}$ as
\begin{eqnarray}
{\bf 2}_{p,q} = \left(
\begin{array}{c}
x_q \\
x_p
\end{array}
\right),
\end{eqnarray}
where we take $p>q$ and $q=0,1,\cdots,N-2$.
Then, each of up and down components, $x_q$ and $x_p$, 
has definite $Z_N \times Z_N'$ charges.
That is, $x_q$ and $x_p$ have $(q,0)$ and $(0,p)$ $Z_N \times Z_N'$ 
charges, respectively.
The characters for doublets are also 
summarized in Table \ref{tab:Sigma-character}.

\begin{table}[t]
\begin{center}
\begin{tabular}{|c|c|c|c|c|}
\hline
   & $h$ & $\chi_{1_{+n}}$ & $\chi_{1_{-n}}$ 
    & $\chi_{2_{p,q}}$  \\ \hline
$C_1$ & 1 &  1 & 1 & 2 \\ \hline
$C_1^{(1)}$ & $N$ & $\rho^{2n}$  &$\rho^{2n}$  & $2\rho^{p+q}$ \\ \hline
$\vdots $ & &       &   &    \\ \hline
$C_1^{(N-1)}$ & $N/gcd(N,N-1)$ & $\rho^{2n(N-1)}$
& $\rho^{2n(N-1)}$ & $2\rho^{(N-1)(p+q)}$ \\ \hline
$ C_{N}'^{(k)}$       & $2N/gcd(N,k)$ & $\rho^{kn}$   & $-\rho^{kn}$ & $0$ \\ \hline
$C_{2}^{(l,m)}$ & $N/gcd(N,l,m)$ & $\rho^{(l+m)n}$ & $\rho^{(l+m)n}$ &
 $\rho^{lq+mp}+\rho^{lp+mq}$ \\ \hline
\end{tabular}
\end{center}
\caption{ Characters of $\Sigma(2N^2)$ representations }
\label{tab:Sigma-character}
\end{table}

\vskip .5cm
{$\bullet$ \bf Tensor products}

Now, let us consider tensor products of doublets ${\bf 2}_{p,q}$.
Because of their $Z_N \times Z_N'$ charges, their 
tensor products can be obtained as 
\begin{eqnarray}
{\2tvec {x_q}  {x_p}}_{{\bf 2}_{q,p}} \otimes 
{\2tvec {y_{q'}}  {y_{p'}}}_{{\bf 2}_{q',p'}} =
{\2tvec {x_q y_{q'}}  {x_p y_{p'}}}_{{\bf 2}_{q+q',p+p'}}
 \oplus
{ \2tvec {x_py_{q'} }  {x_qy_{p'}}}_{{\bf 2}_{q'+p,q+p'}} ,
\end{eqnarray}
for 
$q+q'\neq p+p'~mod(N)$ and $q+p'\neq p+q'~mod(N)$, 
\begin{eqnarray}
{\2tvec {x_q}  {x_p}}_{{\bf 2}_{q,p}} \otimes 
{\2tvec {y_{q'}}  {y_{p'}}}_{{\bf 2}_{q',p'}} &=&
\left({x_q y_{q'}+ x_p y_{p'}}\right)_{{\bf 1}_{+,q+q'}}
\oplus 
\left({x_q y_{q'}- x_p y_{p'}}\right)_{{\bf 1}_{-,q+q'}}   \nonumber \\
& &  \oplus
{ \2tvec {x_py_{q'} }  {x_qy_{p'}}}_{{\bf 2}_{q'+p,q+p'}} ,
\end{eqnarray}
for $q+q'= p+p' ~mod(N)$ and $q+p'\neq p+q'~mod(N)$
\begin{eqnarray}
{\2tvec {x_q}  {x_p}}_{{\bf 2}_{q,p}} \otimes 
{\2tvec {y_{q'}}  {y_{p'}}}_{{\bf 2}_{q',p'}} &=&
\left({x_py_{q'} + x_qy_{p'}}\right)_{{\bf 1}_{+,q+p'}}
\oplus \left({x_py_{q'} - x_qy_{p'}}\right)_{{\bf 1}_{-,q+p'}} \nonumber \\
&\oplus&
{\2tvec {x_q y_{q'}}  {x_p y_{p'}}}_{{\bf 2}_{q+q',p+p'}} ,
\end{eqnarray}
for $q+q'\neq p+p' ~mod(N)$ and $q+p'= p+q'~mod(N)$
\begin{eqnarray}
{\2tvec {x_q}  {x_p}}_{{\bf 2}_{q,p}} \otimes 
{\2tvec {y_{q'}}  {y_{p'}}}_{{\bf 2}_{q',p'}} &=&
\left({x_q y_{q'}+ x_p y_{p'}}\right)_{{\bf 1}_{+,q+q'}}
\oplus 
\left({x_q y_{q'}- x_p y_{p'}}\right)_{{\bf 1}_{-,q+q'}}   \nonumber \\
&\oplus&\left({x_py_{q'} + x_qy_{p'}}\right)_{{\bf 1}_{+,q+p'}}
\oplus \left({x_py_{q'} - x_qy_{p'}}\right)_{{\bf 1}_{-,q+p'}} ,
\end{eqnarray}
for $q+q'= p+p' ~mod(N)$ and $q+p'= p+q'~mod(N)$.
In addition, the tensor products between singlets and 
doublets are obtained as 
\begin{eqnarray}
(y)_{{\bf 1}_{s,n}}\otimes{\2tvec {x_q}  {x_p}}_{{\bf 2}_{q,p}} =
{\2tvec {yx_q }  {yx_p}}_{{\bf 2}_{q+n,p+n}}.
\end{eqnarray}
These tensor products are independent of $s = \pm$.
The tensor products of singlets are simply obtained as 
\begin{eqnarray}
{\bf 1}_{sn} \otimes {\bf 1}_{s'n'} = {\bf 1}_{ss', n+n'} .
\end{eqnarray}

\subsection{$\Sigma(18)$}

The $\Sigma(2)$ group is nothing but the 
Abelian $Z_2$ group.
Furthermore, the $\Sigma(8)$ group is 
isomorphic to $D_4$.
Thus, the simple and non-trivial example is 
$\Sigma(18)$.

The $\Sigma(18)$ has eighteen elements, 
$b^ka^ma'^n$ for $k=0,1$ and $m,n=0,1,2$, 
where $a$, $a'$ and $b$ satisfy 
$b^2=e$, $a^3=a'^3=e$, $aa'=a'a$ and 
$bab=a'$.
These elements are classified into nine 
conjugacy classes, 
\begin{eqnarray}
\begin{array}{ccc}
 C_1:&\{ e \}, &  h=1,\\
 C_1^{(1)}:&\{ aa'\}, &  h=3,\\
 C_1^{(2)}:&\{ a^2a'^2\}, &  h=3,\\
 C_3'^{(0)}:&\{b~,~ba'^2a~,~ba'a^2\},&   h=2,\\
C_3'^{(1)}:&\{ba'~,~ba~,~ba'^2a^2 \},&   h=6, \\
 C_3'^{(2)}:&\{ba'^2~,~ba'a~,~ba^2 \},&   h=6, \\
 C_2^{(1,0)}:&\{ a~,~a' \},&   h=3,\\
  C_2^{(2,0)}:&\{ a^{2}~,~a'^{2} \},&   h=3,\\
   C_2^{(2,1)}:&\{ a^{2}a'~,~aa'^{2} \},&   h=3,\\
\end{array}
\end{eqnarray}
where we have also shown the orders of each element 
in the conjugacy class by $h$.

The $\Sigma(18)$ has six singlets ${\bf 1}_{\pm, n}$ with 
$n=0,1,2$ and three doublets 
${\bf 2}_{p,q}$ with $(p,q)=(1,0), (2,0), (2,1)$.
The characters are shown in Table \ref{18tab:sigma-singlet-character}.

\begin{table}[t]
\begin{center}
\begin{tabular}{|c|c|c|c|c|c|c|c|c|c|c|}
\hline
   & $h$ & $\chi_{1_{+0}}$ & $\chi_{1_{+1}}$  & $\chi_{1_{+2}}$  
   & $\chi_{1_{-0}}$ & $\chi_{1_{-1}}$  & $\chi_{1_{-2}}$ 
   & $\chi_{2_{1,0}}$ & $\chi_{2_{2,0}}$  & $\chi_{2_{2,1}}$    \\ \hline
$C_1$ & 1 &  1 &1 & 1 &1 & 1 & $1$ &  2 & $2$ & $2$   \\ \hline
$C_1^{(1)}$ & $3$ & $1$ & $\rho^2$ & $\rho$ & $1$ & $\rho^2$
& $\rho$ &  $2\rho$ & $2\rho^2$ & $2$  \\ \hline
$C_1^{(2)}$ & $3$ & $1$ & $\rho$ & $\rho^2$ & $1$ & $\rho$
& $\rho^2$ &  $2\rho^2$ & $2\rho$ & $2$  \\ \hline
$C_3'^{(0)}$ & 2 &  1 &1 & 1 & $-1$ & $-1$ & $-1$ &  0 & $0$ & $0$   \\ \hline
$C_3'^{(1)}$ & $6$ & $1$ & $\rho$ & $\rho^2$ & $-1$ & $-\rho$
& $-\rho^2$ &  $0$ & $0$ & $0$  \\ \hline
$C_3'^{(2)}$ & $6$ & $1$ & $\rho^2$ & $\rho$ & $-1$ & $-\rho^2$
& $-\rho$ &  $0$ & $0$ & $0$  \\ \hline
$C_2^{(1,0)}$ & 3 &  $1$ &$\rho$ & $\rho^2$ &$1$ & $\rho$ & $\rho^2$ &  $-\rho^2$ & $-\rho$ & $-1$   \\ \hline
$C_1^{(2,0)}$ & $3$ & $1$ & $\rho^2$ & $\rho$ & $1$ & $\rho^2$
& $\rho$ &  $-\rho$ & $-\rho^2$ & $-1$  \\ \hline
$C_1^{(3,0)}$ & $3$ & $1$ & $1$ & $1$ & $1$ & $1$
& $1$ &  $-1$ & $-1$ & $-1$  \\ \hline
\end{tabular}
\end{center}
\caption{ Characters of $\Sigma(18)$ representations}
\label{18tab:sigma-singlet-character}
\end{table}

The tensor products between doublets are obtained as 
\begin{eqnarray}
&&{\2tvec {x_2}  {x_1}}_{{\bf 2}_{2,1}} \otimes{\2tvec {y_2}  {y_1}}_{{\bf 2}_{2,1}}  =
{(x_1y_2+x_2y_1)}_{{\bf 1}_{+,0}}\oplus {(x_1y_2-x_2y_1)}_{{\bf 1}_{-,0}}
 \oplus
{ \2tvec {x_1y_1 }  {x_2y_2}}_{{\bf 2}_{2,1}}, \nonumber \\
&&{\2tvec {x_2}  {x_0}}_{{\bf 2}_{2,0}} \otimes{\2tvec {y_2}  {y_0}}_{{\bf 2}_{2,0}}  =
{(x_0y_2+x_2y_0)}_{{\bf 1}_{+,2}}\oplus {(x_0y_2-x_2y_0)}_{{\bf 1}_{-,2}}
 \oplus
{ \2tvec {x_2y_2 }  {x_0y_0}}_{{\bf 2}_{1,0}}, \nonumber \\
&&{\2tvec {x_1}  {x_0}}_{{\bf 2}_{1,0}} \otimes{\2tvec {y_1}  {y_0}}_{{\bf 2}_{1,0}}  =
{(x_0y_1+x_1y_0)}_{{\bf 1}_{+,1}}\oplus {(x_0y_1-x_1y_0)}_{{\bf 1}_{-,1}}
 \oplus
{ \2tvec {x_1y_1 }  {x_0y_0}}_{{\bf 2}_{2,0}}, \nonumber \\
&&{\2tvec {x_2}  {x_1}}_{{\bf 2}_{2,1}} \otimes{\2tvec {y_2}  {y_0}}_{{\bf 2}_{2,0}}  =
{(x_2y_2+x_1y_0)}_{{\bf 1}_{+,1}}\oplus {(x_2y_2-x_1y_0)}_{{\bf 1}_{-,1}}
 \oplus
{ \2tvec {x_2y_0 }  {x_1y_2}}_{{\bf 2}_{2,0}}, \nonumber \\
&&{\2tvec {x_2}  {x_1}}_{{\bf 2}_{2,1}} \otimes{\2tvec {y_1}  {y_0}}_{{\bf 2}_{1,0}}  =
{(x_1y_1+x_2y_0)}_{{\bf 1}_{+,2}}\oplus {(x_1y_1-x_2y_0)}_{{\bf 1}_{-,2}}
 \oplus
{ \2tvec {x_1y_0 }  {x_2y_1}}_{{\bf 2}_{1,0}}, \nonumber \\
&&{\2tvec {x_2}  {x_0}}_{{\bf 2}_{2,0}} \otimes{\2tvec {y_1}  {y_0}}_{{\bf 2}_{1,0}}  =
{(x_2y_1+x_0y_0)}_{{\bf 1}_{+,0}}\oplus {(x_2y_1-x_0y_0)}_{{\bf 1}_{-,0}}
 \oplus
{ \2tvec {x_2y_0 }  {x_0y_1}}_{{\bf 2}_{2,1}} . \nonumber \\
\end{eqnarray}
The tensor products between singlets are obtained as
\begin{eqnarray}
&&
{\bf 1}_{\pm,0}\otimes {\bf 1}_{\pm,0}={\bf 1}_{+,0}~,~{\bf 1}_{\pm,1}\otimes {\bf 1}_{\pm,1}={\bf 1}_{+,2}~,~
{\bf 1}_{\pm,2}\otimes {\bf 1}_{\pm,2}={\bf 1}_{+,1}~,~{\bf 1}_{\pm,1}\otimes {\bf 1}_{\pm,0}={\bf 1}_{+,1}~,\nonumber\\
&&
{\bf 1}_{\pm,2}\otimes {\bf 1}_{\pm,0}={\bf 1}_{+,2}~,~{\bf 1}_{\pm,2}\otimes {\bf 1}_{\pm,1}={\bf 1}_{+,0}~,~
{\bf 1}_{\pm,0}\otimes {\bf 1}_{\mp,0}={\bf 1}_{-,0}~,~{\bf 1}_{\pm,1}\otimes {\bf 1}_{\mp,1}={\bf 1}_{-,2}~,~\nonumber\\
&&
{\bf 1}_{\pm,2}\otimes {\bf 1}_{\mp,2}={\bf 1}_{-,1}~,~{\bf 1}_{\pm,1}\otimes {\bf 1}_{\mp,0}={\bf 1}_{-,1}~,~
{\bf 1}_{\pm,2}\otimes {\bf 1}_{\pm,0}={\bf 1}_{-,2}~,~{\bf 1}_{\pm,2}\otimes {\bf 1}_{\mp,1}={\bf 1}_{-,0} ~.\nonumber\\
\end{eqnarray}
The tensor products between singlets and doublets are obtained as 
\begin{eqnarray}
&&(y)_{{\bf 1}_{\pm,0}}\otimes{\2tvec {x_2}  {x_1}}_{{\bf 2}_{2,1}} =
{\2tvec {yx_2 }  {yx_1}}_{{\bf 2}_{2,1}}~,~
(y)_{{\bf 1}_{\pm,1}}\otimes{\2tvec {x_2}  {x_1}}_{{\bf 2}_{2,1}} =
{\2tvec {yx_1 }  {yx_2}}_{{\bf 2}_{2,0}},\nonumber\\
&&(y)_{{\bf 1}_{\pm,2}}\otimes{\2tvec {x_2}  {x_1}}_{{\bf 2}_{2,1}} =
{\2tvec {yx_2 }  {yx_1}}_{{\bf 2}_{1,0}}~,~
(y)_{{\bf 1}_{\pm,0}}\otimes{\2tvec {x_2}  {x_0}}_{{\bf 2}_{2,0}} =
{\2tvec {yx_2 }  {yx_0}}_{{\bf 2}_{2,0}},\nonumber\\
&&(y)_{{\bf 1}_{\pm,1}}\otimes{\2tvec {x_2}  {x_0}}_{{\bf 2}_{2,0}} =
{\2tvec {yx_0 }  {yx_2}}_{{\bf 2}_{1,0}}~,~
(y)_{{\bf 1}_{\pm,2}}\otimes{\2tvec {x_2}  {x_0}}_{{\bf 2}_{2,0}} =
{\2tvec {yx_0 }  {yx_2}}_{{\bf 2}_{2,1}},\nonumber\\
&&(y)_{{\bf 1}_{\pm,0}}\otimes{\2tvec {x_1}  {x_0}}_{{\bf 2}_{1,0}} =
{\2tvec {yx_1 }  {yx_0}}_{{\bf 2}_{1,0}}~,~
(y)_{{\bf 1}_{\pm,1}}\otimes{\2tvec {x_1}  {x_0}}_{{\bf 2}_{1,0}} =
{\2tvec {yx_1 }  {yx_0}}_{{\bf 2}_{2,1}},\nonumber\\
&&(y)_{{\bf 1}_{\pm,2}}\otimes{\2tvec {x_1}  {x_0}}_{{\bf 2}_{1,0}} =
{\2tvec {yx_0 }  {yx_1}}_{{\bf 2}_{2,0}}.
\end{eqnarray}

\subsection{$\Sigma(32)$}

The $\Sigma(32)$ has 
thirty-two elements, 
$b^ka^ma'^n$ for $k=0,1$ and $m,n=0,1,2,3$, 
where $a$, $a'$ and $b$ satisfy 
$b^2=e$, $a^4=a'^4=e$, $aa'=a'a$ and 
$bab=a'$.
These elements are classified into fourteen 
conjugacy classes, 
\begin{eqnarray}
\begin{array}{ccc}
 C_1:&\{ e \}, &  h=1,\\
 C_1^{(1)}:&\{ aa'\}, &  h=4,\\
 C_1^{(2)}:&\{ a^2a'^2\}, &  h=2,\\
 C_1^{(3)}:&\{ a^3a'^3\}, &  h=4,\\
 C_4'^{(0)}:&\{b~,~ba'a^3~,~ba'^2a^2~,~ba'^3a\},&   h=2,\\
C_4'^{(1)}:&\{ba'~,~ba~,~ba'^2a^3~,~ba'^3a^2 \},&   h=8, \\
 C_4'^{(2)}:&\{ba'^2~,~ba'a~,~ba^2~,~ba'^3a^3 \},&   h=4,\\
 C_4'^{(3)}:&\{ba'^3~,~ba'^2a~,~ba'a^2~,~ba^3 \},&   h=8,\\
 C_2^{(1,0)}:&\{ a~,~a' \},&   h=4,\\
  C_2^{(2,0)}:&\{ a^{2}~,~a'^{2} \},&   h=2,\\
   C_2^{(2,1)}:&\{ a^{2}a'~,~aa'^2 \},&   h=4,\\
 C_2^{(3,0)}:&\{ a^3~,~a'^3 \},&   h=4,\\
  C_2^{(3,1)}:&\{ a^{3}a'~,~aa'^3\},&   h=4,\\
   C_2^{(3,2)}:&\{ a^{3}a'^2~,~a^{2}a'^3 \},&   h=4,\\
\end{array}
\end{eqnarray}
where we have also shown the orders of each element 
in the conjugacy class by $h$.

The $\Sigma(32)$ has eight singlets ${\bf 1}_{\pm, n}$ with 
$n=0,1,2,3$ and six doublets 
${\bf 2}_{p,q}$ with $(p,q)=(1,0), (2,0), (3,0), (2,1), 
(3,1), (3,2)$.
The characters are shown in Table \ref{32tab:sigma-singlet-character}.

\begin{table}[t]
\begin{center}
\begin{tabular}{|c|c|c|c|c|c|c|c|c|c|c|c|c|c|c|c|}
\hline
   & $\!\!h\!\!$ & $\!\!\chi_{1_{+0}}\!\!$ & $\!\!\chi_{1_{+1}}\!\!$  & $\!\!\chi_{1_{+2}}\!\!$   & $\!\!\chi_{1_{+3}}\!\!$  
   & $\!\!\chi_{1_{-0}}\!\!$ & $\!\!\chi_{1_{-1}}\!\!$  & $\!\!\chi_{1_{-2}}\!\!$   & $\!\!\chi_{1_{-3}}\!\!$ 
   & $\!\!\chi_{2_{1,0}}\!\!$ & $\!\!\chi_{2_{2,0}}\!\!$  & $\!\!\chi_{2_{2,1}}\!\!$ & $\!\!\chi_{2_{3,0}}\!\!$ & $\!\!\chi_{2_{3,1}}\!\!$  & $\!\!\chi_{2_{3,2}}\!\!$    \\ \hline
$\!\!C_1\!\!$ & $\!\!1\!\!$ &  $\!\!1\!\!$ &$\!\!1\!\!$ & $\!\!1\!\!$ &$\!\!1\!\!$ & $\!\!1\!\!$ & $\!\!1\!\!$ & $\!\!1\!\!$ & $\!\!1\!\!$ &  $\!\!2\!\!$ & $\!\!2\!\!$ & $\!\!2\!\!$  & $\!\!2\!\!$ & $\!\!2\!\!$ & $\!\!2\!\!$    \\ \hline
$\!\!C_1^{(1)}\!\!$ & $\!\!4\!\!$ & $\!\!1\!\!$ & $\!\!-1\!\!$ & $\!\!1\!\!$ & $\!\!-1\!\!$ & $\!\!1\!\!$  & $\!\!-1\!\!$ & $\!\!1\!\!$
& $\!\!-1\!\!$ &  $\!\!2i\!\!$ & $\!\!-2\!\!$ & $\!\!-2i\!\!$  &  $\!\!-2i\!\!$ & $\!\!2\!\!$ & $\!\!2i\!\!$  \\ \hline
$\!\!C_1^{(2)}\!\!$ & $\!\!2\!\!$ & $\!\!1\!\!$ & $\!\!1\!\!$ & $\!\!1\!\!$ & $\!\!1\!\!$ & $\!\!1\!\!$  & $\!\!1\!\!$ & $\!\!1\!\!$
& $\!\!1\!\!$ &  $\!\!-2\!\!$ & $\!\!2\!\!$ & $\!\!-2\!\!$  &  $\!\!-2\!\!$ & $\!\!2\!\!$ & $\!\!-2\!\!$  \\ \hline
$\!\!C_1^{(3)}\!\!$ & $\!\!4\!\!$ &  $\!\!1\!\!$ & $\!\!-1\!\!$ & $\!\!1\!\!$ &  $\!\!-1\!\!$ & $\!\!1\!\!$ & $\!\!-1\!\!$ & $\!\!1\!\!$ & $\!\!-1\!\!$ & $\!\!-2i\!\!$ & $\!\!-2\!\!$ & $\!\!2i\!\!$ &  $\!\!2i\!\!$ & $\!\!2\!\!$ & $\!\!-2i\!\!$   \\ \hline
$\!\!C_4'^{(0)}\!\!$ & $\!\!2\!\!$ &  $\!\!1\!\!$ & $\!\!1\!\!$ &  $\!\!1\!\!$ & $\!\!1\!\!$ &  $\!\!-1\!\!$ & $\!\!-1\!\!$ & $\!\!-1\!\!$ & $\!\!-1\!\!$ &  $\!\!0\!\!$ & $\!\!0\!\!$ & $\!\!0\!\!$  & $\!\!0\!\!$ & $\!\!0\!\!$ & $\!\!0\!\!$    \\ \hline
$\!\!C_4'^{(1)}\!\!$ & $\!\!8\!\!$ & $\!\!1\!\!$ & $\!\!i\!\!$ & $\!\!-1\!\!$ & $\!\!-i\!\!$ & $\!\!-1\!\!$  & $\!\!-i\!\!$ & $\!\!1\!\!$
& $\!\!i\!\!$ &  $\!\!0\!\!$ & $\!\!0\!\!$ & $\!\!0\!\!$  &  $\!\!0\!\!$ & $\!\!0\!\!$ & $\!\!0\!\!$  \\ \hline
$\!\!C_4'^{(2)}\!\!$ & $\!\!4\!\!$ & $\!\!1\!\!$ & $\!\!-1\!\!$ & $\!\!1\!\!$ & $\!\!-1\!\!$ & $\!\!-1\!\!$  & $\!\!1\!\!$ & $\!\!-1\!\!$
& $\!\!1\!\!$ &  $\!\!0\!\!$ & $\!\!0\!\!$ & $\!\!0\!\!$  &  $\!\!0\!\!$ & $\!\!0\!\!$ & $\!\!0\!\!$  \\ \hline
$\!\!C_4'^{(3)}\!\!$ & $\!\!8\!\!$ &  $\!\!1\!\!$ & $\!\!-i\!\!$ & $\!\!-1\!\!$ &  $\!\!i\!\!$ & $\!\!-1\!\!$ & $\!\!i\!\!$ & $\!\!1\!\!$ & $\!\!-i\!\!$ & $\!\!0\!\!$ & $\!\!0\!\!$ & $\!\!0\!\!$ &  $\!\!0\!\!$ & $\!\!0\!\!$ & $\!\!0\!\!$   \\ \hline
$\!\!C_2^{(1,0)}\!\!$ & $\!\!4\!\!$ & $\!\!1\!\!$ & $\!\!i\!\!$ & $\!\!-1\!\!$ & $\!\!-i\!\!$  & $\!\!1\!\!$ & $\!\!i\!\!$ & $\!\!-1\!\!$ & $\!\!-i\!\!$ &
 $\!\! 1+i\!\!$ & $\!\!0\!\!$ & $\!\!-1+i\!\!$  & $\!\!1-i\!\!$ & $\!\!0\!\!$ & $\!\!-1-i\!\!$    \\ \hline
$\!\!C_2^{(2,0)}\!\!$ & $\!\!2\!\!$ & $\!\!1\!\!$ & $\!\!-1\!\!$ & $\!\!1\!\!$ & $\!\!-1\!\!$ &  $\!\!1\!\!$ & $\!\!-1\!\!$ & $\!\!1\!\!$ & $\!\!-1\!\!$ &
$\!\!0\!\!$ & $\!\!2\!\!$ & $\!\!0\!\!$  &  $\!\!0\!\!$ & $\!\!-2\!\!$ & $\!\!0\!\!$  \\ \hline
$\!\!C_2^{(2,1)}\!\!$ & $\!\!4\!\!$ & $\!\!1\!\!$ & $\!\!-i\!\!$ & $\!\!-1\!\!$ & $\!\!i\!\!$  & $\!\!1\!\!$ & $\!\!-i\!\!$ & $\!\!-1\!\!$ & $\!\!i\!\!$ &
$\!\!-1+i\!\!$ & $\!\!0\!\!$ & $\!\!1+i\!\!$  &  $\!\!-1-i\!\!$ & $\!\!0\!\!$ & $\!\!1-i\!\!$  \\ \hline
$\!\!C_2^{(3,0)}\!\!$ & $\!\!4\!\!$ &  $\!\!1\!\!$ & $\!\!-i\!\!$ & $\!\!-1\!\!$ &  $\!\!i\!\!$ & $\!\!1\!\!$ & $\!\!-i\!\!$ & $\!\!-1\!\!$ &  $\!\!i\!\!$ &
$\!\!1-i\!\!$ & $\!\!0\!\!$ & $\!\!-1-i\!\!$ &  $\!\!1+i\!\!$ & $\!\!0\!\!$ & $\!\!-1+i\!\!$   \\ \hline
$\!\!C_2^{(3,1)}\!\!$ & $\!\!4\!\!$ & $\!\!1\!\!$ & $\!\!1\!\!$ & $\!\!1\!\!$ & $\!\!1\!\!$ & $\!\!1\!\!$  & $\!\!1\!\!$ & $\!\!1\!\!$ & $\!\!1\!\!$ &  
$\!\!0\!\!$ & $\!\!-2\!\!$ & $\!\!0\!\!$  &  $\!\!0\!\!$ & $\!\!-2\!\!$ & $\!\!0\!\!$  \\ \hline
$\!\!C_2^{(3,2)}\!\!$ & $\!\!4\!\!$ &  $\!\!1\!\!$ & $\!\!i\!\!$ & $\!\!-1\!\!$ &  $\!\!-i\!\!$ & $\!\!1\!\!$ & $\!\!i\!\!$ & $\!\!-1\!\!$ & $\!\!-i\!\!$ & 
$\!\!-1-i\!\!$ & $\!\!0\!\!$ & $\!\!1-i\!\!$ & $\!\! -1+i\!\!$ & $\!\!0\!\!$ & $\!\!1+i\!\!$   \\ \hline
\end{tabular}
\end{center}
\caption{ Characters of $\Sigma(32)$ representations}
\label{32tab:sigma-singlet-character}
\end{table}

The tensor products between doublets are obtained as 
\begin{equation}
{\2tvec {x_3}  {x_2}}_{{\bf 2}_{3,2}} \otimes{\2tvec {y_3}  {y_2}}_{{\bf 2}_{3,2}}  
=
{(x_2y_3+ x_3y_2)}_{{\bf 1}_{+,1}}\oplus {(x_2y_3-x_3y_2)}_{{\bf 1}_{-,1}}
 \oplus
{ \2tvec {x_3y_3 }  {x_2y_2}}_{{\bf 2}_{2,0}},
\end{equation}
\begin{eqnarray}
{\2tvec {x_3}  {x_1}}_{{\bf 2}_{3,1}} \otimes{\2tvec {y_3}  {y_1}}_{{\bf 2}_{3,1}}  
&=&
{(x_1y_3+x_3y_1)}_{{\bf 1}_{+,0}}\oplus {(x_1y_3-x_3y_1)}_{{\bf 1}_{-,0}}\nonumber\\
 &\oplus&
{(x_3y_3+x_1y_1)}_{{\bf 1}_{+,2}}\oplus {(x_3y_3-x_1y_1)}_{{\bf
    1}_{-,2}} ,
\end{eqnarray}
\begin{equation}
{\2tvec {x_3}  {x_0}}_{{\bf 2}_{3,0}} \otimes{\2tvec {y_3}  {y_0}}_{{\bf 2}_{3,0}}  
=
{(x_0y_3+x_3y_0)}_{{\bf 1}_{+,3}}\oplus {(x_0y_3-x_3y_0)}_{{\bf 1}_{-,3}}
 \oplus
{ \2tvec {x_3y_3 }  {x_0y_0}}_{{\bf 2}_{2,0}},
\end{equation}
\begin{equation}
{\2tvec {x_2}  {x_1}}_{{\bf 2}_{2,1}} \otimes{\2tvec {y_2}  {y_1}}_{{\bf 2}_{2,1}}  
=
{(x_1y_2+x_2y_1)}_{{\bf 1}_{+,3}}\oplus {(x_1y_2-x_2y_1)}_{{\bf 1}_{-,3}}
 \oplus
{ \2tvec {x_1y_1 }  {x_2y_2}}_{{\bf 2}_{2,0}},
\end{equation}
\begin{eqnarray}
{\2tvec {x_2}  {x_0}}_{{\bf 2}_{2,0}} \otimes{\2tvec {y_2}  {y_0}}_{{\bf 2}_{2,0}}  
&=&
{(x_0y_2+x_2y_0)}_{{\bf 1}_{+,2}}\oplus {(x_0y_2-x_2y_0)}_{{\bf 1}_{-,2}}\nonumber\\
 &\oplus&
{(x_2y_2+x_0y_0)}_{{\bf 1}_{+,0}}\oplus {(x_2y_2-x_0y_0)}_{{\bf 1}_{-,0}},
\end{eqnarray}
\begin{equation}
{\2tvec {x_1}  {x_0}}_{{\bf 2}_{1,0}} \otimes{\2tvec {y_1}  {y_0}}_{{\bf 2}_{1,0}}  
=
{(x_0y_1+x_1y_0)}_{{\bf 1}_{+,1}}\oplus {(x_0y_1-x_1y_0)}_{{\bf 1}_{-,1}}
 \oplus
{ \2tvec {x_1y_1 }  {x_0y_0}}_{{\bf 2}_{2,0}},
\end{equation}
\begin{equation}
{\2tvec {x_3}  {x_2}}_{{\bf 2}_{3,2}} \otimes{\2tvec {y_3}  {y_1}}_{{\bf 2}_{3,1}}  
=
{ \2tvec {x_2y_3 }  {x_3y_1}}_{{\bf 2}_{1,0}}
 \oplus
{ \2tvec {x_2y_1 }  {x_3y_3}}_{{\bf 2}_{3,2}},
\end{equation}
\begin{equation}
{\2tvec {x_3}  {x_2}}_{{\bf 2}_{3,2}} \otimes{\2tvec {y_3}  {y_0}}_{{\bf 2}_{3,0}}  
=
{(x_3y_3+x_2y_0)}_{{\bf 1}_{+,2}}\oplus {(x_3y_3-x_2y_0)}_{{\bf 1}_{-,2}}
 \oplus
{ \2tvec {x_3y_0 }  {x_2y_3}}_{{\bf 2}_{3,1}},
\end{equation}
\begin{equation}
{\2tvec {x_3}  {x_2}}_{{\bf 2}_{3,2}} \otimes{\2tvec {y_2}  {y_1}}_{{\bf 2}_{2,1}}  
=
{(x_2y_2+x_3y_1)}_{{\bf 1}_{+,0}}\oplus {(x_2y_2-x_3y_1)}_{{\bf 1}_{-,0}}
 \oplus
{ \2tvec {x_2y_1 }  {x_3y_2}}_{{\bf 2}_{3,1}},
\end{equation}
\begin{equation}
{\2tvec {x_3}  {x_2}}_{{\bf 2}_{3,2}} \otimes{\2tvec {y_2}  {y_0}}_{{\bf 2}_{2,0}}  
=
{ \2tvec {x_3y_0 }  {x_2y_2}}_{{\bf 2}_{3,1}}
 \oplus
{ \2tvec {x_2y_0 }  {x_3y_2}}_{{\bf 2}_{2,1}},
\end{equation}
\begin{equation}
{\2tvec {x_3}  {x_2}}_{{\bf 2}_{3,2}} \otimes{\2tvec {y_1}  {y_0}}_{{\bf 2}_{1,0}}  
=
{(x_2y_1+x_3y_0)}_{{\bf 1}_{+,3}}\oplus {(x_2y_1-x_3y_0)}_{{\bf 1}_{-,3}}
 \oplus
{ \2tvec {x_2y_0 }  {x_3y_1}}_{{\bf 2}_{2,0}},
\end{equation}
\begin{equation}
{\2tvec {x_3}  {x_1}}_{{\bf 2}_{3,1}} \otimes{\2tvec {y_3}  {y_0}}_{{\bf 2}_{3,0}}  
=
{ \2tvec {x_3y_0 }  {x_1y_3}}_{{\bf 2}_{3,0}}
 \oplus
{ \2tvec {x_3y_3 }  {x_1y_0}}_{{\bf 2}_{2,1}},
\end{equation}
\begin{equation}
{\2tvec {x_3}  {x_1}}_{{\bf 2}_{3,1}} \otimes{\2tvec {y_2}  {y_1}}_{{\bf 2}_{2,1}}  
=
{ \2tvec {x_1y_2 }  {x_3y_1}}_{{\bf 2}_{3,0}}
 \oplus
{ \2tvec {x_1y_1 }  {x_3y_2}}_{{\bf 2}_{2,1}},
\end{equation}
\begin{eqnarray}
{\2tvec {x_3}  {x_1}}_{{\bf 2}_{3,1}} \otimes{\2tvec {y_2}  {y_0}}_{{\bf 2}_{2,0}}  
&=&
{(x_1y_2+x_3y_0)}_{{\bf 1}_{+,3}}\oplus {(x_1y_2-x_3y_0)}_{{\bf 1}_{-,3}}\nonumber\\
& \oplus&
{(x_3y_2+x_1y_0)}_{{\bf 1}_{+,1}}\oplus {(x_3y_2-x_1y_0)}_{{\bf 1}_{-,1}},
\end{eqnarray}
\begin{equation}
{\2tvec {x_3}  {x_1}}_{{\bf 2}_{3,1}} \otimes{\2tvec {y_1}  {y_0}}_{{\bf 2}_{1,0}}  
=
{ \2tvec {x_3y_0 }  {x_1y_1}}_{{\bf 2}_{3,2}}
 \oplus
{ \2tvec {x_1y_0 }  {x_3y_1}}_{{\bf 2}_{1,0}},
\end{equation}
\begin{equation}
{\2tvec {x_3}  {x_0}}_{{\bf 2}_{3,0}} \otimes{\2tvec {y_2}  {y_1}}_{{\bf 2}_{2,1}}  
=
{(x_3y_2+x_0y_1)}_{{\bf 1}_{+,1}}\oplus {(x_3y_2-x_0y_1)}_{{\bf 1}_{-,1}}
 \oplus
{ \2tvec {x_0y_2 }  {x_3y_1}}_{{\bf 2}_{2,0}},
\end{equation}
\begin{equation}
{\2tvec {x_3}  {x_0}}_{{\bf 2}_{3,0}} \otimes{\2tvec {y_2}  {y_0}}_{{\bf 2}_{2,0}}  
=
{ \2tvec {x_3y_0 }  {x_0y_2}}_{{\bf 2}_{3,2}}
 \oplus
{ \2tvec {x_3y_2 }  {x_0y_0}}_{{\bf 2}_{1,0}},
\end{equation}
\begin{equation}
{\2tvec {x_3}  {x_0}}_{{\bf 2}_{3,0}} \otimes{\2tvec {y_1}  {y_0}}_{{\bf 2}_{1,0}}  
=
{(x_3y_1+x_0y_0)}_{{\bf 1}_{+,0}}\oplus {(x_3y_1-x_0y_0)}_{{\bf 1}_{-,0}}
 \oplus
{ \2tvec {x_3y_0 }  {x_0y_1}}_{{\bf 2}_{3,1}},
\end{equation}
\begin{equation}
{\2tvec {x_2}  {x_1}}_{{\bf 2}_{2,1}} \otimes{\2tvec {y_2}  {y_0}}_{{\bf 2}_{2,0}}  
=
{(x_2y_2+x_1y_0)}_{{\bf 1}_{+,1}}\oplus {(x_2y_2-x_1y_0)}_{{\bf 1}_{-,1}}
 \oplus
{ \2tvec {x_1y_2 }  {x_2y_0}}_{{\bf 2}_{3,2}},
\end{equation}
\begin{equation}
{\2tvec {x_2}  {x_1}}_{{\bf 2}_{2,1}} \otimes{\2tvec {y_1}  {y_0}}_{{\bf 2}_{1,0}}  
=
{(x_1y_1+x_2y_0)}_{{\bf 1}_{+,2}}\oplus {(x_1y_1-x_2y_0)}_{{\bf 1}_{-,2}}
 \oplus
{ \2tvec {x_2y_1 }  {x_1y_0}}_{{\bf 2}_{3,1}},
\end{equation}
\begin{equation}
{\2tvec {x_2}  {x_0}}_{{\bf 2}_{2,0}} \otimes{\2tvec {y_1}  {y_0}}_{{\bf 2}_{1,0}}  
=
{ \2tvec {x_2y_1 }  {x_0y_0}}_{{\bf 2}_{2,0}}
 \oplus
{ \2tvec {x_2y_0 }  {x_0y_1}}_{{\bf 2}_{2,1}}.
\end{equation}
The tensor products between singlets are obtained as 
\begin{eqnarray}
&&
{\bf 1}_{\pm,0}\otimes {\bf 1}_{\pm,0}={\bf 1}_{+,0}~,~{\bf 1}_{\pm,1}\otimes {\bf 1}_{\pm,1}={\bf 1}_{+,2}~,~
{\bf 1}_{\pm,2}\otimes {\bf 1}_{\pm,2}={\bf 1}_{+,0}~,~{\bf 1}_{\pm,3}\otimes {\bf 1}_{\pm,3}={\bf 1}_{+,2},
\nonumber\\&&
{\bf 1}_{\pm,3}\otimes {\bf 1}_{\pm,2}={\bf 1}_{+,1}~,~{\bf 1}_{\pm,3}\otimes {\bf 1}_{\pm,1}={\bf 1}_{+,0}~,~
{\bf 1}_{\pm,3}\otimes {\bf 1}_{\pm,0}={\bf 1}_{+,3}~,~{\bf 1}_{\pm,2}\otimes {\bf 1}_{\pm,1}={\bf 1}_{+,3},
\nonumber\\&&
{\bf 1}_{\pm,2}\otimes {\bf 1}_{\pm,0}={\bf 1}_{+,2}~,~{\bf 1}_{\pm,1}\otimes {\bf 1}_{\pm,0}={\bf 1}_{+,1},\\
&&
{\bf 1}_{\mp,0}\otimes {\bf 1}_{\pm,0}={\bf 1}_{-,0}~,~{\bf 1}_{\mp,1}\otimes {\bf 1}_{\pm,1}={\bf 1}_{-,2}~,~
{\bf 1}_{\mp,2}\otimes {\bf 1}_{\pm,2}={\bf 1}_{-,0}~,~{\bf 1}_{\mp,3}\otimes {\bf 1}_{\pm,3}={\bf 1}_{-,2},
\nonumber\\&&
{\bf 1}_{\mp,3}\otimes {\bf 1}_{\pm,2}={\bf 1}_{-,1}~,~{\bf 1}_{\mp,3}\otimes {\bf 1}_{\pm,1}={\bf 1}_{-,0}~,~
{\bf 1}_{\mp,3}\otimes {\bf 1}_{\pm,0}={\bf 1}_{-,3}~,~{\bf 1}_{\mp,2}\otimes {\bf 1}_{\pm,1}={\bf 1}_{-,3},
\nonumber\\&&
{\bf 1}_{\mp,2}\otimes {\bf 1}_{\pm,0}={\bf 1}_{-,2}~,~{\bf
  1}_{\mp,1}\otimes {\bf 1}_{\pm,0}={\bf 1}_{-,1} \nonumber.
\end{eqnarray}
The tensor products between singlets and doublets are 
obtained as 
\begin{eqnarray}
&&(y)_{{\bf 1}_{\pm,0}}\otimes{\2tvec {x_3}  {x_2}}_{{\bf 2}_{3,2}} =
{\2tvec {yx_3 }  {yx_2}}_{{\bf 2}_{3,2}}
~,~
(y)_{{\bf 1}_{\pm,1}}\otimes{\2tvec {x_3}  {x_2}}_{{\bf 2}_{3,2}} =
{\2tvec {yx_2 }  {yx_3}}_{{\bf 2}_{3,0}},
\nonumber\\
&&(y)_{{\bf 1}_{\pm,2}}\otimes{\2tvec {x_3}  {x_2}}_{{\bf 2}_{3,2}} =
{\2tvec {yx_3 }  {yx_2}}_{{\bf 2}_{1,0}}
~,~
(y)_{{\bf 1}_{\pm,3}}\otimes{\2tvec {x_3}  {x_2}}_{{\bf 2}_{3,2}} =
{\2tvec {yx_3 }  {yx_2}}_{{\bf 2}_{2,1}},\nonumber
\\
&&(y)_{{\bf 1}_{\pm,0}}\otimes{\2tvec {x_3}  {x_1}}_{{\bf 2}_{3,1}} =
{\2tvec {yx_3 }  {yx_1}}_{{\bf 2}_{3,1}}
~,~
(y)_{{\bf 1}_{\pm,1}}\otimes{\2tvec {x_3}  {x_1}}_{{\bf 2}_{3,1}} =
{\2tvec {yx_1 }  {yx_3}}_{{\bf 2}_{2,0}},
\nonumber\\
&&(y)_{{\bf 1}_{\pm,2}}\otimes{\2tvec {x_3}  {x_1}}_{{\bf 2}_{3,1}} =
{\2tvec {yx_1 }  {yx_3}}_{{\bf 2}_{3,1}}
~,~
(y)_{{\bf 1}_{\pm,3}}\otimes{\2tvec {x_3}  {x_1}}_{{\bf 2}_{3,1}} =
{\2tvec {yx_3 }  {yx_1}}_{{\bf 2}_{2,0}},\nonumber
\\
&&(y)_{{\bf 1}_{\pm,0}}\otimes{\2tvec {x_3}  {x_0}}_{{\bf 2}_{3,0}} =
{\2tvec {yx_3 }  {yx_0}}_{{\bf 2}_{3,0}}
~,~
(y)_{{\bf 1}_{\pm,1}}\otimes{\2tvec {x_3}  {x_0}}_{{\bf 2}_{3,0}} =
{\2tvec {yx_0 }  {yx_3}}_{{\bf 2}_{1,0}},
\nonumber\\
&&(y)_{{\bf 1}_{\pm,2}}\otimes{\2tvec {x_3}  {x_0}}_{{\bf 2}_{3,0}} =
{\2tvec {yx_0 }  {yx_3}}_{{\bf 2}_{2,1}}
~,~
(y)_{{\bf 1}_{\pm,3}}\otimes{\2tvec {x_3}  {x_0}}_{{\bf 2}_{3,0}} =
{\2tvec {yx_0 }  {yx_3}}_{{\bf 2}_{3,2}},\nonumber
\\
&&(y)_{{\bf 1}_{\pm,0}}\otimes{\2tvec {x_2}  {x_1}}_{{\bf 2}_{2,1}} =
{\2tvec {yx_2 }  {yx_1}}_{{\bf 2}_{2,1}}~,~
(y)_{{\bf 1}_{\pm,1}}\otimes{\2tvec {x_2}  {x_1}}_{{\bf 2}_{2,1}} =
{\2tvec {yx_2 }  {yx_1}}_{{\bf 2}_{3,2}},
\\ &&
(y)_{{\bf 1}_{\pm,2}}\otimes{\2tvec {x_2}  {x_1}}_{{\bf 2}_{2,1}} =
{\2tvec {yx_1 }  {yx_2}}_{{\bf 2}_{3,0}}~,~
(y)_{{\bf 1}_{\pm,3}}\otimes{\2tvec {x_2}  {x_1}}_{{\bf 2}_{2,1}} =
{\2tvec {yx_2 }  {yx_1}}_{{\bf 2}_{1,0}},
\nonumber\\
&&(y)_{{\bf 1}_{\pm,0}}\otimes{\2tvec {x_2}  {x_0}}_{{\bf 2}_{2,0}} =
{\2tvec {yx_2 }  {yx_0}}_{{\bf 2}_{2,0}}~,~
(y)_{{\bf 1}_{\pm,1}}\otimes{\2tvec {x_2}  {x_0}}_{{\bf 2}_{2,0}} =
{\2tvec {yx_2 }  {yx_0}}_{{\bf 2}_{3,1}},
\nonumber\\&&
(y)_{{\bf 1}_{\pm,2}}\otimes{\2tvec {x_2}  {x_0}}_{{\bf 2}_{2,0}} =
{\2tvec {yx_0 }  {yx_2}}_{{\bf 2}_{2,0}}~,~
(y)_{{\bf 1}_{\pm,3}}\otimes{\2tvec {x_2}  {x_0}}_{{\bf 2}_{2,0}} =
{\2tvec {yx_0 }  {yx_2}}_{{\bf 2}_{3,1}},
\nonumber\\
&&(y)_{{\bf 1}_{\pm,0}}\otimes{\2tvec {x_1}  {x_0}}_{{\bf 2}_{1,0}} =
{\2tvec {yx_1 }  {yx_0}}_{{\bf 2}_{1,0}}~,~
(y)_{{\bf 1}_{\pm,1}}\otimes{\2tvec {x_1}  {x_0}}_{{\bf 2}_{1,0}} =
{\2tvec {yx_1 }  {yx_0}}_{{\bf 2}_{2,1}},
\nonumber\\&&
(y)_{{\bf 1}_{\pm,2}}\otimes{\2tvec {x_1}  {x_0}}_{{\bf 2}_{1,0}} =
{\2tvec {yx_1 }  {yx_0}}_{{\bf 2}_{3,2}}~,~
(y)_{{\bf 1}_{\pm,3}}\otimes{\2tvec {x_1}  {x_0}}_{{\bf 2}_{1,0}} =
{\2tvec {yx_0 }  {yx_1}}_{{\bf 2}_{3,0}}.\nonumber
\end{eqnarray}


\subsection{$\Sigma(50)$}
The $\Sigma(50)$ has fifty elements, $b^ka^ma^{'n}$ for $k=0,1$ and
$m,~n=0,~1,~2,~3,~4$ where $a$, $a'$ and $b$ satisfy the same
conditions as Eq. (\ref{daisu-kankei}) in the case of 
 $N=5$. These elements are classified into twenty conjugacy classes,
\begin{eqnarray}
\begin{array}{ccc}
 C_1:&\{ e \}, &  h=1,\\
 C_1^{(1)}:&\{ aa'\}, &  h=5,\\
 C_1^{(2)}:&\{ a^2a'^2\}, &  h=5,\\
 C_1^{(3)}:&\{ a^3a'^3\}, &  h=5,\\
  C_1^{(4)}:&\{ a^3a'^3\}, &  h=5,\\
 C_5'^{(0)}:&\{b~,~ba'^2a^3~,~ba'^3a^2~,~ba'^4a~,~ba'a^4\},&   h=2,\\
C_5'^{(1)}:&\{ba'~,~ba~,~ba'^3a^3~,~ba'^4a^2~,~ba'^2a^4 \},&   h=10, \\
 C_5'^{(2)}:&\{ba'^2~,~ba'a~,~ba^2~,~ba'^4a^3~,~ba'^3a^4 \},&   h=10,\\
 C_5'^{(3)}:&\{ba'^3~,~ba'^2a~,~ba'a^2~,~ba^3~,~ba'^4a^4 \},&   h=10,\\
 C_5'^{(4)}:&\{ba'^4~,~ba'^2a^2~,~ba'a^3~,~ba'^3a~,~ba^4 \},&   h=10,\\
 C_2^{(1,0)}:&\{ a~,~a' \},&   h=5,\\
  C_2^{(2,0)}:&\{ a^{2}~,~a'^{2} \},&   h=5,\\
   C_2^{(2,1)}:&\{ a^{2}a'~,~aa'^2 \},&   h=5,\\
 C_2^{(3,0)}:&\{ a^3~,~a'^3 \},&   h=5,\\
  C_2^{(3,1)}:&\{ a^{3}a'~,~aa'^3\},&   h=5,\\
   C_2^{(3,2)}:&\{ a^{3}a'^2~,~a^{2}a'^3 \},&   h=5,\\
   C_2^{(4,0)}:&\{ a^4~,~a'^4 \},&   h=5,\\
 C_2^{(4,1)}:&\{ a^4a'~,~aa'^4 \},&   h=5,\\
  C_2^{(4,2)}:&\{ a^4a'^2~,~a^2a'^4\},&   h=5,\\
   C_2^{(4,3)}:&\{ a^4a'^3~,~a^3a'^4 \},&   h=5,\\
\end{array}
\end{eqnarray}
where we have also shown the orders of each element 
in the conjugacy class by $h$.

The $\Sigma(50)$ has ten singlets ${\bf 1}_{\pm, n}$ with 
$n=0,1,2,3,4$ and ten doublets 
${\bf 2}_{p,q}$ with $(p,q)=(1,0), (2,0), (3,0), (4,0), (2,1), 
(3,1), (4,1), (3,1), (3,2), (4,3)$.
The characters are shown in Tables
 \ref{50tab:sigma-singlet-character} and \ref{50tab:sigma-doublet-character}.
\begin{table}[t]
\begin{center}
\begin{tabular}{|c|c|c|c|c|c|c|}
\hline
& $\!\!h\!\!$ & $\!\!\chi_{1_{\pm0}}\!\!$ & $\!\!\chi_{1_{\pm1}}\!\!$ & $\!\!\chi_{1_{\pm2}}\!\!$ & 
$\!\!\chi_{1_{\pm3}}\!\!$ & $\!\!\chi_{1_{\pm4}}\!\!$
 \\ \hline
$\!\!C_1\!\!$ & $\!\!1\!\!$ &  $\!\!1\!\!$ & $\!\!1\!\!$ & $\!\!1\!\!$ & $\!\!1\!\!$ & $\!\!1\!\!$    \\ \hline
$\!\!C_1^{(1)}\!\!$ & $\!\!5\!\!$  & $\!\!1\!\!$ & $\!\!\rho^2\!\!$ & $\!\!\rho^4\!\!$ & $\!\!\rho\!\!$ & $\!\!\rho^3\!\!$  \\ 
\hline
$\!\!C_1^{(2)}\!\!$ & $\!\!5\!\!$  & $\!\!1\!\!$ & $\!\!\rho^4\!\!$ & $\!\!\rho^3\!\!$ & $\!\!\rho^2\!\!$ & $\!\!\rho\!\!$  \\ 
\hline
$\!\!C_1^{(3)}\!\!$ & $\!\!5\!\!$  & $\!\!1\!\!$ & $\!\!\rho\!\!$ & $\!\!\rho^2\!\!$ & $\!\!\rho^3\!\!$ & $\!\!\rho^4\!\!$  \\ 
\hline
$\!\!C_1^{(4)}\!\!$ & $\!\!5\!\!$  & $\!\!1\!\!$ & $\!\!\rho^3\!\!$ & $\!\!\rho\!\!$ & $\!\!\rho^4\!\!$ & $\!\!\rho^2\!\!$  \\ 
\hline
$\!\!C_5'^{(0)}\!\!$ & $\!\!2\!\!$ & $\!\!\pm1\!\!$ & $\!\!\pm1\!\!$ & $\!\!\pm1\!\!$ & $\!\!\pm1\!\!$ & $\!\!\pm1\!\!$  \\ \hline
$\!\!C_5'^{(1)}\!\!$ & $\!\!10\!\!$ & $\!\!\pm1\!\!$ & $\!\!\pm\rho\!\!$ & $\!\!\pm\rho^2\!\!$ & $\!\!\pm\rho^3\!\!$ & 
$\!\!\pm\rho^4\!\!$  \\ \hline
$\!\!C_5'^{(2)}\!\!$ & $\!\!10\!\!$ & $\!\!\pm1\!\!$ & $\!\!\pm\rho^2\!\!$ & $\!\!\pm\rho^4\!\!$ & $\!\!\pm\rho\!\!$ & 
$\!\!\pm\rho^3\!\!$  \\ \hline
$\!\!C_5'^{(3)}\!\!$ & $\!\!10\!\!$ & $\!\!\pm1\!\!$ & $\!\!\pm\rho^3\!\!$ & $\!\!\pm\rho\!\!$ & $\!\!\pm\rho^4\!\!$ & 
$\!\!\pm\rho^2\!\!$  \\ \hline
$\!\!C_5'^{(4)}\!\!$ & $\!\!10\!\!$ & $\!\!\pm1\!\!$ & $\!\!\pm\rho^4\!\!$ & $\!\!\pm\rho^3\!\!$ & $\!\!\pm\rho^2\!\!$ & 
$\!\!\pm\rho\!\!$  \\ \hline
$\!\!C_2^{(1,0)}\!\!$ & $\!\!5\!\!$   & $\!\!1\!\!$ & $\!\!\rho\!\!$ & $\!\!\rho^2\!\!$ & $\!\!\rho^3\!\!$ & $\!\!\rho^3\!\!$  \\ 
\hline
$\!\!C_2^{(2,0)}\!\!$ & $\!\!5\!\!$   & $\!\!1\!\!$ & $\!\!\rho^2\!\!$ & $\!\!\rho^4\!\!$ & $\!\!\rho\!\!$ & $\!\!\rho^3\!\!$  \\ 
\hline
$\!\!C_2^{(2,1)}\!\!$ & $\!\!5\!\!$ & $\!\!1\!\!$ & $\!\!\rho^3\!\!$ & $\!\!\rho\!\!$ & $\!\!\rho^4\!\!$ & $\!\!\rho^2\!\!$  \\ 
\hline
$\!\!C_2^{(3,0)}\!\!$ & $\!\!5\!\!$   & $\!\!1\!\!$ & $\!\!\rho^3\!\!$ & $\!\!\rho\!\!$ & $\!\!\rho^4\!\!$ & $\!\!\rho^2\!\!$  \\ 
\hline
$\!\!C_2^{(3,1)}\!\!$ & $\!\!5\!\!$   & $\!\!1\!\!$ & $\!\!\rho^4\!\!$ & $\!\!\rho^3\!\!$ & $\!\!\rho^2\!\!$ & $\!\!\rho\!\!$  \\ 
\hline
$\!\!C_2^{(3,2)}\!\!$ & $\!\!5\!\!$   & $\!\!1\!\!$ & $\!\!1\!\!$ & $\!\!1\!\!$ & $\!\!1\!\!$ & $\!\!1\!\!$  \\ \hline
$\!\!C_2^{(4,0)}\!\!$ & $\!\!5\!\!$   & $\!\!1\!\!$ & $\!\!\rho^4\!\!$ & $\!\!\rho^3\!\!$ & $\!\!\rho^2\!\!$ & $\!\!\rho\!\!$  \\ 
\hline
$\!\!C_2^{(4,1)}\!\!$ & $\!\!5\!\!$   & $\!\!1\!\!$ & $\!\!1\!\!$ & $\!\!1\!\!$ & $\!\!1\!\!$ & $\!\!1\!\!$  \\ \hline
$\!\!C_2^{(4,2)}\!\!$ & $\!\!5\!\!$   & $\!\!1\!\!$ & $\!\!\rho\!\!$ & $\!\!\rho^2\!\!$ & $\!\!\rho^3\!\!$ & $\!\!\rho^4\!\!$  \\ 
\hline
$\!\!C_2^{(4,3)}\!\!$ & $\!\!5\!\!$   & $\!\!1\!\!$ & $\!\!\rho^2\!\!$ & $\!\!\rho^4\!\!$ & $\!\!\rho\!\!$ & $\!\!\rho^3\!\!$  \\ 
\hline
\end{tabular}
\end{center}
\caption{ Characters of $\Sigma(50)$ representations,
 where $\rho=e^{2i\pi/5}$
}
\label{50tab:sigma-singlet-character}
\end{table}
\begin{table}[t]
\begin{center}
\begin{tabular}{|c|c|c|c|c|c|c|c|c|c|c|c|}
\hline
& $\!\!h\!\!$  & $\!\!\chi_{2_{1,0}}\!\!$ & $\!\!\chi_{2_{2,0}}\!\!$  & $\!\!\chi_{2_{2,1}}\!\!$ & 
$\!\!\chi_{2_{3,0}}\!\!$ & $\!\!\chi_{2_{3,1}}\!\!$  & $\!\!\chi_{2_{3,2}}\!\!$    & $\!\!\chi_{2_{4,0}}\!\!$  & 
$\!\!\chi_{2_{4,1}}\!\!$   & $\!\!\chi_{2_{4,2}}\!\!$  & $\!\!\chi_{2_{4,3}}\!\!$  \\ \hline
$\!\!C_1\!\!$ & $\!\!1\!\!$ &  $\!\!2\!\!$ & $\!\!2\!\!$ & $\!\!2\!\!$ & $\!\!2\!\!$ & $\!\!2\!\!$ & $\!\!2\!\!$ & $\!\!2\!\!$ & $\!\!2\!\!$ &  $\!\!2\!\!$ & $\!\!2\!\!$    
\\ \hline
$\!\!C_1^{(1)}\!\!$ & $\!\!5\!\!$ & $\!\!2\rho\!\!$ & $\!\!2\rho^2\!\!$ & $\!\!2\rho^3\!\!$  & $\!\!2\rho^3\!\!$  & $\!\!2\rho^4\!\!$ & $\!\!2\!\!$  & $\!\!2\rho^4\!\!$ & $\!\!2\!\!$
& $\!\!2\rho\!\!$ &  $\!\!2\rho^2\!\!$  \\ \hline
$\!\!C_1^{(2)}\!\!$ & $\!\!5\!\!$ & $\!\!2\rho^2\!\!$ & $\!\!2\rho^4\!\!$ & $\!\!2\rho\!\!$  & $\!\!2\rho\!\!$  & $\!\!2\rho^3\!\!$ 
& $\!\!2\!\!$  & $\!\!2\rho^3\!\!$ & $\!\!2\!\!$
& $\!\!2\rho^2\!\!$ &  $\!\!2\rho^4\!\!$   \\ \hline
$\!\!C_1^{(3)}\!\!$ & $\!\!5\!\!$ & $\!\!2\rho^3\!\!$ & $\!\!2\rho\!\!$ & $\!\!2\rho^4\!\!$  & $\!\!2\rho^4\!\!$  & 
$\!\!2\rho^2\!\!$ & $\!\!2\!\!$  & $\!\!2\rho^2\!\!$ & $\!\!2\!\!$
& $\!\!2\rho^3\!\!$ &  $\!\!2\rho\!\!$   \\ \hline   
$\!\!C_1^{(4)}\!\!$ & $\!\!5\!\!$ & $\!\!2\rho^4\!\!$ & $\!\!2\rho^3\!\!$ & $\!\!2\rho^2\!\!$  & $\!\!2\rho^2\!\!$  & 
$\!\!2\rho\!\!$ & $\!\!2\!\!$  & $\!\!2\rho\!\!$ & $\!\!2\!\!$
& $\!\!2\rho^4\!\!$ &  $\!\!2\rho^3\!\!$   \\ \hline   
$\!\!C_5'^{(0)}\!\!$ &  $\!\!2\!\!$  &  $\!\!0\!\!$ & $\!\!0\!\!$ & $\!\!0\!\!$ & $\!\!0\!\!$ & $\!\!0\!\!$ & $\!\!0\!\!$ & $\!\!0\!\!$ & $\!\!0\!\!$ &  $\!\!0\!\!$ & $\!\!0\!\!$    \\ \hline
$\!\!C_5'^{(1-4)}\!\!$ & $\!\!10\!\!$  &  $\!\!0\!\!$ & $\!\!0\!\!$ & $\!\!0\!\!$ & $\!\!0\!\!$ & $\!\!0\!\!$ & $\!\!0\!\!$ & $\!\!0\!\!$ & $\!\!0\!\!$ &  $\!\!0\!\!$ & $\!\!0\!\!$    \\ \hline
$\!\!C_2^{(1,0)}\!\!$ & $5\!\!$ & $\!\!1+\rho\!\!$ & $\!\!1+\rho^2\!\!$ & $\!\!\rho+\rho^2\!\!$ & $\!\!1+\rho^3\!\!$  & $\!\!\rho+\rho^3\!\!$ & $\!\!\rho^2+\rho^3\!\!$ & $\!\!1+\rho^4\!\!$ & $\!\!\rho+\rho^4\!\!$ & $\!\!\rho^2+\rho^4\!\!$ & $\!\!\rho^3+\rho^4\!\!$  \\ \hline
$\!\!C_2^{(2,0)}\!\!$ & $\!\!5\!\!$ & $\!\!1+\rho^2\!\!$ & $\!\!1+\rho^4\!\!$ & $\!\!\rho^2+\rho^4\!\!$ & $\!\!1+\rho\!\!$  & $\!\!\rho+\rho^2\!\!$ & $\!\!\rho+\rho^4\!\!$ & $\!\!1+\rho^3\!\!$ & $\!\!\rho^2+\rho^3\!\!$ & $\!\!\rho^3+\rho^4\!\!$ & $\!\!\rho+\rho^3\!\!$  \\ \hline
$\!\!C_2^{(2,1)}\!\!$ & $\!\!5\!\!$ & $\!\!\rho+\rho^2\!\!$ & $\!\!\rho^2+\rho^4\!\!$ & $\!\!1+\rho^4\!\!$ & 
$\!\!\rho+\rho^3\!\!$  & $\!\!1+\rho^2\!\!$ & $\!\!\rho^2+\rho^3\!\!$ & $\!\!\rho^3+\rho^4\!\!$ & $\!\!\rho+\rho^4\!\!$ & 
$\!\!1+\rho^3\!\!$ & $\!\!1+\rho\!\!$  \\ \hline
$\!\!C_2^{(3,0)}\!\!$ & $\!\!5\!\!$ & $\!\!1+\rho^3\!\!$ & $\!\!1+\rho\!\!$ & $\!\!\rho+\rho^3\!\!$ & $\!\!1+\rho^4\!\!$  & $\!\!\rho^3+\rho^4\!\!$ & $\!\!\rho+\rho^4\!\!$ & $\!\!1+\rho^2\!\!$ & $\!\!\rho^2+\rho^3\!\!$ & 
$\!\!\rho+\rho^2\!\!$ & $\!\!\rho^2+\rho^4\!\!$  \\ \hline
$\!\!C_2^{(3,1)}\!\!$ & $\!\!5\!\!$ & $\!\!\rho+\rho^3\!\!$ & $\!\!\rho+\rho^2\!\!$ & $\!\!1+\rho^2\!\!$ & 
$\!\!\rho^3+\rho^4\!\!$  & $\!\!1+\rho\!\!$ & $\!\!\rho+\rho^4\!\!$ & $\!\!\rho^2+\rho^4\!\!$ & $\!\!\rho^2+\rho^3\!\!$ & 
$\!\!1+\rho^4\!\!$ & $\!\!1+\rho^3\!\!$  \\ \hline
$\!\!C_2^{(3,2)}\!\!$ & $\!\!5\!\!$ & $\!\!\rho^2+\rho^3\!\!$ & $\!\!\rho+\rho^4\!\!$ & $\!\!\rho^2+\rho^3\!\!$ & $\!\!\rho+\rho^4\!\!$  & $\!\!\rho+\rho^4\!\!$ & $\!\!\rho^2+\rho^3\!\!$ & $\!\!\rho^2+\rho^3\!\!$ & 
$\!\!\rho+\rho^4\!\!$ & $\!\!\rho+\rho^4\!\!$ & $\!\!\rho^2+\rho^3\!\!$  \\ \hline
$\!\!C_2^{(4,0)}\!\!$ & $\!\!5\!\!$ & $\!\!1+\rho^4\!\!$ & $\!\!1+\rho^3\!\!$ & $\!\!\rho^3+\rho^4\!\!$ & $\!\!1+\rho^2\!\!$  & $\!\!\rho^2+\rho^4\!\!$ & $\!\!\rho^2+\rho^3\!\!$ & $\!\!1+\rho\!\!$ & $\!\!\rho+\rho^4\!\!$ & $\!\!\rho+\rho^3\!\!$ & $\!\!\rho+\rho^2\!\!$  \\ \hline
$\!\!C_2^{(4,1)}\!\!$ & $\!\!5\!\!$ & $\!\!\rho+\rho^4\!\!$ & $\!\!\rho^2+\rho^3\!\!$ & $\!\!\rho+\rho^4\!\!$ & $\!\!\rho^2+\rho^3\!\!$  & $\!\!\rho^2+\rho^3\!\!$ & $\!\!\rho+\rho^4\!\!$ & $\!\!\rho+\rho^4\!\!$ & 
$\!\!\rho^2+\rho^3\!\!$ & $\!\!\rho^2+\rho^3\!\!$ & $\!\!\rho+\rho^4\!\!$  \\ \hline
$\!\!C_2^{(4,2)}\!\!$ & $\!\!5\!\!$ & $\!\!\rho^2+\rho^4\!\!$ & $\!\!\rho^3+\rho^4\!\!$ & $\!\!1+\rho^3\!\!$ & $\!\!\rho+\rho^2\!\!$  & $\!\!1+\rho^4\!\!$ & $\!\!1+\rho\!\!$ & $\!\!\rho+\rho^3\!\!$ & $\!\!\rho^2+\rho^3\!\!$ & 
$\!\!1+\rho\!\!$ & $\!\!1+\rho^2\!\!$  \\ \hline
$\!\!C_2^{(4,3)}\!\!$ & $\!\!5\!\!$ & $\!\!\rho^3+\rho^4\!\!$ & $\!\!\rho+\rho^3\!\!$ & $\!\!1+\rho\!\!$ & 
$\!\!\rho^2+\rho^4\!\!$  & $\!\!1+\rho^3\!\!$ & $\!\!\rho^2+\rho^3\!\!$ & $\!\!\rho+\rho^2\!\!$ & $\!\!\rho+\rho^4\!\!$ & 
$\!\!1+\rho^2\!\!$ & $\!\!1+\rho^4\!\!$  \\ \hline
\end{tabular}
\end{center}
\caption{ Characters of $\Sigma(50)$ representations,
 where $\rho=e^{2i\pi/5}$}
\label{50tab:sigma-doublet-character}
\end{table}
Since the tensor products are obtained in the same ways as the cases of lower order, as can been seen from the previous ones, we omit the explicit expressions.

\clearpage


\section{$\Delta (3N^2)$}
\label{sec:Delta-3n}

\subsection{Generic aspects}

The discrete group $\Delta (3N^2)$ is isomorphic 
to $(Z_N \times Z_N')\rtimes Z_3$.
(See also Ref.~\cite{Luhn:2007uq}.)
We denote the generators of $Z_N$ and $Z_N'$ by $a$ and $a'$, 
respectively, and the $Z_3$ generator is written by $b$.
They satisfy
\begin{eqnarray}
& & a^N = {a'}^N = b^3 = e, \qquad aa' = a'a, \nonumber \\
& &  bab^{-1}=a^{-1}(a')^{-1}, \qquad ba'b^{-1}=a.
\end{eqnarray}
Using them, all of $\Delta (3N^2)$ elements are written as 
\begin{eqnarray}
 & & g=b^{k}a^{m}a'^{n},
\end{eqnarray}
for $k=0,1,2$ and $m,n=0,1,2,\cdots ,N-1$.

The generators, $a$, $a'$ and $b$, are represented, e.g. as 
\begin{eqnarray}\label{eq:delta-3n-3}
b=\Mat3{0}{1}{0}{0}{0}{1}{1}{0}{0},\quad a=
\Mat3{\rho}{0}{0} {0}{1}{0} {0}{0}{\rho^{-1}},\quad 
a'=\Mat3{\rho^{-1}}{0}{0} {0}{\rho}{0} {0}{0}{1},
\end{eqnarray}
where $\rho=e^{2\pi i/N}$.
Then, all elements of $\Delta(3N^2)$ are represented as 
\begin{eqnarray}
\Mat3{\rho^m}{0}{0}{0}{p^n}{0}{0}{0}{\rho^{-m-n}},\quad
\Mat3{0}{\rho^m}{0}{0}{0}{\rho^n}{\rho^{-m-n}}{0}{0},\quad
\Mat3{0}{0}{\rho^m}{\rho^n}{0}{0}{0}{\rho^{-m-n}}{0}, 
\end{eqnarray}
for $m,n=0,1,2,\cdots ,N-1$.

\vskip .5cm
{$\bullet$ \bf Conjugacy classes}

Now, let us study the conjugacy classes.
It is found that 
\begin{eqnarray}
ba^{\ell}a'^{m}b^{-1} = a^{-\ell+m}a'^{-\ell}, 
\qquad
b^2a^{\ell}a'^{m}b^{-2}=a^{-m}a'^{\ell- m}.
\end{eqnarray}
Thus, these elements,
$a^{\ell}a'^{m},~a^{-\ell+m}a'^{-\ell},~a^{-m}a'^{\ell -m}$, 
must belong to the same conjugacy class.
These are independent elements of $\Delta(3N^2)$ 
unless $N/3=$ integer and $3\ell = \ell + m = 0$ (mod $N$).
On the other hand, if 
$N/3=$ integer and $3\ell = \ell + m = 0$ (mod $N$), 
the above elements are the same, i.e. $a^{\ell}a'^{-\ell}$. 
As a result, the elements $a^\ell a'^m$ are classified 
into the following conjugacy classes,
\begin{eqnarray}
C^{(\ell,m)}_3=\{a^{\ell}a'^{m},~a^{-\ell+m}a'^{-\ell},~a^{-m}a'^{\ell-m} \},
\end{eqnarray}
for $N/3 \neq$ integer,
\begin{eqnarray}
\hspace{-7mm}&& C^{\ell}_1=\{a^{\ell}a'^{-\ell}\},\quad \quad \quad 
\ell =\frac{N}{3},~\frac{2N}{3}, \nonumber \\
\hspace{-7mm}&&
C^{(\ell,m)}_3
=\{a^{\ell}a'^{m},~a^{-\ell +m}a'^{-\ell},~a^{-m}a'^{\ell -m} \}
,\quad(\ell,m)\neq\L(\frac{N}{3},~\frac{2N}{3}\R),\L(\frac{2N}{3},~\frac{N}{3}\R), 
\end{eqnarray}
for $N/3=$ integer.

Similarly, we can obtain conjugacy classes including 
$ba^{\ell}a'^{m}$.
Let us consider the conjugates of the simplest element $b$ among 
$ba^{\ell}a'^{m}$.
It is found that 
\begin{eqnarray}
a^pa'^q (b) a^{-p}a'^{-q} = ba^{-p-q}a'^{p-2q} = ba^{-n+3q}a'^{n},
\end{eqnarray}
where we have written for convenience by using $n \equiv p-2q$.
We also obtain 
\begin{eqnarray}
& & b (ba^{-n+3q}a'^{n})b^{-1} =  b a^{2n-3q}a'^{n-3q}, \\
& & b^2 (ba^{-n+3q}a'^{n})b^{-2} = b a^{-n} a'^{-2n+3q}  .
\end{eqnarray}
The important property is that $q$ appears only in the form of $3q$.
Thus, if $N/3 \neq $ integer,  
the element of $b$ is conjugate to all of $ba^{\ell}a'^{m}$. 
That is, all of them belong to the same conjugacy class 
$C_{N^2}^{1}$.
Similarly, all of $b^2a^{\ell}a'^{m}$ belong to the same 
conjugacy class $C_{N^2}^{2}$ for $N/3 \neq $ integer.

On the other hand, if $N/3 =$ integer, the situation is different.
Among the above elements conjugate to $b$, 
there does not appear $ba$.
Its conjugates are also obtained as 
\begin{eqnarray}
& & a^pa'^q (ba) a^{-p}a'^{-q} = ba^{1-p-q}a'^{p-2q} = ba^{1-n+3q}a'^{n}, \\
& & b (ba^{1-n+3q}a'^{n})b^{-1} =  b a^{-1+2n-3q}a'^{-1+n-3q}, \\
& & b^2 (ba^{1-n+3q}a'^{n})b^{-2} = b a^{-n} a'^{1-2n+3q}  .
\end{eqnarray}
It is found that these elements conjugate to $ba$ as well as  
conjugates of $b$ do not include $ba^2$ when $N/3 =$ integer.
As a result, it is found that  for $N/3 =$ integer, 
the elements  $ba^{\ell}a'^{m}$ are classified into three 
conjugacy classes, $C^{(\ell)}_{N^2/3}$ for $\ell=0,1,2$, i.e., 
\begin{eqnarray}
C^{(\ell)}_{N^2/3}=\{ba^{\ell-n-3m}a'^{n}|
m=0,1,...,\frac{N-3}{3};~n=0,...,N-1\}.
\end{eqnarray}
Similarly, the $b^2a^{\ell}a'^{m}$ are classified into three 
conjugacy classes, $C^{(\ell)}_{N^2/3}$ for $\ell=0,1,2$, i.e., 
\begin{eqnarray}
C^{(\ell)}_{N^2/3}=\{b^2a^{\ell-n-3m}a'^{n}|
m=0,1,...,\frac{N-3}{3};~n=0,...,N-1\}.
\end{eqnarray}

Here, we summarize the conjugacy classes of $\Delta(3N^2)$.
For $N/3 \neq$ integer, the $\Delta(3N^2)$ has the following 
conjugacy classes,
\begin{eqnarray}
\begin{array}{ccc}
C_1:&\{e\},  & h=1, \\
C_3^{(\ell,m)}
:&\{a^\ell a'^m,a^{-\ell+m} a'^{-\ell},
a^{-\sigma} a'^{\ell-m}\},  & h=N/\gcd(N,\ell,m), \\
C^1_{N^2}:& \{ba^\ell a'^m|\ell,m=0,1,\cdots,N-1\}, & h=N/\gcd(N,3,\ell,m), \\
C^2_{N^2}:& \{b^2a^\ell a'^m|\ell,m=0,1,\cdots,N-1\}, & h=N/\gcd(N,3,\ell,m) .
\end{array}
\end{eqnarray}
The number of the conjugacy classes $C_3^{(\ell,m)}$ is equal to 
$(N^2 -1)/3$.
Then, the total number of conjugacy classes is equal to $3 + (N^2 -1)/3$.
The relations (\ref{eq:character-2-e}) and (\ref{eq:sum-dim})
for $\Delta (3n^2)$ with  $N/3 \neq$ integer 
lead to
\begin{eqnarray}
&&m_1+2^2m_2+3^2m_3 + \cdots =3N^2,\\
&&m_1+m_2+m_3 + \cdots =3+ (N^2 -1)/3.
\end{eqnarray}
The solution is found as $(m_1,m_3)=(3,(N^2-1)/3)$.
That is, there are three singlets and $(N^2-1)/3$ triplets.

On the other hand, for $N/3=$ integer, the $\Delta(3N^2)$ has the following 
conjugacy classes,
\begin{eqnarray}
\hspace{-20mm}
\begin{array}{cc}
C_1 :&\{e\},    \\
C_1^{(k)} :&\{a^{k} a'^{-k}\},~~k=\frac{N}{3}, \frac{2N}{3}, \\
C_3^{(\ell,m)} :& 
\{a^\ell a'^m,a^{-\ell +m} a'^{-\ell},
a^{-m} a'^{\ell -m}\},~~(\ell,m)\neq 
\left( \frac{N}{3}, \frac{2N}{3}\right), 
\left( \frac{2N}{3}, \frac{N}{3}\right),
\\
C_{N^2/3}^{(1,p)}
: & \{ba^{p-n-3m}a'^n|
m=0,1,\cdots,\frac{N-3}{3},n=0,1,\cdots,N-1\},~~p=0,1,2,
\\
C_{N^2/3}^{(2,p)}
:& \{b^2a^{p-n-3m}a'^n|
m=0,1,\cdots,\frac{N-3}{3},n=0,1,\cdots,N-1\},~~p=0,1,2.
\end{array}
\end{eqnarray}
The orders of each element in the conjugacy classes, i.e. $g^h=e$ are 
obtained as follows,
\begin{eqnarray}
\hspace{-20mm}
\begin{array}{cc}
C_1 :   & h=1, \\
C_1^{(k)} :& h=N/\gcd(N,N/3,2N/3), \\
C_3^{(\ell,m)} :
& h=N/\gcd(N,\ell,m), \\
C_{N^2/3}^{(1,p)}
: 
& h = N/\gcd(N,3,p-n-3m,n),\\
C_{N^2/3}^{(2,p)}
: &h=  N/\gcd(N,3,p-n-3m,n).
\end{array}
\end{eqnarray}
The numbers of the conjugacy classes $C_1^{(k)}$, 
$C_3^{(\ell,m)}$, $C_{N^2/3}^{(1,p)}$ and $C_{N^2/3}^{(2,p)}$, 
are equal to 2, $(N^2-3)/3$, 3 and 3, respectively.
The total number of conjugacy classes is equal to 
$9 + (N^2-3)/3$.
The relations (\ref{eq:character-2-e}) and (\ref{eq:sum-dim})
for $\Delta (3N^2)$ with $N/3=$ integer  lead to
\begin{eqnarray}
&&m_1+2^2m_2+3^2m_3 + \cdots =3N^2,\\
&&m_1+m_2+m_3  + \cdots =9+ (N^2 -3)/3.
\end{eqnarray}
The solution is found as $(m_1,m_3)=(9,(N^2-3)/3)$.
That is, there are nine singlets and $(N^2-3)/3$ triplets.

\vskip .5cm
{$\bullet$ \bf Characters and representations}

Now, we study characters.
We start with $N/3 \neq $ integer.
In this case, there are 3 singlets.
Because $b^3=e$, characters of three singlets 
have three possible values $\chi_{1k}(b) = \omega^k$ with 
$k=0,1,2$ and 
they correspond to three singlets, ${\bf 1}_{k}$.
Note that $\chi_{1k}(a) = \chi_{1k}(a') = 1$, 
because $\chi_{1k}(b) = \chi_{1k}(ba) = \chi_{1k}(ba')$.
These characters are shown in Table \ref{tab:Delta-3N-character}.

Next, let us consider triplets for $N/3 \neq $ integer.
Indeed, the matrices (\ref{eq:delta-3n-3}) correspond to 
one of triplet representations.
Similarly, we can obtain $(3 \times 3)$ matrix representations 
for generic triplets, e.g. by replacing 
\begin{eqnarray}\label{eq:delta-3n-3rep}
a \rightarrow a^\ell a'^m, \qquad a' \rightarrow
b^2ab^{-2}=a^{-m}a'^{\ell -m}.
\end{eqnarray}
However, note that the following two types of replacing 
\begin{eqnarray}
 & & (a,a') \rightarrow (a^{-\ell+m} a'^{-\ell}, 
a^{\ell}a'^{m}), \nonumber \\
 & & (a,a') \rightarrow (a^{-m} a'^{\ell -m}, 
a^{-\ell+m}a'^{-\ell}),
\end{eqnarray}
also lead to the representation equivalent to 
the above (\ref{eq:delta-3n-3rep}), because 
the three elements, $a^\ell a'^m$, $a^{-\ell+m} a'^{-\ell}$ and 
$a^{-m} a'^{\ell -m}$, belong to the same conjugacy class, 
$C_3^{(\ell,m)}$.
Thus, the $\Delta(3N^2)$ group with $N/3 \neq $ integer 
is represented as 
\begin{eqnarray}
b= \begin{pmatrix}0   & 1 & 0 \\ 
                   0    & 0   &1    \\
                   1 & 0  & 0    \\
 \end{pmatrix},
 \quad
a= \begin{pmatrix}\rho^\ell   & 0 & 0 \\ 
                   0    & \rho^k   &0    \\
                   0 & 0  & \rho^{-k-\ell}    \\
 \end{pmatrix},
 \quad
a'= \begin{pmatrix}\rho^{-k-\ell}   & 0 & 0 \\ 
                   0    & \rho^\ell   &0    \\
                   0 & 0  & \rho^k    \\
 \end{pmatrix},
\end{eqnarray}
on the triplet ${\bf 3}_{[k][\ell]}$, where 
$[k][\ell]$ denotes \footnote{The notation 
$[k][\ell]$ corresponds to $\widetilde{(k,\ell)}$ in Ref.~\cite{Luhn:2007uq}.}
\begin{eqnarray}
[k][\ell]
=(k,\ell),\ \
(-k-\ell,k) \ \text{or}\
(\ell,-k-\ell) .
\end{eqnarray}
We also denote the vector of ${\bf 3}_{[k][\ell]}$ as 
\begin{eqnarray}
{\bf 3}_{[k][\ell]}
= \begin{pmatrix}x_{\ell,-k-\ell} \\ 
                  x_{k,\ell}   \\
                  x_{-k-\ell,k}    \\
 \end{pmatrix},
\end{eqnarray}
for $k,\ell =0,1,\cdots, N-1$,
where $k$ and $\ell$ correspond to $Z_N$ and $Z_N'$ charges, 
respectively.
When $(k,\ell)=(0,0)$, the matrices $a$ and $a'$ are 
proportional to the identity matrix.
Thus, we exclude the case with $(k,\ell)=(0,0)$.
The characters are shown in Table \ref{tab:Delta-3N-character}.
\begin{table}[t]
\begin{center}
\begin{tabular}{|c|c|c|c|c|c|c|c|c|}
\hline
&$h$&$\chi_{1_0}$&$\chi_{1_1}$&$\chi_{1_2}$&$\chi_{3_{{[0][1]}}}$&$\chi_{3_{[0][2]}}$&$\cdots$&$\chi_{3_{[n-1][n-1]}}$\\
\hline
$C_1$&1&1 & 1&1&3&3& &3 \\ 
\hline
$C_3^{(\ell,m)}$&$\frac{N}{\gcd(N,\ell,m)}$&1 & 1&1&$ \rho^{\ell-m}$ 
&$\rho^{2\ell-2m}$ && 
$ \rho^{n-1}(\rho^{\ell-2m}$\\ 
&& & &&$+\rho^{m}+\rho^{-\ell}$ 
&$+\rho^{2m}+\rho^{-2\ell}$ && 
$ +\rho^{\ell+m}+\rho^{-2\ell+m})$\\ 
\hline
$C^1_{N^2}$&$\frac{N}{\gcd(N,3,\ell,m)}$&1 & $\omega$&$\omega^2$     &0&0&&0 \\ 
\hline
$C^2_{N^2}$ &$\frac{N}{\gcd(N,3,\ell,m)}$& 1                  &$\omega^2$ &$\omega$&0&0&&0\\
\hline
\end{tabular}
\end{center}
\caption{Characters of $\Delta(3N^2)$ for $N/3 \neq $ integer}
\label{tab:Delta-3N-character}
\end{table}

Similarly, we study the characters and representations 
for $N/3=$  integer.
In this case, there are nine singlets.
Their characters must satisfy $\chi_\alpha(b)=\omega^k$ 
($k=0,1,2$) similarly to the above case.
In addition, it is found that
$\chi_\alpha(a)=\chi_\alpha(a')=\omega^\ell$ ($\ell=0,1,2$).
Thus, nine singlets can be specified by combinations of 
$\chi_\alpha(b)$ and $\chi_\alpha(a)$, i.e. 
${\bf 1}_{k,\ell}$ ($k,\ell = 0,1,2$) with $\chi_\alpha(b)=\omega^k$ 
and  $\chi_\alpha(a)=\chi_\alpha(a')=\omega^\ell$.
These characters are shown in Table \ref{tab:Delta-3N-chara-2}.

The triplet representations are also given similarly to 
the case with $N/3 \neq $  integer.
That is, the $\Delta(3N^2)$ group with $N/3 = $ integer 
is represented as 
\begin{eqnarray}
b= \begin{pmatrix}0   & 1 & 0 \\ 
                   0    & 0   &1    \\
                   1 & 0  & 0    \\
 \end{pmatrix},
 \quad
a= \begin{pmatrix}\rho^\ell   & 0 & 0 \\ 
                   0    & \rho^k   &0    \\
                   0 & 0  & \rho^{-k-\ell}    \\
 \end{pmatrix},
 \quad
a'= \begin{pmatrix}\rho^{-k-\ell}   & 0 & 0 \\ 
                   0    & \rho^\ell   &0    \\
                   0 & 0  & \rho^k    \\
 \end{pmatrix},
\end{eqnarray}
on the triplet ${\bf 3}_{[k][\ell]}$.
However, note that when 
$(k,\ell) =(0,0),(N/3,N/3),(2N/3,2N/3)$ 
the matrices, $a$ and $a'$ are trivial.
Thus, we exclude such values of $(k,\ell)$.
These characters are shown in Table \ref{tab:Delta-3N-chara-2}.

\begin{table}[t]
\begin{center}
\begin{tabular}{|c|c|c|c|c|c|c|}
\hline
& $\!\!h\!\!$ & $\!\!\chi_{1_{(r,s)}}\!\!$ & $\!\!\chi_{3_{[0][1]}}\!\!$ & $\!\!\chi_{3_{[0][2]}}\!\!$
& $\!\!\cdots\!\!$ & $\!\!\chi_{3_{[n-1][n-1]}}\!\!$\\\hline
$\!\!1C_1\!\!$ & $\!\!1\!\!$  & $\!\!1\!\!$ & $\!\!3\!\!$ &$\!\!3\!\!$ & & $\!\!3\!\!$\\ 
\hline
$\!\!1C_1^{(k)}\!\!$ & $\!\!\frac{N}{\gcd(N,N/3,2N/3)}\!\!$ & $\!\!1\!\!$ & $\!\!\rho^{2k}+2\rho^{-k}\!\!$ & 
$\!\!\rho^{4k}+2\rho^{-2k}\!\!$ & & $\!\!1+ \rho^{3(n-1)\rho}+\rho^{-3(n-1)\rho}\!\!$ \\ 
\hline
$\!\!3C_1^{(\ell,m)}\!\!$ & $\!\!\frac{N}{\gcd(N,\ell,m)}\!\!$ & $\!\!\omega^{s(\ell+m)}\!\!$ & 
$\!\!\rho^{\ell-m}\!\!$ & $\!\!\rho^{2\ell-2m}\!\!$ & & $\!\!\rho^{n-1}(\rho^{\ell-2m}\!\!$\\ 
& & & $\!\!+\rho^{m}+\rho^{-\ell}\!\!$ 
& $\!\!+\rho^{2m}+\rho^{-2\ell}\!\!$ && 
$\!\!+\rho^{\ell+m}+\rho^{-2\ell+m})\!\!$\\ 
\hline
$\!\!C_{N^2/3}^{(1,p)}\!\!$ & $\!\!\frac{N}{\gcd(N,3,p-n-3m,n)}\!\!$ & $\!\!\omega^{r+s p}\!\!$ 
& $\!\!0\!\!$ & $\!\!0\!\!$ & & $\!\!0\!\!$ \\ 
\hline
$\!\!C_{N^2/3}^{(2,p)}\!\!$ & $\!\!\frac{N}{\gcd(N,3,p-n-3m,n)}\!\!$ & $\!\!\omega^{2r+s p}\!\!$      
& $\!\!0\!\!$ & $\!\!0\!\!$ & & $\!\!0\!\!$\\
\hline
\end{tabular}
\end{center}
\caption{Characters of $\Delta(3N^2)$ for $N/3=$ integer}
\label{tab:Delta-3N-chara-2}
\end{table}

\vskip .5cm
{$\bullet$ \bf Tensor products}

Now, we study tensor products.
First, we consider the $\Delta(3N^2)$ with $N/3 \neq $ integer.
Because of their $Z_N \times Z_N'$ charges, 
tensor products of triplets ${\bf 3}$ can be obtained as 
\begin{eqnarray}\label{eq:delta3n-3-3}
\begin{pmatrix}x_{\ell,-k-\ell} \\ 
                  x_{k,\ell}   \\
                  x_{-k-\ell,k}    \\
 \end{pmatrix}_{{\bf 3}_{[k][\ell] }} &\otimes& 
\begin{pmatrix}y_{ \ell',-k'-\ell'} \\ 
                  y_{k', \ell'}   \\
                  y_{-k'-\ell', k'}    \\
 \end{pmatrix}_{{\bf 3}_{[k'][\ell'] }} 
=
 \begin{pmatrix}x_{\ell,-k-\ell}y_{ \ell',-k'-\ell'} \\ 
                  x_{k,\ell}y_{  k', \ell'}    \\
                  x_{-k-\ell,k}y_{-k'-\ell', k'}    \\
 \end{pmatrix}_{{\bf 3}_{([k+k'][\ell+\ell']}}   \nonumber \\
&\oplus&
\begin{pmatrix}x_{-k-\ell,k}y_{ \ell',-k'-\ell'} \\ 
                  x_{\ell,-k-\ell}y_{  k', \ell'}    \\
                  x_{k,\ell}y_{-k'-\ell', k'}    \\
 \end{pmatrix}_{{\bf 3}_{[\ell+k'][-k-\ell+\ell']}}   
\oplus
 \begin{pmatrix}x_{\ell,-k-\ell}y_{-k'-\ell',k'} \\ 
                  x_{k,\ell}y_{\ell',-k'-\ell'}    \\
                  x_{-k-\ell,k}y_{k',\ell'}    \\
 \end{pmatrix}_{{\bf 3}_{[k+\ell'][\ell-k'-\ell']}}
\end{eqnarray}
 
for $-(k,\ell) \neq [k'][\ell']$,
\begin{eqnarray}
\begin{pmatrix} x_{\ell,-k-\ell} \\ 
                  x_{k,\ell}   \\
                  x_{-k-\ell,k}    \\
 \end{pmatrix}_{{\bf 3}_{[k][\ell] }} &\otimes& 
\begin{pmatrix}y_{-\ell,k+\ell} \\ 
                   y_{-k,-\ell}   \\
                   y_{k+\ell,-k}    \\
\end{pmatrix}_{{\bf 3}_{-[k][\ell]}} 
 = 
\left(
x_{\ell,-k-\ell}y_{-\ell,k+\ell}
+x_{k,\ell}y_{-k,-\ell}
+x_{-k-\ell,k}y_{k+\ell,-k}
\right)_{{\bf 1}_0}  \nonumber \\
&\oplus&
\left(x_{\ell,-k-\ell}y_{-\ell,k+\ell}
+\omega^2x_{k,\ell}y_{-k,-\ell}
+\omega x_{-k-\ell,k}y_{k+\ell,-k}
\right)_{{\bf 1}_1} \nonumber \\
&\oplus&
\left(
x_{\ell,-k-\ell}y_{-\ell,k+\ell}
+\omega x_{k,\ell}y_{-k,-\ell}
+\omega^2x_{-k-\ell,k}y_{k+\ell,-k}
\right)_{{\bf 1}_2}   \nonumber \\
&\oplus&
\begin{pmatrix}x_{-k-\ell,k}y_{-\ell,k+\ell} \\ 
                  x_{\ell,-k-\ell}y_{-k,-\ell}    \\
                   x_{k,\ell}y_{k+\ell,-k}    \\
 \end{pmatrix}_{{\bf 3}_{[-k+\ell][-k-2\ell]}}
\oplus 
\begin{pmatrix}x_{\ell,-k-\ell}y_{k+\ell,-k} \\ 
                  x_{k,\ell}y_{-\ell,k+\ell}    \\
                  x_{-k-\ell,k}y_{-k,-\ell}    \\
 \end{pmatrix}_{{\bf 3}_{[k-\ell][k+2\ell]}}.
\end{eqnarray}
A product of ${\bf 3}_{[k][\ell]}$ and ${\bf 1}_r$ is obtained as 
\begin{eqnarray}
\begin{pmatrix}
x_{(\ell,-k-\ell)}\\
x_{(k,\ell)}\\
x_{(-k-\ell,k)}
\end{pmatrix}_{[k][\ell]}
\otimes
(z_{r})_{{\bf 1}_{r}}
=
\begin{pmatrix}
x_{(\ell,-k-\ell)}z_{r}\\
\omega^{r}x_{(k,\ell)}z_{r}\\
\omega^{2r}x_{(-k-\ell,k)}z_{r}
\end{pmatrix}_{[k][\ell]}.
\end{eqnarray}
The tensor products of singlets ${\bf 1}_k$ and ${\bf 1}_{k'}$ 
are obtained simply as 
\begin{eqnarray}
{\bf 1}_k \otimes {\bf 1}_{k'} = {\bf 1}_{k+k'}.
\end{eqnarray}

Next, we study tensor products of $N/3=$ integer.
We consider the tensor products of two triplets,
\begin{eqnarray}
\begin{pmatrix}x_{\ell,-k-\ell} \\ 
                  x_{k,\ell}   \\
                  x_{-k-\ell,k}    \\
 \end{pmatrix}_{{\bf 3}_{[k][\ell] }} ~~~{\rm and}~~~
\begin{pmatrix}y_{ \ell',-k'-\ell'} \\ 
                  y_{k', \ell'}   \\
                  y_{-k'-\ell', k'}    \\
 \end{pmatrix}_{{\bf 3}_{[k'][\ell'] }} . 
\end{eqnarray}
Unless $(k',\ell') = -[k+mN/3][\ell+mN/3]$ for 
$m=0,1,2$, their tensor products are the same as (\ref{eq:delta3n-3-3}).
Thus, we do not repeat them.
For $(k',\ell') = -[-k+mN/3][-\ell+mN/3]$ ($m=0,1$ or 2), tensor 
products of the above triplets are obtained as 
\begin{eqnarray}
& & 
\begin{pmatrix}x_{\ell,-k-\ell} \\ 
                  x_{k,\ell}   \\
                  x_{-k-\ell,k}    \\
 \end{pmatrix}_{{\bf 3}_{[k][\ell] }} \otimes 
\begin{pmatrix}y_{-\ell+mN/3,k+\ell-2mN/3} \\ 
                  y_{-k+mN/3,-\ell+mN/3}   \\
                   y_{k+\ell-2mN/3,-k+mN/3}    \\
\end{pmatrix}_{{\bf 3}_{[-k+mN/3][-\ell+mN/3]}} \nonumber \\
& & = 
\left(
x_{\ell,-k-\ell}y_{-\ell,k+\ell-2mN/3}
+x_{k,\ell}y_{-k+mN/3,-\ell+mN/3}
+x_{-k-\ell,k}y_{k+\ell-2mN/3,-k+mN/3}
\right)_{{\bf 1}_{0,m}}  \nonumber \\
&\oplus&
\!\!\!\!\!\!\left(x_{\ell,-k-\ell}y_{-\ell+mN/3,k+\ell-2mN/3}
+\omega^2x_{k,\ell}y_{-k+mN/3,-\ell+mN/3}
+\omega x_{-k-\ell,k}y_{k+\ell-2mN/3,-k+mN/3}
\right)_{{\bf 1}_{1,m}}\!\!\!\!\!\! \nonumber \\
&\oplus&
\!\!\!\!\!\!\left(
x_{\ell,-k-\ell}y_{-\ell+mN/3,k+\ell-2mN/3}
+\omega x_{k,\ell}y_{-k+mN/3,-\ell+mN/3}
+\omega^2x_{-k-\ell,k}y_{k+\ell-2mN/3,-k+mN/3}
\right)_{{\bf 1}_{2,m}}\!\!\!\!\!\!   \nonumber \\
&\oplus&
\!\!\!\!\!\begin{pmatrix}x_{-k-\ell,k}y_{-\ell+mN/3,k+\ell-2mN/3} \\ 
                   x_{\ell,-k-\ell}y_{-k+mN/3,-\ell+mN/3}    \\
                   x_{k,\ell}\phi'_{k+\ell-2mN/3,-k+mN/3}    \\
 \end{pmatrix}_{{\bf 3}_{[-k+\ell+mN/3][-k-2\ell+mN/3]}}\!\!\!\!\! \nonumber \\
&\oplus&
\!\!\!\!\!\begin{pmatrix}x_{\ell,-k-\ell}y_{k+\ell-2mN/3,-k+mN/3} \\ 
                   x_{k,\ell}y_{-\ell+mN/3,k+\ell-2mN/3}    \\
                   x_{-k-\ell,k}y_{-k+mN/3,-\ell+mN/3}    \\
 \end{pmatrix}_{{\bf 3}_{[k-\ell+mN/3][k+2\ell-2mN/3]}}.\!\!\!\!\!
\end{eqnarray}
A product of ${\bf 3}_{[k][\ell]}$ and ${\bf 1}_{r,s}$ is
\begin{eqnarray}
\begin{pmatrix}
x_{(\ell,-k-\ell)}\\
x_{(k,\ell)}\\
x_{(-k-\ell,k)}
\end{pmatrix}_{{\bf 3}_{[k][\ell]}}
\otimes
(z_{r,s})_{{\bf 1}_{r,s}}
=
\begin{pmatrix}
x_{(\ell,-k-\ell)}z_{r,s}\\
\omega^{r}x_{(k,\ell)}z_{r,s}\\
\omega^{2r}x_{(-k-\ell,k)}z_{r,s}
\end{pmatrix}_{{\bf 3}_{[k+sN/3][\ell+sN/3]}}.
\end{eqnarray}
The tensor products of singlets ${\bf 1}_{k,\ell}$ and ${\bf 1}_{k',\ell'}$ 
are obtained simply as 
\begin{eqnarray}\label{eq:tensor-27-1-1}
{\bf 1}_{k,\ell} \otimes {\bf 1}_{k',\ell'} = {\bf 1}_{k+k',\ell+\ell'}.
\end{eqnarray}

\subsection{$\Delta(27)$}

The $\Delta(3)$ is nothing but the $Z_3$ group and 
$\Delta(12)$ is isomorphic to $A_4$.
Thus, the simple and non-trivial example is $\Delta(27)$.

The conjugacy classes of $\Delta(27)$ are obtained as 
\begin{eqnarray}\label{eq:conjugacy-27}
\begin{array}{ccc}
 C_1:&\{ e \}, &  h=1,\\
 C_1^{(1)}:&\{ a, a'^2 \}, &  h=3,\\
 C_1^{(2)}:&\{ a^2, a' \}, &  h=3,\\
 C_3^{(0,1)}: &\{ a',a,a^2a'^2 \}, & h=3,\\
 C_3^{(0,2)}: &\{ a'^2,a^2,aa' \}, & h=3,\\
 C_3^{(1,p)}: &\{ ba^p,ba^{p-1}a',ba^{p-2}a'^2 \}, & h=3,\\
 C_3^{(2,p)}: &\{ ba^p,ba^{p-1}a',ba^{p-2}a'^2 \}, & h=3.\\
\end{array}
\end{eqnarray}

The $\Delta(27)$ has nine singlets ${\bf 1}_{r,s}$ 
($r,s=0,1,2$) and two triplets, ${\bf 3}_{[0][1]}$ and 
${\bf 3}_{{[0][2]}}$.
The characters are shown in Table \ref{tab:delta-27}.

\begin{table}[t]
\begin{center}
\begin{tabular}{|c|c|c|c|c|}
\hline
&h&$\chi_{1_{(r,s)}}$&$\chi_{3_{[0,1]}}$&$\chi_{3_{[0,2]}}$\\
\hline
$1C_1$&1&1 & 3&3\\ 
\hline
$1C_1^{(1)}$&1&1 & $3\omega^2$&$3\omega$\\ 
\hline
$1C_1^{(2)}$&1&1 & $3\omega$&$3\omega^2$\\ 
\hline
$3C_1^{(0,1)}$&$3$&$\omega^{s}$ & 
$ 0$ &$0$ \\ 
\hline
$3C_1^{(0,2)}$&$3$&$\omega^{2s}$ & 
$ 0$ &$0$ \\ 
\hline
$C_3^{(1,p)}$&$3$&$\omega^{r+s p}$      &0&0 \\ 
\hline
$C_3^{(2,p)}$&$3$ & $\omega^{2r+s p}$      &0&0\\
\hline
\end{tabular}
\end{center}
\caption{Characters of $\Delta(27)$}
\label{tab:delta-27}
\end{table}

Tensor products between triplets are obtained as 
\begin{equation}\label{eq:delta27-3-3}
\begin{pmatrix} x_{1,-1} \\ 
                   x_{0,1}   \\
                   x_{-1,0}    \\
 \end{pmatrix}_{{\bf 3}_{[0][1] }} \otimes 
\begin{pmatrix}y_{ 1,-1} \\ 
                   y_{0, 1}   \\
                   y_{-1, 0}    \\
 \end{pmatrix}_{{\bf 3}_{[0][1] }} 
=
 \begin{pmatrix}x_{1,-1}y_{1,-1} \\ 
                   x_{0,1}y_{  0, 1}    \\
                   x_{-1,0}y_{-1, 0}    \\
 \end{pmatrix}_{{\bf 3}_{[0][2]}} 
\oplus
\begin{pmatrix}x_{-1,0}y_{ 1,-1} \\ 
                   x_{1,-1}y_{  0,1}    \\
                   x_{0,1}y_{-1, 0}    \\
 \end{pmatrix}_{{\bf 3}_{[0][2]}}   
\oplus
 \begin{pmatrix}x_{1,-1}y_{-1,0} \\ 
                   x_{0,1}y_{1,-1}    \\
                   x_{-1,0}y_{0,1}    \\
 \end{pmatrix}_{{\bf 3}_{[0][2]}} ,
\end{equation}

\begin{equation}\label{eq:delta27-3-3}
\begin{pmatrix} x_{2,-2} \\ 
                   x_{0,2}   \\
                   x_{-2,0}    \\
 \end{pmatrix}_{{\bf 3}_{[0][2] }} \otimes 
\begin{pmatrix}y_{ 2,-2} \\ 
                   y_{0, 2}   \\
                   y_{-2, 0}    \\
 \end{pmatrix}_{{\bf 3}_{[0][2] }} 
=
 \begin{pmatrix}x_{2,-2}y_{ 2,-2} \\ 
                   x_{0,2}y_{  0, 2}    \\
                   x_{-2,0}y_{-2, 0}    \\
 \end{pmatrix}_{{\bf 3}_{[0][1]}} 
\oplus
\begin{pmatrix}x_{-2,0}y_{ 2,-2} \\ 
                   x_{2,-2}y_{  0, 2}    \\
                   x_{0,2}y_{-2, 0}    \\
 \end{pmatrix}_{{\bf 3}_{[0][1]}}   
\oplus
 \begin{pmatrix}x_{2,-2}y_{-2,0} \\ 
                   x_{0,2}y_{2,-2}    \\
                   x_{-2,0}y_{0,2}    \\
 \end{pmatrix}_{{\bf 3}_{[0][1]}} ,
\end{equation}

\begin{eqnarray}\label{eq:delta27-3-3d}
\begin{pmatrix} x_{1,-1} \\ 
                   x_{0,1}   \\
                   x_{-1,0}    \\
 \end{pmatrix}_{{\bf 3}_{[0][1] }} \otimes 
\begin{pmatrix} y_{ -1,1} \\ 
                   y_{0, -1}   \\
                   y_{1, 0}    \\
 \end{pmatrix}_{{\bf 3}_{[0][2] }} 
=&&
\sum_r(x_{1,-1}y_{ -1,1}
+
\omega^{2r}x_{0,1}y_{0, -1}
+
\omega^{r}x_{-1,0}y_{1, 0})_{{\bf 1}_{(r,0)}}
\nonumber\\
&\oplus&
\sum_r(x_{1,-1}y_{ 0,-1}
+\omega^{2r}x_{0,1}y_{1,0}
+\omega^{r}x_{-1,0}y_{-1,1})_{{\bf 1}_{(r,1)}}
\nonumber\\
&\oplus&
\sum_r(x_{1,-1}y_{ 1,0}
+\omega^{2r}x_{0,1}y_{-1,1}
+\omega^{r}x_{-1,0}y_{0, -1})_{{\bf 1}_{(r,2)}}.\nonumber\\
\end{eqnarray}

The tensor products between singlets and triplets are obtained as 

\begin{eqnarray}
&&\begin{pmatrix}
x_{(1,-1)}\\
x_{(0,1)}\\
x_{(-1,0)}
\end{pmatrix}_{{\bf 3}_{[0][1]}}
\otimes
(z_{r,s})_{1_{r,s}}
=
\begin{pmatrix}
x_{(1,-1)}z_{r,s}\\
\omega^{r}x_{(0,1)}z_{r,s}\\
\omega^{2r}x_{(-1,0)}z_{r,s}
\end{pmatrix}_{{\bf 3}_{[s][1+s]}},
\nonumber\\
&&\begin{pmatrix}
x_{(2,-2)}\\
x_{(0,2)}\\
x_{(-2,0)}
\end{pmatrix}_{{\bf 3}_{[0][2]}}
\otimes
(z_{r,s})_{1_{r,s}}
=
\begin{pmatrix}
x_{(2,-2)}z_{r,s}\\
\omega^{r}x_{(0,2)}z_{r,s}\\
\omega^{2r}x_{(-2,0)}z_{r,s}
\end{pmatrix}_{{\bf 3}_{[s][2+s]}}.
\end{eqnarray}
The tensor products of singlets are obtained as Eq.~(\ref{eq:tensor-27-1-1}).

\clearpage

\section{$ T_7$}
\label{sec:T7}

It is useful to construct a discrete group as a subgroup of 
known groups.
Through such a procedure, we can obtain 
group-theoretical aspects such as representations 
of the subgroup from those of larger groups.
As such an example, here we study $T_7$, which is 
isomorphic to $Z_7\rtimes Z_3$ and a subgroup of 
$\Delta(3N^2)$ with $N=7$.
The discrete group $T_7$\cite{Hagedorn:2008bc} is known as the minimal
non-Abelian discrete group with respect to having a complex triplet.

We denote the generators of $Z_7$ by $a$ and $Z_3$ generator is written by $b$. They satisfy 
\begin{equation}
a^7=1, \quad ab=ba^4 .
\end{equation}
Using them, all of $T_7$ elements are written as
\begin{equation}
g=b^{m}a^{n} ,
\end{equation}
with $m=0,1,2$ and $n=0,\cdots,6$.

The generators, $a$ and $b$, are represented e.g. as 
\begin{equation}
b=\Mat3{0}{1}{0} {0}{0}{1} {1}{0}{0},\quad 
a=\Mat3{\rho}{0}{0} {0}{\rho^2}{0} {0}{0}{\rho^4},
\end{equation}
where $\rho=e^{2i\pi/7}$.
These elements are classified into five conjugacy classes,
\begin{eqnarray}
\begin{array}{ccc}
 C_1:&\{ e \}, &  h=1,\\
 C_7^{(1)}:&\{b~,~ ba~,~ba^{2}~,~ba^{3}~,~ ba^{4}~,~ba^{5}~,~ba^{6}\}, &  
h=3,\\
  C_7^{(2)}:&\{b^2~,~ b^2a~,~b^2a^{2}~,~b^2a^{3}~,~ 
b^2a^{4}~,~b^2a^{5}~,~b^2a^{6}\}, &  h=3,\\
 C_{3}:&\{ a~,~a^{2}~,~a^{4} \},&   
 h=7,\\
 C_{\bar3}:&\{ a^{3}~,~a^{5}~,~a^{6} \},&   
 h=7.\\
\end{array}
\end{eqnarray}
The $T_7$ group has three singlets ${\bf 1}_k$ with $k=0,~1,~2$ and two triplets ${\bf 3}$ and ${\bf \bar3}$.
The characters are shown in Table \ref{T7}, where 
$\xi=\frac{-1+i\sqrt7}{2}$.

\begin{table}[t]
\begin{center}
\begin{tabular}{|c|c|c|c|c|c|c|c|}
\hline
        & $n$ & 
$h$&$\chi_{\bf 1_{0}}$&$\chi_{\bf 1_{1}}$&$\chi_{\bf1_{2}}$&$\chi_{\bf3}$&$\chi_{\bf\bar3}$ \\ \hline
$C^{(0)}_1$  & $1$ &$1$&   $1$  &    $1$    &    $1$     &$3$ & $3$   \\ 
\hline
$C^{(1)}_{7}$  & $7$ &$3$&   $1$  &    $\omega$    &    $\omega^2$     
&$0$&$0$  \\ \hline
$C^{(2)}_{7}$ & $7$  &$3$&   $1$  & $\omega^2$  & $\omega$ &$0$ & $0$   \\ 
\hline
$C_{3}$ &  $3$  &$7$&   $1$  & $1$  & $1$ &$\xi$ &$\bar\xi$  \\ \hline
$C_{\bar3}$ & $3$ &$7$&   $1$  & $1$&  $1$  &$\bar\xi$ &$\xi$  \\
\hline
\end{tabular}
\end{center}
\caption{Characters of $T_7$}
\label{T7}
\end{table}

Using the order of $\rho$ in $a$, 
we define the triplet ${\bf3}$( ${\bf\bar3}$) as
\begin{eqnarray}
{\bf3}\equiv\3tvec{x_1}{x_2}{x_4},\quad
{\bf\bar3}\equiv\3tvec{x_{-1}}{x_{-2}}{x_{-4}}=\3tvec{x_6}{x_5}{x_3}.
\end{eqnarray}

The tensor products between triplets are obtained as 
\begin{eqnarray}
\3tvec{x_1}{x_2}{x_4}_{{\bf 3}}\otimes\3tvec{y_1}{y_2}{y_4}_{{\bf 3}}
&=&
\3tvec{x_2y_{4}}{x_4y_{1}}{x_1y_{2}}_{{\bf \bar3}}\oplus
\3tvec{x_4y_{2}}{x_1y_{4}}{x_2y_{1}}_{{\bf \bar3}}\oplus
\3tvec{x_4y_{4}}{x_1y_{1}}{x_2y_{2}}_{{\bf 3}},
\\
\3tvec{x_6}{x_5}{x_3}_{{\bf \bar3}}\otimes\3tvec{y_6}{y_5}{y_3}_{{\bf 
\bar3}}
&=&
\3tvec{x_5y_{3}}{x_3y_{6}}{x_6y_{5}}_{{\bf 3}}\oplus
\3tvec{x_3y_{5}}{x_6y_{3}}{x_5y_{6}}_{{\bf 3}}\oplus
\3tvec{x_3y_{3}}{x_6y_{6}}{x_5y_{5}}_{{\bf \bar3}},
\\
\3tvec{x_1}{x_2}{x_4}_{{\bf 3}}\otimes\3tvec{y_6}{y_5}{y_3}_{{\bf \bar3}}
&=&
\3tvec{x_2y_{6}}{x_4y_{5}}{x_1y_{3}}_{{\bf 3}}\oplus
\3tvec{x_1y_{5}}{x_2y_{3}}{x_4y_{6}}_{{\bf \bar3}} \nonumber \\
& &\oplus\sum_{k=0,1,2} (x_1y_6+\omega^k x_2y_5+\omega^{2k} x_4y_3)_{{\bf1}_k}
.
\end{eqnarray}
The tensor products between singlets are obtained as 
\begin{eqnarray}
& & (x)_{{\bf1}_0}(y)_{{\bf1}_0}=(x)_{{\bf1}_1}(y)_{{\bf1}_2}=(x)_{{\bf1}_2}
(y)_{{\bf1}_1}=(xy)_{{\bf1}_0},~ \nonumber \\
& & (x)_{{\bf1}_1}(y)_{{\bf1}_1}=(xy)_{{\bf1}_2},~
(x)_{{\bf1}_2}(y)_{{\bf1}_2}=(xy)_{{\bf1}_1}.
\end{eqnarray}
\\
The tensor products between triplets and singlets are obtained as 
\begin{eqnarray}
(y)_{{\bf 1}_k}\otimes \3tvec{x_{1(6)}}{x_{2(5)}}{x_{4(3)}}_{{\bf 3(\bar3)}}= 
\3tvec{yx_{1(6)}}{yx_{2(5)}}{yx_{4(3)}}_{{\bf 3(\bar3)}} .
\end{eqnarray}

\clearpage


\section{$\Sigma(3N^3)$}

\subsection{Generic aspects}

The discrete group $\Sigma (3N^3)$ is defined as a 
closed algebra of 
three Abelian symmetries, $Z_N$, $Z_N'$ and $Z_N''$, 
which commute each other, and their $Z_3$ permutations.
That is, when we denote the generators of $Z_N$, $Z_N'$ 
and $Z_N''$ by $a$, $a'$ and $a''$, 
respectively, and the $Z_3$ generator is written by $b$, 
all of $\Sigma(3N^3)$ elements are written as 
\begin{equation}
g=b^ka^ma'^na''^{\ell} ,
\end{equation}
with $k=0,1,2$, and $m,n,\ell=0,...,N-1$, 
where $a$, $a'$, $a''$ and $b$ satisfy the following relations:
\begin{eqnarray}
&&a^N=a'^N=a''^N=1,
\quad
aa'=a'a, ~~aa''=a''a, ~~a''a'=a'a'',
\quad
b^3=1 ,
\nonumber\\
&&b^2ab=a'',\quad b^2a'b=a, \quad  b^2a''b=a'.
\end{eqnarray}
These generators, $a$, $a'$, $a''$ and $b$, are represented, e.g, as
\begin{eqnarray}
b=\Mat3{0}{1}{0} {0}{0}{1} {1}{0}{0},~~ 
a=\Mat3{1}{0}{0} {0}{1}{0} {0}{0}{\rho} ,~~
a'=\Mat3{1}{0}{0} {0}{\rho}{0} {0}{0}{1} ,~~ a''=\Mat3{\rho}{0}{0} 
{0}{1}{0} {0}{0}{1} ,
\end{eqnarray}
where $\rho=e^{2i\pi/N}$. Then, all of $\Sigma(3N^3)$ elements are written 
as 
\begin{eqnarray}
\Mat3{0}{\rho^n}{0} {0}{0}{\rho^m} {\rho^{\ell}}{0}{0},\quad 
\Mat3{\rho^{\ell}}{0}{0} {0}{\rho^m}{0} {0}{0}{\rho^n},
\quad
\Mat3{0}{0}{\rho^m} {\rho^{\ell}}{0}{0} {0}{\rho^n}{0}.
\end{eqnarray}

When $N=2$, the element $aa'a''$ commutes with all of the elements.
In addition, when we define $\tilde a=aa''$ and $\tilde a'=a'a''$, 
the closed algebra among $\tilde a$, $\tilde a'$ and $b$ 
corresponds to $\Delta(12)$, where the element $aa'a''$ is not 
included.
That is, this group is isomorphic to $Z_2 \times \Delta(12)$.

The situation for $N=3$ is different.
It is the same as the fact that the element $aa'a''$ commutes with all of the elements.
Furthermore, when we define 
$\tilde a=a^2a''$ and $\tilde a'=a'a''^{2}$, 
the closed algebra among $\tilde a$, $\tilde a'$ and $b$ 
corresponds to $\Delta(27)$.
However, since the element $aa'a''$ is written by 
$aa'a'' = \tilde a^2 \tilde a' $ in this case, 
the element is inside of $\Delta(27)$.
Thus, the group $\Sigma(81)$ is not $Z_3 \times \Delta(27)$, 
but isomorphic to $(Z_3 \times Z_3' \times Z_3'') \rtimes Z_3$.

Similarly, for generic value of $N$, the element 
$aa'a''$ commutes with all of the elements.
When we define $\tilde a=a^{N-1}a''$ and $\tilde a'=a'a''^{N-1}$, 
the closed algebra among $\tilde a$, $\tilde a'$ and $b$ 
corresponds to $\Delta(3N^2)$.
When $N/3 \neq $ integer, the element $aa'a''$ is not 
included in $\Delta(3N^2)$.
Thus, we find that this group is isomorphic to $Z_N\times \Delta(3N^2)$. 
On the other hand, when $N/3 = $ integer, 
the element  $aa'a''$ is included in $\Delta(3N^2)$.
That is, the group $\Sigma (3N^3)$ can not be 
$Z_N\times \Delta(3N^2)$. 
The group $\Sigma (3N^3)$ with $N/3 = $ integer 
 has $N(N^2+8)/3$ conjugacy classes, $3N$ singlets, 
and $N(N^2-1)/3$ triplets. \\

 
\subsection{$\Sigma(81)$}
$\Sigma(81)$ has eighty-one elements and 
those  
are written as $b^ka^ma'^na''^{\ell} $for $k=0,1,2$ and 
$m,n,\ell=0,~1,~2$, where $a,~a',~a''$, 
and $b$ satisfy $a^3=a'^3=a''^3=1, ~aa'=a'a, ~aa''=a''a,
~a''a'=a'a'',~b^3=1$, $b^2ab=a''$, $b^2a'b=a$ and $b^2a''b=a'$. 
These elements are classified into seventeen conjugacy classes,

\begin{eqnarray}
\begin{array}{ccc}
 C_1:&\{ e \}, &  h=1,\\
 C_1^{(1)}:&\{ aa'a''\}, &  h=3,\\
  C_1^{(2)}:&\{ (aa'a'')^2\}, &  h=3,\\
 C_{3}^{0)}:&\{ 
a^{0}a'^{1}a''^{2}~,~a''^{0}a^{1}a'^{2}~,~a'^{0}a''^{1}a^{2} \},&   
 h=3,\\
 C_{3}^{'(0)}:&\{ 
a^{1}a'^{0}a''^{2}~,~a''^{1}a^{0}a'^{2}~,~a'^{1}a''^{0}a^{2} \},&   
 h=3,\\
 C_{3}^{(1)}:&\{ 
a^{0}a'^{0}a''^{1}~,~a''^{0}a^{0}a'^{1}~,~a'^{0}a''^{0}a^{1} \},&   
 h=3, \\
  C_{3}^{'(1)}:&\{ 
a^{0}a'^{2}a''^{2}~,~a''^{0}a^{2}a'^{2}~,~a'^{0}a''^{2}a^{2} \},&   
 h=3,\\
  C_{3}^{''(1)}:&\{ 
a^{1}a'^{1}a''^{2}~,~a''^{1}a^{1}a'^{2}~,~a'^{1}a''^{1}a^{2} \},&   
 h=3,\\
 C_{3}^{(2)}:&\{ 
a^{0}a'^{0}a''^{2}~,~a''^{0}a^{0}a'^{2}~,~a'^{0}a''^{0}a^{2} \},&   
 h=3,\\
  C_{3}^{'(2)}:&\{ 
a^{0}a'^{1}a''^{1}~,~a''^{0}a^{1}a'^{1}~,~a'^{0}a''^{1}a^{1} \},&   
 h=3,\\
 C_{3}^{''(2)}:&\{ 
a^{1}a'^{2}a''^{2}~,~a''^{1}a^{2}a'^{2}~,~a'^{1}a''^{2}a^{2} \},&   
 h=3,\\
 C_{9}^{0}:&\{ba^aa'^ba''^c\}& 
 h=3,\\
  C_{9}^{1}:&\{ba^aa'^ba''^c\}& 
 h=9,\\
   C_{9}^{2}:&\{ba^aa'^ba''^c\}& 
 h=9,\\
 C_{9}^{'(0)}:&\{b^2a^{a}a'^{b}a''^{c}\},& 
 h=3,\\
 C_{9}^{'(1)}:&\{b^2a^{a}a'^{b}a''^{c}\},& 
 h=9,\\
  C_{9}^{''(2)}:&\{b^2a^{a}a'^{b}a''^{c}\},& 
 h=9,\\
\end{array}
\end{eqnarray}
where we have shown also 
 the orders of each element in the conjugacy class by $h$.

The relations (\ref{eq:character-2-e}) and (\ref{eq:sum-dim})
for $\Sigma(81)$
lead to
\begin{eqnarray}
&&m_1+2^2m_2+3^2m_3 + \cdots =81,\\
&&m_1+m_2+m_3 + \cdots =17.
\end{eqnarray}
The solution is found as $(m_1,m_3)=(9,8)$.
That is, there are nine singlets ${\bf 1}^k_\ell$ with $k,\ell=0,1,2 $ 
and eight triplets, ${\bf 3}_A$, ${\bf 3}_B$, ${\bf 3}_C$, ${\bf
  3}_D$, ${\bf \bar3}_{A}$, ${\bf\bar 3}_{B}$, ${\bf\bar 3}_{C}$ and 
${\bf \bar 3}_{D}$.
The character tables are given by Tables \ref{81tab:sigma-singlet-character} 
and  \ref{81tab:sigma-triplet-character}.
\begin{table}[t]
\begin{center}
\begin{tabular}{|c|c|c|c|c|c|c|c|c|c|c|}
\hline
   & $h$ & $\chi_{1^{0}_{0}}$ & $\chi_{1^{0}_{1}}$  & 
$\chi_{1^{0}_{2}}$  
   & $\chi_{1^{1}_{0}}$ & $\chi_{1^{1}_{1}}$  & $\chi_{1^{1}_{2}}$ 
    & $\chi_{1^{2}_{0}}$ & $\chi_{1^{2}_{1}}$  & $\chi_{1^{2}_{2}}$     
\\ \hline
$C_1$ & 1 &  1 &1 & 1 &1 & 1 & $1$ &  1 & $1$ & $1$   \\ \hline
$C_1^{(1)}$  & 1 &  1 &1 & 1 &1 & 1 & $1$ &  1 & $1$ & $1$  \\ \hline
$C_1^{(2)}$  & 1 &  1 &1 & 1 &1 & 1 & $1$ &  1 & $1$ & $1$  \\ \hline
$C_{3}^{(0)}$ & $3$ &  $1$ & $1$ & $1$ & $1$ & $1$ & $1$ &  $1$ & 
$1$ & $1$   \\ \hline
$C_{3}^{'(0)}$ & $3$ &  $1$ & $1$ & $1$ & $1$ & $1$ & $1$ &  $1$ & 
$1$ & $1$   \\ \hline
$C_{3}^{(1)}$ & $3$ &  $1$ & $1$ & $1$ & $\omega$ & $\omega$ & 
$\omega$ &  $\omega^2$ & $\omega^2$ & $\omega^2$    \\ \hline
$C_{3}^{'(1)}$ & $3$ &  $1$ & $1$ & $1$ & $\omega$ & $\omega$ & 
$\omega$ &  $\omega^2$ & $\omega^2$ & $\omega^2$    \\ \hline
$C_{3}^{''(1)}$ & $3$ &  $1$ & $1$ & $1$ & $\omega$ & $\omega$ & 
$\omega$ &  $\omega^2$ & $\omega^2$ & $\omega^2$    \\ \hline
$C_{3}^{(2)}$ & $3$ &  $1$ & $1$ & $1$ & $\omega^2$ & $\omega^2$ & 
$\omega^2$ &  $\omega$ & $\omega$ & $\omega$    \\ \hline
$C_{3}^{'(2)}$  & $3$ &  $1$ & $1$ & $1$ & $\omega^2$ & $\omega^2$ & 
$\omega^2$ &  $\omega$ & $\omega$ & $\omega$    \\ \hline
$C_{3}^{''(2)}$  & $3$ &  $1$ & $1$ & $1$ & $\omega^2$ & $\omega^2$ & 
$\omega^2$ &  $\omega$ & $\omega$ & $\omega$    \\ \hline
$C_{9}^{(0)}$ & $3$ &  $1$ & $\omega$ & $\omega^2$ & $1$ & $\omega$ & 
$\omega^2$ &  $1$ & $\omega$ & $\omega^2$    \\ \hline
$C_{9}^{(1)}$ & $9$ & $1$ & $\omega$ & $\omega^2$ & $\omega$ & $\omega^2$ & 
$1$ &  $\omega^2$ & $1$ & $\omega$    \\ \hline
$C_{9}^{(2)}$ & $9$ &  $1$ & $\omega$ & $\omega^2$ & $\omega^2$ & $1$ & 
$\omega$ &  $\omega$ & $\omega^2$ & $1$    \\ \hline
$C'^{(0)}_{9}$ & $3$ &  $1$ & $\omega^2$ & $\omega$ & $1$ & $\omega^2$ & 
$\omega$ &  $1$ & $\omega^2$ & $\omega$    \\ \hline
$C'^{(1)}_{9}$  & $9$ & $1$ & $\omega^2$ & $\omega$ & $\omega^2$ & $\omega$ 
& $1$ &  $\omega$ & $1$ & $\omega^2$    \\ \hline
$C'^{(2)}_{9}$   & $9$ &  $1$ & $\omega^2$ & $\omega$ & $\omega$ & $1$ & 
$\omega^2$ &  $\omega^2$ & $\omega$ & $1$    \\ \hline

\end{tabular}
\end{center}
\caption{Characters of $\Sigma(81)$ for the 9 one-dimensional 
representations.}
\label{81tab:sigma-singlet-character}
\end{table}
\begin{table}[htb]
\begin{center}
\begin{tabular}{|c|c|c|c|c|c|c|c|c|c|c|}
\hline 
Class & $n$ & $h$ & 
$\chi_{ 3_{A}}$ & $\chi_{ \bar3_{A}}$ &
$\chi_{ 3_{B}}$ & $\chi_{ \bar3_{B}}$ & 
$\chi_{ 3_{C}}$ & $\chi_{ \bar3_{C}}$ & 
$\chi_{ 3_{D}}$ & $\chi_{ \bar3_{D}}$ \\ 
\hline
$C_1$ & 1 & 1 & 3 & 3 & 3 & 3 & 3 & 3 & 3 & 3 \\ 
$C^{(1)}_1$ & 1 & 3 & 3$\omega$ & 3$\omega^2$ & 3$\omega$ & 3$\omega^2$ & 
3$\omega$ 
& 3$\omega^2$ & 3 & 3 \\ 
$C^{(2)}_1$ & 1 & 3 & 3$\omega^2$ & 3$\omega$ & 3$\omega^2$ & 3$\omega$ & 
3$\omega^2$ & 3$\omega$ & 3 & 3 \\ 
$C^{(0)}_{3}$ & 3 & 3 & 0 & 0 & 0 & 0 & 0 & 0 & 3$\omega$ & 
3$\omega^2$ \\ 
$C^{'(0)}_{3}$ & 3 & 3 & 0 & 0 & 0 & 0 & 0 & 0 & 3$\omega^2$ & 
3$\omega$ \\ 
$C^{(1)}_{3}$ & 3 & 3 & $-i\sqrt{3}$ & $i\sqrt{3}$ & 
$-i\sqrt{3}\omega^2$ & 
$i\sqrt{3}\omega$ & $-i\sqrt{3}\omega$ & $i\sqrt{3}\omega^2$ & 0 & 0 \\ 
$C^{(2)}_{3}$ & 3 & 3 & $i\sqrt{3}$ & $-i\sqrt{3}$ & 
$i\sqrt{3}\omega$ & 
$-i\sqrt{3}\omega^2$ & $i\sqrt{3}\omega^2$ & $-i\sqrt{3}\omega$ & 0 & 0 \\ 
$C^{'(1)}_{3}$ & 3 & 3 & $-i\sqrt{3}\omega^2$ & $i\sqrt{3}\omega$ & 
$-i\sqrt{3}\omega$ & $i\sqrt{3}\omega^2$ & $-i\sqrt{3}$ & $i\sqrt{3}$ & 
0 & 0 \\ 
$C^{'(2)}_{3}$ & 3 & 3 & $i\sqrt{3}\omega$ & $-i\sqrt{3}\omega^2$ & 
$i\sqrt{3}\omega^2$ & $-i\sqrt{3}\omega$ & $i\sqrt{3}$ & $-i\sqrt{3}$ & 
0 & 0 \\ 
$C^{''(1)}_{3}$ & 3 & 3 & $-i\sqrt{3}\omega$ & $i\sqrt{3}\omega^2$ & 
$-i\sqrt{3}$ & 
$i\sqrt{3}$ & $-i\sqrt{3}\omega^2$ & $i\sqrt{3}\omega$ & 0 & 0 \\ 
$C^{''(2)}_{3}$ & 3 & 3 & $i\sqrt{3}\omega^2$ & $-i\sqrt{3}\omega$ & 
$i\sqrt{3}$ & 
$-i\sqrt{3}$ & $i\sqrt{3}\omega$ & $-i\sqrt{3}\omega^2$ & 0 & 0 \\ 
$C^{(0)}_9$ & 9 & 3 & 0 & 0 & 0 & 0 & 0 & 0 & 0 & 0 \\ 
$C^{(1)}_{9}$ & 9 & 9 & 0 & 0 & 0 & 0 & 0 & 0 & 0 & 0 \\ 
$C^{(2)}_{9}$ & 9 & 9 & 0 & 0 & 0 & 0 & 0 & 0 & 0 & 0 \\ 
$C'^{(0)}_9$ & 9 & 3 & 0 & 0 & 0 & 0 & 0 & 0 & 0 & 0 \\
$C'^{(1)}_{9}$ & 9 & 9 & 0 & 0 & 0 & 0 & 0 & 0 & 0 & 0 \\ 
$C'^{(2)}_{9}$ & 9 & 9 & 0 & 0 & 0 & 0 & 0 & 0 & 0 & 0 \\ 
\hline
\end{tabular}
\end{center}
\caption{Characters of $\Sigma(81)$ for the 8 three-dimensional 
representations. }
\label{81tab:sigma-triplet-character}
\end{table}

On all of the triplets, the generator $b$ is represented as 
\begin{eqnarray}
b=\Mat3{0}{1}{0} {0}{0}{1} {1}{0}{0}.
\end{eqnarray}
The generators, $a$, $a'$ and $a''$, are represented 
on each triplet as 
\begin{eqnarray}
a=\Mat3{\omega}{0}{0} {0}{1}{0} {0}{0}{1} ,\quad
a'=\Mat3{1}{0}{0} {0}{\omega}{0} {0}{0}{1} ,\quad 
a''=\Mat3{1}{0}{0} {0}{1}{0} {0}{0}{\omega} ,
\end{eqnarray}
on ${\bf 3}_A$,
\begin{eqnarray}
a=\Mat3{1}{0}{0} {0}{\omega^2}{0} {0}{0}{\omega^2} ,\quad
a'=\Mat3{\omega^2}{0}{0} {0}{1}{0} {0}{0}{\omega^2} ,\quad 
a''=\Mat3{\omega^2}{0}{0} {0}{\omega^2}{0} {0}{0}{1} ,
\end{eqnarray}
on ${\bf 3}_B$,
\begin{eqnarray}
a=\Mat3{\omega^2}{0}{0} {0}{\omega}{0} {0}{0}{\omega} ,\quad
a'=\Mat3{\omega}{0}{0} {0}{\omega^2}{0} {0}{0}{\omega} ,\quad 
a''=\Mat3{\omega}{0}{0} {0}{\omega}{0} {0}{0}{\omega^2} ,
\end{eqnarray}
on ${\bf 3}_C$,
\begin{eqnarray}
a=\Mat3{\omega^2}{0}{0} {0}{1}{0} {0}{0}{\omega} ,\quad
a'=\Mat3{\omega}{0}{0} {0}{\omega^2}{0} {0}{0}{1} ,\quad 
a''=\Mat3{1}{0}{0} {0}{\omega}{0} {0}{0}{\omega^2} ,
\end{eqnarray}
on ${\bf 3}_D$.
The representations of $a$, $a'$ and $a''$ on 
${\bf 3}_{\bar A}$, ${\bf 3}_{\bar B}$, ${\bf 3}_{\bar C}$ 
and ${\bf 3}_{\bar D}$ are obtained as complex conjugates of 
the representations on 
${\bf 3}_A$, ${\bf 3}_B$, ${\bf 3}_C$ and ${\bf 3}_D$, 
respectively.

On the other hand, these generators are represented on 
the singlet ${\bf 1}^k_\ell$ as 
$b=\omega^\ell$ and $a=a'=a''=\omega^k$.

The tensor products between triplets are obtained as 
\begin{eqnarray}
{\3tvec {x_1}  {x_2} {x_3}}_{{\bf 3}_A} 
\otimes 
{\3tvec {y_1}  {y_2} {y_3}}_{{\bf 3}_A}
&=&
{ \3tvec {x_1y_1  }  {x_2y_2 }  {x_3y_3}}_{{\bf \bar3}_{A}}
 \oplus 
{ \3tvec {x_2y_3  }  {x_3y_1 }  {x_1y_2}}_{{\bf \bar3}_{B}}
 \oplus 
{ \3tvec {x_3y_2  }  {x_1y_3 }  {x_2y_1}}_{{\bf \bar3}_{B}},
\\
{\3tvec {x_1}  {x_2} {x_3}}_{{\bf 3}_A} 
\otimes 
{\3tvec {y_1}  {y_2} {y_3}}_{{\bf \bar3}_{A}}
&=&
\left( \sum_{\ell =0,1,2} 
(x_1y_1+\omega^{2\ell}x_2y_2+\omega^\ell x_3y_3)_{{\bf 1}^0_\ell} \right)
\nonumber \\ & & 
 \oplus 
{ \3tvec {x_3y_1  }  {x_1y_2 }  {x_2y_3}}_{{\bf 3}_{D}}
 \oplus 
{ \3tvec {x_1y_3  }  {x_2y_1 }  {x_3y_2}}_{{\bf\bar 3}_{D}},
\\
{\3tvec {x_1}  {x_2} {x_3}}_{{\bf 3}_A} 
\otimes 
{\3tvec {y_1}  {y_2} {y_3}}_{{\bf 3}_{B}}
&=&
{ \3tvec {x_1y_1  }  {x_2y_2 }  {x_3y_3}}_{{\bf\bar 3}_{C}}
 \oplus 
{ \3tvec {x_3y_2  }  {x_1y_3 }  {x_2y_1}}_{{\bf \bar3}_{A}}
 \oplus 
{ \3tvec {x_2y_3  }  {x_3y_1 }  {x_1y_2}}_{{\bf \bar3}_{A}},
\\
{\3tvec {x_1}  {x_2} {x_3}}_{{\bf 3}_A} 
\otimes 
{\3tvec {y_1}  {y_2} {y_3}}_{{\bf \bar3}_{B}}
&=&
\left( \sum_{\ell =0,1,2}
(x_1y_1+\omega^{2\ell}x_2y_2+\omega^\ell x_3y_3)_{{\bf 1}^2_\ell} \right)
\nonumber \\ & & 
 \oplus 
{ \3tvec {x_2y_3  }  {x_3y_1 }  {x_1y_2}}_{{\bf 3}_{D}}
 \oplus 
{ \3tvec {x_2y_1  }  {x_3y_2 }  {x_1y_3}}_{{\bf \bar3}_{D}},
\\
{\3tvec {x_1}  {x_2} {x_3}}_{{\bf 3}_A} 
\otimes 
{\3tvec {y_1}  {y_2} {y_3}}_{{\bf 3}_{C}}
&=&
{ \3tvec {x_1y_1  }  {x_2y_2 }  {x_3y_3}}_{{\bf\bar 3}_{B}}
 \oplus 
{ \3tvec {x_2y_3  }  {x_3y_1 }  {x_1y_2}}_{{\bf \bar3}_{C}}
 \oplus 
{ \3tvec {x_3y_2  }  {x_1y_3 }  {x_2y_1}}_{{\bf \bar3}_{C}},
\\
{\3tvec {x_1}  {x_2} {x_3}}_{{\bf 3}_A} 
\otimes 
{\3tvec {y_1}  {y_2} {y_3}}_{{\bf \bar3}_{C}}
&=&
\left( \sum_{\ell =0,1,2} 
(x_1y_1+\omega^{2\ell }x_2y_2+\omega^\ell x_3y_3)_{{\bf 1}^2_\ell} \right)
\nonumber \\ & & 
 \oplus 
{ \3tvec {x_2y_3  }  {x_3y_1 }  {x_1y_2}}_{{\bf 3}_{D}}
 \oplus 
{ \3tvec {x_2y_1  }  {x_1y_2 }  {x_3y_3}}_{{\bf \bar3}_{D}},
\end{eqnarray}
\begin{eqnarray}
{\3tvec {x_1}  {x_2} {x_3}}_{{\bf \bar3}_{A}} 
\otimes 
{\3tvec {y_1}  {y_2} {y_3}}_{{\bf 3}_{C}}
&=& \left(
\sum_{\ell =0,1,2} 
(x_1y_1+\omega^{2\ell}x_2y_2+\omega^\ell x_3y_3)_{{\bf 1}^1_\ell} \right)
\nonumber \\ & & 
 \oplus
{ \3tvec {x_1y_2 }  {x_2y_3 }  {x_3y_1}}_{{\bf 3}_{D}}
 \oplus 
{ \3tvec {x_3y_2  }  {x_1y_3 }  {x_2y_1}}_{{\bf \bar3}_{D}},
\\
{\3tvec {x_1}  {x_2} {x_3}}_{{\bf 3}_D} 
\otimes 
{\3tvec {y_1}  {y_2} {y_3}}_{{\bf 3}_{D}}
&=&
{ \3tvec {x_1y_1  }  {x_2y_2 }  {x_3y_3}}_{{\bf \bar3}_{D}}
 \oplus 
{ \3tvec {x_2y_3  }  {x_3y_1 }  {x_1y_2}}_{{\bf \bar3}_{D}}
 \oplus 
{ \3tvec {x_3y_2  }  {x_1y_3 }  {x_2y_1}}_{{\bf \bar3}_{D}},
\\
{\3tvec {x_1}  {x_2} {x_3}}_{{\bf 3}_{D}} 
\otimes 
{\3tvec {y_1}  {y_2} {y_3}}_{{\bf \bar3}_{D}}
&=& \sum_{\ell =0,1,2}[
(x_1y_1+\omega^{2\ell }x_2y_2+\omega^\ell x_3y_3)_{{\bf 1}^0_\ell}
\nonumber \\ & & \oplus
(x_2y_3+\omega^{2\ell}x_3y_1+\omega^\ell x_1y_2)_{{\bf 1}^1_\ell}
\nonumber \\ & & \oplus 
(x_3y_2+\omega^{2\ell }x_1y_3+\omega^\ell x_2y_1)_{{\bf 1}^2_\ell}].
\end{eqnarray}

The tensor products between singlets are obtained as 
\begin{eqnarray}
{\bf 1}^{k}_{\ell}\otimes {\bf 1}^{k'}_{\ell'}=  {\bf
  1}^{k+k'~({\rm mod ~3})}_{\ell+\ell'~({\rm mod ~3})} .
\end{eqnarray}
\\
The tensor products between singlets and triplets are obtained as 
\begin{eqnarray}
&&
(x)_{{\bf 1}^{0}_0}\otimes \3tvec{y_1}{y_2}{y_3}_{{\bf 3(\bar3)}_{A}}=
 \3tvec{xy_1}{xy_2}{xy_3}_{{\bf 3(\bar3)}_{A}},
~
 (x)_{{\bf 1}^{0}_1}\otimes \3tvec{y_1}{y_2}{y_3}_{{\bf 3(\bar3)}_{A}}=
 \3tvec{xy_1}{\omega xy_2}{\omega^2 xy_3}_{{\bf 3(\bar3)}_{A}},
 \nonumber\\
&&
( x)_{{\bf 1}^{0}_2}\otimes \3tvec{y_1}{y_2}{y_3}_{{\bf 3(\bar3)}_{A}}=
 \3tvec{xy_1}{\omega^2 xy_2}{\omega xy_3}_{{\bf 3(\bar3)}_{A}},
\\
 &&
(x)_{{\bf 1}^{1}_0}\otimes \3tvec{y_1}{y_2}{y_3}_{{\bf 3(\bar3)}_{A}}=
 \3tvec{xy_1}{xy_2}{xy_3}_{{\bf 3}_{C},({\bf \bar3}_B) },
~
 (x)_{{\bf 1}^{1}_1}\otimes \3tvec{y_1}{y_2}{y_3}_{{\bf 3(\bar3)}_{A}}=
 \3tvec{xy_1}{\omega xy_2}{\omega^2 xy_3}_{{\bf 3}_{C},({\bf \bar3}_B) },
 \nonumber\\
&&
( x)_{{\bf 1}^{1}_2}\otimes \3tvec{y_1}{y_2}{y_3}_{{\bf 3(\bar3)}_{A}}=
 \3tvec{xy_1}{\omega^2 xy_2}{\omega xy_3}_{{\bf 3}_{C},({\bf \bar3}_B) },
 \\
 &&
(x)_{{\bf 1}^{2}_0}\otimes \3tvec{y_1}{y_2}{y_3}_{{\bf 3(\bar3)}_{A}}=
 \3tvec{xy_1}{xy_2}{xy_3}_{{\bf 3}_{B},({\bf \bar3}_C) },
~
 (x)_{{\bf 1}^{2}_1}\otimes \3tvec{y_1}{y_2}{y_3}_{{\bf 3(\bar3)}_{A}}=
 \3tvec{xy_1}{\omega xy_2}{\omega^2 xy_3}_{{\bf 3}_{B},({\bf \bar3}_C) },
 \nonumber\\
&&
( x)_{{\bf 1}^{2}_2}\otimes \3tvec{y_1}{y_2}{y_3}_{{\bf \bf 3(\bar3)}_{A}}=
 \3tvec{xy_1}{\omega^2 xy_2}{\omega xy_3}_{{\bf \bf 3}_{B},({\bf \bar3}_C) }
, 
\end{eqnarray}

\begin{eqnarray}
&&
(x)_{{\bf 1}^{0}_0}\otimes \3tvec{y_1}{y_2}{y_3}_{{\bf 3(\bar3)}_{B}}=
 \3tvec{xy_1}{xy_2}{xy_3}_{{\bf 3(\bar3)}_{B}},
~
 (x)_{{\bf 1}^{0}_1}\otimes \3tvec{y_1}{y_2}{y_3}_{{\bf 3(\bar3)}_{B}}=
 \3tvec{xy_1}{\omega xy_2}{\omega^2 xy_3}_{{\bf 3(\bar3)}_{B}},
 \nonumber\\
&&
( x)_{{\bf 1}^{0}_2}\otimes \3tvec{y_1}{y_2}{y_3}_{{\bf 3(\bar3)}_{B}}=
 \3tvec{xy_1}{\omega^2 xy_2}{\omega xy_3}_{{\bf 3(\bar3)}_{B}},
\\
 &&
(x)_{{\bf 1}^{1}_0}\otimes \3tvec{y_1}{y_2}{y_3}_{{\bf 3(\bar3)}_{B}}=
 \3tvec{xy_1}{xy_2}{xy_3}_{{\bf 3}_{A},({\bf \bar 3}_C)},
~
 (x)_{{\bf 1}^{1}_1}\otimes \3tvec{y_1}{y_2}{y_3}_{{\bf 3(\bar3)}_{B}}=
 \3tvec{xy_1}{\omega xy_2}{\omega^2 xy_3}_{{\bf 3}_{A},({\bf \bar 3}_C)},
 \nonumber\\
&&
( x)_{{\bf 1}^{1}_2}\otimes \3tvec{y_1}{y_2}{y_3}_{{\bf 3(\bar 3)}_{B}}=
 \3tvec{xy_1}{\omega^2 xy_2}{\omega xy_3}_{{\bf 3}_{A},({\bf \bar 3}_C)},
 \\
 &&
(x)_{{\bf 1}^{2}_0}\otimes \3tvec{y_1}{y_2}{y_3}_{{\bf 3(\bar 3)}_{B}}=
 \3tvec{xy_1}{xy_2}{xy_3}_{{\bf 3}_{C},({\bf \bar 3}_A)},
~
 (x)_{{\bf 1}^{2}_1}\otimes \3tvec{y_1}{y_2}{y_3}_{{\bf 3(\bar 3)}_{B}}=
 \3tvec{xy_1}{\omega xy_2}{\omega^2 xy_3}_{{\bf 3}_{C},({\bf \bar 3}_A)},
 \nonumber\\
&&
( x)_{{\bf 1}^{2}_2}\otimes \3tvec{y_1}{y_2}{y_3}_{{\bf 3(\bar 3)}_{B}}=
 \3tvec{xy_1}{\omega^2 xy_2}{\omega xy_3}_{{\bf 3}_{C},({\bf \bar 3}_A)}
,
\end{eqnarray}
\begin{eqnarray}
&&
(x)_{{\bf 1}^{0}_0}\otimes \3tvec{y_1}{y_2}{y_3}_{{\bf 3(\bar 3)}_{C}}=
 \3tvec{xy_1}{xy_2}{xy_3}_{{\bf 3(\bar 3)}_{C}},
~
 (x)_{{\bf 1}^{0}_1}\otimes \3tvec{y_1}{y_2}{y_3}_{{\bf 3(\bar 3)}_{C}}=
 \3tvec{xy_1}{\omega xy_2}{\omega^2 xy_3}_{{\bf 3(\bar 3)}_{C}},
 \nonumber\\
&&
( x)_{{\bf 1}^{0}_2}\otimes \3tvec{y_1}{y_2}{y_3}_{{\bf 3(\bar 3)}_{C}}=
 \3tvec{xy_1}{\omega^2 xy_2}{\omega xy_3}_{{\bf 3(\bar 3)}_{C}},
\\
 &&
(x)_{{\bf 1}^{1}_0}\otimes \3tvec{y_1}{y_2}{y_3}_{{\bf 3(\bar 3)}_{C}}=
 \3tvec{xy_1}{xy_2}{xy_3}_{{\bf 3}_{B},({\bf \bar3}_A) },
~
 (x)_{{\bf 1}^{1}_1}\otimes \3tvec{y_1}{y_2}{y_3}_{{\bf 3(\bar 3)}_{C}}=
 \3tvec{xy_1}{\omega xy_2}{\omega^2 xy_3}_{{\bf 3}_{B},({\bf \bar3}_A) },
 \nonumber\\
&&
( x)_{{\bf 1}^{1}_2}\otimes \3tvec{y_1}{y_2}{y_3}_{{\bf 3(\bar 3)}_{C}}=
 \3tvec{xy_1}{\omega^2 xy_2}{\omega xy_3}_{{\bf 3}_{B},({\bf \bar3}_A) },
 \\
 &&
(x)_{{\bf 1}^{2}_0}\otimes \3tvec{y_1}{y_2}{y_3}_{{\bf 3(\bar 3)}_{C}}=
 \3tvec{xy_1}{xy_2}{xy_3}_{{\bf 3}_{A},({\bf \bar3}_B) },
~
 (x)_{{\bf 1}^{2}_1}\otimes \3tvec{y_1}{y_2}{y_3}_{{\bf 3(\bar 3)}_{C}}=
 \3tvec{xy_1}{\omega xy_2}{\omega^2 xy_3}_{{\bf 3}_{A},({\bf \bar3}_B) },
 \nonumber\\
&&
( x)_{{\bf 1}^{2}_2}\otimes \3tvec{y_1}{y_2}{y_3}_{{\bf 3(\bar 3)}_{C}}=
 \3tvec{xy_1}{\omega^2 xy_2}{\omega xy_3}_{{\bf 3}_{A},({\bf \bar3}_B) },
\end{eqnarray} 
\begin{equation}
\hspace{-7.5cm}
(x)_{{\bf 1}^{k}_\ell}\otimes \3tvec{y_1}{y_2}{y_3}_{{\bf 3(\bar 3)}_{D}}=
 \3tvec{xy_1}{xy_2}{xy_3}_{{\bf 3(\bar 3)}_{D}}, 
\end{equation}
where $k,~\ell=0,~1,~2$.

\clearpage


\section{$\Delta (6N^2)$}
\label{sec:Delta-6N}

\subsection{Generic aspects}

The discrete group $\Delta (6N^2)$ is isomorphic 
to $(Z_N \times Z_N')\rtimes S_3$.
(See also Ref.~\cite{Escobar:2008vc}.)
Its order is equal to $6N^2$.
We denote the generators of $Z_N$ and $Z_N'$ by $a$ and $a'$, 
respectively, and the $Z_3$ and $Z_2$ generators of $S_3$ are 
written by $b$ and $c$, respectively.
They satisfy 
\begin{eqnarray}\label{eq:delta-6n}
& & a^N={a'}^N=b^3=c^2=(bc)^2=e, \nonumber \\
& & aa'= a'a, \nonumber \\
& & bab^{-1} = a^{-1} a'^{-1}, \qquad ba'b^{-1} = a, \\
& & cac^{-1} = a'^{-1}, \qquad ca'c^{-1} = a^{-1}. \nonumber 
\end{eqnarray}
Using them, all of $\Delta (6N^2)$ elements are written as 
\begin{eqnarray}
b^kc^\ell a^ma'^n,
\end{eqnarray}
where $k=0,1,2$, $\ell = 0,1$ and $m,n=0,\cdots, (N-1)$.
Similarly to the previous sections, we can find 
conjugacy classes, characters, representations 
and tensor products of generic $\Delta (6N^2)$.
Instead of showing those aspects for generic $\Delta (6N^2)$, 
we concentrate on an example, i.e. $\Delta(54)$.

\subsection{$\Delta(54)$}

The $\Delta(6)$ is nothing but $S_3$ and 
the $\Delta(24)$ is isomorphic to the $S_4$ group.
Thus, the simple and non-trivial example is 
$\Delta(54)$.

\vskip .5cm
{$\bullet$ \bf Conjugacy classes}

All of the $\Delta(54)$ elements are written as 
$b^kc^\ell a^ma'^n$, where $k,m,n=0,1,2$ and $\ell = 0,1$.
Half of them are the elements of $\Delta(27)$, whose 
conjugacy classes are shown in (\ref{eq:conjugacy-27}).
Because of $cac^{-1} = a'^{-1}$ and $ca'c^{-1} = a^{-1}$, 
the conjugacy classes $C_1^{(1)}$ and $C_1^{(2)}$ of 
$\Delta(27)$ correspond to the conjugacy classes of 
$\Delta(54)$, still.
However, the conjugacy classes $C_3^{(0,1)}$ and 
$C_3^{(0,1)}$ of $\Delta(27)$ are combined to 
a conjugacy class of $\Delta(54)$.
Similarly, since $cba^ka'^\ell c^{-1} = b^2ca^ka'^\ell c^{-1}$,
 the conjugacy classes $C_3^{(1,p)}$ and 
$C_3^{(2,p')}$ of $\Delta(27)$ for $p+p'=0$ (mod 3) are combined to 
a conjugacy class of $\Delta(54)$.


Next, let us consider the conjugacy classes of elements 
including $c$.
For example, we obtain 
\begin{eqnarray}
a^ka'^\ell (ca^m) a^{-k}a'^{-\ell} = ca^{m+p}a'^p,
\end{eqnarray}
where $p=-k-\ell$.
Thus, the element $ca^m$ is conjugate to $ca^{m+p}a'^p$ with 
$p=0,1,2$.
Furthermore, it is found that 
\begin{eqnarray}
b (ca^{m+p}a'^p) b^{-1} = b^2ca^{-m}a'^{-m-p}, \\
b (ca^{-m}a'^{-m-p}) b^{-1} = bca^{-p}a'^{m}.
\end{eqnarray}
Then, these elements belong to the same conjugacy class.

Using the above results, the $\Delta(54)$ elements are 
classified into the following conjugacy classes,
\begin{eqnarray}\label{eq:conjugacy-54}
\begin{array}{ccc}
 C_1:&\{ e \}, &  h=1, \\
 C_1^{(1)}:&\{ a a'^2 \}, &  h=3,  \\
 C_1^{(2)}:&\{ a^2 a' \}, &  h=3,  \\
 C_6^{(0)}: &\{ a',a,a^2a'^2,  a'^2,a^2,aa'\}, & h=3, \\
 C_6^{(1,0)}: &\{ b,ba^{2}a',ba^{1}a'^2,  b^2,b^2a^{2}a',b^2a^{1}a'^2 \}, & 
h=3, \\
 C_6^{(1,1)}: &\{ ba,ba',ba^{2}a'^2, b^2a,b^2a',b^2a^{2}a'^2\}, & h=3, 
 \\
 C_6^{(1,2)}: &\{ ba^2,ba^{1}a',ba'^2, b^2a^2,b^2a^{1}a',b^2a'^2\}, &
 h=3, \\
C_9^{(m)}: &\{ca^{m+p}a'^p,b^2ca^{-m}a'^{-m-p},bca^{-p}a'^{m} ~|~
p=0,1,2 \}, &
\end{array}
\end{eqnarray}
where $m=0,1,2$.
The total number of conjugacy classes is equal to ten.
The relations (\ref{eq:character-2-e}) and (\ref{eq:sum-dim}) 
for $\Delta(54)$ lead to 
\begin{eqnarray}
&&m_1+2^2m_2+3^2m_3 +\cdots =54,\\
&&m_1+m_2+m_3 + \cdots =10.
\end{eqnarray}
The solution is found as 
$(m_1,m_2,m_3)=(2,4,4)$.
That is, there are two singlets, four doublets 
and four triplets.

\vskip .5cm
{$\bullet$ \bf Characters and representations}

Now, let us study characters and representations.
We start with two singlets.
It is straightforward to find 
$\chi_{1\alpha}(a)=\chi_{1\alpha}(a')=\chi_{1\alpha}(b)=1$ 
for two singlets from the above structure of conjugacy classes.
In addition, because of $c^2=e$, the two values $\pm 1$ for 
$\chi_{1 \pm}(c)$ are possible.
They correspond to two singlets, ${\bf 1}_\pm$.

Next, we study triplets.
For example, the generators, $a$, $a'$, $b$ and $c$, are 
represented by 
\begin{eqnarray}
& & 
a=
\begin{pmatrix}
\omega^k & 0 & 0\\
0 & \omega^{2k} & 0 \\
0 & 0 & 1 
\end{pmatrix}, \qquad 
a'=
\begin{pmatrix}
1 & 0 & 0\\
0 & \omega^{k} & 0 \\
0 & 0 & \omega^{2k} 
\end{pmatrix}, \nonumber \\
& & 
b=
\begin{pmatrix}
0 & 1 & 0\\
0 & 0 & 1 \\
1 & 0 & 0 
\end{pmatrix}, \qquad 
c=
\begin{pmatrix}
0 & 0 & 1\\
0 & 1 & 0 \\
1 & 0 & 0 
\end{pmatrix}, 
\end{eqnarray}
on ${\bf 3}_{1(k)}$ for $k=1,2$.
Obviously, the $\Delta(54)$ algebra (\ref{eq:delta-6n}) 
is satisfied when we replace $c$ by $-c$.
That is, the generators,  $a$, $a'$, $b$ and $c$, are 
represented by 
\begin{eqnarray}
& & 
a=
\begin{pmatrix}
\omega^k & 0 & 0\\
0 & \omega^{2k} & 0 \\
0 & 0 & 1 
\end{pmatrix}, \qquad 
a'=
\begin{pmatrix}
1 & 0 & 0\\
0 & \omega^{k} & 0 \\
0 & 0 & \omega^{2k} 
\end{pmatrix}, \nonumber \\
& & 
b=
\begin{pmatrix}
0 & 1 & 0\\
0 & 0 & 1 \\
1 & 0 & 0 
\end{pmatrix}, \qquad 
c=
\begin{pmatrix}
0 & 0 & -1\\
0 & -1 & 0 \\  -1 & 0 & 0 
\end{pmatrix}, 
\end{eqnarray}
on ${\bf 3}_{2(k)}$ for $k=1,2$.
Then, characters $\chi_3$ for ${\bf 3}_{1(k)}$ and ${\bf 3}_{2(k)}$ 
are shown in Table \ref{tab:delta-54}.

There are four doublets and 
the generators, $a$, $a'$, $b$ and $c$, are 
represented by 
\begin{eqnarray}
& & 
a=a'=
\begin{pmatrix}
1 & 0 \\
0 & 1   
\end{pmatrix}, \qquad 
b=
\begin{pmatrix}
\omega & 0 \\
0 & \omega^2
\end{pmatrix}, \qquad 
c=
\begin{pmatrix}
0 & 1 \\
1 & 0   
\end{pmatrix}, \qquad {\rm on~~} {\bf 2}_1,
\end{eqnarray}
\begin{eqnarray}
& & 
a=a'=
\begin{pmatrix}
\omega^2 & 0 \\
0 & \omega   
\end{pmatrix}, \qquad 
b=
\begin{pmatrix}
\omega & 0 \\
0 & \omega^2
\end{pmatrix}, \qquad 
c=
\begin{pmatrix}
0 & 1 \\
1 & 0   
\end{pmatrix}, \qquad {\rm on~~} {\bf 2}_2,
\end{eqnarray}
\begin{eqnarray}
& & 
a=a'=
\begin{pmatrix}
\omega & 0 \\
0 & \omega^2   
\end{pmatrix}, \qquad 
b=
\begin{pmatrix}
\omega & 0 \\
0 & \omega^2
\end{pmatrix}, \qquad 
c=
\begin{pmatrix}
0 & 1 \\
1 & 0   
\end{pmatrix}, \qquad {\rm on~~} {\bf 2}_3,
\end{eqnarray}
\begin{eqnarray}
& & 
a=a'=
\begin{pmatrix}
\omega & 0 \\
0 & \omega^2   
\end{pmatrix}, \qquad 
b=
\begin{pmatrix}
1 & 0 \\
0 & 1
\end{pmatrix}, \qquad 
c=
\begin{pmatrix}
0 & 1 \\
1 & 0   
\end{pmatrix}, \qquad {\rm on~~} {\bf 2}_4.
\end{eqnarray}
Then, characters $\chi_2$ for ${\bf 2}_{1,2,3,4}$ 
are shown in Table \ref{tab:delta-54}.

\begin{table}[t]
\begin{center}
\begin{tabular}{|c|c|c|c|c|c|c|c|c|}
\hline
&$\chi_{1_+}$& $\chi_{1_-}$& $\chi_{3_{1(k)}}$& $\chi_{3_{2(k)}}$ 
& $\chi_{2_1}$ & $\chi_{2_2}$ & $\chi_{2_3}$ & $\chi_{2_4}$\\
\hline
$C_1$ & 1 & 1 & 3 & 3 & 2 & 2& 2 & 2 \\ \hline
$C_1^{(1)}$  & 1 & 1 & $3 \omega^k$ & $3 \omega^{2k}$ 
& 2 & 2& 2 & 2 \\ \hline
$C_1^{(2)}$  & 1 & 1 & $3 \omega^{2k}$ & $3 \omega^{k}$ 
& 2 & 2& 2 & 2 \\ \hline
$C_6^{(0)}$ & 1 & 1 & 0 & 0 & 2 & $-1$ & $-1$ & $-1$ \\ \hline
$C_6^{(1,0)}$ & 1 & 1 & 0 & 0 & $-1$ & $-1$ & $-1$ & 2 \\ \hline
$C_6^{(1,1)}$ & 1 & 1 & 0 & 0 & $-1$ & $2$ & $-1$ & $-1$ \\ \hline
$C_6^{(1,2)}$ & 1 & 1 & 0 & 0 & $-1$ & $-1$ & $2$ & $-1$ \\ \hline
$C_9^{(0)}$ & 1 & $-1$ & 1 & $-1$ & 0 & 0 & 0 & 0 \\ \hline
$C_9^{(1)}$ & 1 & $-1$ & $\omega^{2k}$  & $-\omega^{2k}$ 
& 0 & 0 & 0 & 0 \\ \hline
$C_9^{(2)}$ & 1 & $-1$ & $\omega^{k}$  & $-\omega^{k}$ 
& 0 & 0 & 0 & 0 \\
\hline
\end{tabular}
\end{center}
\caption{Characters of $\Delta(54)$}
\label{tab:delta-54}
\end{table}

\vskip .5cm
{$\bullet$ \bf Tensor products}

The tensor products between triplets are obtained as 
\begin{eqnarray}
\begin{pmatrix}
x_1 \\ x_2 \\ x_3 \\
\end{pmatrix}_{{\bf 3}_{1(1)}} \otimes
\begin{pmatrix}
y_1 \\ y_2 \\ y_3 \\
\end{pmatrix}_{{\bf 3}_{1(1)}} 
=\begin{pmatrix}
x_1y_1 \\ x_2y_2 \\ x_3y_3 \\
\end{pmatrix}_{{\bf 3}_{1(2)}} \oplus
\begin{pmatrix}
x_2y_3+x_3y_2 \\ x_3y_1+x_1y_3 \\ x_1y_2+x_2y_1 \\
\end{pmatrix}_{{\bf 3}_{1(2)}} 
\oplus
\begin{pmatrix}
x_2y_3-x_3y_2 \\ x_3y_1-x_1y_3 \\ x_1y_2-x_2y_1 \\
\end{pmatrix}_{{\bf 3}_{2(2)}} ,
\end{eqnarray}
\begin{eqnarray}
\begin{pmatrix}
x_1 \\ x_2 \\ x_3 \\
\end{pmatrix}_{{\bf 3}_{1(2)}} \otimes
\begin{pmatrix}
y_1 \\ y_2 \\ y_3 \\
\end{pmatrix}_{{\bf 3}_{1(2)}} 
=\begin{pmatrix}
x_1y_1 \\ x_2y_2 \\ x_3y_3 \\
\end{pmatrix}_{{\bf 3}_{1(1)}} \oplus
\begin{pmatrix}
x_2y_3+x_3y_2 \\ x_3y_1+x_1y_3 \\ x_1y_2+x_2y_1 \\
\end{pmatrix}_{{\bf 3}_{1(1)}} 
\oplus
\begin{pmatrix}
x_2y_3-x_3y_2 \\ x_3y_1-x_1y_3 \\ x_1y_2-x_2y_1 \\
\end{pmatrix}_{{\bf 3}_{2(1)}} ,
\end{eqnarray}
\begin{eqnarray}
\begin{pmatrix}
x_1 \\ x_2 \\ x_3 \\
\end{pmatrix}_{{\bf 3}_{2(1)}} \otimes
\begin{pmatrix}
y_1 \\ y_2 \\ y_3 \\
\end{pmatrix}_{{\bf 3}_{2(1)}} 
=\begin{pmatrix}
x_1y_1 \\ x_2y_2 \\ x_3y_3 \\
\end{pmatrix}_{{\bf 3}_{1(2)}} \oplus
\begin{pmatrix}
x_2y_3+x_3y_2 \\ x_3y_1+x_1y_3 \\ x_1y_2+x_2y_1 \\
\end{pmatrix}_{{\bf 3}_{1(2)}} 
\oplus
\begin{pmatrix}
x_2y_3-x_3y_2 \\ x_3y_1-x_1y_3 \\ x_1y_2-x_2y_1 \\
\end{pmatrix}_{{\bf 3}_{2(2)}} ,
\end{eqnarray}
\begin{eqnarray}
\begin{pmatrix}
x_1 \\ x_2 \\ x_3 \\
\end{pmatrix}_{{\bf 3}_{2(2)}} \otimes
\begin{pmatrix}
y_1 \\ y_2 \\ y_3 \\
\end{pmatrix}_{{\bf 3}_{2(2)}} 
=\begin{pmatrix}
x_1y_1 \\ x_2y_2 \\ x_3y_3 \\
\end{pmatrix}_{{\bf 3}_{1(1)}} \oplus 
\begin{pmatrix}
x_2y_3+x_3y_2 \\ x_3y_1+x_1y_3 \\ x_1y_2+x_2y_1 \\
\end{pmatrix}_{{\bf 3}_{1(1)}} 
\oplus
\begin{pmatrix}
x_2y_3-x_3y_2 \\ x_3y_1-x_1y_3 \\ x_1y_2-x_2y_1 \\
\end{pmatrix}_{{\bf 3}_{2(1)}} ,
\end{eqnarray}
\begin{eqnarray}
\begin{pmatrix}
x_1 \\ x_2 \\ x_3 \\
\end{pmatrix}_{{\bf 3}_{1(1)}} \otimes
\begin{pmatrix}
y_1 \\ y_2 \\ y_3 \\
\end{pmatrix}_{{\bf 3}_{1(2)}} 
&=&\begin{pmatrix}
x_1y_1 + x_2y_2 + x_3y_3
\end{pmatrix}_{{\bf 1}_{+}} \oplus
\begin{pmatrix}
x_1y_1+\omega ^2x_2y_2+\omega x_3y_3 \\ \omega x_1y_1+\omega 
^2x_2y_2+x_3y_3
\end{pmatrix}_{{\bf 2}_1} \nonumber \\ & & \oplus
\begin{pmatrix}
x_1y_2+\omega ^2x_2y_3+\omega x_3y_1 \\ \omega x_1y_3+\omega 
^2x_2y_1+x_3y_2
\end{pmatrix}_{{\bf 2}_2} \oplus
\begin{pmatrix}
x_1y_3+\omega ^2x_2y_1+\omega x_3y_2 \\ \omega x_1y_2+\omega 
^2x_2y_3+x_3y_1
\end{pmatrix}_{{\bf 2}_3} \nonumber \\ & & \oplus  
\begin{pmatrix}
x_1y_3+x_2y_1+x_3y_2 \\ x_1y_2+x_2y_3+x_3y_1
\end{pmatrix}_{{\bf 2}_4} ,
\end{eqnarray}
\begin{eqnarray}
\begin{pmatrix}
x_1 \\ x_2 \\ x_3 \\
\end{pmatrix}_{{\bf 3}_{1(1)}} \otimes
\begin{pmatrix}
y_1 \\ y_2 \\ y_3 \\
\end{pmatrix}_{{\bf 3}_{2(2)}} 
&=&\begin{pmatrix}
x_1y_1 + x_2y_2 + x_3y_3
\end{pmatrix}_{{\bf 1}_{-}} \oplus
\begin{pmatrix}
x_1y_1+\omega ^2x_2y_2+\omega x_3y_3 \\ -\omega x_1y_1-\omega 
^2x_2y_2-x_3y_3\end{pmatrix}_{{\bf 2}_1} 
\nonumber \\ & & \oplus
\!\!\begin{pmatrix}
x_1y_2+\omega ^2x_2y_3+\omega x_3y_1 \\ -\omega x_1y_3-\omega 
^2x_2y_1-x_3y_2\end{pmatrix}_{{\bf 2}_2} \!\!
\oplus
\!\!\begin{pmatrix}
x_1y_3+\omega ^2x_2y_1+\omega x_3y_2 \\ -\omega x_1y_2-\omega ^2x_2y_3-x_3y_1
\end{pmatrix}_{{\bf 2}_3}\!\!
\nonumber \\ & & 
\oplus
\begin{pmatrix}
x_1y_3+x_2y_1+x_3y_2 \\ -x_1y_2-x_2y_3-x_3y_1
\end{pmatrix}_{{\bf 2}_4} ,
\end{eqnarray}
\begin{eqnarray}
\begin{pmatrix}
x_1 \\ x_2 \\ x_3 \\
\end{pmatrix}_{{\bf 3}_{1(2)}} \otimes
\begin{pmatrix}
y_1 \\ y_2 \\ y_3 \\
\end{pmatrix}_{{\bf 3}_{2(1)}} 
&=&\begin{pmatrix}
x_1y_1 + x_2y_2 + x_3y_3
\end{pmatrix}_{{\bf 1}_{-}} \oplus
\begin{pmatrix}
x_1y_1+\omega ^2x_2y_2+\omega x_3y_3 \\ -\omega x_1y_1-\omega 
^2x_2y_2-x_3y_3
\end{pmatrix}_{{\bf 2}_1} 
\nonumber \\ & & \oplus
\!\!\begin{pmatrix}
x_1y_3+\omega ^2x_2y_1+\omega x_3y_2 \\ -\omega x_1y_2-\omega 
^2x_2y_3-x_3y_1\end{pmatrix}_{{\bf 2}_2}\!\! 
\oplus
\!\!\begin{pmatrix}
x_1y_2+\omega ^2x_2y_3+\omega x_3y_1 \\ -\omega x_1y_3-\omega 
^2x_2y_1-x_3y_2\end{pmatrix}_{{\bf 2}_3}\!\!
\nonumber \\ & & \oplus
\begin{pmatrix}
x_1y_2+x_2y_3+x_3y_1 \\ -x_1y_3-x_2y_1-x_3y_2
\end{pmatrix}_{{\bf 2}_4} .
\end{eqnarray}

The tensor products between doublets are obtained as 
\begin{eqnarray}
\begin{pmatrix}
x_1 \\ x_2 \\
\end{pmatrix}_{{\bf 2}_k} \otimes 
\begin{pmatrix}
y_1 \\ y_2 \\
\end{pmatrix}_{{\bf 2}_k}   = 
\begin{pmatrix}
x_1y_2+x_2y_1 \\
\end{pmatrix}_{{\bf 1}_+} \oplus
\begin{pmatrix}
x_1y_2 -x_2y_1  \\
\end{pmatrix}_{{\bf 1}_-} \oplus
\begin{pmatrix}
x_2y_2 \\ x_1y_1  \\
\end{pmatrix}_{{\bf 2}_k} ,
\end{eqnarray}
for $k=1,2,3,4$,
\begin{eqnarray}
\begin{pmatrix}
x_1 \\ x_2 \\
\end{pmatrix}_{{\bf 2}_1} \otimes 
\begin{pmatrix}
y_1 \\ y_2 \\
\end{pmatrix}_{{\bf 2}_2}   = 
\begin{pmatrix}
x_2y_2 \\x_1y_1 \\
\end{pmatrix}_{{\bf 2}_3} \oplus
\begin{pmatrix}
x_2y_1 \\x_1y_2 \\
\end{pmatrix}_{{\bf 2}_4} , 
\end{eqnarray}
\begin{eqnarray}
\begin{pmatrix}
x_1 \\ x_2 \\
\end{pmatrix}_{{\bf 2}_1} \otimes 
\begin{pmatrix}
y_1 \\ y_2 \\
\end{pmatrix}_{{\bf 2}_3}   = 
\begin{pmatrix}
x_2y_2 \\x_1y_1 \\
\end{pmatrix}_{{\bf 2}_2} \oplus
\begin{pmatrix}
x_2y_1 \\x_1y_2 \\
\end{pmatrix}_{{\bf 2}_4} , 
\end{eqnarray}
\begin{eqnarray}
\begin{pmatrix}
x_1 \\ x_2 \\
\end{pmatrix}_{{\bf 2}_1} \otimes 
\begin{pmatrix}
y_1 \\ y_2 \\
\end{pmatrix}_{{\bf 2}_4}   = 
\begin{pmatrix}
x_1y_2 \\x_2y_1
\end{pmatrix}_{{\bf 2}_2} \oplus
\begin{pmatrix}
x_1y_1 \\x_2y_2
\end{pmatrix}_{{\bf 2}_3} ,
\end{eqnarray}
\begin{eqnarray}
\begin{pmatrix}
x_1 \\ x_2 \\
\end{pmatrix}_{{\bf 2}_2} \otimes 
\begin{pmatrix}
y_1 \\ y_2 \\
\end{pmatrix}_{{\bf 2}_3}   = 
\begin{pmatrix}
x_2y_2 \\ x_1y_1
\end{pmatrix}_{{\bf 2}_1} \oplus
\begin{pmatrix}
x_1y_2 \\ x_2y_1
\end{pmatrix}_{{\bf 2}_4} ,
\end{eqnarray}
\begin{eqnarray}
\begin{pmatrix}
x_1 \\ x_2 \\
\end{pmatrix}_{{\bf 2}_2} \otimes 
\begin{pmatrix}
y_1 \\ y_2 \\
\end{pmatrix}_{{\bf 2}_4}   = 
\begin{pmatrix}
x_1y_1 \\ x_2y_2
\end{pmatrix}_{{\bf 2}_1} \oplus
\begin{pmatrix}
x_1y_2 \\ x_2y_1
\end{pmatrix}_{{\bf 2}_3} ,
\end{eqnarray}
\begin{eqnarray}
\begin{pmatrix}
x_1 \\ x_2 \\
\end{pmatrix}_{{\bf 2}_3} \otimes 
\begin{pmatrix}
y_1 \\ y_2 \\
\end{pmatrix}_{{\bf 2}_4}   = 
\begin{pmatrix}
x_1y_2 \\ x_2y_1
\end{pmatrix}_{{\bf 2}_1} \oplus
\begin{pmatrix}
x_1y_1 \\ x_2y_2
\end{pmatrix}_{{\bf 2}_2} .
\end{eqnarray}

The tensor products between doublets and triplets 
are obtained as 
\begin{eqnarray}
\begin{pmatrix}
x_1 \\ x_2 \\
\end{pmatrix}_{{\bf 2}_1} \otimes 
\begin{pmatrix}
y_1 \\ y_2 \\ y_3 
\end{pmatrix}_{{\bf 3}_{1(k)}}   = 
\begin{pmatrix}
x_1y_1 + \omega^2 x_2 y_1 \\
\omega x_1y_2 + \omega x_2y_2 \\
\omega^2 x_1 y_3 + x_2 y_3
\end{pmatrix}_{{\bf 3}_{1(k)}} \oplus
\begin{pmatrix}
x_1y_1 - \omega^2 x_2 y_1 \\
\omega x_1y_2 - \omega x_2y_2 \\
\omega^2 x_1 y_3 - x_2 y_3
\end{pmatrix}_{{\bf 3}_{2(k)}} ,
\end{eqnarray}
\begin{eqnarray}
\begin{pmatrix}
x_1 \\ x_2 \\
\end{pmatrix}_{{\bf 2}_1} \otimes 
\begin{pmatrix}
y_1 \\ y_2 \\ y_3 
\end{pmatrix}_{{\bf 3}_{2(k)}}   = 
\begin{pmatrix}
x_1y_1 + \omega^2 x_2 y_1 \\
\omega x_1y_2 + \omega x_2y_2 \\
\omega^2 x_1 y_3 + x_2 y_3
\end{pmatrix}_{{\bf 3}_{2(k)}} \oplus
\begin{pmatrix}
x_1y_1 - \omega^2 x_2 y_1 \\
\omega x_1y_2 - \omega x_2y_2 \\
\omega^2 x_1 y_3 - x_2 y_3
\end{pmatrix}_{{\bf 3}_{1(k)}} ,
\end{eqnarray}
\begin{eqnarray}
\begin{pmatrix}
x_1 \\ x_2 \\
\end{pmatrix}_{{\bf 2}_2} \otimes 
\begin{pmatrix}
y_1 \\ y_2 \\ y_3 
\end{pmatrix}_{{\bf 3}_{1(1)}}   = 
\begin{pmatrix}
x_1y_2 + \omega x_2 y_3 \\
\omega x_1y_3 + \omega^2 x_2 y_1 \\
\omega^2 x_1 y_2 + x_2 y_2
\end{pmatrix}_{{\bf 3}_{1(1)}} \oplus
\begin{pmatrix}
x_1y_2 - \omega x_2 y_3 \\
\omega x_1y_3 - \omega^2 x_2 y_1 \\
\omega^2 x_1 y_2 - x_2 y_2
\end{pmatrix}_{{\bf 3}_{2(1)}} ,
\end{eqnarray}
\begin{eqnarray}
\begin{pmatrix}
x_1 \\ x_2 \\
\end{pmatrix}_{{\bf 2}_2} \otimes 
\begin{pmatrix}
y_1 \\ y_2 \\ y_3 
\end{pmatrix}_{{\bf 3}_{2(1)}}   = 
\begin{pmatrix}
x_1y_2 + \omega x_2 y_3 \\
\omega x_1y_3 + \omega^2 x_2 y_1 \\
\omega^2 x_1 y_2 + x_2 y_2
\end{pmatrix}_{{\bf 3}_{2(1)}} \oplus
\begin{pmatrix}
x_1y_2 - \omega x_2 y_3 \\
\omega x_1y_3 - \omega^2 x_2 y_1 \\
\omega^2 x_1 y_2 - x_2 y_2
\end{pmatrix}_{{\bf 3}_{1(1)}} ,
\end{eqnarray}
\begin{eqnarray}
\begin{pmatrix}
x_1 \\ x_2 \\
\end{pmatrix}_{{\bf 2}_2} \otimes 
\begin{pmatrix}
y_1 \\ y_2 \\ y_3 
\end{pmatrix}_{{\bf 3}_{1(2)}}   = 
\begin{pmatrix}
x_1y_3 + \omega x_2 y_2 \\
\omega x_1y_1 + \omega^2 x_2 y_3 \\
\omega^2 x_1 y_2 + x_2 y_1
\end{pmatrix}_{{\bf 3}_{1(2)}} \oplus
\begin{pmatrix}
x_1y_3 - \omega x_2 y_2 \\
\omega x_1y_1 - \omega^2 x_2 y_2 \\
\omega^2 x_1 y_2 - x_2 y_1
\end{pmatrix}_{{\bf 3}_{2(2)}} ,
\end{eqnarray}
\begin{eqnarray}
\begin{pmatrix}
x_1 \\ x_2 \\
\end{pmatrix}_{{\bf 2}_2} \otimes 
\begin{pmatrix}
y_1 \\ y_2 \\ y_3 
\end{pmatrix}_{{\bf 3}_{2(2)}}   = 
\begin{pmatrix}
x_1y_3 + \omega x_2 y_2 \\
\omega x_1y_1 + \omega^2 x_2 y_3 \\
\omega^2 x_1 y_2 + x_2 y_1
\end{pmatrix}_{{\bf 3}_{2(2)}} \oplus
\begin{pmatrix}
x_1y_3 - \omega x_2 y_2 \\
\omega x_1y_1 - \omega^2 x_2 y_2 \\
\omega^2 x_1 y_2 - x_2 y_1
\end{pmatrix}_{{\bf 3}_{1(2)}} ,
\end{eqnarray}
\begin{eqnarray}
\begin{pmatrix}
x_1 \\ x_2 \\
\end{pmatrix}_{{\bf 2}_3} \otimes 
\begin{pmatrix}
y_1 \\ y_2 \\ y_3 
\end{pmatrix}_{{\bf 3}_{1(1)}}   = 
\begin{pmatrix}
x_1y_3 + \omega x_2 y_2 \\
\omega x_1y_1 + \omega^2 x_2 y_3 \\
\omega^2 x_1 y_2 + x_2 y_1
\end{pmatrix}_{{\bf 3}_{1(1)}} \oplus
\begin{pmatrix}
x_1y_3 - \omega x_2 y_2 \\
\omega x_1y_1 - \omega^2 x_2 y_2 \\
\omega^2 x_1 y_2 - x_2 y_1
\end{pmatrix}_{{\bf 3}_{2(1)}} ,
\end{eqnarray}
\begin{eqnarray}
\begin{pmatrix}
x_1 \\ x_2 \\
\end{pmatrix}_{{\bf 2}_3} \otimes 
\begin{pmatrix}
y_1 \\ y_2 \\ y_3 
\end{pmatrix}_{{\bf 3}_{2(1)}}   = 
\begin{pmatrix}
x_1y_3 + \omega x_2 y_2 \\
\omega x_1y_1 + \omega^2 x_2 y_3 \\
\omega^2 x_1 y_2 + x_2 y_1
\end{pmatrix}_{{\bf 3}_{2(1)}} \oplus
\begin{pmatrix}
x_1y_3 - \omega x_2 y_2 \\
\omega x_1y_1 - \omega^2 x_2 y_2 \\
\omega^2 x_1 y_2 - x_2 y_1
\end{pmatrix}_{{\bf 3}_{1(1)}} ,
\end{eqnarray}
\begin{eqnarray}
\begin{pmatrix}
x_1 \\ x_2 \\
\end{pmatrix}_{{\bf 2}_3} \otimes 
\begin{pmatrix}
y_1 \\ y_2 \\ y_3 
\end{pmatrix}_{{\bf 3}_{1(2)}}   = 
\begin{pmatrix}
x_1y_2 + \omega x_2 y_3 \\
\omega x_1y_3 + \omega^2 x_2 y_1 \\
\omega^2 x_1 y_1 + x_2 y_2
\end{pmatrix}_{{\bf 3}_{1(2)}} \oplus
\begin{pmatrix}
x_1y_2 - \omega x_2 y_3 \\
\omega x_1y_3 - \omega^2 x_2 y_1 \\
\omega^2 x_1 y_1 - x_2 y_2
\end{pmatrix}_{{\bf 3}_{2(2)}} ,
\end{eqnarray}
\begin{eqnarray}
\begin{pmatrix}
x_1 \\ x_2 \\
\end{pmatrix}_{{\bf 2}_3} \otimes 
\begin{pmatrix}
y_1 \\ y_2 \\ y_3 
\end{pmatrix}_{{\bf 3}_{2(2)}}   = 
\begin{pmatrix}
x_1y_2 + \omega x_2 y_3 \\
\omega x_1y_3 + \omega^2 x_2 y_1 \\
\omega^2 x_1 y_1 + x_2 y_2
\end{pmatrix}_{{\bf 3}_{2(2)}} \oplus
\begin{pmatrix}
x_1y_2 - \omega x_2 y_3 \\
\omega x_1y_3 - \omega^2 x_2 y_1 \\
\omega^2 x_1 y_1 - x_2 y_2
\end{pmatrix}_{{\bf 3}_{1(2)}} ,
\end{eqnarray}
\begin{eqnarray}
\begin{pmatrix}
x_1 \\ x_2 \\
\end{pmatrix}_{{\bf 2}_4} \otimes 
\begin{pmatrix}
y_1 \\ y_2 \\ y_3 
\end{pmatrix}_{{\bf 3}_{1(1)}}   = 
\begin{pmatrix}
x_1y_3 +  x_2 y_2 \\
 x_1y_1 +  x_2 y_3 \\
x_1 y_2 + x_2 y_1
\end{pmatrix}_{{\bf 3}_{1(1)}} \oplus
\begin{pmatrix}
x_1y_3 -  x_2 y_2 \\
 x_1y_1 -  x_2 y_3 \\
x_1 y_2 - x_2 y_1
\end{pmatrix}_{{\bf 3}_{2(1)}} ,
\end{eqnarray}
\begin{eqnarray}
\begin{pmatrix}
x_1 \\ x_2 \\
\end{pmatrix}_{{\bf 2}_4} \otimes 
\begin{pmatrix}
y_1 \\ y_2 \\ y_3 
\end{pmatrix}_{{\bf 3}_{2(1)}}   = 
\begin{pmatrix}
x_1y_3 +  x_2 y_2 \\
 x_1y_1 +  x_2 y_3 \\
x_1 y_2 + x_2 y_1
\end{pmatrix}_{{\bf 3}_{2(1)}} \oplus
\begin{pmatrix}
x_1y_3 -  x_2 y_2 \\
 x_1y_1 -  x_2 y_3 \\
x_1 y_2 - x_2 y_1
\end{pmatrix}_{{\bf 3}_{1(1)}} ,
\end{eqnarray}
\begin{eqnarray}
\begin{pmatrix}
x_1 \\ x_2 \\
\end{pmatrix}_{{\bf 2}_4} \otimes 
\begin{pmatrix}
y_1 \\ y_2 \\ y_3 
\end{pmatrix}_{{\bf 3}_{1(2)}}   = 
\begin{pmatrix}
x_1y_2 +  x_2 y_3 \\
 x_1y_3 +  x_2 y_1 \\
x_1 y_1 + x_2 y_2
\end{pmatrix}_{{\bf 3}_{1(2)}} \oplus
\begin{pmatrix}
x_1y_2 -  x_2 y_3 \\
 x_1y_3 -  x_2 y_1 \\
x_1 y_1 - x_2 y_2
\end{pmatrix}_{{\bf 3}_{2(2)}} ,
\end{eqnarray}
\begin{eqnarray}
\begin{pmatrix}
x_1 \\ x_2 \\
\end{pmatrix}_{{\bf 2}_4} \otimes 
\begin{pmatrix}
y_1 \\ y_2 \\ y_3 
\end{pmatrix}_{{\bf 3}_{2(2)}}   = 
\begin{pmatrix}
x_1y_2 +  x_2 y_3 \\
 x_1y_3 +  x_2 y_1 \\
x_1 y_1 + x_2 y_2
\end{pmatrix}_{{\bf 3}_{2(2)}} \oplus
\begin{pmatrix}
x_1y_2 -  x_2 y_3 \\
 x_1y_3 -  x_2 y_1 \\
x_1 y_1 - x_2 y_2
\end{pmatrix}_{{\bf 3}_{1(2)}} .
\end{eqnarray}

Furthermore, the tensor products of the non-trivial singlet 
${\bf 1}_-$ with other representations are obtained as 
\begin{eqnarray}
{\bf 2}_k\otimes {\bf 1}_- = {\bf 2}_k, \qquad 
{\bf 3}_{1(k)} \otimes {\bf 1}_- = {\bf 3}_{2(k)}, 
\qquad 
{\bf 3}_{2(k)} \otimes {\bf 1}_- = {\bf 3}_{1(k)}.
\end{eqnarray}

\clearpage

\section{Subgroups and decompositions of multiplets}

In the section \ref{sec:application}, we see 
interesting applications of non-Abelian discrete symmetries 
for particle physics.
For such applications, breaking of discrete symmetries 
is quite important, that is, 
breaking patterns of discrete groups and 
decompositions of multiplets.
In this section, we study decompositions 
of multiplets for groups, which are studied 
in the previous sections.
Suppose that a finite group $G$ has the order $N$ 
and $M$ is a divisor of $N$.
Then, Lagrange's theorem implies fine group $H$ with the order $M$ is a 
candidate for subgroups of $G$.
(See Appendix A.)

An irreducible representation $\mbox{\boldmath $r$}_G$ of $G$ 
can be decomposed in terms of irreducible representations 
$\mbox{\boldmath $r$}_{H,m}$ of its subgroup $H$ as 
$\mbox{\boldmath $r$}_G = \sum_m \mbox{\boldmath $r$}_{H,m}$.
If the trivial singlet of $H$ is included in 
such a decomposition, $\sum_m \mbox{\boldmath $r$}_{H,m}$, and 
a scalar field with such a trivial singlet develops its 
vacuum expectation value (VEV), the group $G$ breaks to $H$.
On the other hand, if a scalar field in a multiplet 
$\mbox{\boldmath $r$}_G$ develops its VEV and it does not correspond to 
the trivial singlet of $H$, 
the group $G$ breaks not to $H$, but to another group.

Furthermore, when we know group-theoretical aspects such as 
representations of $G$, it would be useful to use them 
to study those for subgroups of $G$.

In what follows, we show decompositions of multiplets 
 of $G$ into multiplets of subgroups.
For a finite group $G$, there are several chains of 
subgroups, $G \rightarrow G_1 \rightarrow \cdots \rightarrow G_k 
\rightarrow Z_N \rightarrow \{ e \}$, 
$G \rightarrow G'_1 \rightarrow \cdots \rightarrow G'_m \rightarrow
Z_M\rightarrow \{ e \}$, 
etc.
It would be obvious that the smallest non-trivial subgroup in those 
chains is an Abelian  group such as $Z_N$ or $Z_M$.
In most of cases, we concentrate on subgroups, 
which are shown explicitly in the previous sections.
Then, we show the largest subgroup such as 
$G_1$ and $G_1'$ in each chain of subgroups.

\subsection{$S_3$}

Here, we start with $S_3$, because 
$S_3$ is the minimal non-Abelian discrete group.
Its order is equal to $2\times 3 =6$.
Thus, there are two candidates for subgroups.
One is a group with the order two, and 
the other has the order three.
The former corresponds to $Z_2$ and the latter 
corresponds to $Z_3$.
As in section \ref{subsec:S3}, 
the $S_3$ consists of $\{e,a,b,ab,ba,bab\}$, 
where $a^2=e$ and $(ab)^3$=e.
Indeed, the subgroup $Z_2$ consists of e.g. $\{e,a \}$, 
while the other combinations such as $\{e,b \}$  and $\{e,bab \}$ also 
correspond to $Z_2$.
The subgroup $Z_3$ consists of  $ \{e, ab, ba=(ab)^2  \}$.
The $S_3$ has two singlets, ${\bf 1}$ and ${\bf 1}'$ and 
one doublet ${\bf 2}$.
Both of subgroups, $Z_2$ and $Z_3$, are Abelian.
Thus, decompositions of multiplets under $Z_2$ and $Z_3$ are 
rather simple.
We show such decompositions in what follows.


\vskip .5cm
{$\bullet$ $S_3 \to Z_3$}

The following elements 
\[ \{e, ab, ba  \}, \]
of $S_3$ construct the $Z_3$ subgroup, which is the normal 
subgroup.
There is no other choice to make a $Z_3$ subgroup.
There are three singlet representations, 
${\bf 1}_k$ $k=0,1,2$ for $Z_3$, that is, 
$ab=\omega^k$ on ${\bf 1}_k$.
Recall that $\chi_1(ab)=\chi_{1'}(ab)=1$ for 
both ${\bf 1}$ and ${\bf 1}'$ of $S_3$.
Thus, both ${\bf 1}$ and ${\bf 1}'$ of $S_3$ correspond 
to ${\bf 1}_0$ of $Z_3$.
On the other hand, the doublet ${\bf 2}$ of $S_3$ 
decomposes into two singlets of $Z_3$.
Since $\chi_2(ab)=-1$, 
the $S_3$ doublet ${\bf 2}$ decomposes into ${\bf 1}_1$ 
and  ${\bf 1}_2$ of $Z_3$.

In order to see this, we use the 
two dimensional representations of the group element $ab$ 
(\ref{eq:s3-2-rep}),
\begin{eqnarray}
ab=\mat2{-\frac12}{-\frac{\sqrt{3}}{2}}{\frac{\sqrt{3}}{2}}{-\frac12}.
\end{eqnarray}
Then the doublet $(x_1,x_2)$ decompose into two non-trivial singlets as
\begin{eqnarray}
{\bf 1}_1: x_1-ix_2 \quad \quad
{\bf 1}_2: x_1+ix_2.
\end{eqnarray}

\vskip .5cm
{$\bullet$ $S_3 \to Z_2$}


We consider the $Z_2$ subgroup of $S_3$, which consists of e.g. 
\[ \{e, a\}.\]
There are two singlet representations ${\bf 1}_k$ $k=0,1$, 
for $Z_2$, that is, $a=(-1)^k$ on ${\bf 1}_k$.
Recall that $\chi_1(a)=1$ and  $\chi_{1'}(a)=-1$ for 
${\bf  1}$ and ${\bf  1}'$ of $S_3$.
Thus, ${\bf  1}$ and ${\bf  1}'$ of $S_3$ correspond to 
${\bf 1}_0$ and ${\bf 1}_1$ of $Z_2$, respectively.
On the other hand, the doublet ${\bf 2}$ of $S_3$ decomposes into 
two singlets of $Z_2$.
Since $\chi_2(a)=-1$, the $S_3$ doublet ${\bf 2}$ decomposes 
into ${\bf 1}_0$ and ${\bf 1}_1$ of $Z_2$.
Indeed, the element $a$ is represented on ${\bf 2}$ in 
(\ref{eq:s3-2-rep}) as 
\begin{eqnarray}
a=\mat2{1}{0}{0}{-1}.
\end{eqnarray}
Then, for the doublet $(x_1,x_2)$, the elements $x_1$ and $x_2$ 
correspond to $x_1={\bf 1}_0$ and $x_2={\bf 1}_1$, respectively.

In addition to $\{e, a\}$, there are other $Z_2$ subgroups, 
$\{e, b\}$ and $\{e, aba\}$.
In both cases, the same results are obtained when we choose a 
proper basis.
This is an example of Abelian subgroups.
In non-Abelian subgroups, the same situation happens.
That is, different elements of a finite group $G$ 
construct the same subgroup.
A simple example is $D_6$. 
All of the $D_6$ elements 
are written by $a^mb^k$ for $m=0,1,\cdots,5$ and $k=0,1$, 
where $a^6=e$ and $bab=a^{-1}$.
Here, we denote $\tilde a=a^2$.
Then, the elements $\tilde a^mb^k$ for $m=0,1,2$ and $k=0,1$ 
correspond to the subgroup $D_3 \simeq S_3$.
On the other hand, we denote $\tilde b=ab$.
Then, the elements $\tilde a^m \tilde b^k$ for $m=0,1,2$ and $k=0,1$ 
correspond to another $D_3$ subgroup.
The decompositions of $D_6$ multiplets into $D_3$ multiplets 
are the same between both $D_3$ subgroups when we change 
a proper basis.

\subsection{$S_4$}

As mentioned in section \ref{sec:Delta-6N}, the 
$S_4$ group is isomorphic to $\Delta(24)$ and $(Z_2 \times Z_2)
\rtimes S_3$.
It would be convenient to use the terminology of $(Z_2 \times Z_2)
\rtimes S_3$.
That is, all of the elements are written as 
$b^kc^\ell a^m a'^{~n}$ with $k=0,1,2$ and $\ell,m,n=0,1$.
(See section \ref{sec:Delta-6N}.)
The generators, $a$, $a'$, $b$ and $c$, are related with 
the notation in section \ref{subsec:S4} as follows,
\begin{eqnarray}
b=c_1, \qquad c=f_1, \qquad a=a_4, \qquad a'=a_2.
\end{eqnarray}
They satisfy the following algebraic relations 
\begin{eqnarray}
&&b^3=c^2=(bc)^2=a^2=a'^2=e, \quad aa'=a'a,
\nonumber\\
&&
bab^{-1}=a^{-1}a'^{-1},
\quad
ba'b^{-1}=a,
\quad
cac^{-1}=a'^{-1},
\quad
ca'a^{-1}=a^{-1}.
\end{eqnarray}
Furthermore, their representations on ${\bf 1}$,  ${\bf 1}'$, 
 ${\bf 2}$,   ${\bf 3}$ and   ${\bf 3}'$ are shown in Table 
\ref{tab:S4-rep}. 
As subgroups, the $S_4$ includes non-Abelian groups, $S_3$, $A_4$ and 
$\Sigma(8)$, which is $(Z_2 \times Z_2) \rtimes Z_2$.
Thus, the decompositions of $S_4$ are non-trivial compared with 
those of $S_3$.

\begin{table}[t]
\begin{center}
\begin{tabular}{|c|c|c|c|c|c|c|}
\hline
&${\bf 1}$&${\bf 1}'$&${\bf 2}$&${\bf 3}$&${\bf 3}'$\\
\hline
$b$&1&1&$\mat2{\omega}{0}{0}{\omega^2}$&$\Mat3{0}{1}{0} {0}{0}{1} 
{1}{0}{0}$&$\Mat3{0}{1}{0} {0}{0}{1} {1}{0}{0}$\\
\hline
$c$&1&-1&$\mat2{0}{1}{1}{0}$&$\Mat3{0}{0}{1} {0}{1}{0} 
{1}{0}{0}$&$\Mat3{0}{0}{-1} {0}{-1}{0} {-1}{0}{0}$\\
\hline
$a$&1&1&$\mat2{1}{0}{0}{1}$&$\Mat3{-1}{0}{0} {0}{-1}{0} 
{0}{0}{1}$&$\Mat3{-1}{0}{0} {0}{-1}{0} {0}{0}{1}$\\
\hline
$a'$&1&1&$\mat2{1}{0}{0}{1}$&$\Mat3{1}{0}{0} {0}{-1}{0} 
{0}{0}{-1}$&$\Mat3{1}{0}{0} {0}{-1}{0} {0}{0}{-1}$\\
\hline
\end{tabular}
\end{center}
\caption{Representations of $S_4$ elements}
\label{tab:S4-rep}
\end{table}

\vskip .5cm
{$\bullet$ $S_4 \to S_3$}

The subgroup $S_3$ elements are $\{a_1,b_1,d_1,d_1,e_1,f_1\}$. 
Alternatively, they are denoted by $b^kc^\ell$ with $k=0,1,2$ and 
$\ell=0,1$, i.e., 
$\{e, b,b^2,c,bc,b^2c \}$.
Among them, Table \ref{tab:S3-rep-for-S3} shows 
the representations of the generators $b$ and $c$ 
on ${\bf 1}$, ${\bf 1}'$ and  ${\bf 2}$ of $S_3$.
Then each representation of $S_4$ is decomposed as 
\begin{eqnarray}
\begin{array}{cccccc}
S_4 &{\bf 1}&{\bf 1}'&{\bf 2}&{\bf 3} &{\bf 3}'   
 \\
 &\downarrow &\downarrow &\downarrow  &\downarrow  &\downarrow 
\\
S_3 &{\bf 1}&{\bf 1}'&{\bf 2}&{\bf 1}+{\bf 2} &{\bf 1}'+{\bf 2}  
\end{array} .
\end{eqnarray}
The components of ${\bf 3}$ $(x_1,x_2,x_3)$ are decomposed to 
${\bf 1}$ and ${\bf 2}$ as 
\begin{eqnarray}
{\bf 1}: (x_1+x_2+x_3), \qquad 
{\bf 2}:
\left(
\begin{array}{c}
x_1+\omega^2 x_2+\omega x_3  \\   
\omega(x_1+\omega^2 x_2+\omega x_3 ) \\   
\end{array}
\right) ,
\end{eqnarray}
and components of ${\bf 3}'$ are decomposed to 
${\bf 1}'$ and ${\bf 2}$ as 
\begin{eqnarray}
{\bf 1}': ( x_1+x_2+x_3), \qquad 
{\bf 2}: 
\left(
\begin{array}{c}
x_1+\omega^2 x_2+\omega x_3  \\   
-\omega(x_1+\omega^2 x_2+\omega x_3 ) \\   
\end{array}
\right) .
\end{eqnarray}

\begin{table}[t]
\begin{center}
\begin{tabular}{|c|c|c|c|c|c|c|}
\hline
&${\bf 1}$&${\bf 1}'$&${\bf 2}$\\
\hline
$b$&1&1&$\mat2{\omega}{0}{0}{\omega^2}$\\
\hline
$c$&1&-1&$\mat2{0}{1}{1}{0}$\\
\hline
\end{tabular}
\end{center}
\caption{Representations of $S_3$ elements}
\label{tab:S3-rep-for-S3}
\end{table}

\vskip .5cm
{$\bullet$ $S_4 \to A_4$}

The $A_4$ subgroup consists of $b^ka^ma'^n$ with $k=0,1,2$ and 
$m,n=0,1$. 
Recall that the $A_4$ is isomorphic to $\Delta(12)$.
Table \ref{tab:S4-rep-for-A4} shows the representations of 
the generators $b$, $a$ and $a'$ on ${\bf 1}$, ${\bf 1}'$, ${\bf 1}''$
and ${\bf 3}$ of $A_4$.
Then each representation of $S_4$ is decomposed as 
\begin{eqnarray}
\begin{array}{cccccc}
S_4 &{\bf 1}&{\bf 1}'&{\bf 2}&{\bf 3} &{\bf 3}'   
\\
 &\downarrow &\downarrow &\downarrow  &\downarrow  &\downarrow 
\\
A_4 &{\bf 1}&{\bf 1}&{\bf 1}'+{\bf 1}''&{\bf 3} &{\bf 3} 
\end{array} .
\end{eqnarray}

\begin{table}[t]
\begin{center}
\begin{tabular}{|c|c|c|c|c|c|c|c|c|}
\hline
&${\bf 1}$&${\bf 1}'$&${\bf 1}''$&${\bf 3}$\\
\hline
$b$&1&$\omega$&$\omega^2$&$\Mat3{0}{1}{0} {0}{0}{1} {1}{0}{0}$\\
\hline
$a$&1&1&1&$\Mat3{-1}{0}{0} {0}{-1}{0} {0}{0}{1}$\\
\hline
$a'$&1&1&1&$\Mat3{1}{0}{0} {0}{-1}{0} {0}{0}{-1}$\\
\hline
\end{tabular}
\end{center}
\caption{Representations of $A_4$ elements}
\label{tab:S4-rep-for-A4}
\end{table}

\vskip .5cm
{$\bullet$ $S_4 \to \Sigma (8)$}

The subgroup $\Sigma(8)$, i.e. $(Z_2 \times Z_2)\rtimes Z_2$, 
consists of $c^\ell a^ma'^n$ with $\ell,m,n=0,1$.
Table \ref{tab:S4-rep-for-Sigama8} shows the representations 
of the generators $c$, $a$ and $a'$ on 
${\bf 1}_{+0}$, ${\bf 1}_{+1}$, ${\bf 1}_{-0}$, ${\bf 1}_{-1}$ 
and ${\bf 2}_{1,0}$ of $\Sigma(8)$.
\begin{table}[t]
\begin{center}
\begin{tabular}{|c|c|c|c|c|c|c|c|c|}
\hline
&${\bf 1}_{+0}$&${\bf 1}_{+1}$&${\bf 1}_{-0}$&${\bf 1}_{-1}$&${\bf 
2}_{1,0}$\\
\hline
$c$&1&1&-1&-1&$\mat2{0}{1} {1}{0}$\\
\hline
$a$&1&-1&1&-1&$\mat2{1}{0} {0}{-1}$\\
\hline
$a'$&1&-1&1&-1&$\mat2{-1}{0} {0}{1}$\\
\hline
\end{tabular}
\end{center}
\caption{Representations of $\Sigma(8)$ elements}
\label{tab:S4-rep-for-Sigama8}
\end{table}
Then each representation of $S_4$ is decomposed as 
\begin{eqnarray}
\begin{array}{cccccc}
S_4 &{\bf 1}&{\bf 1}'&{\bf 2}&{\bf 3} &{\bf 3}'   
\\
 &\downarrow &\downarrow &\downarrow  &\downarrow  &\downarrow 
\\
\Sigma(8) &{\bf 1}_{+0}&{\bf 1}_{-0}&{\bf 1}_{+0}+{\bf 1}_{-0}
&{\bf 1}_{+1}+{\bf 2} &{\bf 1}_{-1}+{\bf 2} 
\end{array} .
\end{eqnarray}
The components of ${\bf 3}$ $(x_1,x_2,x_3)$ are decomposed 
to ${\bf 1}_{+1}$ and ${\bf 2} $ as 
\begin{eqnarray}
{\bf 1}_{+1}:x_2, \qquad 
{\bf 2}:
\left(
\begin{array}{c}
x_3  \\   
x_1 \\   
\end{array}
\right),
\end{eqnarray}
and the components of ${\bf 3}'$ are decomposed to ${\bf 1}_{-1}$ and
{\bf 2} as 
\begin{eqnarray}
{\bf 1}_{-1}:x_2, \qquad 
{\bf 2}:\quad
\left(
\begin{array}{c}
x_3  \\   
-x_1 \\   
\end{array}
\right) .
\end{eqnarray}

\subsection{$A_5$}

All of the $A_5$ elements are written 
by products of $s=a$ and $t=bab$ 
as shown in section \ref{sec:A5}.

\vskip .5cm
{$\bullet$ $A_5 \to A_4$}

The subgroup $A_4$ elements are 
$\{e, b, \tilde{a} , b\tilde{a}b^2, b^2\tilde{a}b, b\tilde{a}, 
\tilde{a}b, \tilde{a}b\tilde{a}, b^2 \tilde{a}, b^2\tilde{a}
b\tilde{a}b \}$ where $\tilde{a}=ab^2aba$. 
We denote $\tilde t =b$ and $\tilde s =  \tilde{a}$.
They satisfy the following relations 
\begin{eqnarray}
\tilde s^2=\tilde t^3=(\tilde s \tilde t)^3=e,
\end{eqnarray}
and correspond to the generators, $s$ and $t$, of the $A_4$ group 
in section \ref{sec:A4}. 
Each representation of $A_5$
is decomposed as
\begin{eqnarray}
\begin{array}{cccccc}
A_5 &{\bf 1}&{\bf 3}&{\bf 3}'&{\bf 4} &{\bf 5}   
\\
 &\downarrow &\downarrow &\downarrow  &\downarrow  &\downarrow 
\\
A_4 &{\bf 1}&{\bf 3}&{\bf 3}&{\bf 1}+{\bf 3} &
{\bf 1'}+{\bf 1''}+{\bf 3} 
\end{array} .
\end{eqnarray}

\vskip .5cm
{$\bullet$ $A_5 \to D_{5}$}

The $D_{5}$ subgroup consists of 
$a^k\tilde{a}^m$ with $k=0,1$ and $m=0,1,2,3,4$
where $\tilde{a} \equiv bab^2a$.
They satisfy $a^2=\tilde{a}^5=e$ and  $a\tilde{a}a=\tilde{a}^4$.   
In order to identify the $D_5$ basis 
used in section \ref{sec:DN}, 
we define $\tilde{b}=abab^2a$.  
Table \ref{tab:D5-rep-for-A5} shows the representations of 
these generators $\tilde{a}$, $\tilde{b}$ 
on ${\bf 1_+}$, ${\bf 1_-}$, ${\bf 2_1}$
and ${\bf 2_2}$ of $D_5$.
Then each representation of $A_5$ is decomposed as 
\begin{eqnarray}
\begin{array}{cccccc}
A_5 &{\bf 1}&{\bf 3}&{\bf 3}'&{\bf 4} &{\bf 5}   
\\
 &\downarrow &\downarrow &\downarrow  &\downarrow  &\downarrow 
\\
D_5 &{\bf 1_+}&{\bf 1_-}+{\bf 2_1} & {\bf 1_-}+{\bf 2_2} &
{\bf 2_1}+{\bf 2_2} &{\bf 1_+}+{\bf 2_1}+{\bf 2_2} 
\end{array} .
\end{eqnarray}

\begin{table}[t]
\begin{center}
\begin{tabular}{|c|c|c|c|c|c|}
\hline
&{${\bf 1}_+$}&{${\bf 1}_-$}&{${\bf 2}_1$}&{${\bf 2}_2$}\\
\hline
$\tilde{a}$&1&1&$\mat2{\exp{2\pi i/5}}{0}{0}{-\exp{2\pi i/5}} $
&$\mat2{\exp{4\pi i/5}}{0}{0}{-\exp{4\pi i/5}} $  \\
\hline
$\tilde{b}$&1&-1&$\mat2{0}{1}{1}{0} $
&$\mat2{0}{1}{1}{0} $  \\
\hline
\end{tabular}
\end{center}
\caption{Representations of $D_5$ elements.}
\label{tab:D5-rep-for-A5}
\end{table}

\vskip .5cm
{$\bullet$ $A_5 \to S_3 \simeq D_3$}

Recall that the $S_3$ group is isomorphic to the $D_3$ group. 
The subgroup $D_3$ consists of  $b^k\tilde{a}^m$ 
with $k=0,1,2$ and $m=0,1$
where we define $\tilde{a} = ab^2ab^2ab$.
These generators satisfy  $\tilde{a}^2=e$ and 
$\tilde{a}b\tilde{a}=b^2$.   
Then each representation of $A_5$ is decomposed as 
\begin{eqnarray}
\begin{array}{cccccc}
A_5 &{\bf 1}&{\bf 3}&{\bf 3}'&{\bf 4} &{\bf 5}   
\\
 &\downarrow &\downarrow &\downarrow  &\downarrow  &\downarrow 
\\
D_3 &{\bf 1_+}&{\bf 1_-}+{\bf 2} & {\bf 1_-}+{\bf 2} &
{\bf 1_+}+{\bf 1_-}+{\bf 2} &{\bf 1_+}+{\bf 2}+{\bf 2} 
\end{array} .
\end{eqnarray}

\subsection{$T'$}

All of the $T'$ elements are written 
in terms of the generators, $s$ and $t$ as well as $r$, which 
satisfy the algebraic relations, 
$s^2=r$,  $r^2=t^3=(st)^3=e$ and $rt=tr$.
Table \ref{tab:T'-rep} shows the representations of $s$, $t$ and $r$ on 
each representation.

\begin{table}[t]
\begin{center}
\begin{tabular}{|c|c|c|c|c|c|c|c|c|c|c|}
\hline
&\!\!{\bf 1}\!\!&\!\!{\bf 1}'\!\!&\!\!{\bf 1}''\!\!&\!\!{\bf 2}\!\!&\!\!{\bf 2}'\!\!&\!\!{\bf 2}''\!\!&\!\!{\bf 3}\!\!
\\\hline
$\!\!s\!\!$&\!\!1\!\!&\!\!1\!\!&\!\!1\!\!&\!\!$\frac{i}{\sqrt3}\mat2{1}{\sqrt2}{\sqrt2}{-1}$\!\!
&\!\!$\frac{i}{\sqrt3}\mat2{1}{\sqrt2}{\sqrt2}{-1}$\!\!&\!\!$\frac{i}{\sqrt3}\mat2{1}{\sqrt2}{\sqrt2}{-1}$\!\!
&\!\!$\frac13\Mat3{-1}{2}{2} {2}{-1}{2} {2}{2}{-1}$\!\! 
\\\hline
$\!\!r\!\!$&\!\!1\!\!&\!\!1\!\!&\!\!1\!\!&\!\!$\mat2{-1}{0}{0}{-1}$\!\!
&\!\!$\mat2{-1}{0}{0}{-1}$\!\!&\!\!$\mat2{-1}{0}{0}{-1}$\!\!
&\!\!$\Mat3{1}{0}{0} {0}{1}{0} {0}{0}{1}$\!\!
\\\hline
$\!\!t\!\!$&\!\!1\!\!&\!\!$\omega$\!\!&\!\!$\omega^2$\!\!
&\!\!$\mat2{\omega}{0}{0}{\omega^2}$\!\!&\!\!$\mat2{\omega^2}{0}{0}{1}$\!\!
&\!\!$\mat2{1}{0}{0}{\omega}$\!\!&\!\!$\frac13\Mat3{1}{0}{0} {0}{\omega}{0} {0}{0}{\omega^2}$\!\!
\\\hline
\end{tabular}
\end{center}
\caption{Representations of $T'$}
\label{tab:T'-rep}
\end{table}

\vskip .5cm
{$\bullet$ $T' \to Z_6$}


The subgroup $Z_6$ consists of  $a^m$, with $m=0,\cdots, 5$, 
where $a=rt$ and $a^6=e$.
The $Z_6$ group has six singlet representations, 
${\bf 1}_n$ with $n=0,\cdots,5$.
On the singlet  ${\bf 1}_n$, the generator $a$ is 
represented as $a=e^{2\pi in/6}$.
Thus, each representation of $T'$ is decomposed as 
\begin{eqnarray}
\begin{array}{ccccccccc}
T' &{\bf 1}&{\bf 1}'&{\bf 1}''&{\bf 2} &{\bf 2}'&{\bf 2}''&{\bf 3}   
\\
 &\downarrow &\downarrow &\downarrow &\downarrow &\downarrow  &\downarrow  
&\downarrow 
\\
Z_6 &{\bf 1}_0&{\bf 1}_2&{\bf 1}_4&{\bf 1}_5+{\bf 1}_1&{\bf 1}_5+{\bf
  1}_3
&{\bf 1}_3+{\bf 1}_5&{\bf 1}_0+{\bf 1}_2+{\bf 1}_4 
\end{array} .
\end{eqnarray}

\vskip .5cm
{$\bullet$ $T' \to Z_4$}

The subgroup $Z_4$ consists of $\{e,s,s^2,s^3\}$. 
The $Z_4$ group has two singlet representations, 
${\bf 1}_m$ with $m=0,1,2,3$.
On the singlet  ${\bf 1}_m$, the generator $s$ is 
represented as $s=e^{\pi im/2}$.
All of the doublets of $T'$, ${\bf 2}$, ${\bf 2}'$, ${\bf 2}''$, are 
decomposed to two singlets of $Z_4$ ${\bf 1}_1$ and ${\bf 1}_3$ 
as  ${\bf 1}_1:\frac{1+\sqrt3}{\sqrt2}x_1+x_2$ and 
${\bf 1}_3:-\frac{-1+\sqrt3}{\sqrt2}ix_1+x_2$, where 
$(x_1,x_2)$ correspond to the doublets.
In addition, the triplet ${\bf 3}:(x_1,x_2,x_3)$ is 
decomposed to singlets,  ${\bf 1}_0+{\bf 1}_2+{\bf 1}_2$ as 
${\bf 1}_0: (x_1+x_2+x_3)$, 
${\bf 1}_2:(-x_1+x_3)$ and 
${\bf 1}_2:(-x_1+x_2)$. 
The results are summarized as 
\begin{eqnarray}
\begin{array}{ccccccccc}
T' &{\bf 1}&{\bf 1}'&{\bf 1}''&{\bf 2}&{\bf 2}'&{\bf 2}''&{\bf 3}   
\\
 &\downarrow &\downarrow &\downarrow &\downarrow &\downarrow  &\downarrow  
&\downarrow 
\\
Z_4 &{\bf 1}_0&{\bf 1}_0&{\bf 1}_0&{\bf 1}_1+{\bf 1}_3&{\bf 1}_1+{\bf 1}_3
&{\bf 1}_1+{\bf 1}_3&{\bf 1}_0+{\bf 1}_2+{\bf 1}_2 
\end{array}  .
\end{eqnarray}

\vskip .5cm
{$\bullet$ $T' \to Q_4$}


We consider the subgroup $Q_4$, which 
consists of $s^mb^k$ with $m=0,1,2,3$ and $k=0,1$. 
The generator $b$ is defined by $b=tst^2$. 
Then, each representation of $T'$ is decomposed as 
\begin{eqnarray}
\begin{array}{ccccccccc}
T' &{\bf 1}&{\bf 1}'&{\bf 1}''&{\bf 2}&{\bf 2}'&{\bf 2}''&{\bf 3}   
\\
 &\downarrow &\downarrow &\downarrow &\downarrow &\downarrow  &\downarrow  
&\downarrow 
\\
Q_4 &{\bf 1}_{++}&{\bf 1}_{++}&{\bf 1}_{++}&{\bf 2}&{\bf 2}
  &{\bf 2}&{\bf 1}_{+-}+{\bf 1}_{-+}+{\bf 1}_{--} 
\end{array} .
\end{eqnarray}

\subsection{$D_4$}
Here, we study $D_4$, which is the second minimum discrete symmetry.
All of the $D_4$ elements are written by $a^mb^k$ with $m=0,1,2,3$ 
and $k=0,1$.
Since the order of $D_4$ is 8, it contains the order 2 and 4 subgroups. 
There are two types of the order $4$ groups which are corresponding to  
$Z_2 \times Z_2$ and $Z_4$ groups.  
All of subgroups are Abelian.
Thus, decompositions are rather simple.

\vskip .5cm
{$\bullet$ $D_4 \to Z_4$}

The subgroup $Z_4$ is consist of the elements $\{e,a,a^2,a^3\}$.
Obviously it is the normal subgroups of $D_4$ and 
there are four types of irreducible singlet representations 
${\bf 1}_m$ with $m=0,1,2,3$, where $a$ is represented as $a=e^{\pi im/2}$.  
{}From the characters of $D_4$ groups, 
it is found that ${\bf 1}_{++}$ and ${\bf  1}_{--}$ 
of $D_4$ correspond to ${\bf 1}_0$ of $Z_4$ and 
${\bf 1}_{+-}$ and ${\bf 1}_{-+}$ of $D_4$ correspond to ${\bf 1}_2$ of 
$D_4$.
For the $D_4$ doublet ${\bf 2}$, it is convenient to 
use the diagonal base of matrix $a$ as 
\begin{eqnarray}
a=\mat2{i}{0}{0}{-i}. 
\end{eqnarray}
Then we can read the doublet ${\bf 2}:(x_1,x_2)$ decomposes 
to two singlets as ${\bf 1}_1: x_1$ and ${\bf 1}_3: x_2$.

\vskip .5cm
{$\bullet$ $D_4 \to Z_2 \times Z_2$}

We denote $\tilde a = a^2$.
Then, the subgroup $Z_2 \times Z_2$ consists 
of $\{e ,\tilde{a} ,{b} ,\tilde{a}{b} \}$, where  
$\tilde a b = b \tilde a$ and  $\tilde{a}^2={b}^2=e$. 
Obviously, their representations are quite simple, 
that is, ${\bf 1}_{\pm \pm}$, whose $Z_2 \times Z_2$ charges 
are determined by $\tilde a=\pm 1$ and $ b=\pm 1$.
We use the notation that the first (second) subscript of  ${\bf 1}_{\pm 
\pm}$ 
denotes the $Z_2$ charge for $\tilde a$ ($b$).
%
Then, the singlets ${\bf 1}_{++}$ and ${\bf 1}_{+-}$ of $D_4$ 
correspond to ${\bf 1}_{++}$ of $Z_2 \times Z_2$
and ${\bf 1}_{-+}$ and ${\bf 1}_{--}$ of $D_4$ 
correspond to ${\bf 1}_{+-}$ of $Z_2 \times Z_2$.
The doublet {\bf 2} of $D_4$ decomposes to 
${\bf 1}_{-+}$ and ${\bf 1}_{--}$ of $Z_2$.

In addition to the above, 
there is another choice of $Z_2 \times Z_2$ subgroup, 
which consists of  $\{e, a^2, ab, a^3b\}$.
In this case, we can obtain the same decomposition of $D_4$.

\vskip .5cm
{$\bullet$ $D_4 \to Z_2$}

Furthermore, both $Z_4$ and $Z_2\times Z_2$ include 
$Z_2$ subgroup.
The decomposition of $D_4$ to $Z_2$ is rather straightforward.

\subsection{general $D_N$}

Since the group $D_N$ is isomorphic to $Z_N \rtimes Z_2$, 
$D_M$ and $Z_N$ as well as $Z_2$ appear as subgroups of $D_N$, 
where $M$ is a divisor of $N$.
Recall that all of $D_N$ elements are written by 
$a^mb^k$ with $m=0,\cdots,N-1$ and $k=0,1$.
There are singlets and doublets ${\bf 2}_k$, where 
$k=1,\cdots,N/2-1$ for $N=$ even and 
$k=1,\cdots,(N -1)/2$ for $N=$ odd.
On the doublet ${\bf 2}_k$, the generators $a$ and $b$ are 
represented as  
\begin{eqnarray}
a=\mat2{\rho^k}{0}{0}{\rho^{-k}}, 
\quad
b=\mat2{0}{1}{1}{0}, 
\end{eqnarray}
where $\rho=e^{2\pi i /N}$. 
For $N=$ even, there are four singlets ${\bf 1}_{\pm \pm}$.
The generator $b$ is represented as $b=1$ on ${\bf 1}_{+\pm}$, 
while $b=-1$ on  ${\bf 1}_{-\pm}$.
The generator $a$ is represented as $a=1$ on ${\bf 1}_{+ +}$ 
and ${\bf 1}_{--}$, while $a=-1$ on ${\bf 1}_{+ -}$ 
and ${\bf 1}_{+-}$.
For $N=$ odd, there are two singlets ${\bf 1}_\pm$.
The generator $b$ is represented as $b=1$ on ${\bf 1}_+$ 
and $b=-1$ on ${\bf 1}_-$, while $a=1$ on 
both singlets.

\vskip .5cm
{$\bullet$ $D_N \to Z_2$}

The two elements $e$ and $b$ construct the 
$Z_2$ subgroup.
Obviously,
there are two singlet representations
${\bf 1}_0, {\bf 1}_1$, 
where the subscript denotes the $Z_2$ charge.
That is, we have $b=1$ on ${\bf 1}_0$ and 
$b=-1$ on ${\bf 1}_1$.

When $N$ is even, the singlets ${\bf 1}_{++}$ and ${\bf 1}_{+-}$ of $D_N$ 
become ${\bf 1}_0$ of $Z_2$ and the singlets  
${\bf 1}_{-+}$ and ${\bf 1}_{--}$ of $D_N$ become ${\bf 1}_1$ of $Z_2$. 
The doublets ${\bf 2}_k$ of $D_N$, $(x_1,x_2)$, decompose 
two singlets as ${\bf 1}_0: x_1+x_2$ and ${\bf 1}_1: x_1-x_2$.
These results are summarized as follows,
\begin{eqnarray}
\begin{array}{cccccc}
D_N &{\bf 1}_{++}&{\bf 1}_{+-}&{\bf 1}_{-+}&{\bf 1}_{--} &{\bf 2}_{k}   
\\
 &\downarrow &\downarrow &\downarrow  &\downarrow  &\downarrow 
\\
Z_2 &{\bf 1}_{0}&{\bf 1}_{0}&{\bf 1}_{1}&{\bf 1}_{1} &{\bf 1}_0+{\bf 1}_1 
\end{array} .
\end{eqnarray}

When $N$ is odd, the singlet ${\bf 1}_{+}$ of $D_N$ 
becomes ${\bf 1}_0$ of $Z_2$ and the singlet 
${\bf 1}_{-}$ of $D_N$ becomes ${\bf 1}_1$ of $Z_2$. 
The decompositions of doublets ${\bf 2}_k$ are the same as 
those for $N=$ even.
These results are summarized as 
\begin{eqnarray}
\begin{array}{cccc}
D_N &{\bf 1}_{+}&{\bf 1}_{-} &{\bf 2}_{k}   
\\
 &\downarrow   &\downarrow  &\downarrow 
\\
Z_2 &{\bf 1}_{0}&{\bf 1}_{1} &{\bf 1}_0+{\bf 1}_1 
\end{array} .
\end{eqnarray}

\vskip .5cm
{$\bullet$ $D_N \to Z_N$}

The subgroup $Z_N$  consists of the elements $\{e,a,\cdots,a^{N-1}\}$.
Obviously it is the normal subgroups of $D_N$ and 
there are $N$ types of irreducible singlet representations
${\bf 1}_0, {\bf 1}_1,\cdots,{\bf 1}_{N-1}$.
On the ${\bf 1}_k$, the generator $a$ is represented as $a=\rho^k$.

When $N$ is even, the singlets ${\bf 1}_{++}$ and ${\bf 1}_{--}$ of $D_N$ 
become ${\bf 1}_0$ of $Z_N$ and the singlets  
${\bf 1}_{+-}$ and ${\bf 1}_{-+}$ of $D_N$ become ${\bf 1}_{N/2}$ of $Z_N$. 
The doublets  ${\bf 2}_k$, $(x_1,x_2)$,   
decompose to two singlets as ${\bf 1}_{k}: x_1$ and ${\bf 1}_{N-k}: x_2$.
These results are summarized as follows,
\begin{eqnarray}
\begin{array}{cccccc}
D_N &{\bf 1}_{++}&{\bf 1}_{+-}& {\bf 1}_{-+}& {\bf 1}_{--} &{\bf 2}_{k}   
\\
 &\downarrow &\downarrow &\downarrow  &\downarrow  &\downarrow 
\\
Z_N &{\bf 1}_{0}&{\bf 1}_{N/2}&{\bf 1}_{N/2}&{\bf 1}_{0} &
{\bf 1}_{k}+{\bf 1}_{N-k} 
\end{array}.
\end{eqnarray}

When $N$ is odd, both ${\bf 1}_{+}$ and ${\bf 1}_-$ of $D_N$ 
become ${\bf 1}_0$ of $Z_N$. 
The decompositions of doublets ${\bf 2}_k$ are the same as 
those for $N=$ even.

\vskip .5cm
{$\bullet$ $D_N \to D_M$}

The above decompositions of $D_N$ are rather straightforward, 
because subgroups are Abelian.
Here we consider the $D_M$ subgroup, where $M$ is a divisor of $N$.
The decompositions of $D_N$ to $D_M$ would be non-trivial.
We denote $\tilde a=a^\ell$ with $\ell = N/M$, where $\ell$ is integer.
The subgroup $D_M$ consists of $\tilde a^m b^k$ with 
$m=0,\cdots,M-1$ and $k=0,1$.
There are three combinations of $(N,M)$, 
i.e. $(N,M)$=(even,even), (even,odd) and (odd,odd).

We start with the combination $(N,M)=$(even,even).
Recall that $(ab)$ of $D_N$ is represented as $ab=1$ on ${\bf 1}_{\pm+}$ 
and $ab=-1$ on ${\bf 1}_{\pm -}$.
Thus, the representations of $(a^\ell b)$ depend on whether $\ell$ 
is even or odd.
When $\ell$ is odd, $(ab)$ and $(a^\ell b)$ are represented 
in the same way on each of the above singlets.
On the other hand, when $\ell$ = even, we always have 
the singlet representations with $a^\ell=1$.
The doublets ${\bf 2}_k$ of $D_N$ correspond to 
the doublets ${\bf 2}_{k'}$ of $D_M$ when 
$k=k'$ (mod $M$).
In addition, when  $k=-k'$ (mod $M$), 
doublets ${\bf 2}_k$ $(x_1,x_2)$ of $D_N$ correspond to 
the doublets ${\bf 2}_{M-k'}$ $(x_2,x_1)$ of $D_M$.
That is, the components are exchanged each other 
and we denote it by $\tilde {\bf 2}_{M-k'}$.
Furthermore, the other doublets ${\bf 2}_k$ of $D_N$ 
decompose to two singlets of $D_M$ as 
${\bf 1}_{+-}+{\bf 1}_{-+}$ 
with ${\bf 1}_{+-}:x_1+x_2$ and ${\bf 1}_{-+}:x_1-x_2$ for $k=(M/2)$
(mod $M$) and 
${\bf 1}_{++}+{\bf 1}_{--}$ with ${\bf 1}_{++}:x_1+x_2$ 
and ${\bf 1}_{--}:x_1-x_2$ for $k=0$ (mod $M$). 
These results are summarized as follows,
\begin{equation}
\begin{array}{ccccccccccc}
&D_N &{\bf 1}_{++}&{\bf 1}_{+-}&{\bf 1}_{-+}&{\bf 1}_{--} &
{\bf 2}_{k'+Mn} &{\bf 2}_{\frac{M}{2}-k'+Mn} &{\bf 2}_{\frac{M}{2}(2n+1)}& {\bf 2}_{Mn}   
\\
&&\downarrow &\downarrow &\downarrow  &\downarrow  &\downarrow  
&\downarrow  &\downarrow&\downarrow 
\\
(\ell=\textrm{odd})&D_M &{\bf 1}_{++}&{\bf 1}_{+-}& {\bf 1}_{-+}&{\bf 1}_{--} 
&{\bf 2}_{k'} &\tilde {\bf 2}_{\frac{M}{2}-k'} &{\bf 1}_{+-}+{\bf 1}_{-+}
&{\bf 1}_{++}+{\bf 1}_{--}  
\\
(\ell=\textrm{even})\!\!&\!\!D_M &{\bf 1}_{++}&{\bf 1}_{++}&{\bf 1}_{--}&{\bf 1}_{--} 
&{\bf 2}_{k'} &\tilde {\bf 2}_{\frac{M}{2}-k'} &{\bf 1}_{+-}+{\bf 1}_{-+}&{\bf 1}_{++}+{\bf 1}_{--}  
\end{array} ,
\end{equation}
where $n$ is integer.

Next we consider the case with $(N,M)=$(even,odd).
In this case, the singlet ${\bf 1}_{++}$, ${\bf 1}_{+-}$, 
${\bf 1}_{-+}$ and ${\bf 1}_{--}$ 
of $D_N$ become ${\bf 1}_{+}$, ${\bf 1}_{+}$, ${\bf 1}_{-}$ 
and ${\bf 1}_{-}$ of $D_M$, respectively.  
The doublets ${\bf 2}_k$ of $D_N$ correspond to 
the doublets ${\bf 2}_{k'}$ of $D_M$ when 
$k=k'$ (mod $M$).
In addition, when  $k=-k'$ (mod $M$), 
the doublets ${\bf 2}_k$ $(x_1,x_2)$ of $D_N$ correspond to 
the doublets ${\bf 2}_{M-k'}$ $(x_2,x_1)$ of $D_M$.
Furthermore, when $k=0$ (mod $M$),
other doublets ${\bf 2}_k$ of $D_N$ 
decompose to two singlets of $D_M$ as 
${\bf 1}_{+}+{\bf 1}_{-}$, where ${\bf 1}_{+}:x_1+x_2$ 
and ${\bf 1}_{-}:x_1-x_2$.
These results are summarized as follows,
\begin{eqnarray}
\begin{array}{cccccccc}
D_N &{\bf 1}_{++}&{\bf 1}_{-+}&{\bf 1}_{+-}& {\bf 1}_{--} 
&{\bf 2}_{k'+Mn} &{\bf 2}_{Mn-k'} &{\bf 2}_{Mn}   
\\
 &\downarrow &\downarrow &\downarrow  &\downarrow  &\downarrow  &\downarrow  
&\downarrow 
\\
D_M &{\bf 1}_{+}&{\bf 1}_{+}&{\bf 1}_{-}&{\bf 1}_{-} 
&{\bf 2}_{k'} &{\tilde {\bf 2}}_{M-k'} &{\bf 1}_{+}+{\bf 1}_{-}  
\end{array} ,
\end{eqnarray}
where $n$ is integer.

Now, let us consider the case with $(N,M)=$(odd,odd). 
In this case, the singlets ${\bf 1}_{+}$ and ${\bf 1}_{-}$ 
of $D_N$ become ${\bf 1}_{+}$ and ${\bf 1}_{-}$ of $D_M$.
The doublets ${\bf 2}_k$ of $D_N$ correspond to 
the doublets ${\bf 2}_{k'}$ of $D_M$ when 
$k=k'$ (mod $M$).
In addition, when  $k=-k'$ (mod $M$), 
doublets ${\bf 2}_k$ $(x_1,x_2)$ of $D_N$ correspond to 
the doublets ${\bf 2}_{M-k'}$ $(x_2,x_1)$ of $D_M$.
Furthermore, when $k=0$ (mod $M$),
other doublets ${\bf 2}_k$ of $D_N$ 
decompose to two singlets of $D_M$ as 
${\bf 1}_{+}+{\bf 1}_{-}$, where ${\bf 1}_{+}:x_1+x_2$ 
and ${\bf 1}_{-}:x_1-x_2$.
These results are summarized as follows,
\begin{eqnarray}
\begin{array}{cccccc}
D_N &{\bf 1}_{+}&{\bf 1}_{-} &{\bf 2}_{k'+Mn} &{\bf 2}_{Mn-k'} &{\bf 
2}_{Mn}   
\\
 &\downarrow &\downarrow &\downarrow  &\downarrow  &\downarrow 
\\
D_M &{\bf 1}_{+}&{\bf 1}_{-} &{\bf 2}_{k'} &{\tilde {\bf 2}}_{M-k'} 
&{\bf 1}_{+}+{\bf 1}_{-}  
\end{array} ,
\end{eqnarray}
where $n$ is integer.

\subsection{$Q_4$}

Here, we study $Q_4$.
All of the $Q_4$ elements are written by $a^mb^k$ with $m=0,1,2,3$ 
and $k=0,1$. 
Since the order of $Q_4$ is equal to 8, 
it contains the order 2 and 4 subgroups. 
There are a few types of the order $4$ groups which correspond to  
$Z_4$ groups.

\vskip .5cm
{$\bullet$ $Q_4 \to Z_4$}

For example, the elements $\{e,a,a^2,a^3\}$ construct 
the $Z_4$ subgroup.
Obviously it is the normal subgroups of $Q_4$ and 
there are four types of irreducible singlet representations 
${\bf 1}_m$ with $m=0,1,2,3$, where $a$ is represented as $a=e^{\pi im/2}$. 
{}From the characters of $Q_4$ groups, 
it is found that ${\bf 1}_{++}$ and ${\bf
  1}_{--}$ of $Q_4$ correspond to ${\bf 1}_0$ of $Z_4$ and 
${\bf 1}_{-+}$ and ${\bf 1}_{+-}$ of $Q_4$ correspond to ${\bf 1}_2$ of 
$Z_4$. 
For the doublets of $Q_4$, it is convenient to 
use the diagonal base of matrix $a$ as 
\begin{eqnarray}
a=\mat2{i}{0}{0}{-i}. 
\end{eqnarray}
Then we find that the doublet {\bf 2} $(x_1,x_2)$ decomposes 
to  two singlets as ${\bf 1}_1: x_1$ and ${\bf 1}_3: x_2$.
These results are summarized as follows,
\begin{eqnarray}
\begin{array}{cccccc}
Q_4 &{\bf 1}_{++}&{\bf 1}_{+-}&{\bf 1}_{-+}&{\bf 1}_{--} &{\bf 2}   
\\
 &\downarrow &\downarrow &\downarrow  &\downarrow  &\downarrow 
\\
Z_4 &{\bf 1}_{0}&{\bf 1}_{2}&{\bf 1}_{2}&{\bf 1}_{2} &{\bf 1}_1+{\bf 1}_3 
\end{array} .
\end{eqnarray}

In addition, other $Z_4$ subgroups consist of 
$\{e,b,b^2,b^3\}$ and $\{e,ab,(ab)^2,(ab)^3\}$.
For those $Z_4$ subgroups, we obtain the same results 
when we choose proper basis.
Furthermore, subgroups of $Z_2$ can appear from 
the above $Z_4$ groups.
The decomposition of $Z_4$ to $Z_2$ is rather straightforward.

\subsection{general $Q_N$}

Recall that all of the $Q_N$ elements are written as 
$a^mb^k$ with $m=0,\cdots,N-1$ and $k=0,1$, where 
$a^N=e$ and $b^2=a^{N/2}$.
Similarly to $D_N$ with $N=$ even, 
there are four singlets ${\bf 1}_{\pm \pm}$ and 
doublets ${\bf 2}_k$ with $k=1,\cdots, N/2-1$.
Tables \ref{tab:QN-4N} and \ref{tab:QN-4N+2} 
show the representations of $a$ and $b$ on 
these representations for $N=4n$ and $N=4n+2$.

\begin{table}[t]
\begin{center}
\begin{tabular}{|c|c|c|c|c|c|c|}
\hline
${(N=4n)}$&${\bf 1}_{++}$&${\bf 1}_{+-}$&${\bf 1}_{-+}$&${\bf 1}_{--}$ 
&${\bf 2}_{k=\text{odd}}$&${\bf 2}_{k=\text{even}}$\\
\hline
 $a$&1&-1&-1&1&$\mat2{\rho^{k}}{0} {0}{\rho^{-k}}$&$\mat2{\rho^{k}}{0} 
{0}{\rho^{-k}}$\\
\hline
 $b$&1&1&-1&-1&$\mat2{0}{i} {i}{0}$&$\mat2{0}{1} {1}{0}$\\
\hline
\end{tabular}
\end{center}
\caption{Representations of $Q_N$ for $N=4n$}
\label{tab:QN-4N}
\end{table}

\begin{table}[t]
\begin{center}
\begin{tabular}{|c|c|c|c|c|c|c|}
\hline
${(N=4n+2)}$&${\bf 1}_{++}$&${\bf 1}_{+-}$&${\bf 1}_{-+}$&${\bf 1}_{--}$ 
&${\bf 2}_{k=\text{odd}}$&${\bf 2}_{k=\text{even}}$\\
\hline
 $a$&1&-1&-1&1&$\mat2{\rho^{k}}{0} {0}{\rho^{-k}}$&$\mat2{\rho^{k}}{0} 
{0}{\rho^{-k}}$\\
\hline
 $b$&1&$i$&$-i$&-1&$\mat2{0}{i} {i}{0}$&$\mat2{0}{1} {1}{0}$ \\
\hline
\end{tabular}
\end{center}
\caption{Representations of $Q_N$ for $N=4n+2$}
\label{tab:QN-4N+2}
\end{table}

\vskip .5cm
{$\bullet$ $Q_N \to Z_4$}

First, we consider the subgroup $Z_4$, which consists of 
the elements $\{e,b,b^2,b^3\}$. 
Obviously, 
there are four singlet representations ${\bf 1}_m$ for $Z_4$ 
and the generator $b$ is represented as $b=e^{\pi im/2}$ on 
${\bf 1}_m$.

When $N=4n$, ${\bf 1}_{++}$ and ${\bf 1}_{+-}$ of $Q_N$ 
correspond to ${\bf 1}_0$ of $Z_4$ and 
${\bf 1}_{-+}$ and ${\bf 1}_{--}$ of $Q_N$ correspond to ${\bf 1}_2$ of 
$Z_4$. 
The doublets ${\bf 2}_k$ of $Q_N$, $(x_1,x_2)$ decompose to two singlets 
as ${\bf 1}_1: (x_1-ix_2)$ and ${\bf 1}_3: (x_1+ix_2)$.
These results are summarized as follows,
\begin{eqnarray}
\begin{array}{cccccc}
Q_N &{\bf 1}_{++}&{\bf 1}_{+-}&{\bf 1}_{-+}&{\bf 1}_{--} &{\bf 2}_{k}   
\\
 &\downarrow &\downarrow &\downarrow  &\downarrow  &\downarrow 
\\
Z_4 &{\bf 1}_{0}&{\bf 1}_{0}&{\bf 1}_{2}&{\bf 1}_{2} &{\bf 1}_{1}+{\bf
  1}_{3} 
\end{array}.
\end{eqnarray}

When $N=4n+2$, ${\bf 1}_{++}$, ${\bf 1}_{+-}$, ${\bf 1}_{-+}$ 
 and ${\bf 1}_{--}$ of $Q_N$ 
correspond to ${\bf 1}_0$, ${\bf 1}_1$, ${\bf 1}_2$ and 
${\bf 1}_3$ of $Z_4$, respectively. 
The decompositions of doublets ${\bf 2}_k$ are the same as 
those for $N=4n$.
Then, these results are summarized as follows,
\begin{eqnarray}
\begin{array}{cccccc}
Q_N &{\bf 1}_{++}&{\bf 1}_{+-}&{\bf 1}_{-+}&{\bf 1}_{--} &{\bf 2}_{k}   
\\
 &\downarrow &\downarrow &\downarrow  &\downarrow  &\downarrow 
\\
Z_4 &{\bf 1}_{0}&{\bf 1}_{1}&{\bf 1}_{2}&{\bf 1}_{3} &{\bf 1}_{1}+{\bf
  1}_{3} 
\end{array} .
\end{eqnarray}

\vskip .5cm
{$\bullet$ $Q_N \to Z_N$}

Next, we consider the subgroup $Z_N$, which 
consist of the elements $\{e,a,\cdots,a^{N-1}\}$.
Obviously it is the normal subgroups of $Q_N$ and 
there are $N$ types of irreducible singlet representations
${\bf 1}_0, {\bf 1}_1,\cdots,{\bf 1}_{N-1}$.
On the singlet ${\bf 1}_m$ of $Z_N$, 
the generator $a$ is represented as $a=\rho^m$. 
The singlets, ${\bf 1}_{++}$ and ${\bf 1}_{--}$ of $Q_N$ 
correspond to  ${\bf 1}_0$ of $Z_N$ and the singlets 
${\bf 1}_{+-}$ and ${\bf 1}_{-+}$ of $Q_N$ 
correspond to  ${\bf 1}_{N/2}$ of $Z_N$. 
The doublets ${\bf 2}_k$, $(x_1,x_2)$, of $Q_N$ 
decompose to two singlets as 
${\bf 1}_{k}: x_1$ and ${\bf 1}_{N-k}: x_2$.
These results are summarizes as follows,
\begin{eqnarray}
\begin{array}{cccccc}
Q_N &{\bf 1}_{++}&{\bf 1}_{+-}&{\bf 1}_{-+}&{\bf 1}_{--} 
&{\bf 2}_{k}   
\\
 &\downarrow &\downarrow &\downarrow  &\downarrow  &\downarrow 
\\
Z_2 &{\bf 1}_{0}&{\bf 1}_{N/2}&{\bf 1}_{N/2}&{\bf 1}_{0} 
&{\bf 1}_{k} +{\bf 1}_{N-k} 
\end{array} .
\end{eqnarray}

\vskip .5cm
{$\bullet$ $Q_N \to Q_M$}

We consider the $Q_M$ subgroup, 
where $M$ is a divisor of $N$.
We denote $\tilde a = a^\ell$ with $\ell = N/M$, 
where $\ell =$ integer.
The subgroup $Q_M$ consists of $\tilde a^m b^k$ with 
$m=0,\cdots,M-1$ and $k=0,1$.
There are three combinations of $(N,M)$, i.e. 
$(N,M)=(4n,4m)$, $(4n,4m+2)$ and $(4n+2,4m+2)$.

We start with the combination $(N,M)=(4n,4m)$, 
where $\ell=N/M$ can be even or odd.
Recall that $(ab)$ of $Q_N$ is represented as $ab=1$ on ${\bf 1}_{\pm+}$ 
and $ab=-1$ on ${\bf 1}_{\pm -}$.
Thus, the representations of $(a^\ell b)$ depend on whether $\ell$ 
is even or odd.
When $\ell$ is odd, $(ab)$ and $(a^\ell b)$ are represented 
in the same way as on each of the above singlets.
On the other hand, when $\ell$ = even, we always have 
the singlet representations with $a^\ell=1$.
The doublets ${\bf 2}_k$ of $Q_N$ correspond to 
the doublets ${\bf 2}_{k'}$ of $Q_M$ when 
$k=k'$ (mod $M$).
In addition, when  $k=-k'$ (mod $M$), 
doublets ${\bf 2}_k$ $(x_1,x_2)$ of $Q_N$ correspond to 
the doublets ${\bf 2}_{M-k'}$ $(x_2,x_1)$ of $Q_M$.
Furthermore, other doublets ${\bf 2}_k$ of $Q_N$ 
decompose to two singlets of $Q_M$ as 
${\bf 1}_{+-}+{\bf 1}_{-+}$ 
with ${\bf 1}_{+-}:x_1+x_2$ and ${\bf 1}_{-+}:x_1-x_2$ for $k=(M/2)$
(mod $M$) and 
${\bf 1}_{++}+{\bf 1}_{--}$ with ${\bf 1}_{++}:x_1+x_2$ 
and ${\bf 1}_{--}:x_1-x_2$ for $k=0$ (mod $M$). 
These results are summarized as follows,
\begin{equation}
\begin{array}{cccccccccc}
(N=4n)&Q_N &{\bf 1}_{++}&{\bf 1}_{+-}&{\bf 1}_{-+}&{\bf 1}_{--} 
&{\bf 2}_{k+Mk'} &{\bf 2}_{Mk'-k} &{\bf 2}_{\frac{M}{2}(2k'+1)}&{\bf 2}_{Mk'}   
\\
(M=4m) &&\downarrow &\downarrow &\downarrow  &\downarrow  &\downarrow  
&\downarrow  &\downarrow&\downarrow 
\\
(\ell=\textrm{odd})&Q_M &{\bf 1}_{++}&{\bf 1}_{+-}&{\bf 1}_{-+}&{\bf 1}_{--} 
&{\bf 2}_{k} &{\tilde {\bf 2}}_{M-k} &{\bf 1}_{+-}+{\bf 1}_{-+}&{\bf 1}_{++}+{\bf 1}_{--}  
\\
(\ell=\textrm{even})&Q_M &{\bf 1}_{++}&{\bf 1}_{++}&{\bf 1}_{--}&{\bf 1}_{--} 
&{\bf 2}_{k} &{\tilde {\bf 2}}_{M-k} &{\bf 1}_{+-}+{\bf 1}_{-+}&{\bf 1}_{++}+{\bf 1}_{--}  
\end{array} ,
\end{equation}
where $k'$ is integer.

Next we consider the case with $(N,M)=(4n,4m+2)$, 
where $\ell$ must be even.
Similarly to the above case with $\ell =$ even, 
the singlets ${\bf 1}_{++}$, ${\bf 1}_{+-}$, ${\bf 1}_{-+}$ 
and ${\bf 1}_{--}$ of $Q_N$ correspond to 
${\bf 1}_{++}$, ${\bf 1}_{++}$, ${\bf 1}_{--}$ and ${\bf 1}_{--}$ 
of $Q_M$.
The results on decompositions of doublets are also 
the same as the above case with   $(N,M)=(4n,4m)$ and $\ell=N/M=$ even.
These results are summarized as follows,
\begin{equation}
\begin{array}{cccccccccc}
(N=4n)&Q_N &{\bf 1}_{++}&{\bf 1}_{+-}&{\bf 1}_{-+}&{\bf 1}_{--} 
&{\bf 2}_{k+Mk'} &{\bf 2}_{Mk'-k} &{\bf 2}_{\frac{M}{2}(2k'+1)}&{\bf 
2}_{Mk'}   
\\
(M=4m+2)  &&\downarrow &\downarrow &\downarrow  &\downarrow  &\downarrow  
&\downarrow  &\downarrow&\downarrow 
\\
(\ell=\textrm{even})&Q_M &{\bf 1}_{++}&{\bf 1}_{++}&{\bf 1}_{--}&{\bf
  1}_{--}
 &{\bf 2}_{k} &{\tilde {\bf 2}}_{M-k} &{\bf 1}_{+-}+{\bf 1}_{-+}&
{\bf 1}_{++}+{\bf 1}_{--}  
\end{array} ,
\end{equation}
where $k'$ is integer.

Next, we consider the case with $(N,M)=(4n+2,4m+2)$, 
where $\ell$ must be odd.
In this case, the results on decompositions are the same 
as the case with $(N,M)=(4n,4m)$ and $\ell=N/M=$ odd.
These results are summarized as follows,
\begin{equation}
\begin{array}{cccccccccc}
(N=4n+2)&Q_N &{\bf 1}_{++}&{\bf 1}_{-+}&{\bf 1}_{+-}&{\bf 1}_{--} 
&{\bf 2}_{k+Mk'} &{\bf 2}_{Mk'-k} &{\bf 2}_{\frac{M}{2}(2k'+1)}&{\bf 
2}_{Mk'}   
\\
(M=4m+2)& &\downarrow &\downarrow &\downarrow  &\downarrow  &\downarrow  
&\downarrow  &\downarrow&\downarrow 
\\
(\ell=\textrm{odd})&Q_M &{\bf 1}_{++}&{\bf 1}_{-+}&{\bf 1}_{+-}&{\bf
  1}_{--} 
&{\bf 2}_{k} &{\tilde {\bf 2}}_{M-k} &{\bf 1}_{+-}+{\bf 1}_{-+}&
{\bf 1}_{++}+{\bf 1}_{--}  
\end{array} ,
\end{equation}
where $k'=$ integer.

\subsection{general $\Sigma(2N^2)$}

Recall that all of the $\Sigma (2N^2)$ elements 
are written by $b^ka^ma'^n$ with $k=0,1$ and $m,n=0,1,\cdots,N-1$. 
The generators, $a$, $a'$ and $b$, satisfy 
$a^N=a'^N=b^2=e$, $aa'=a'a$ and $bab=a'$, 
that is, $a$, $a'$ and $b$ correspond to 
$Z_N$, $Z_N'$ and $Z_2$ of $(Z_N \times Z_N') \rtimes Z_2$, 
respectively.
Table \ref{tab:Sigma2N-rep} shows the representations of these generators 
on each representation.
The number of doublets ${\bf 2}_{p,q}$ is 
equal to $N(N-1)/2$ with the relation 
$p>q$.

\begin{table}[t]
\begin{center}
\begin{tabular}{|c|c|c|c|}
\hline
&${\bf 1}_{+n}$&${\bf 1}_{-n}$&${\bf 2}_{p,q}$\\
\hline
$a$&$\rho^{n}$&$\rho^{n}$&$\mat2{\rho^{q}}{0}{0}{\rho^{p}}$\\
\hline
$a'$&$\rho^{n}$&$\rho^{n}$&$\mat2{\rho^{p}}{0}{0}{\rho^{q}}$\\
\hline
$b$&1&-1&$\mat2{0}{1}{1}{0}$\\
\hline
\end{tabular}
\end{center}
\caption{Representations of $\Sigma(2N^2)$}
\label{tab:Sigma2N-rep}
\end{table}

\vskip .5cm
{$\bullet$ $\Sigma(2N^2) \to Z_2$}


The subgroup $Z_2$ consists of the elements 
$\{e,b\}$. 
There are two singlet representations 
${\bf 1}_{0},{\bf 1}_{1}$ for $Z_2$ 
and the generator $b$ is represented as 
$b=(-1)^m$ on ${\bf 1}_m$.
Then, each representation of $\Sigma(2N^2)$ is 
decomposed as 
\begin{eqnarray}
\begin{array}{cccc}
\Sigma(2N^2) &{\bf 1}_{+n}&{\bf 1}_{-n}&{\bf 2}_{\ell,m}   
\\
 &\downarrow &\downarrow &\downarrow 
\\
Z_2 &{\bf 1}_{0}&{\bf 1}_{1}&{\bf 1}_{0}+{\bf 1}_{1}  
\end{array} ,
\end{eqnarray}
where the components of doublets $(x_1,x_2)$ 
correspond to  ${\bf 1}_0:(x_1+x_2)$ and ${\bf 1}_1:(x_1-x_2)$.

\vskip .5cm
{$\bullet$ $\Sigma(2N^2) \to Z_N \times Z_N$}

The subgroup $Z_N\times Z_N$ consists of the elements 
$a^ma'^n$ with $m,n=0,\cdots,N-1$.
Obviously it is the normal subgroups of $\Sigma(2N^2)$.
There are $N^2$ singlet representations 
${\bf 1}_{m,n}$ and the generators $a$ and $a'$ are 
represented as $a=\rho^m$ and $a'=\rho^n$ on 
${\bf 1}_{m,n}$.
Then, each representation of $\Sigma(2N^2)$ is 
decomposed as 
\begin{eqnarray}
\begin{array}{cccc}
\Sigma(2N^2) &{\bf 1}_{+n}&{\bf 1}_{-n}&{\bf 2}_{\ell,m}   
\\
 &\downarrow &\downarrow &\downarrow 
\\
Z_N\times Z_N &{\bf 1}_{n,n}&{\bf 1}_{n,n}&{\bf 1}_{\ell,m} +{\bf 
1}_{m,\ell}
\end{array} .
\end{eqnarray}

\vskip .5cm
{$\bullet$ $\Sigma(2N^2) \to  D_N$}


We consider  $D_N$ as a subgroup of $\Sigma(2N^2)$.
We denote $\tilde a=a^{-1}a'$.
Then, the subgroup $D_N$ consists of 
the elements $\tilde a^m b^k$ with $k=0,1$ and $m=0,\cdots, N-1$.
Table \ref{tab:sigma2N-DN} shows the representations of 
the generators, $\tilde a$ and $b$, on 
each representation of $\Sigma(2N^2)$.

\begin{table}[t]
\begin{center}
\begin{tabular}{|c|c|c|c|}
\hline
&${\bf 1}_{+n}$&${\bf 1}_{-n}$&${\bf 2}_{p,q}$\\
\hline
$\tilde{a}$&1&1&$\mat2{\rho^{p-q}}{0}{0}{\rho^{-(p-q)}}$\\
\hline
$b$&1&-1&$\mat2{0}{1}{1}{0}$\\
\hline
\end{tabular}
\end{center}
\caption{Representations of $\tilde a$ and $b$ in $\Sigma(2N^2)$}
\label{tab:sigma2N-DN}
\end{table}

At first, we consider the case that $N$ is even. 
The doublets ${\bf 2}_{p,q}$ of $\Sigma(2N^2)$ are still doublets of 
$D_N$ except $p-q=\frac{N}{2}$.
On the other hand, when $p-q=\frac{N}{2}$, 
the doublets decompose to two singlets of $D_N$.
Then, each representation of $\Sigma(2N^2)$ is 
decomposed as 
\begin{eqnarray}
\begin{array}{ccccccc}
\Sigma(2N^2) &{\bf 1}_{+n}&{\bf 1}_{-n}&
{\bf 2}_{q+k',q}  &{\bf 2}_{q-k',q}   &{\bf 2}_{q+\frac{N}{2},q}   
\\
 &\downarrow &\downarrow &\downarrow &\downarrow &\downarrow 
\\
D_N &{\bf 1}_{++}&{\bf 1}_{--}&{\bf 2}_{k'} &{\tilde {\bf 2}}_{k'}  
&{\bf 1}_{+-}+{\bf 1}_{-+}
\end{array} .
\end{eqnarray}

Next, we consider the case that $N$ is odd. 
In this case, each representation of $\Sigma(2N^2)$ is 
decomposed as 
\begin{eqnarray}
\begin{array}{ccccccc}
\Sigma(2N^2) &{\bf 1}_{+n}&{\bf 1}_{-n}&{\bf 2}_{q+k',q}  &{\bf 2}_{q-k',q}    
\\
 &\downarrow &\downarrow &\downarrow &\downarrow  
\\
D_N &{\bf 1}_{+}&{\bf 1}_{-}&{\bf 2}_{k'} &{\tilde {\bf 2}}_{N-k'} 
\end{array} .
\end{eqnarray}

\vskip .5cm
{$\bullet$ $\Sigma(2N^2) \to  Q_N$}


We consider $Q_N$ as a subgroup of $\Sigma(2N^2)$ 
with $N=$ even.
We denote $\tilde a = a^{-1}a'$ and $\tilde b=ba'^{N/2}$.
Then, the subgroup $Q_N$ consists of 
$\tilde a^m \tilde b^k$ with $m=0,\cdots,N-1$ and $k=0,1$.
Table \ref{tab:sigma2N-QN} shows the representations of these generators 
$\tilde a$ and $\tilde b$ on each representation of 
$\Sigma(2N^2)$.
Then the singlets of $\Sigma(2N^2)$ become singlets of $Q_N$ as follows
\begin{eqnarray}
\begin{array}{cccc}
\Sigma(2N^2) &{\bf 1}_{+n}&{\bf 1}_{-n} &
\\
 &\downarrow &\downarrow &
\\
Q_N &{\bf 1}_{++}, &{\bf 1}_{--},& (n:\rm{even}) 
\\
    &{\bf 1}_{--}, &{\bf 1}_{++},& (n:\rm{odd}) 
\end{array} .
\end{eqnarray}
The decompositions of doublets are obtained in a way 
similar to the decomposition, $\Sigma(2N^2) \rightarrow D_N$, as 
follows, 
\begin{eqnarray}
\begin{array}{ccc}
\Sigma(2N^2) &{\bf 2}_{q+k',q}  & {\bf 2}_{q+\frac{N}{2} }
\\
 &\downarrow &\downarrow 
\\
Q_N & {\bf 2}_{k'} &{\bf 1}_{+-}+{\bf 1}_{-+}
\end{array} .
\end{eqnarray}

\begin{table}[t]
\begin{center}
\begin{tabular}{|c|c|c|c|}
\hline
&${\bf 1}_{+n}$&${\bf 1}_{-n}$&${\bf 2}_{p,q}$\\
\hline
$\tilde{a}$&1&1&$\mat2{\rho^{p-q}}{0}{0}{\rho^{-(p-q)}}$\\
\hline
$\tilde b$&1&-1&$\mat2{0}{(-1)^q}{(-1)^p}{0}$\\
\hline
\end{tabular}
\end{center}
\caption{Representations of $\tilde a$ and $\tilde b$ in $\Sigma(2N^2)$}
\label{tab:sigma2N-QN}
\end{table}

\vskip .5cm
{$\bullet$ $\Sigma(2N^2) \to \Sigma(2M^2)$}

We consider the subgroup $\Sigma(2M^2)$, where 
$M$ is a divisor of $N$.
We denote $\tilde a=a^\ell$ and $\tilde a' = a'^\ell$ 
with $\ell =N/M$, where $\ell =$ integer.
The subgroup  $\Sigma(2M^2)$ consists of 
$b^k \tilde a^m \tilde a'^n$ with $k=0,1$ 
and $m,n=0,\cdots,M-1$.
Table \ref{tab:sigma2N-M} shows the representations of 
$\tilde a$, $\tilde a'$ and $\tilde b$ on 
each representation of $\Sigma(2N^2)$.
Then, representations of $\Sigma(2N^2)$ correspond to representations 
of  $\Sigma(2M^2)$ as follows, 
\begin{eqnarray}
\begin{array}{ccccc}
\Sigma(2N^2) &{\bf 1}_{+n}&{\bf 1}_{-n}&{\bf 2}_{p+Mn,q+Mn'}   
\\
 &\downarrow &\downarrow &\downarrow 
\\
\Sigma(2m^2) &{\bf 1}_{+n}&{\bf 1}_{-n}&{\bf 2}_{p,q}   
\end{array} ,
\end{eqnarray}
where $n,n'$ are integers.

\begin{table}[t]
\begin{center}
\begin{tabular}{|c|c|c|c|}
\hline
&${\bf 1}_{+n}$&${\bf 1}_{-n}$&${\bf 2}_{p,q}$\\
\hline
$\tilde{a}$&$\rho^{n\ell}$&$\rho^{n\ell}$&$\mat2{\rho^{q\ell}}{0}{0}{\rho^{p\ell}}$\\
\hline
$\tilde{a}'$&$\rho^{n\ell}$&$\rho^{n\ell}$&$\mat2{\rho^{p\ell}}{0}{0}{\rho^{q\ell}}$\\
\hline
$b$ &1&-1&$\mat2{0}{1}{1}{0}$\\
\hline
\end{tabular}
\end{center}
\caption{Representations of $\tilde a$, $\tilde a'$ and $\tilde b$ in 
$\Sigma(2N^2)$}
\label{tab:sigma2N-M}
\end{table}

\subsection{$\Sigma(32)$}

The $\Sigma(32)$ group includes subgroups, 
$D_4$, $Q_4$ and $\Sigma(8)$ as well as Abelian groups, 
as shown in the previous section.
In addition, the $\Sigma(32)$ group includes the subgroup, 
which has not been studied in the previous sections.
It is useful to construct a discrete group as 
a subgroup of known groups, as 
explained in section \ref{sec:T7}, where the example $T_7$ was 
shown.
Here, we show another example 
$(Z_4 \times Z_2) \rtimes Z_2$ as a subgroup of 
$\Sigma(32) \simeq (Z_4 \times Z_4) \rtimes Z_2$.

All of the $\Sigma(32)$ elements are written by 
$b^ka^ma'^n$ with $k=0,1$ and $m,n=0,1,2,3$.
The generators, $a$, $a'$ and $b$, satisfy 
$a^4=a'^4=b^2=e$, $aa'=a'a$ and $bab=a'$.
Here we define $\tilde a=aa'$ and $\tilde a' = a^2$,
where $\tilde a^4=e$ and $\tilde a'^2 =e$.
Then, the elements 
$b^k\tilde a^m \tilde a'^n$ with $k,n=0,1$ and $m=0,1,2,3$ 
construct a closed subalgebra, i.e. 
$(Z_4 \times Z_2) \rtimes Z_2$.
It has ten conjugacy classes 
and eight singlets, ${\bf 1}_{\pm 0}$,  ${\bf 1}_{\pm 1}$,  
${\bf 1}_{\pm 2}$ and  ${\bf 1}_{\pm 3}$,  and 
two doublets, ${\bf 2}_1$ and ${\bf 2}_2$.
These conjugacy classes and characters are shown in Table  
\ref{tab:Z4-Z2-Z2}.
{}From this table, we can find decompositions of $\Sigma(32)$ 
representations to representation of 
$(Z_4\times Z_2) \rtimes Z_2$ as follows,
\begin{eqnarray}
\begin{array}{ccccccc}
\Sigma(32) &{\bf 1}_{\pm0,\pm1,\pm2,\pm3}& {\bf 2}_{1,0}, {\bf
  2}_{3,2} 
& {\bf 2}_{3,0},
{\bf 2}_{2,1} & {\bf 2}_{2,0} & {\bf 2}_{3,1} \\            
\\
 &\downarrow &\downarrow &\downarrow  &\downarrow  &\downarrow 
\\
(Z_4\times Z_2) \rtimes Z_2 
&{\bf 1}_{\pm0,\pm1,\pm0,\pm1} & {\bf 2}_1 & {\bf 2}_2 & {\bf
  1}_{+3}+{\bf 1}_{-3} & {\bf 1}_{+2}+{\bf 1}_{-2} .
\end{array} .
\end{eqnarray}

\begin{table}[t]
\begin{center}
\begin{tabular}{ccc|ccccccc}
 & & $h$ & $\chi_{\pm0}$& $\chi_{\pm1}$&
 $\chi_{\pm2}$ & $\chi_{\pm3}$ & $\chi_{2_1}$ & $\chi_{2_2}$\\
\hline 
 $C_1$:&$\{ e \}$, &  1 & 1 & 1 & 1 & 1 & 2 & 2 \\
 $C_1^{(1)}$:&$\{ \tilde a\tilde a'\},$ &  4 & 1 & $-1$ & 1 & $-1$ & $2i$ & 
$-2i$    \\
 $C_1^{(2)}$:&$\{ \tilde a^2\tilde a'^2\}$, &  2 &  1 & 1 & 1 & 1 & $-2$ & 
$-2$    \\
 $C_1^{(3)}$:&$\{ \tilde a^3\tilde a'^3\},$ &  4 & 1 & $-1$ & 1 & $-1$ & 
$-2i$ & $2i$     \\
 $C_2^{(0)}$:&$\{ b, b\tilde a^2\tilde a'^2\},$ &  2  & $\pm1$ & $\pm1$ & 
$\pm1$ & $\pm1$ & 0 & 0  \\
 $C_2^{(0)}$:&$\{ b\tilde a\tilde a', b\tilde a^3\tilde a'^3\},$ &  4  & 
$\pm1$ & $\mp1$ & $\pm1$ & $\mp1$ & 0 & 0  
\\
 $C_2^{(0)}$:&$\{ b\tilde a^2, b\tilde a'^2\},$ &  4  & $\pm1$ & $\mp1$ & 
$\mp1$ & $\pm1$ & 0 & 0  \\
 $C_2^{(0)}$:&$\{ b\tilde a\tilde a'^3, b\tilde a^3\tilde a'\},$ &  2 & 
$\pm1$ & $\pm1$ & $\mp1$ & $\mp1$ & 0 & 0  
\\
 $C_2^{(2,0)}$:&$\{ \tilde a^2, \tilde a'^2\},$ &  2 & 1 & $-1$ & $-1$ & 1 
& 0 & 0  \\ 
 $C_2^{(3,1)}$:&$\{ \tilde a\tilde a'^3, \tilde a^3\tilde a'\},$ &  2  & 1 
& 1 & $-1$ & $-1$ & 0 & 0  \\ 
\end{tabular}
\end{center}
\caption{Conjugacy classes and characters of $(Z_4\times Z_2) \rtimes Z_2$}
\label{tab:Z4-Z2-Z2}
\end{table}

\subsection{General $\Delta(3N^2)$}

All of the $\Delta(3N^2)$ elements are written by 
$b^ka^ma'^n$ with $k=0,1,2$ and $m,n=0,\cdots,N-1$, 
where the generators, $b$, $a$ and $a'$, correspond to 
$Z_3$, $Z_N$ and $Z_N'$ of $(Z_N \times Z_N') \rtimes Z_3$, 
respectively.
Table \ref{tab:delta3N-1} shows the representations of 
generators, $b$, $a$ and $a'$ 
on each representation of $\Delta(3N^2)$ for $N/3 \neq$ integer.
Also Table \ref{tab:delta3N-2} shows the same for $N/3 =$ integer.

\begin{table}[t]
\begin{center}
\begin{tabular}{|c|c|c|}
\hline
&${\bf 1}_{k}$&${\bf 3}_{[k][\ell]}$\\
\hline
${a}$&1&$\Mat3{\rho^\ell}{0}{0}{0}{\rho^k}{0}{0}{0}{\rho^{-k-\ell}}$\\
\hline
${a}'$&1&$\Mat3{\rho^{-k-\ell}}{0}{0}{0}{\rho^\ell}{0}{0}{0}{\rho^{k}}$\\
\hline
$b$&$\omega^{k}$&$\Mat3{0}{1}{0}{0}{0}{1}{1}{0}{0}$\\
\hline
\end{tabular}
\end{center}
\caption{Representations of $a$, $ a'$ and $ b$ 
in $\Delta(3N^2)$ for $N/3 \neq$ integer}
\label{tab:delta3N-1}
\end{table}

\begin{table}[t]
\begin{center}
\begin{tabular}{|c|c|c|}
\hline
&${\bf 1}_{k,\ell}$&${\bf 3}_{[k][\ell]}$\\
\hline
${a}$&$\omega^\ell$&$\Mat3{\rho^\ell}{0}{0}{0}{\rho^k}{0}{0}{0}{\rho^{-k-\ell}}$\\
\hline
${a}'$&$\omega^\ell$&$\Mat3{\rho^{-k-\ell}}{0}{0}{0}{\rho^\ell}{0}{0}{0}{\rho^{k}}$\\
\hline
$b$&$\omega^{k}$&$\Mat3{0}{1}{0}{0}{0}{1}{1}{0}{0}$\\
\hline
\end{tabular}
\end{center}
\caption{Representations of $ a$, $ a'$ and $ b$ 
in $\Delta(3N^2)$ for $N/3 =$ integer}
\label{tab:delta3N-2}
\end{table}

\vskip .5cm
{$\bullet$ $\Delta(3N^2) \to Z_3$}


The subgroup $Z_3$ consists of $\{e,b,b^2\}$. 
There are three singlet representations 
${\bf 1}_{m}$ with $m=0,1,2$ for $Z_3$ 
and the generator $b$ is represented as 
$b=\omega^m$ on ${\bf 1}_m$.
When $N/3\neq$ integer, each representation of $\Delta(3N^2)$ is 
decomposed as 
\begin{eqnarray}
\begin{array}{ccc}
\Delta(3N^2) &{\bf 1}_{k}&{\bf 3}_{[k][\ell]}   
\\
 &\downarrow &\downarrow
\\
Z_3 &{\bf 1}_k&{\bf 1}_0+{\bf 1}_1+{\bf 1}_2 
\end{array} .
\end{eqnarray}
On the other hand, when $N/3=$ integer, 
each representation of $\Delta(3N^2)$ is 
decomposed as 
\begin{eqnarray}
\begin{array}{ccc}
\Delta(3N^2) &{\bf 1}_{k,\ell}&{\bf 3}_{[k][\ell]}   
\\
 &\downarrow &\downarrow
\\
Z_3 &{\bf 1}_k&{\bf 1}_0+{\bf 1}_1+{\bf 1}_2 
\end{array} .
\end{eqnarray}
In both cases, the triplet components $(x_1,x_2,x_3)$
of $\Delta(3N^2)$ are decomposed to singlets of $Z_3$ as 
${\bf 1}_0:x_1+x_2+x_3$, ${\bf 1}_1:x_1+\omega^2 x_2+\omega x_3$
and ${\bf 1}_2:x_1+\omega x_2+\omega^2 x_3$.

\vskip .5cm
{$\bullet$ $\Delta(3N^2) \to Z_N \times Z_N$}


The subgroup $Z_N\times Z_N$ consists of $\{a^ma'^n\}$ 
with $m, n=0,1,\cdots N-1$. 
There are $N^2$ singlet representations 
${\bf 1}_{m,n}$ and the generators $a$ and $a'$ are 
represented as $a=\rho^m$ and $a'=\rho^n$ on 
${\bf 1}_{m,n}$.
When $N/3\neq$ integer, each representation of $\Delta(3N^2)$ is 
decomposed as 
\begin{eqnarray}
\begin{array}{ccc}
\Delta(3N^2) &{\bf 1}_{k}&{\bf 3}_{[k][\ell]}   
\\
 &\downarrow &\downarrow
\\
Z_N\times Z_N &{\bf 1}_{0,0}&{\bf 1}_{\ell,-k-\ell}+{\bf 1}_{k,\ell}
+{\bf 1}_{-k-\ell,k} 
\end{array} .
\end{eqnarray}
In addition, when $N/3=$ integer, we have the following 
decompositions
\begin{eqnarray}
\begin{array}{ccc}
\Delta(3N^2) &{\bf 1}_{k,\ell}&{\bf 3}_{[k][\ell]}   
\\
 &\downarrow &\downarrow
\\
Z_N\times Z_N &{\bf 1}_{N\ell/3,N\ell/3}&{\bf 1}_{\ell,-k-\ell}+
{\bf 1}_{k,\ell}+{\bf 1}_{-k-\ell,k} 
\end{array} .
\end{eqnarray}

\vskip .5cm
{$\bullet$ $\Delta(3N^2) \to \Delta(3M^2)$}


We consider the subgroup $\Delta(3M^2)$, where 
$M$ is a divisor of $N$.
We denote $\tilde a=a^p$ and $\tilde a' = a'^p$ 
with $p =N/M$, where $p =$ integer.
The subgroup $\Delta(3M^2)$ consists of 
$b^k \tilde a^m \tilde a'^n$ with $k=0,1,2$ 
and $m,n=0,\cdots,M-1$.
Table \ref{tab:DeltaN-M-1} shows the representations of 
$\tilde a$, $\tilde a'$ and $\tilde b$ on 
each representation of $\Delta(3N^2)$ for $N/3=$ integer.
In addition, Table \ref{tab:DeltaN-M-2} shows the representations of 
$\tilde a$, $\tilde a'$ and $\tilde b$ on 
each representation of $\Delta(3N^2)$ for $N/3 \neq$ integer.
There are three types of combinations $(N,M)$, i.e. 
(1) both $N/3$ and $M/3$ are integers, (2) 
$N/3$ is integer, but $M/3$ is not integer, 
(3) either $N/3$ or $M/3$ is not integer.

\begin{table}[t]
\begin{center}
\begin{tabular}{|c|c|c|}
\hline
&${\bf 1}_{k,\ell}$&${\bf 3}_{[k][\ell]}$\\
\hline
$\tilde{a}$&$\omega^{p\ell}$&$\Mat3{\rho^{p\ell}}{0}{0}{0}{\rho^{p 
k}}{0}{0}{0}{\rho^{-p(k+\ell)}}$\\
\hline
$\tilde{a}'$&$\omega^{p\ell}$&$\Mat3{\rho^{-p(k+\ell)}}{0}{0}{0}{\rho^{p\ell}}{0}{0}{0}{\rho^{p 
k}}$\\
\hline
$b$&$\omega^{k}$&$\Mat3{0}{1}{0}{0}{0}{1}{1}{0}{0}$\\
\hline
\end{tabular}
\end{center}
\caption{Representations of $\tilde a$, $\tilde a'$ and $\tilde b$ 
in $\Delta(3N^2)$ for $N/3=$ integer}
\label{tab:DeltaN-M-1}
\end{table}

\begin{table}[t]
\begin{center}
\begin{tabular}{|c|c|c|}
\hline
&${\bf 1}_{k}$&${\bf 3}_{[k][\ell]}$\\
\hline
$\tilde{a}$&$1$&$\Mat3{\rho^{p\ell}}{0}{0}{0}{\rho^{p 
k}}{0}{0}{0}{\rho^{-p(k+\ell)}}$\\
\hline
$\tilde{a}'$&1&$\Mat3{\rho^{-p(k+\ell)}}{0}{0}{0}{\rho^{p\ell}}{0}{0}{0}{\rho^{p 
k}}$\\
\hline
$b$&$\omega^{k}$&$\Mat3{0}{1}{0}{0}{0}{1}{1}{0}{0}$\\
\hline
\end{tabular}
\end{center}
\caption{Representations of $\tilde a$, $\tilde a'$ and $\tilde b$ 
in $\Delta(3N^2)$ for $N/3 \neq $ integer}
\label{tab:DeltaN-M-2}
\end{table}

When both $N/3$ and $M/3$ are integers, 
each representation of $\Delta(3N^2)$ is 
decomposed to representations of $\Delta(3M^2)$ 
as follows, 
\begin{eqnarray}
\begin{array}{ccc}
\Delta(3N^2) &{\bf 1}_{k,\ell}&{\bf 3}_{[k+Mn][\ell+Mn']}   
\\
 &\downarrow &\downarrow
\\
\Delta(3M^2) &{\bf 1}_{k,p\ell}&{\bf 3}_{[k][\ell]} 
\end{array} ,
\end{eqnarray}
where $n$ and $n'$ are integers.

Next we consider the case that $N/3=$ integer and $M/3 \neq $ integer, 
where $p=N/M$ must be $3n$.
In this case, each representation of $\Delta(3N^2)$ is 
decomposed to representations of $\Delta(3M^2)$ 
as follows, 
\begin{eqnarray}
\begin{array}{ccc}
\Delta(3N^2) &{\bf 1}_{k,\ell}&{\bf 3}_{[k+Mn][\ell+Mn']}   
\\
 &\downarrow &\downarrow
\\
\Delta(3M^2) &{\bf 1}_{k}&{\bf 3}_{[k][\ell]} 
\end{array} ,
\end{eqnarray}
where $n$ and $n'$ are integers.

The last case is that either $N/3$ or $M/3$ is not integer. 
In this case, each representation of $\Delta(3N^2)$ is 
decomposed to representations of $\Delta(3M^2)$ 
as follows, 
\begin{eqnarray}
\begin{array}{ccc}
\Delta(3N^2) &{\bf 1}_{k}&{\bf 3}_{[k+Mn][\ell+Mn']}   
\\
 &\downarrow &\downarrow
\\
\Delta(3M^2) &{\bf 1}_{k}&{\bf 3}_{[k][\ell]} 
\end{array}   ,
\end{eqnarray}
where $n$ and $n'$ are integers.

\subsection{$A_4$}

The $A_4$ group is isomorphic to $\Delta(12)$.
Here, we apply the above generic results to the $A_4$ group.
All of the $\Delta(12)$ elements are written by 
$b^ka^ma'^n$ with $k=0,1,2$ and $m,n=0,1$.
Table \ref{tab:delta-12} shows the representations of generators 
$a$, $a'$ and $b$ on each representation.

\begin{table}[t]
\begin{center}
\begin{tabular}{|c|c|c|}
\hline
&${\bf 1}_{k}$&{\bf 3} \\
\hline
${a}$&1&$\Mat3{-1}{0}{0} {0}{1}{0} {0}{0}{-1}$\\
\hline
${a}'$&1&$\Mat3{-1}{0}{0} {0}{-1}{0} {0}{0}{1}$\\
\hline
$b$&$\omega^{k}$&$\Mat3{0}{1}{0}{0}{0}{1}{1}{0}{0}$\\
\hline
\end{tabular}
\end{center}
\caption{Representations of $\tilde a$, $\tilde a'$ and $\tilde b$ 
in $\Delta(12)$}
\label{tab:delta-12}
\end{table}

\vskip .5cm
{$\bullet$ $A_4 \to Z_3$}


The $Z_3$ group consists of $\{e,b,b^2\}$. 
Each  representation of $\Delta(12)$ is 
decomposed as 
\begin{eqnarray}
\begin{array}{ccc}
A_4 \simeq \Delta(12) &{\bf 1}_{k}&{\bf 3}   
\\
 &\downarrow &\downarrow
\\
Z_3 &{\bf 1}_k&{\bf 1}_0+{\bf 1}_1+{\bf 1}_2 
\end{array} .
\end{eqnarray}
Decomposition of triplet $(x_1,x_2,x_3)$ is obtained as  
${\bf 1}_0:x_1+x_2+x_3$, ${\bf 1}_1:x_1+\omega^2 x_2+\omega x_3$
and ${\bf 1}_2:x_1+\omega x_2+\omega^2 x_3$.

\vskip .5cm
{$\bullet$ $A_4 \to Z_2 \times Z_2$}


The subgroup $Z_2\times Z_2$ consists of $\{e, a, a',aa'\}$. 
Each  representation of $\Delta(12)$ is 
decomposed as 
\begin{eqnarray}
\begin{array}{ccc}
A_4 \simeq \Delta(12) &{\bf 1}_{k}&{\bf 3}   
\\
 &\downarrow &\downarrow
\\
Z_2\times Z_2 &{\bf 1}_{0,0}&{\bf 1}_{1,1}+{\bf 1}_{0,1}+{\bf 1}_{1,0} 
\end{array} .
\end{eqnarray}

\subsection{$T_7$}

All of the $T_7$ elements are written as 
$b^ma^n$ with $m=0,1,2$ and $n=0,\cdots,6$, 
where $b^3=e$ and $a^7=e$.
Table \ref{tab:T7} shows the representation of generators $a$ and $b$ on 
each representation of $T_7$.

\begin{table}[t]
\begin{center}
\begin{tabular}{|c|c|c|c|c|c|c|c|c|c|}
\hline
        & ${\bf 1}_0$ & ${\bf 1}_1$&${\bf 1}_2$&${\bf 3}$&$\bar {\bf 3}$\\
 \hline
$a$ & 1&1&1&$\Mat3{\rho}{0}{0} {0}{\rho^2}{0} {0}{0}{\rho^4}$ 
&$\Mat3{\rho^{-1}}{0}{0} {0}{\rho^{-2}}{0} {0}{0}{\rho^{-4}}$\\
\hline
$b$ & 1&$\omega$&$\omega^2$&$\Mat3{0}{1}{0} {0}{0}{1} {1}{0}{0}$ 
&$\Mat3{0}{1}{0} {0}{0}{1} {1}{0}{0}$\\
\hline
\end{tabular}
\end{center}
\caption{Representations of $a$ and $ b$ 
in $T_7$}
\label{tab:T7}
\end{table}

\vskip .5cm
{$\bullet$ $T_7 \to Z_3$}

The subgroup $Z_3$ consists of $\{e,b,b^2\}$.
The three singlet representations ${\bf 1}_m$ of $Z_3$ with 
$m=0,1,2$  
are specified such that $b=\omega^m$ on ${\bf 1}_m$.
Then, each representation of $T_7$ is decomposed as 
\begin{eqnarray}
\begin{array}{cccccc}
T_7 &{\bf 1}_0&{\bf 1}_1&{\bf 1}_2&{\bf 3} &\bar {\bf 3}   
\\
 &\downarrow &\downarrow &\downarrow  &\downarrow  &\downarrow 
\\
Z_3 &{\bf 1}_0&{\bf 1}_1&{\bf 1}_2&{\bf 1}_0+{\bf 1}_1+{\bf 1}_2 
&{\bf 1}_0+{\bf 1}_1+{\bf 1}_2 
\end{array} .
\end{eqnarray}
Here the $T_7$ triplet ${\bf 3}:(x_1,x_2,x_3)$ decomposes 
to three singlets, ${\bf 1}_0+{\bf 1}_1+{\bf 1}_2 $, 
and their components correspond to 
\begin{eqnarray}
&{\bf 1}_0~:~x+y+z, \qquad 
{\bf 1}_1~:~x+\omega^2 y+\omega z, \qquad 
{\bf 1}_2~:~x+\omega y+\omega^2 z  .
\end{eqnarray}

\vskip .5cm
{$\bullet$ $T_7 \to Z_7$}

The subgroup $Z_7$ consists of $a^n$ with $n=0,\cdots,6$.
The seven singlets ${\bf 1}_m$ of $Z_ 7$ with 
$m=0,\cdots,6$  are specified such that $b=\rho^m$ on ${\bf 1}_m$.
Then, each representation of $T_7$ is decomposed as 
\begin{eqnarray}
\begin{array}{cccccc}
T_7 &{\bf 1}_0&{\bf 1}_1&{\bf 1}_2&{\bf 3} &\bar {\bf 3}   
\\
 &\downarrow &\downarrow &\downarrow  &\downarrow  &\downarrow 
\\
Z_3 &{\bf 1}_0&{\bf 1}_0&{\bf 1}_0&{\bf 1}_1+{\bf 1}_2+{\bf 1}_4 
&{\bf 1}_3+{\bf 1}_5+{\bf 1}_6 
\end{array} .
\end{eqnarray}

\subsection{$\Sigma(81)$}

All of the $\Sigma(81)$ elements are written as 
$b^ka^\ell a'^ma''^n$ with $k,\ell,m,n=0,1,2$, 
where these generators satisfy 
$a^3=a'^3=a''^3=1, ~aa'=a'a, ~aa''=a''a, ~a''a'=a'a'',~b^3=1$, 
$b^2ab=a''$, $b^2a'b=a$ and $b^2a''b=a'$.
Table \ref{tab:sigma81} shows the representations of 
generators, $b$, $a$, $a'$ and $a''$ 
on each representation of $\Sigma(81)$.

\begin{table}[t]
\begin{center}
\begin{tabular}{|c|c|c|c|c|c|c|c|c|c|c|}
\hline
&${\bf 1}_{\ell}^{k}$&${\bf 3}_{A}$&${\bf 3}_{B}$&${\bf
  3}_{C}$&${\bf 3}_{D}$\\
\hline
${a}$&$\omega^k$ 
&$\Mat3{\omega}{0}{0} {0}{1}{0} {0}{0}{1}$
&$\Mat3{\omega^2}{0}{0} {0}{\omega}{0} {0}{0}{\omega}$
&$\Mat3{1}{0}{0} {0}{\omega^2}{0} {0}{0}{\omega^2}$
&$\Mat3{\omega^2}{0}{0} {0}{1}{0} {0}{0}{\omega}$
\\
\hline
${a'}$&$\omega^k$ 
&$\Mat3{1}{0}{0} {0}{\omega}{0} {0}{0}{1}$
&$\Mat3{\omega}{0}{0} {0}{\omega^2}{0} {0}{0}{\omega}$
&$\Mat3{\omega^2}{0}{0} {0}{1}{0} {0}{0}{\omega^2}$
&$\Mat3{\omega}{0}{0} {0}{\omega^2}{0} {0}{0}{1}$\\
\hline
${a''}$&$\omega^k$ 
&$\Mat3{1}{0}{0} {0}{1}{0} {0}{0}{\omega}$
&$\Mat3{\omega}{0}{0} {0}{\omega}{0} {0}{0}{\omega^2}$
&$\Mat3{\omega^2}{0}{0} {0}{\omega^2}{0} {0}{0}{1}$
&$\Mat3{1}{0}{0} {0}{\omega}{0} {0}{0}{\omega^2}$\\
\hline
${b}$&$\omega^\ell$ &$\Mat3{0}{1}{0} {0}{0}{1} {1}{0}{0}$
&$\Mat3{0}{1}{0} {0}{0}{1} {1}{0}{0}$
&$\Mat3{0}{1}{0} {0}{0}{1} {1}{0}{0}$&$\Mat3{0}{1}{0} {0}{0}{1} 
{1}{0}{0}$\\
\hline
\end{tabular}
\end{center}
\caption{Representations of $a$, $a'$, $ a''$ 
and $b$ 
in $\Sigma(81)$}
\label{tab:sigma81}
\end{table}

\vskip .5cm
{$\bullet$ $\Sigma(81) \to Z_3 \times Z_3 \times Z_3$}

The subgroup $Z_3 \times Z_3 \times Z_3$ consists of 
$\{e,a,a^2,a',a'^2,a'',a''^2,\cdots \}$.
There are $3^3$ singlets ${\bf 1}_{k,\ell,m}$ of 
$Z_3 \times Z_3 \times Z_3$ and 
the generators, $a$, $a'$ and $a''$, are 
represented on ${\bf 1}_{k,\ell,m}$  as 
$a=\omega^k$, $a'=\omega^\ell$ and $a''=\omega^m$.
Then, each representation of $\Sigma(81)$ is decomposed 
 as follows,
\begin{eqnarray}
&&\begin{array}{cccccccccc}
\Sigma(81) &{\bf 1}^{k}_\ell &{\bf 3}_{A}&{\bf 3}_{B}\\
 &\downarrow &\downarrow 
&\downarrow
\\
Z_3 \times Z_3 \times Z_3 &{\bf 1}_{k,k,k}
&{\bf 1}_{1,0,0}+{\bf 1}_{0,1,0}+{\bf 1}_{0,0,1} 
&{\bf 1}_{2,1,1}+{\bf 1}_{1,2,1}+{\bf 1}_{1,1,2} 
\end{array}   ,
\nonumber\\
\nonumber \\
\nonumber \\
&&\begin{array}{cccccccccc}
\Sigma(81) &{\bf 3}_{C}&{\bf
  3}_{D}\\
 &\downarrow &\downarrow 
\\
Z_3 \times Z_3 \times Z_3 
&{\bf 1}_{0,2,2}+{\bf 1}_{2,0,2}+{\bf 1}_{2,2,0} 
&{\bf 1}_{2,1,0}+{\bf 1}_{0,2,1}+{\bf 1}_{1,0,2} 
\end{array}   ,
\nonumber   \\
~~~~~~~~~~\nonumber \\
\nonumber \\
&&\begin{array}{cccccccccc}
\Sigma(81) 
&{\bar {\bf 3}}_{A}&{\bar {\bf 3}}_{B} \\
 &\downarrow &\downarrow 
\\
Z_3 \times Z_3 \times Z_3 
&{\bf 1}_{2,0,0}+{\bf 1}_{0,2,0}+{\bf 1}_{0,0,2} 
&{\bf 1}_{1,2,2}+{\bf 1}_{2,1,2}+{\bf 1}_{2,2,1} 
\end{array}  ,
 \\
~~~~~~~~~~\nonumber \\
\nonumber \\
&&\begin{array}{cccccccccc}
\Sigma(81) 
&{\bar {\bf 3}}_{C}&{\bar {\bf 3}}_{D}\\
&\downarrow &\downarrow
\\
Z_3 \times Z_3 \times Z_3 
&{\bf 1}_{0,1,1}+{\bf 1}_{1,0,1}+{\bf 1}_{1,1,0} 
&{\bf 1}_{1,2,0}+{\bf 1}_{0,1,2}+{\bf 1}_{2,0,1} 
\end{array}   .\nonumber
\end{eqnarray}

\vskip .5cm
{$\bullet$ $\Sigma(81) \to \Delta(27)$}

The subgroup $\Delta(27)$ consists of 
$b^k\tilde a^m \tilde a'^n$, where 
$\tilde a=a^2a''$ and $\tilde a'=a'a''^2$.
Table \ref{tab:sigma81-delta27} shows 
the representations of 
the generators, $b$, $\tilde a$ and $\tilde a'$ 
on each representation of $\Sigma(81)$.
Then, each representation of $\Sigma(81)$ is decomposed 
to representations of $\Delta(27)$ as 
\begin{eqnarray}
\begin{array}{ccccccccccccc}
\Sigma(81) &{\bf 1}_{\ell}^{k}&{\bf 3}_{A}&{\bf 3}_{B}&{\bf 3}_{C}
&{\bf 3}_{D}
\\
 &\downarrow &\downarrow &\downarrow &\downarrow
  &\downarrow 
\\
\Delta(27) &{\bf 1}_{\ell,0} 
&{\bf 3}_{[0][1]}&{\bf 3}_{[0][1]}&{\bf 3}_{[0][1]} 
&{\bf 1}_{0,2}+{\bf 1}_{1,2}+{\bf 1}_{2,2}
\end{array}   ,  \nonumber 
\end{eqnarray}
\begin{eqnarray}
\begin{array}{ccccccccccccc}
\Sigma(81) 
&{\bar {\bf 3}}_{A}&{\bar {\bf 3}}_{B}&{\bar {\bf 3}}_{C}&{\bar {\bf 
3}}_{D}   
\\
 &\downarrow &\downarrow &\downarrow &\downarrow
\\
\Delta(27) 
&{\bf 3}_{[0][2]}&{\bf 3}_{[0][2]}&{\bf 3}_{[0][2]} 
&{\bf 1}_{0,1}+{\bf 1}_{1,1}+{\bf 1}_{2,1}
\end{array} .
\end{eqnarray}

\begin{table}[t]
\begin{center}
\begin{tabular}{|c|c|c|c|c|c|c|c|c|c|c|}
\hline
&${\bf 1}_{\ell}^{k}$&${\bf 3}_{A}$&${\bf 3}_{B}$&${\bf 3}_{C}$
&${\bf 3}_{D}$\\
\hline
${\tilde a}$&1
&$\Mat3{\omega}{0}{0} {0}{1}{0} {0}{0}{\omega^2}$
&$\Mat3{\omega}{0}{0} {0}{1}{0} {0}{0}{\omega^2}$
&$\Mat3{\omega}{0}{0} {0}{1}{0} {0}{0}{\omega^2}$
&$\Mat3{\omega^2}{0}{0} {0}{\omega^2}{0} {0}{0}{\omega^2}$
\\
\hline
${{\tilde a'}}$&1 
&$\Mat3{\omega^2}{0}{0} {0}{\omega}{0} {0}{0}{1}$
&$\Mat3{\omega^2}{0}{0} {0}{\omega}{0} {0}{0}{1}$
&$\Mat3{\omega^2}{0}{0} {0}{\omega}{0} {0}{0}{1}$
&$\Mat3{\omega^2}{0}{0} {0}{\omega^2}{0} {0}{0}{\omega^2}$\\
\hline
${b}$&$\omega^\ell$ 
&$\Mat3{0}{1}{0} {0}{0}{1} {1}{0}{0}$
&$\Mat3{0}{1}{0} {0}{0}{1} {1}{0}{0}$ 
&$\Mat3{0}{1}{0} {0}{0}{1} {1}{0}{0}$ 
&$\Mat3{0}{1}{0} {0}{0}{1} {1}{0}{0}$\\ 
\hline
\end{tabular}
\end{center}
\caption{Representations of $ b$, $\tilde a $ and $\tilde a'$ of 
$\Delta(27)$
in $\Sigma(81)$}
\label{tab:sigma81-delta27}
\end{table}

\subsection{$\Delta(54)$}

All of the $\Delta(54)$ elements are written as 
$b^kc^\ell a^m a'^n$ with $k,m,n=0,1,2$  and $\ell=0,1$.
Here, the generators $a$ and $a'$ correspond to 
$Z_3$ and $Z_3'$ of $(Z_3 \times Z_3') \rtimes S_3$, 
respectively, while 
$b$ and $c$ correspond to $Z_3$ and $Z_2$ in $S_3$ of 
$(Z_3 \times Z_3') \rtimes S_3$, respectively.
Table \ref{tab:delta54} shows the representations of 
generators, $b$, $c$, $a$ and $a'$ 
on each representation of $\Delta(54)$.

\begin{table}[t]
\begin{center}
\begin{tabular}{|c|c|c|c|c|c|c|c|c|c|c|}
\hline
&$\!\!{\bf 1}_{+}\!\!$&$\!\!{\bf 1}_{-}\!\!$&$\!\!{\bf 2}_1\!\!$&$\!\!{\bf 2}_2\!\!$
&$\!\!{\bf 2}_3\!\!$&$\!\!{\bf 2}_4\!\!$&$\!\!{\bf 3}_{1(k)}\!\!$&$\!\!{\bf 3}_{2(k)}\!\!$\\
\hline
$\!\!{a}\!\!$&$\!\!1\!\!$& $\!\!1\!\!$
&$\!\!\mat2{1}{0} {0}{1}\!\!$&$\!\!\mat2{\!\!\omega^2\!\!}{0} {0}{\!\!\omega\!\!}\!\!$
&$\!\!\mat2{\!\!\omega\!\!}{0} {0}{\!\!\omega^2\!\!}\!\!$
&$\!\!\mat2{\!\!\omega\!\!}{0} {0}{\!\!\omega^2\!\!}\!\!$
&$\!\!\Mat3{\!\!\omega^k\!\!}{0}{0} {0}{\!\!\omega^{2k}\!\!}{0} {0}{0}{1}
\!\!$&$\!\!\Mat3{\!\!\omega^k\!\!}{0}{0} {0}{\!\!\omega^{2k}\!\!}{0} 
{0}{0}{1}\!\!$\\\hline
$\!\!{a'}\!\!$&$\!\!1\!\!$&$\!\!1\!\!$
&$\!\!\mat2{1}{0} {0}{1}\!\!$ & $\!\!\mat2{\!\!\omega^2\!\!}{0} {0}{\!\!\omega\!\!}\!\!$
&$\!\!\mat2{\!\!\omega\!\!}{0} {0}{\!\!\omega^2\!\!}\!\!$
&$\!\!\mat2{\!\!\omega\!\!}{0} {0}{\!\!\omega^2\!\!}\!\!$
&$\!\!\Mat3{1}{0}{0} {0}{\omega^{k}}{0} {0}{0}{\!\!\omega^{2k}\!\!}
\!\!$&$\!\!\Mat3{1}{0}{0}{0}{\!\!\omega^{k}\!\!}{0} 
{0}{0}{\!\!\omega^{2k}\!\!}\!\!$\\\hline
$\!\!{b}\!\!$&$\!\!1\!\!$&$\!\!1\!\!$
&$\!\!\mat2{\!\!\omega\!\!}{0} {0}{\!\!\omega^2\!\!}\!\!$
&$\!\!\mat2{\!\!\omega\!\!}{0} {0}{\!\!\omega^2\!\!}\!\!$
&$\!\!\mat2{\!\!\omega\!\!}{0} {0}{\!\!\omega^2\!\!}\!\!$
&$\!\!\mat2{1}{0} {0}{1}\!\!$
&$\!\!\Mat3{0}{1}{0} {0}{0}{1} {1}{0}{0}\!\!$
&$\!\!\Mat3{0}{1}{0} {0}{0}{1} {1}{0}{0}\!\!$
\\
\hline
$\!\!{c}\!\!$&$\!\!1\!\!$&$\!\!-1\!\!$
&$\!\!\mat2{0}{1} {1}{0}$&$\mat2{0}{1} {1}{0}\!\!$
&$\!\!\mat2{0}{1} {1}{0}$&$\mat2{0}{1} {1}{0}\!\!$
&$\!\!\Mat3{0}{0}{1} {0}{1}{0} {1}{0}{0}\!\!$
&$\!\!\Mat3{0}{0}{\!\!-1\!\!} {0}{\!\!-1\!\!}{0} {\!\!-1\!\!}{0}{0}\!\!$
\\
\hline
\end{tabular}
\end{center}
\caption{Representations of $a$, $a'$, $ b$ 
and $c$ 
in $\Delta(54)$}
\label{tab:delta54}
\end{table}

\vskip .5cm
{$\bullet$ $\Delta(54) \to S_3 \times Z_3$}

The $\Delta(54)$ group includes $S_3 \times Z_3$ as a subgroup.
The subgroup $S_3$ consists of $\{ e,b,c,b^2,bc,b^2c \}$.
The $Z_3$ part of $S_3 \times Z_3$ consists of 
$\{ e,aa'^2, a^2a' \}$, where 
$(aa'^2)^3=e$ and the element $aa'^2$ commutes with all of 
the $S_3$ elements.
Representations, $\mbox{\boldmath $r$}_k$, for  $S_3 \times Z_3$ 
are specified by representations $\mbox{\boldmath $r$}$ of $S_3$ and 
the $Z_3$ charge $k$, where $\mbox{\boldmath $r$}={\bf 1}, {\bf 1}', {\bf 2}$ 
and $k=0,1,2$.
That is, the element  $aa'^2$ is represented as $aa'^2=\omega^k$ 
on $\mbox{\boldmath $r$}_k$ for $k=0,1,2$.
For the decomposition of $\Delta(54)$ to $S_3 \times Z_3$, 
it would be convenient to use the basis for $S_3$ representations, 
${\bf 1}$, ${\bf 1}'$ and ${\bf 2}$, 
which is shown in Table \ref{tab:delta54-S3}.
Then, each representation of $\Delta(54)$ is decomposed 
to representations of $S_3 \times Z_3$ as follows,
\begin{eqnarray}
\begin{array}{ccccccccccccc}
\Delta(54) 
&{\bf 1}_{+}&{\bf 1}_{-}&{\bf 2}_1&{\bf 2}_2&{\bf 2}_3&
{\bf 2}_4&{\bf 3}_{1(k)}&{\bf 3}_{2(k)}\\
 &\downarrow &\downarrow &\downarrow &\downarrow
 &\downarrow &\downarrow &\downarrow &\downarrow
\\
S_3 &{\bf 1}_0&{\bf 1}'_0&{\bf 2}_0&{\bf 2}_0&{\bf 2}_0 
&{\bf 1}_0+{\bf 1}'_0&{\bf 1}_k+{\bf 2}_k&{\bf 1}'_k+{\bf 2}_k
\end{array} ,
\end{eqnarray}
for $k=1,2$.
Components of $S_3$ doublets and singlets obtained from 
$\Delta(54)$ triplets 
are the same ones as considered in 
the decomposition for $S_4\to S_3$.

\begin{table}[t]
\begin{center}
\begin{tabular}{|c|c|c|c|c|c|c|c|c|c|c|}
\hline
&{\bf 1}&{\bf 1}'&{\bf 2}\\
\hline
${b}$&1&1
&$\mat2{\omega}{0} {0}{\omega^2}$
\\
\hline
${c}$&1&-1
&$\mat2{0}{1} {1}{0}$
\\
\hline
\end{tabular}
\end{center}
\caption{Representations of $ b$ 
and $c$ of $S_3$
in $\Delta(54)$}
\label{tab:delta54-S3}
\end{table}

\vskip .5cm
{$\bullet$ $\Delta(54) \to \Sigma(18)$}

We consider the subgroup $\Sigma(18)$, which 
consists of $\tilde b^\ell \tilde a^ma'^n$ with $\ell =0,1$ and 
$m,n=0,1,2$, where $\tilde b=c$ and $\tilde a=a^2$.
Table \ref{tab:delta54-18} shows the representations of the generators, 
$\tilde a$, $a'$ and $\tilde b$ on each representation 
of $\Delta(54)$.
Then, each representation of $\Delta(54)$ is decomposed 
to representations of $\Sigma(18)$ as follows,
\begin{eqnarray}
\begin{array}{ccccccccccccc}
\Delta(54) 
&{\bf 1}_{+}&{\bf 1}_{-}&{\bf 2}_1&{\bf 2}_2&{\bf 2}_3&{\bf 2}_4
&{\bf 3}_{1(k)}&{\bf 3}_{2(k)}\\
 &\downarrow &\downarrow &\downarrow &\downarrow
 &\downarrow &\downarrow &\downarrow &\downarrow
\\
\Sigma(18) &{\bf 1}_{+0}&{\bf 1}_{-0}
&{\bf 2}_{0,0}&{\bf 2}_{2,1}
&{\bf 2}_{1,2}&{\bf 2}_{1,2} 
&{\bf 1}_{+k}+{\bf 2}_{0,2k}&{\bf 1}_{-k}+{\bf 2}_{2k,0}
\end{array} .
\end{eqnarray}
The decomposition of triplet components is obtained as follows,
\begin{eqnarray}
\3tvec{x_1}{x_2}{x_3}_{{\bf 3}_{1(k)}}
\to
(x_2)_{{\bf 1}_{+k}}
\oplus\2tvec{x_1}{x_3}_{{\bf 2}_{0,2k}},
\quad
\3tvec{x_1}{x_2}{x_3}_{{\bf 3}_{2(k)}}
\to
(x_2)_{{\bf 1}_{-k}}
\oplus\2tvec{x_3}{-x_1}_{{\bf 2}_{2k,0}}.
\end{eqnarray}

\begin{table}[t]
\begin{center}
\begin{tabular}{|c|c|c|c|c|c|c|c|c|c|c|}
\hline
&$\!\!{\bf 1}_{+}\!\!$&$\!\!{\bf 1}_{-}\!\!$&$\!\!{\bf 2}_1\!\!$&$\!\!{\bf 2}_2\!\!$&$\!\!{\bf 2}_3\!\!$&
$\!\!{\bf 2}_4\!\!$&$\!\!{\bf 3}_{1(k)}\!\!$&$\!\!{\bf 3}_{2(k)}\!\!$\\
\hline
$\!\!\tilde{a}\!\!$&$\!\!1\!\!$&$\!\!1\!\!$
&$\!\!\mat2{1}{0} {0}{1}\!\!$&$\!\!\mat2{\!\!\omega\!\!}{0} {0}{\!\!\omega^2\!\!}\!\!$
&$\!\!\mat2{\!\!\omega^2\!\!}{0} {0}{\!\!\omega\!\! }\!\!$
&$\!\!\mat2{\!\!\omega^2\!\!}{0} {0}{\!\!\omega \!\!}\!\!$
&$\!\!\Mat3{\!\!\omega^{2k}\!\!}{0}{0} {0}{\!\!\omega^{k}\!\!}{0}{0}{0}{1}\!\!$
&$\!\!\Mat3{\!\!\omega^{2k}\!\!}{0}{0}{0}{\!\!\omega^{k}\!\!}{0}{0}{0}{\!\!1\!\!}\!\!$
\\\hline
${\!\! {a}'\!\!}$&$\!\!1\!\!$&$\!\!1\!\!$
&$\!\!\mat2{1}{0} {0}{1}\!\!$
&$\!\!\mat2{\!\!\omega^2\!\!}{0} {0}{\!\!\omega\!\!}\!\!$
&$\!\!\mat2{\!\!\omega\!\!}{0} {0}{\!\!\omega^2\!\!}\!\!$
&$\!\!\mat2{\!\!\omega\!\!}{0} {0}{\!\!\omega^2\!\!}\!\!$
&$\!\!\Mat3{1}{0}{0}{0}{\!\!\omega^{k}\!\!}{0}{0}{0}{\!\!\omega^{2k}\!\!}\!\!$
&$\!\!\Mat3{\!\!1\!\!}{0}{0}{0}{\!\!\omega^{k}\!\!}{0}{0}{0}{\!\!\omega^{2k}\!\!}\!\!$\\\hline
$\!\!{\tilde b}\!\!$&$\!\!1\!\!$&$\!\!-1\!\!$
&$\!\!\mat2{0}{1} {1}{0}\!\!$&$\!\!\mat2{0}{1} {1}{0}\!\!$
&$\!\!\mat2{0}{1} {1}{0}$&$\mat2{0}{1} {1}{0}\!\!$
&$\!\!\Mat3{0}{0}{1} {0}{1}{0} {1}{0}{0}\!\!$
&$\!\!\Mat3{0}{0}{\!\!-1\!\!}{0}{\!\!-1\!\!}{0} {\!\!-1\!\!}{0}{0}\!\!$
\\
\hline
\end{tabular}
\end{center}
\caption{Representations of $ \tilde b$, $\tilde a$ and $a'$ 
of $\Sigma(18)$ in $\Delta(54)$}
\label{tab:delta54-18}
\end{table}

\vskip .5cm
{$\bullet$ $\Delta(54) \to \Delta(27)$}

We consider the subgroup $\Delta(27)$, which 
consists of $b^ka^ma'^n$ with 
$k,m,n=0,1,2$.
By use of Table \ref{tab:delta54}, it is found 
that each representation of $\Delta(54)$ is decomposed 
to representations of $\Delta(27)$ as follows,
\begin{equation}
\begin{array}{ccccccccccccc}
\Delta(54) 
&{\bf 1}_{+}&{\bf 1}_{-}&{\bf 2}_1&{\bf 2}_2&
{\bf 2}_3&{\bf 2}_4&{\bf 3}_{1(k)}&{\bf 3}_{2(k)}\\
 &\downarrow &\downarrow &\downarrow &\downarrow
 &\downarrow &\downarrow &\downarrow &\downarrow
\\
\Delta(27) &{\bf 1}_{0,0}&{\bf 1}_{0,0}
&{\bf 1}_{1,0}+{\bf 1}_{2,0}&{\bf 1}_{1,1}+{\bf 1}_{2,2}
&{\bf 1}_{1,2}+{\bf 1}_{2,1}&{\bf 1}_{0,2}+{\bf 1}_{0,1} 
&{\bf 3}_{{[0][k]}}&{\bf 3}_{{[0][k]}}
\end{array} .
\end{equation}

\clearpage

\clearpage





\section{Anomalies}

\subsection{Generic aspects}

In section \ref{sec:application}, some phenomenological 
applications of 
non-Abelian discrete symmetries are shown as 
flavor symmetries.
In general, symmetries at the tree-level can be 
broken by quantum effects, i.e. anomalies.
Anomalies of continuous symmetries, in particular 
gauge symmetries, have been studied well.
Here we review about anomalies of non-Abelian discrete symmetries.
For our purpose, the path integral approach is convenient.
Thus, we use Fujikawa's method \cite{Fujikawa:1979ay,Fujikawa:1980eg} 
to derive anomalies of discrete symmetries.
(See e.g. Ref.~\cite{Araki:2008ek}.)

Let us consider a gauge theory with a (non-Abelian) gauge group $G_g$ 
and a set of fermions 
$\Psi = [\psi^{(1)}, \cdots, \psi^{(M)}]$.
Then, we assume that their Lagrangian is invariant under 
the following chiral transformation,
\begin{eqnarray}\label{eq:U-trans}
\Psi(x) \rightarrow U \Psi(x),
\end{eqnarray}
with $U=\exp(i\alpha P_L)$ and $\alpha = \alpha^A T_A$, 
where $T_A$ denote the generators of the transformation and 
$P_L$ is the left-chiral projector.
Here, the above transformation is not necessary a gauge transformation. 
The fermions $\Psi(x)$ are the (irreducible) $M$-plet 
representation $\boldsymbol{R}^M$.
For the moment, we suppose that $\Psi(x)$ correspond to a (non-trivial) 
singlet under the flavor symmetry while they correspond to 
the $\boldsymbol{R}^M$ 
representation under the gauge group $G_g$.
Since the generator $T_A$ as well as $\alpha$ is represented on 
$\boldsymbol{R}^M$ as a $(M \times M)$ matrix, 
we use the notation, $T_A(\boldsymbol{R}^M)$ 
and $\alpha(\boldsymbol{R}^M)=\alpha^AT_A(\boldsymbol{R}^M)$.

The anomaly appears in Fujikawa's method from 
the transformation of the path integral measure as the 
Jacobian, $J(\alpha)$, i.e.,
\begin{equation}
\mathcal{D}\Psi \mathcal{D}\bar \Psi \rightarrow \mathcal{D}\Psi 
\mathcal{D}\bar \Psi J(\alpha),
\end{equation}
where 
\begin{eqnarray}
J(\alpha)=\exp \left( i \int d^4x \mathcal{A}(x;\alpha)  \right).
\end{eqnarray}
The anomaly function  $\mathcal{A}$ decomposes into a gauge part 
and a gravitational  part
\cite{AlvarezGaume:1983ig,AlvarezGaume:1984dr,Fujikawa:1986hk}
\begin{equation}\label{eq:AnomalyFunctionA}
 \mathcal{A}~=~
  \mathcal{A}_\mathrm{gauge}+\mathcal{A}_\mathrm{grav}
\;.
\end{equation}
The gauge part is given by 
\begin{equation}
 \mathcal{A}_\mathrm{gauge}(x;\alpha)~=~\frac{1}{32\,\pi^2}
 \mathrm{Tr}\left[\alpha(\boldsymbol{R}^M)\,F^{\mu\nu}(x)\,
\widetilde{{F}}_{\mu\nu}(x)\right]\;,
\end{equation}
where ${F}^{\mu\nu}$ denotes the field strength of 
the gauge fields, ${F}_{\mu\nu}=[D_\mu,D_\nu]$, and 
$\widetilde{{F}}_{\mu\nu}$ denotes its dual, 
$\widetilde{{F}}^{\mu\nu}=
\varepsilon^{\mu\nu\rho\sigma}{F}_{\rho\sigma}$.
The trace `Tr' runs over all internal indices. 
When the transformation corresponds to a continuous symmetry, 
this anomaly can be calculated by the triangle diagram with 
external lines of two gauge bosons and one current corresponding 
to the symmetry for Eq.~(\ref{eq:U-trans}).

Similarly, the gravitation part is obtained as 
\cite{AlvarezGaume:1983ig,AlvarezGaume:1984dr,Fujikawa:1986hk}
\begin{equation}
 \mathcal{A}_\mathrm{grav}
 ~=~
 -\mathcal{A}_\mathrm{grav}^\mathrm{Weyl\:fermion} 
   \mathrm{tr} \left[\alpha(\boldsymbol{R}^{(M)})\right]\;, 
\label{eq:Agrav}
\end{equation}
where 
`$\mathrm{tr}$' is the trace for the matrix $(M \times M)$
$T_A(\boldsymbol{R}^M)$. 
The contribution  of a single Weyl fermion to the
gravitational anomaly is given by 
\cite{AlvarezGaume:1983ig,AlvarezGaume:1984dr,Fujikawa:1986hk}  
\begin{equation}
 \mathcal{A}_\mathrm{grav}^\mathrm{Weyl\:fermion}  ~=~
 \frac{1}{384 \pi^2}\,\frac{1}{2}\,
 \varepsilon^{\mu\nu\rho\sigma}\, R_{\mu\nu}{}^{\lambda\gamma}\,
 R_{\rho\sigma\lambda\gamma}\;.
\end{equation}
When other sets of $M_i$-plet fermions $\Psi_{M_i}$ 
are included in a theory, 
the total gauge and gravity anomalies are obtained as their summations, 
$\sum_{\Psi_{M_i}} \mathcal{A}_\mathrm{gauge}$ and 
$\sum_{\Psi_{M_i}} \mathcal{A}_\mathrm{grav}$.

For the evaluation of these anomalies, it is useful to recall the
index theorems~\cite{AlvarezGaume:1983ig,AlvarezGaume:1984dr}, which
imply
\begin{subequations}\label{eq:indexTheorems}
\begin{eqnarray}
 \int d^4 x\, \frac{1}{32\pi^2}\,\varepsilon^{\mu\nu\rho\sigma}\,
 	F_{\mu\nu}^a\,F_{\rho\sigma}^b\,
	\mathrm{tr}\left[\mathsf{t}_a\,\mathsf{t}_b\right]
 & \in & \mathbbm{Z}
 \;, \label{eq:index1}\\*
 \frac{1}{2}\,\int d^4 x\, \frac{1}{384 \pi^2}\,\frac{1}{2}\,
 \varepsilon^{\mu\nu\rho\sigma}\, R_{\mu\nu}{}^{\lambda\gamma}\,
 R_{\rho\sigma\lambda\gamma}
 & \in & \mathbbm{Z}
 \;,\label{eq:index2}
\end{eqnarray}
\end{subequations}
where $\mathsf{t}_a$ are generators of $G_g$ 
in the fundamental representation.
We use the convention that 
$\mathrm{tr}[\mathsf{t}_a\,\mathsf{t}_b]=\frac{1}{2}\delta_{ab}$.
The factor $\frac12$ in Eq.~(\ref{eq:index2}) follows from Rohlin's theorem~\cite{Rohlin:1959}, as
discussed in~\cite{Csaki:1997aw}.
Of course, these indices are independent of each other.
The path integral includes all possible configurations 
corresponding to different index numbers.

First of all, we study anomalies of the continuous $U(1)$ symmetry.
We consider a theory with a (non-Abelian) gauge symmetry $G_g$ as well
as the continuous $U(1)$ symmetry, which may be gauged.
This theory include fermions with $U(1)$ charges, $q^{(f)}$ and 
representations $\boldsymbol{R}^{(f)}$.
Those anomalies vanish if and only if the Jacobian is trivial, 
i.e. $J(\alpha)=1$ for an arbitrary value of $\alpha$. 
Using the index theorems, one can find that 
the anomaly-free conditions require
\begin{equation}\label{eq:A[U(1)-G-G]}
 A_{U(1)-G_g-G_g}~\equiv~
  \sum_{\boldsymbol{R}^{(f)}} q^{(f)}\,T_2(\boldsymbol{R}^{(f)})~=~0,
\end{equation}
for the mixed $U(1)- G_g-G_g$ anomaly, and 
\begin{equation}
 A_{U(1)-\mathrm{grav}-\mathrm{grav}}
 ~\equiv~ 
 \sum_{f} q^{(f)}~=~0,
\end{equation}
for the $U(1)$--gravity--gravity anomaly.
Here, $T_2(\boldsymbol{R}^{(f)})$ is the Dynkin index of 
the $\boldsymbol{R}^f$ representation, i.e.
\begin{equation}
\mathrm{tr}\left[\mathsf{t}_a\left(\boldsymbol{R}^{(f)}\right)\,
 	\mathsf{t}_b\left(\boldsymbol{R}^{(f)}\right)\right]
      ~=~\delta_{ab}
T_2(\boldsymbol{R}^{(f)})
 \;.
\end{equation}

Next, let us study anomalies of the Abelian discrete symmetry, 
i.e. the $Z_N$ symmetry.
For the $Z_N$ symmetry, we write $\alpha =2 \pi Q_N/N$, 
where $Q_N$ is the $Z_N$ charge operator and its eigenvalues are 
integers.
Here we denote $Z_N$ charges of fermions as $q^{(f)}_N$.
Then we can evaluate the $Z_N - G_g-G_g$ and $Z_N$-gravity-gravity 
anomalies as the above $U(1)$ anomalies.
However, the important difference is that $\alpha$ takes a discrete 
value.
Then, the anomaly-free conditions, i.e., $J(\alpha)=1$ for a discrete 
transformation, require 
\begin{eqnarray}\label{eq:A_Z_N-G-G}
 A_{Z_{N}-G_g-G_g} & = & \frac{1}{N}\sum_{\boldsymbol{R}^{(f)}} q^{(f)_N} \,
 \big(2\,T_2(\boldsymbol{R}^{(f)})\big) \in \mathbbm{Z}
 \;,
\end{eqnarray}
for the $Z_N - G_g-G_g$  anomaly, and 
\begin{eqnarray}\label{eq:A_Z_N-grav-grav}
 A_{Z_{N}-\mathrm{grav}-\mathrm{grav}} &= & 
 \frac{2}{N}
 \sum_{f} q^{(f)}_N \,
 \dim\boldsymbol{R}^{(f)} \in \mathbbm{Z} \; ,
\end{eqnarray}
for the $Z_N$-gravity-gravity anomaly.
These anomaly-free conditions reduce to 
\begin{subequations}\label{eq:ZNconditions}
\begin{eqnarray}
& &  \sum_{\boldsymbol{R}^{(f)}} q^{(f)}_N \,
 T_2(\boldsymbol{R}^{(f)}) ~=~ 0\mod N/2\;,\label{eq:condition-gauge}\\
& & 
 \sum_{f}
 q^{(f)}_N \,\dim\boldsymbol{R}^{(f)} ~=~0\mod N/2\;.\label{eq:condition-grav}
\end{eqnarray}
\end{subequations}
Note  that the $Z_2$ symmetry is always free from 
the $Z_2$-gravity-gravity anomaly.

Finally, we study anomalies of non-Abelian discrete symmetries $G$ 
\cite{Araki:2006mw,Araki:2008ek}.
A discrete group $G$ consists of the finite number of elements, $g_i$.
Hence, the non-Abelian discrete symmetry is anomaly-free 
if and only if the Jacobian is vanishing for the transformation 
corresponding to each element $g_i$.
Furthermore, recall that $(g_i)^{N_i} =1$. 
That is, each element $g_i$ in the non-Abelian discrete group 
generates a $Z_{N_i}$ symmetry.
Thus, the analysis on non-Abelian discrete anomalies reduces to 
one on Abelian discrete anomalies.
One can take the field basis such that $g_i$ is represented in 
a diagonal form.
In such a basis, each field has a definite $Z_{N_i}$ charge, 
$q^{(f)}_{N_i}$.
The anomaly-free conditions for the $g_i$ transformation 
are written as 
\begin{subequations}\label{eq:ZNi-conditions}
\begin{eqnarray}
& &  \sum_{\boldsymbol{R}^{(f)}} q^{(f)}_{N_i} \,
 T_2(\boldsymbol{R}^{(f)}) ~=~ 0\mod N_i/2\;,\label{eq:condition-gauge-i}\\
& & 
 \sum_{f}
 q^{(f)}_{N_i} \,\dim\boldsymbol{R}^{(f)} ~=~0\mod N_i/2\;.
\label{eq:condition-grav-i}
\end{eqnarray}
\end{subequations}
If these conditions are satisfied for all of $g_i \in G$, 
there are no anomalies of the full non-Abelian symmetry $G$.
Otherwise, the non-Abelian symmetry is broken completely 
or partially to its subgroup by quantum effects.

In principle, we can investigate anomalies of non-Abelian 
discrete symmetries $G$ following the above procedure.
However, we give a practically simpler way to analyze 
those anomalies \cite{Araki:2006mw,Araki:2008ek}.
Here, we consider again the transformation similar to
\eqref{eq:U-trans} 
for a set of fermions 
$\Psi = [\psi^{(1)}, \cdots, \psi^{(Md_\alpha)}]$, 
which correspond to the $\boldsymbol{R}^M$ irreducible representation 
of the gauge group $G_g$ 
and the $\mbox{\boldmath $r$}^\alpha$ irreducible representation of 
the non-Abelian discrete symmetry $G$ with the dimension $d_\alpha$.
Let $U$ correspond to one of group elements $g_i \in G$, 
which is represented by the matrix $D_\alpha(g_i)$ on $\mbox{\boldmath $r$}^\alpha$.
Then, the Jacobian is proportional to its determinant, $\det D(g_i)$. 
Thus, the representations with $\det D_\alpha(g_i) = 1$ 
do not contribute to anomalies.
Therefore, the non-trivial Jacobian, i.e. anomalies 
are originated from representations with $\det D_\alpha(g_i) \neq 1$.
Note that $\det D_\alpha(g_i) = \det D_\alpha(gg_ig^{-1})$ 
for $g \in G$, that is, 
the determinant is constant in a conjugacy class.
Thus, it would be useful to calculate the determinants 
of elements on each irreducible representation.
Such a determinant for the  conjugacy class $C_i$ can be written by 
\begin{eqnarray}
\det(C_i)_\alpha= e^{2\pi i q_{\hat N_i}^\alpha/\hat N_i},
\end{eqnarray}  
on the 
irreducible representation $\mbox{\boldmath $r$}^\alpha$.
Note that $\hat N_i$ is a divisor of $N_i$, where 
$N_i$ is the order of $g_i$ in the conjugacy class $C_i$, i.e.
$g^{N_i}=e$, such that $q_{\hat N_i}^\alpha$ are normalized 
to be integers for all of the 
irreducible representations $\mbox{\boldmath $r$}^\alpha$.
We consider the $Z_{\hat N_i}$ symmetries and 
their anomalies.
Then, we obtain the anomaly-free conditions similar to 
\eqref{eq:ZNi-conditions}.
That is, the anomaly-free conditions for the 
conjugacy classes $C_i$ are written as 
\begin{subequations}\label{eq:ZhatNi-conditions}
\begin{eqnarray}
& &  \sum_{\boldsymbol{r}^{(\alpha)},\boldsymbol{R}^{(f)}} 
q^{\alpha(f)}_{\hat N_i} \,
 T_2(\boldsymbol{R}^{(f)}) 
~=~ 0\mod \hat N_i/2\;,\label{eq:condition-gauge-hati}\\
& & 
 \sum_{\alpha. f}
 q^{\alpha(f)}_{\hat N_i} \,\dim\boldsymbol{R}^{(f)} ~=~0\mod \hat
 N_i/2\; ,
\label{eq:condition-grav-hati}
\end{eqnarray}
\end{subequations}
for the theory including fermions with the $\boldsymbol{R}^{(f)}$ 
representations of the gauge group $G_g$ and the $\mbox{\boldmath $r$}^{\alpha(f)}$
representations of the flavor group $G$, which correspond to 
the $Z_{\hat N_i}$ charges, $q^{\alpha (f)}_{\hat N_i}$.
Note that the fermion fields with the $d_\alpha$-dimensional 
representation $\mbox{\boldmath $r$}^\alpha$ contribute to these anomalies, 
$q^{\alpha(f)}_{\hat N_i} \,
 T_2(\boldsymbol{R}^{(f)})$ and 
$q^{\alpha(f)}_{\hat N_i} \,\dim\boldsymbol{R}^{(f)}$, but not 
$d_\alpha q^{\alpha(f)}_{\hat N_i} \,
 T_2(\boldsymbol{R}^{(f)})$ and 
$d_\alpha q^{\alpha(f)}_{\hat N_i} \,\dim\boldsymbol{R}^{(f)}$.
If these conditions are satisfied for all of conjugacy classes of 
$G$, the full non-Abelian symmetry $G$ is free from anomalies.
Otherwise, the non-Abelian symmetry is broken by 
quantum effects.
As we will see below, in concrete examples, the above 
anomaly-free conditions often lead to the same conditions 
between different conjugacy classes. 
Note, when $\hat N_i=2$, the symmetry is always free from 
the mixed gravitational anomalies.  
We study explicitly more for concrete groups in what follows.

\subsection{Explicit calculations}

Here, we apply the above studies on anomalies for concrete groups.

\vskip .5cm
{$\bullet$ ${\bf S_3}$}

\vskip .2cm
We start with $S_3$.
As shown in section \ref{subsec:S3},
the $S_3$ group has the three conjugacy classes, $C_1=\{ e \}$,  
$C_2=\{ ab, ba \}$ and   $C_3=\{ a,b,bab \}$,  
and three irreducible representations, ${\bf 1}$, ${\bf 1}'$ and 
${\bf 2}$.
Note that the determinants of elements are constant in a 
conjugacy class.
The determinants of elements in singlet representations 
are equal to characters.
Obviously, the determinants of elements in a trivial singlet 
representation ${\bf 1}$ are always equal to $1$.
On the doublet representation ${\bf 2}$, the determinants of 
representation matrices in $C_1$, $C_2$ and $C_3$ are 
obtained as $1$, $1$ and $-1$, respectively.
These determinants are shown in Table \ref{tab:S3-determinants}.

\begin{table}[t]
\begin{center}
\begin{tabular}{|c|c|c|c|}
\hline
     &${\bf 1}$&${\bf 1}'$& ${\bf 2}$\\ \hline
$\det(C_1)$&  $1$   &   $1$     &   $1$   \\ \hline
$\det(C_2)$&  $1$   &   $1$     &   $1$  \\ \hline
$\det(C_3)$&  $1$   &   $-1$    &   $-1$   \\ 
\hline
\end{tabular}
\end{center}
\caption{Determinants on $S_3$ representations}
\label{tab:S3-determinants}
\end{table}

{}From these results, it is found that only the conjugacy class $C_3$
is relevant to anomalies and the only $Z_2$ symmetry can be anomalous.
Under such a $Z_2$ symmetry, the trivial singlet has vanishing 
$Z_2$ charge, while the other representations, 
 ${\bf 1}'$ and ${\bf 2}$ have the $Z_2$ charges $q_2=1$, 
that is, 
\begin{eqnarray}
Z_2 ~{\rm even} &:& {\bf 1}, \nonumber \\                   
Z_2 ~{\rm odd} &:& {\bf 1}', \quad {\bf 2}.
\end{eqnarray}
Thus, the anomaly-free conditions for the $Z_2-G_g-G_g$ mixed anomaly 
(\ref{eq:ZhatNi-conditions}) are written as 
\begin{eqnarray}
\sum_{{\bf 1}'}\sum_{\boldsymbol{R}^{(f)}}  T_2(\boldsymbol{R}^{(f)})
+ \sum_{{\bf 2}}\sum_{\boldsymbol{R}^{(f)}} \,
 T_2(\boldsymbol{R}^{(f)}) ~=~ 0\mod 1\;  .
\end{eqnarray}
Note that a doublet ${\bf 2}$ contributes on the 
anomaly coefficient by not $2T_2(\boldsymbol{R}^{(f)})$ but 
$ T_2(\boldsymbol{R}^{(f)})$, which is the same as 
${\bf 1}'$.
To show this explicitly, we have written the summations 
on ${\bf 1}'$ and ${\bf 2}$ separately.

\vskip .5cm
{$\bullet$ ${\bf S_4}$}

\vskip .2cm
Similarly, we can study anomalies of $S_4$.
As seen in section 3.2, 
the $S_4$ group has five the conjugacy classes, 
$C_1$, $C_3$, $C_6$, $C_6'$ and $C_8$ 
and the five irreducible representations, 
${\bf 1}$, ${\bf 1}'$, ${\bf 2}$, ${\bf 3}$ and ${\bf 3}'$.
The determinants of group elements in each representation 
are shown in Table \ref{tab:S4-determinants}.
These results imply that only the $Z_2$ symmetry can be anomalous.
Under such a $Z_2$ symmetry, each representation has the following 
behaviors,
\begin{eqnarray}
Z_2 ~{\rm even} &:& {\bf 1}, \quad {\bf 3}', \nonumber \\                   
Z_2 ~{\rm odd} &:& {\bf 1}', \quad {\bf 2},  \quad {\bf 3}.
\end{eqnarray}
Then, the anomaly-free conditions for the $Z_2-G_g-G_g$ mixed anomaly 
(\ref{eq:ZNi-conditions}) are written as 
\begin{eqnarray}
\sum_{{\bf 1}'}\sum_{\boldsymbol{R}^{(f)}}  T_2(\boldsymbol{R}^{(f)})
+ \sum_{{\bf 2}}\sum_{\boldsymbol{R}^{(f)}} \,
 T_2(\boldsymbol{R}^{(f)}) + \sum_{{\bf 3}}\sum_{\boldsymbol{R}^{(f)}} \,
 T_2(\boldsymbol{R}^{(f)}) ~=~ 0\mod 1\;  .
\end{eqnarray}

\begin{table}[t]
\begin{center}
\begin{tabular}{|c|c|c|c|c|c|}
\hline
     &${\bf 1}$&${\bf 1}'$& ${\bf 2}$&${\bf 3}$&${\bf 3}'$\\ \hline
$\det(C_1)$&  $1$   &   $1$     &   $1$    &  $1$  & $1$ \\ \hline
$\det(C_3)$&  $1$   &   $1$     &   $1$    &  $1$ & $1$ \\ \hline
$\det(C_6)$&  $1$   &   $-1$    &   $-1$   &  $-1$ & $1$ \\ \hline
$\det(C_6')$& $1$   &   $-1$   &    $-1$   &  $-1$& $1$ \\ \hline
$\det(C_8)$& $1$    & $1$     &    $1$   &  $1$ & $1$ \\
\hline
\end{tabular}
\end{center}
\caption{Determinants on $S_4$ representations}
\label{tab:S4-determinants}
\end{table}

\vskip .5cm
{$\bullet$ ${\bf A_4}$}

\vskip .2cm
We study anomalies of $A_4$.
As shown in section \ref{sec:A4}, there are four conjugacy classes, 
$C_1, C_3, C_4$ and $C_4'$, and 
four irreducible representations, 
${\bf 1}, {\bf 1}', {\bf 1}''$ and ${\bf 3}$.
The determinants of group elements in each representation 
are shown in Table \ref{tab:A4-determinants}, where $\omega = e^{2\pi i/3}$.
These results imply that only the $Z_3$  symmetry can be anomalous.
Under such a $Z_3$ symmetry, each representation 
has the following $Z_3$ charge $q_3$,
\begin{eqnarray}
q_3=0 &:& {\bf 1}, \quad {\bf 3},\nonumber\\                   
q_3=1 &:& {\bf 1}', \\
q_3=2 &:& {\bf 1}''.\nonumber
\end{eqnarray}
This corresponds to the $Z_3$ symmetry for the conjugacy class $C_4$.
There is another $Z_3$ symmetry for the conjugacy class $C_4'$, 
but it is not independent of the former $Z_3$.
Then, the anomaly-free conditions are written as 
\begin{eqnarray}
\sum_{{\bf 1}'}\sum_{\boldsymbol{R}^{(f)}}  T_2(\boldsymbol{R}^{(f)})
+ 2\sum_{{\bf 1}''}\sum_{\boldsymbol{R}^{(f)}} \,
 T_2(\boldsymbol{R}^{(f)}) ~=~ 0\mod 3/2\;  ,
\end{eqnarray}
for the $Z_3-G_g-G_g$ anomaly and 
\begin{eqnarray}
\sum_{{\bf 1}'}\sum_{\boldsymbol{R}^{(f)}}  \dim \boldsymbol{R}^{(f)}
+ 2\sum_{{\bf 1}''}\sum_{\boldsymbol{R}^{(f)}} \,
 \dim \boldsymbol{R}^{(f)} ~=~ 0\mod 3/2\;  ,
\end{eqnarray}
for the $Z_3$-gravity-gravity anomaly.

\begin{table}[t]
\begin{center}
\begin{tabular}{|c|c|c|c|c|}
\hline
     &${\bf 1}$&${\bf 1}'$& ${\bf 1}''$&${\bf 3}$ \\ \hline
$\det(C_1)$&  $1$   &   $1$     &   $1$    &  $1$ \\ \hline
$\det(C_3)$&  $1$   &   $1$     &   $1$    &  $1$ \\ \hline
$\det(C_4)$&  $1$   &   $\omega$    &   $\omega^2$   &  $1$\\ \hline
$\det(C_4')$& $1$   &   $\omega^2$   &    $\omega$   &  $1$ \\ 
\hline
\end{tabular}
\end{center}
\caption{Determinants on $A_4$ representations}
\label{tab:A4-determinants}
\end{table}

\vskip .5cm
{$\bullet$ ${\bf A_5}$}

\vskip .2cm
We study anomalies of $A_5$.
As shown in section \ref{sec:A5}, 
there are five conjugacy classes, 
$C_1, C_{15}, C_{20}, C_{12}$ and $C_{12}'$, and 
five irreducible representations, 
${\bf 1}, {\bf 3}, {\bf 3}', {\bf 4}$ and ${\bf 5}$.
The determinants of group elements in each representation 
are shown in Table \ref{tab:A4-determinants}.
That is, the determinants of all  the $A_5$ elements are 
equal to one on any representation.
This result can be understood as follows.
All of the $A_5$ elements are written by products of 
$s=a$ and $t=bab$.
The generators, $s$ and $t$, are written as real matrices 
on all of representations, 
${\bf 1}, {\bf 3}, {\bf 3}', {\bf 4}$ and ${\bf 5}$.
Thus, it is found $\det (t)=1$, because $t^5=e$.
Similarly, since $s^2=b^3=e$, the possible values are 
obtained as $\det(s)=\pm 1 $ and $\det(b)=\omega^k$ with 
$k=0,1,2$.
By imposing $\det(bab)=\det (t)=1$, we find 
$\det (s)=\det(b)=1$.
Thus, it is found that $\det(g)=1$ for all of 
the $A_5$ elements on any representation.
Therefore, the $A_5$ symmetry is always anomaly-free.

\begin{table}[t]
\begin{center}
\begin{tabular}{|c|c|c|c|c|c|}
\hline
     &${\bf 1}$&${\bf 3}$& ${\bf 3}'$&${\bf 4}$&${\bf 5}$ \\ \hline
$\det(C_1)$&     $1$  &  $1$  &  $1$   &  $1$ & $1$ \\ \hline
$\det(C_{15})$&  $1$  &  $1$  &  $1$   &  $1$ & $1$  \\ \hline
$\det(C_{20})$&  $1$  &  $1$  &  $1$   &  $1$ & $1$ \\ \hline
$\det(C_{12})$&  $1$  &  $1$  &  $1$   &  $1$ & $1$ \\ \hline
$\det(C_{12}')$& $1$  &  $1$  &  $1$   &  $1$ & $1$ \\ 
\hline
\end{tabular}
\end{center}
\caption{Determinants on $A_5$ representations}
\label{tab:A5-determinants}
\end{table}

\vskip .5cm
{$\bullet$ ${\bf T}'$}

\vskip .2cm
We study anomalies of $T'$.
As shown in section \ref{sec:T'}, 
the $T'$ group has seven conjugacy classes, 
$C_1$, $C_1'$, $C_4$, $C_4'$, $C_4''$, $C_4'''$ and $C_6$, 
and seven irreducible representations, 
${\bf 1}$, ${\bf 1}'$,  ${\bf 1}''$, ${\bf 2}$, ${\bf 2}'$, 
${\bf 2}''$ and ${\bf 3}$.
The determinants of group elements on each representation 
are shown in Table \ref{tab:T'-determinants}.
These results imply that only the $Z_3$  symmetry can be anomalous.
Under such a $Z_3$ symmetry, each representation 
has the following $Z_3$ charge $q_3$,
\begin{eqnarray}
q_3=0 &:& {\bf 1}, \quad {\bf 2},\quad {\bf 3},\nonumber\\                   
q_3=1 &:& {\bf 1}', \quad {\bf 2}'',\\
q_3=2 &:& {\bf 1}''\quad {\bf 2}'.\nonumber
\end{eqnarray}
This corresponds to the $Z_3$ symmetry for the conjugacy class $C_4$.
There is other $Z_3$ symmetries for the conjugacy classes $C_4'$, 
$C_4''$ and $C_4'''$,  
but those are not independent of the former $Z_3$.
Then, the anomaly-free conditions are written as 
\begin{eqnarray}
& & \sum_{{\bf 1}'}\sum_{\boldsymbol{R}^{(f)}}  T_2(\boldsymbol{R}^{(f)})
+ 2\sum_{{\bf 1}''}\sum_{\boldsymbol{R}^{(f)}} \,
 T_2(\boldsymbol{R}^{(f)}) + 
\sum_{{\bf 2}''}\sum_{\boldsymbol{R}^{(f)}}  T_2(\boldsymbol{R}^{(f)}) 
\nonumber \\
& & ~~~~~~~~~~~+ 2\sum_{{\bf 2}'}\sum_{\boldsymbol{R}^{(f)}} \,
 T_2(\boldsymbol{R}^{(f)})
~=~ 0\mod 3/2\;  ,
\end{eqnarray}
for the $Z_3-G_g-G_g$ anomaly and 
\begin{eqnarray}
& & \sum_{{\bf 1}'}\sum_{\boldsymbol{R}^{(f)}}  \dim \boldsymbol{R}^{(f)}
+ 2\sum_{{\bf 1}''}\sum_{\boldsymbol{R}^{(f)}} \,
 \dim \boldsymbol{R}^{(f)} + 
\sum_{{\bf 2}''}\sum_{\boldsymbol{R}^{(f)}}  \dim \boldsymbol{R}^{(f)} 
\nonumber \\
& & ~~~~~~~~~~~+ 2\sum_{{\bf 2}'}\sum_{\boldsymbol{R}^{(f)}} \,
 \dim \boldsymbol{R}^{(f)}
~=~ 0\mod 3/2\;  ,
\end{eqnarray}
for the $Z_3$-gravity-gravity anomaly.

\begin{table}[t]
\begin{center}
\begin{tabular}{|c|c|c|c|c|c|c|c|}
\hline
     &${\bf 1}$&${\bf 1}'$& ${\bf 1}''$
      &${\bf 2}$&${\bf 2}'$& ${\bf 2}''$&${\bf 3}$ \\ \hline
$\det(C_1)$&  $1$   &   $1$     &   $1$    
           &  $1$   &   $1$     &   $1$  &  $1$ \\ \hline
$\det(C_1')$&  $1$   &   $1$     &   $1$  
            &  $1$   &   $1$     &   $1$    &  $1$ \\ \hline
$\det(C_4)$&  $1$   &   $\omega$     &   $\omega^2$    
             &  $1$ & $\omega^2$     &   $\omega$  &$1$   \\ \hline
$\det(C_4')$&  $1$   &   $\omega^2$    &   $\omega$   
           &  $1$ & $\omega$     &   $\omega^2$  &$1$    \\ \hline
$\det(C_4'')$&  $1$   &   $\omega$     &   $\omega^2$    
             &  $1$ & $\omega^2$     &   $\omega$  &$1$   \\ \hline
$\det(C_4''')$&  $1$   &   $\omega^2$    &   $\omega$   
           &  $1$ & $\omega$     &   $\omega^2$  &$1$    \\ \hline
$\det(C_6)$&  $1$   &   $1$     &   $1$    
           &  $1$   &   $1$     &   $1$  &  $1$ \\
\hline
\end{tabular}
\end{center}
\caption{Determinants on $T'$ representations}
\label{tab:T'-determinants}
\end{table}

\vskip .5cm
{$\bullet$ \bf  ${\bf D_{N}}$ (${\bf N=}$even)}

\vskip .2cm
We study anomalies of $D_N$ with $N=$ even.
As shown in section \ref{sec:DN}, 
the $D_N$ group with $N=$ even has the four 
singlets ${\bf 1}_{\pm \pm}$ and $(N/2-1)$ 
doublets ${\bf 2}_k$.
All of the $D_N$ elements can be written as 
products of two elements, $a$ and $b$.
Their determinants on ${\bf 2}_k$ are obtained as 
$\det (a) =1$ and $\det (b) = -1$.
Similarly, we can obtain determinants of $a$ and $b$ on 
four singlets, ${\bf 1}_{\pm \pm}$.
Indeed, four singlets are classified by values of 
$\det (b)$ and $\det (ab)$, that is, 
$\det (b) =1$ for  ${\bf 1}_{+ \pm}$, 
$\det (b) =-1$ for  ${\bf 1}_{- \pm}$,
$\det (ab) =1$ for  ${\bf 1}_{\pm +}$ and 
$\det (ab) =-1$ for  ${\bf 1}_{\pm -}$.
Thus, the determinants of $b$ and $ab$ are essential for anomalies.
Those determinants are summarized in Table
\ref{tab:DN-even-determinants}.
This implies that two $Z_2$ symmetries can be anomalous.
One $Z_2$ corresponds to $b$ and the other $Z_2'$ corresponds to 
$ab$.
Under these $Z_2 \times Z_2'$ symmetry, each representation 
has the following behavior,
\begin{eqnarray}
Z_2 ~{\rm even} &:& {\bf 1}_{+\pm},\nonumber\\                   
Z_2 ~{\rm odd} &:& {\bf 1}_{-\pm}, \quad {\bf 2}_k,
\end{eqnarray}
\begin{eqnarray}
Z'_2 ~{\rm even} &:& {\bf 1}_{\pm +},\nonumber\\                   
Z'_2 ~{\rm odd} &:& {\bf 1}_{\pm -}, \quad {\bf 2}_k.
\end{eqnarray}
Then, the anomaly-free conditions are written as 
\begin{eqnarray}
\sum_{{\bf 1}_{- \pm}}\sum_{\boldsymbol{R}^{(f)}}  T_2(\boldsymbol{R}^{(f)})
+ \sum_{{\bf 2}_k}\sum_{\boldsymbol{R}^{(f)}} \,
 T_2(\boldsymbol{R}^{(f)}) 
~=~ 0\mod 1\;  ,
\end{eqnarray}
for the $Z_2-G_g-G_g$ anomaly and 
\begin{eqnarray}
\sum_{{\bf 1}_{\pm -}}\sum_{\boldsymbol{R}^{(f)}}  T_2(\boldsymbol{R}^{(f)})
+ \sum_{{\bf 2}_k}\sum_{\boldsymbol{R}^{(f)}} \,
 T_2(\boldsymbol{R}^{(f)}) 
~=~ 0\mod 1\;  ,
\end{eqnarray}
for the $Z_2'-G_g-G_g$ anomaly.

\begin{table}[t]
\begin{center}
\begin{tabular}{|c|c|c|c|c|c|}
\hline
     &${\bf 1}_{++}$&${\bf 1}_{+-}$& ${\bf 1}_{-+}$& ${\bf 1}_{--}$
      &${\bf 2}_k$ \\ \hline
$\det(b)$&  $1$   &   $1$     &   $-1$    &  $-1$  & $-1$ \\ \hline
$\det(ab)$&  $1$   &   $-1$     &   $1$    &  $-1$ & $-1$\\ 
\hline
\end{tabular}
\end{center}
\caption{Determinants on $D_N$ representations for $N=$ even}
\label{tab:DN-even-determinants}
\end{table}

\vskip .5cm
{$\bullet$ \bf  ${\bf D_{N}}$ (${\bf N=}$odd)}

\vskip .2cm
Similarly, we study anomalies of $D_N$ with $N=$ odd.
As shown in section \ref{sec:DN}, the $D_N$ group with $N=$ odd 
has the two singlets ${\bf 1}_{\pm }$ and $(N-1)/2$ 
doublets ${\bf 2}_k$.
Similarly to $D_N$ with $N=$ even, all elements of 
$D_N$ with $N=$ odd are written by products of two elements, $a$ and
$b$.
The determinants of $a$ are obtained as $\det(a)=1$ on 
all of representations, ${\bf 1}_{\pm }$ and ${\bf 2}_k$.
The determinants of $b$ are 
obtained as $\det b = 1$ on ${\bf 1}_{+ }$  and 
$\det(b)=-1$ on ${\bf 1}_{- }$ and ${\bf 2}_k$.
These are shown in Table \ref{tab:DN-odd-determinants}.
Thus, only the $Z_2$ symmetry corresponding to $b$ 
can be anomalous.
Under such a $Z_2$ symmetry, each representation has 
the following behavior,
\begin{eqnarray}
Z_2 ~{\rm even} &:& {\bf 1}_{+}, \nonumber \\                   
Z_2 ~{\rm odd} &:& {\bf 1}_{-}, \quad {\bf 2}_k .
\end{eqnarray}
Then, the anomaly-free condition is  written as 
\begin{eqnarray}
\sum_{{\bf 1}_{- }}\sum_{\boldsymbol{R}^{(f)}}  T_2(\boldsymbol{R}^{(f)})
+ \sum_{{\bf 2}_k}\sum_{\boldsymbol{R}^{(f)}} \,
 T_2(\boldsymbol{R}^{(f)}) 
~=~ 0\mod 1\;  ,
\end{eqnarray}
for the $Z_2-G_g-G_g$ anomaly.

\begin{table}[t]
\begin{center}
\begin{tabular}{|c|c|c|c|}
\hline
     &${\bf 1}_{+}$&${\bf 1}_{-}$
      &${\bf 2}_k$ \\ \hline
$\det(b)$&  $1$   &    $-1$     & $-1$ \\ \hline
$\det(a)$&  $1$   &   $1$     &   $1$   \\ 
\hline
\end{tabular}
\end{center}
\caption{Determinants on $D_N$ representations for $N=$ odd}
\label{tab:DN-odd-determinants}
\end{table}

\vskip .5cm
{$\bullet$ \bf  ${\bf Q_{N}}$ (${\bf N=4n}$)}

\vskip .2cm
We study anomalies of $Q_N$ with $N=4n$.
As shown in section \ref{sec:QN}, the $Q_N$ group 
with $N=4n$ has four singlets  ${\bf 1}_{\pm \pm}$
and  $(N/2-1)$ 
doublets ${\bf 2}_k$.
All elements of $Q_N$ are written by products of 
$a$ and $b$.
The determinant of $a$ is obtained as 
$\det(a) =1$  on all of doublets, ${\bf 2}_k$.
On the other hand, the determinant of $b$ is obtained as 
$\det(b)=1$ on the doublets ${\bf 2}_k$ with $k=$ odd 
and $\det(b)=-1$ on the doublets ${\bf 2}_k$ with $k=$ even.
Similarly to $D_N$ with $N=$ even, 
the four singlets ${\bf 1}_{\pm \pm}$ are classified by values of 
$\det (b)$ and $\det (ab)$, that is, 
$\det (b) =1$ for  ${\bf 1}_{+ \pm}$, 
$\det (b) =-1$ for  ${\bf 1}_{- \pm}$,
$\det (b) =1$ for  ${\bf 1}_{\pm +}$ and 
$\det (b) =-1$ for  ${\bf 1}_{\pm -}$.
Thus, the determinants of $b$ and $ab$ are essential for anomalies.
Those determinants are summarized in Table
\ref{tab:QN-4n-determinants}.
Similarly to $D_N$ with $N=$ even, 
two $Z_2$ symmetries can be anomalous.
One $Z_2$ corresponds to $b$ and the other $Z_2'$ corresponds to 
$ab$.
Under these $Z_2 \times Z_2'$ symmetry, each representation 
has the following behavior,
\begin{eqnarray}
Z_2 ~{\rm even} &:& {\bf 1}_{+\pm},\quad {\bf 2}_{k={\rm ~odd}}, \nonumber \\                   
Z_2 ~{\rm odd} &:& {\bf 1}_{-\pm} ,\quad {\bf 2}_{k={\rm ~even}}. 
\end{eqnarray}
\begin{eqnarray}
Z'_2 ~{\rm even} &:& {\bf 1}_{\pm +}, \quad {\bf 2}_{k={\rm ~odd}}, \nonumber \\                   
Z'_2 ~{\rm odd} &:& {\bf 1}_{\pm -},\quad {\bf 2}_{k={\rm ~even}}  .
\end{eqnarray}
Then, the anomaly-free conditions are written as 
\begin{eqnarray}
\sum_{{\bf 1}_{- \pm}}\sum_{\boldsymbol{R}^{(f)}}  T_2(\boldsymbol{R}^{(f)})
+\sum_{{\bf 2}_{k={\rm ~even}}}\sum_{\boldsymbol{R}^{(f)}}  T_2(\boldsymbol{R}^{(f)})
~=~ 0\mod 1\;  ,
\end{eqnarray}
for the $Z_2-G_g-G_g$ anomaly and 
\begin{eqnarray}
\sum_{{\bf 1}_{\pm -}}\sum_{\boldsymbol{R}^{(f)}}  T_2(\boldsymbol{R}^{(f)})
+\sum_{{\bf 2}_{k={\rm ~even}}}\sum_{\boldsymbol{R}^{(f)}}  T_2(\boldsymbol{R}^{(f)})
~=~ 0\mod 1\;  ,
\end{eqnarray}
for the $Z_2'-G_g-G_g$ anomaly.

\begin{table}[t]
\begin{center}
\begin{tabular}{|c|c|c|c|c|c|c|}
\hline
     &${\bf 1}_{++}$&${\bf 1}_{+-}$& ${\bf 1}_{-+}$& ${\bf 1}_{--}$
      &${\bf 2}_{k= {\rm ~odd}}$ & ${\bf 2}_{k= {\rm ~even}}$ \\ \hline
$\det(b)$&  $1$   &   $1$     &   $-1$    &  $-1$  & $1$ & $-1$ \\ \hline
$\det(ab)$&  $1$   &   $-1$     &   $1$    &  $-1$ & $1$ & $-1$ \\ 
\hline
\end{tabular}
\end{center}
\caption{Determinants on $Q_N$ representations for $N/2=$ even}
\label{tab:QN-4n-determinants}
\end{table}

\vskip .5cm
{$\bullet$ \bf  ${\bf Q_{N}}$ (${\bf N=4n+2}$)}

\vskip .2cm
Similarly, we study anomalies of $Q_N$ with $N=4n$+2.
As shown in section \ref{sec:QN}, the $Q_N$ group 
with $N=4n+2$ has four singlets  ${\bf 1}_{\pm \pm}$
and  $(N/2-1)$ 
doublets ${\bf 2}_k$.
All elements of $Q_N$ are written by products of 
$a$ and $b$.
The determinants of $a$ are obtained as 
$\det(a) =1$ on all of doublets, ${\bf 2}_k$.
On the other hand, the determinants of $b$ are obtained as 
$\det(b)=1$ on the doublets ${\bf 2}_k$ with $k=$ odd 
and $\det(b)=-1$ on the doublets ${\bf 2}_k$ with $k=$ even.
For all of singlets, it is found that 
$\chi_\alpha(a) = \chi_\alpha(b^2)$, i.e. $\det(a)= \det(b^2)$.
This implies that the determinants of $b$ are more essential 
for anomalies than $a$.
Indeed, the determinants of $b$ are obtained as 
$\det (b) =1$ on ${\bf 1}_{++}$,
$\det (b) =i$ on ${\bf 1}_{+-}$,
$\det (b) =-i$ on ${\bf 1}_{-+}$ and 
$\det (b) =-1$ on ${\bf 1}_{--}$.
Those determinants are summarized in Table 
\ref{tab:QN-4n+2-determinants}.
This result implies that only the $Z_4$ symmetry corresponding to $b$ 
can be anomalous.
Under such a $Z_4$ symmetry, each representation has 
the following $Z_4$ charge $q_4$,
\begin{eqnarray}
q_4=0 &:& {\bf 1}_{++}, \quad {\bf 2}_{k={\rm ~odd}} , \nonumber \\                   
q_4=1 &:& {\bf 1}_{+-}, \nonumber\\
q_4=2 &:& {\bf 1}_{--}, \quad {\bf 2}_{k={\rm ~even}} ,\\
q_4=3 &:& {\bf 1}_{-+}. \nonumber
\end{eqnarray}
That includes the $Z_2$ symmetry corresponding to 
$a$ and the $Z_2$ charge $q_2$ for each representation is defined as 
$q_2=q_4$ mod $2$.
The anomaly-free conditions are written as 
\begin{eqnarray}
& & \sum_{{\bf 1}_{+-}}\sum_{\boldsymbol{R}^{(f)}}  T_2(\boldsymbol{R}^{(f)})
+ 2\sum_{{\bf 1}_{--}}\sum_{\boldsymbol{R}^{(f)}} \,
 T_2(\boldsymbol{R}^{(f)}) + 
 3\sum_{{\bf 1}_{-+}}\sum_{\boldsymbol{R}^{(f)}}  T_2(\boldsymbol{R}^{(f)})
\nonumber \\
& &  \qquad \qquad  
+ 2\sum_{{\bf 2}_{k={\rm ~ even}}}\sum_{\boldsymbol{R}^{(f)}} \,
 T_2(\boldsymbol{R}^{(f)}) 
~=~ 0\mod 2\;  ,
\end{eqnarray}
for the $Z_4-G_g-G_g$ anomaly and 
\begin{eqnarray}
& & \sum_{{\bf 1}_{+-}}\sum_{\boldsymbol{R}^{(f)}}  \dim \boldsymbol{R}^{(f)}
+ 2\sum_{{\bf 1}_{--}}\sum_{\boldsymbol{R}^{(f)}} \,
 \dim \boldsymbol{R}^{(f)} + 
 3\sum_{{\bf 1}_{-+}}\sum_{\boldsymbol{R}^{(f)}}  \dim \boldsymbol{R}^{(f)}
\nonumber \\
& &   \qquad \qquad  
+ 2\sum_{{\bf 2}_{k={\rm ~ even}}}\sum_{\boldsymbol{R}^{(f)}} \,
 \dim \boldsymbol{R}^{(f)} 
~=~ 0\mod 2\;  ,
\end{eqnarray}
for the $Z_4$-gravity-gravity anomaly.
Similarly, we can obtain the anomaly-free condition on 
the $Z_2$ symmetry corresponding to $a$ as 
\begin{eqnarray}
\sum_{{\bf 1}_{+-}}\sum_{\boldsymbol{R}^{(f)}}  T_2(\boldsymbol{R}^{(f)})
+
 \sum_{{\bf 1}_{-+}}\sum_{\boldsymbol{R}^{(f)}}  T_2(\boldsymbol{R}^{(f)})
~=~ 0\mod 1\;  ,
\end{eqnarray}
for the $Z_2-G_g-G_g$ anomaly.

\begin{table}[t]
\begin{center}
\begin{tabular}{|c|c|c|c|c|c|c|}
\hline
     &${\bf 1}_{++}$&${\bf 1}_{+-}$& ${\bf 1}_{-+}$& ${\bf 1}_{--}$
      &${\bf 2}_{k={\rm ~odd}}$ &${\bf 2}_{k={\rm ~even}}$ \\ \hline
$\det(b)$&  $1$   &   $i$     &   $-i$    &  $-1$  & $1$  & $-1$ \\ \hline
$\det(a)$&  $1$   &   $-1$     &   $-1$    &  $1$ & $1$ & $1$\\ 
\hline
\end{tabular}
\end{center}
\caption{Determinants on $Q_N$ representations for $N/2=$ odd}
\label{tab:QN-4n+2-determinants}
\end{table}

\vskip .5cm
{$\bullet$ \bf  ${\bf \Sigma (2N^2)}$ }

\vskip .2cm
We study anomalies of $\Sigma(2N^2)$.
As shown in section \ref{sec:sigma-2n}, 
the $\Sigma(2N^2)$ group has 
$2N$ singlets, ${\bf 1}_{\pm n}$, and 
$N(N-1)/2$ doublets, ${\bf 2}_{p,q}$.
All elements of  $\Sigma(2N^2)$ can be 
written by products of $a$, $a'$ and $b$.
Their determinants for each representation are 
shown in Table \ref{tab:sigma-2n-determinants}, 
where $\rho = e^{2 \pi i/N}$.
Then, it is found that only the $Z_2$ symmetry 
corresponding to $b$ and the $Z_N$ symmetry corresponding 
to $a$ can be anomalous.
Another $Z_N$ symmetry corresponding to $a'$ is not independent of 
the $Z_N$ symmetry for $a$.
Under such $Z_2$ symmetry, each representation has 
the following behavior,
\begin{eqnarray}
Z_2 ~{\rm even} &:& {\bf 1}_{+ n}, \nonumber\\                   
Z_2 ~{\rm odd} &:& {\bf 1}_{- n} ,\quad {\bf 2}_{p,q},
\end{eqnarray}
and under the $Z_N$ symmetry corresponding to $a$ 
each representation has the following $Z_N$ charge $q_N$,
\begin{eqnarray}
q_N=n &:& {\bf 1}_{\pm n},\nonumber\\                   
q_N=p+q &:& {\bf 2}_{p,q}. 
\end{eqnarray}
Then, the anomaly-free condition is obtained as 
\begin{eqnarray}
\sum_{{\bf 1}_{-n}}\sum_{\boldsymbol{R}^{(f)}}  T_2(\boldsymbol{R}^{(f)})
+
 \sum_{{\bf 2}_{p,q}}\sum_{\boldsymbol{R}^{(f)}}  T_2(\boldsymbol{R}^{(f)})
~=~ 0\mod 1\;  ,
\end{eqnarray}
for the $Z_2-G_g-G_g$ anomaly.
Similarly, the anomaly-free conditions for the $Z_N$ symmetry 
are obtained as 
\begin{eqnarray}
\sum_{{\bf 1}_{\pm n}}\sum_{\boldsymbol{R}^{(f)}}  nT_2(\boldsymbol{R}^{(f)})
+ \sum_{{\bf 2}_{p,q}}\sum_{\boldsymbol{R}^{(f)}} \,
 (p+q) T_2(\boldsymbol{R}^{(f)}) 
~=~ 0\mod N/2\;  ,
\end{eqnarray}
for the $Z_N-G_g-G_g$ anomaly and 
\begin{eqnarray}
\sum_{{\bf 1}_{\pm n}}\sum_{\boldsymbol{R}^{(f)}}  n \dim \boldsymbol{R}^{(f)}
+ \sum_{{\bf 2}_{p,q}}\sum_{\boldsymbol{R}^{(f)}} \,
 (p+q) \dim \boldsymbol{R}^{(f)}
~=~ 0\mod N/2\;  ,
\end{eqnarray}
for the $Z_N$-gravity-gravity anomaly.

\begin{table}[t]
\begin{center}
\begin{tabular}{|c|c|c|c|}
\hline
     &${\bf 1}_{+n}$&${\bf 1}_{-n}$& ${\bf 2}_k$ \\ \hline
$\det(b)$&  $1$     &   $-1$        &   $-1$    \\ \hline
$\det(a)$&  $\rho^n$     &   $\rho^n$       &   $\rho^{p+q}$   \\  \hline
$\det(a')$&  $\rho^n$     &   $\rho^n$       &  $\rho^{p+q}$     \\ 
\hline
\end{tabular}
\end{center}
\caption{Determinants on $\Sigma(2N^2)$ representations }
\label{tab:sigma-2n-determinants}
\end{table}

\vskip .5cm
{$\bullet$ \bf  ${\bf \Delta (3N^2)}$ (${\bf N/3\neq}$integer) }

\vskip .2cm
We study anomalies of $\Delta(3N^2)$ with $N/3\neq$ integer.
As shown in section \ref{sec:Delta-3n}, 
the $\Delta(3N^2)$ group with $N/3\neq$ integer has 
three singlets, ${\bf 1}_0$, ${\bf 1}_1$ and ${\bf 1}_2$, 
and $(N^2 -1)/3$ triplets, ${\bf 3}_{[k][\ell]}$.
All elements of $\Delta(3N^2)$  can be written 
by products of $a$, $a'$ and $b$.
It is found that $\det(a)=\det(a')=1$ on all of representations.
Thus, these elements are irrelevant to anomalies.
On the other hand, the determinant of $b$ is obtained as 
$\det (b) = 1$ for all ${\bf 3}_{[k][\ell]}$ and 
${\bf 1}_0$, $\det (b) = \omega$  for ${\bf 1}_1$ and 
$\det (b) = \omega^2$  for ${\bf 1}_2$, 
with $\omega = e^{2 \pi i/3}$,
as shown in Table \ref{tab:delta-3n-determinants}.
This implies that only the $Z_3$ symmetry corresponding to 
$b$ can be anomalous.
Under such a $Z_3$ symmetry, 
each representation has the following $Z_3$ charge $q_3$,
\begin{eqnarray}
q_3=0 &:& {\bf 1}_0, \quad {\bf 3}_{[k][\ell]},\nonumber\\                   
q_3=1 &:& {\bf 1}_1, \\
q_3=2 &:& {\bf 1}_2 .\nonumber
\end{eqnarray}
Then, the anomaly-free conditions are written as 
\begin{eqnarray}
\sum_{{\bf 1}_1}\sum_{\boldsymbol{R}^{(f)}}  T_2(\boldsymbol{R}^{(f)})
+ 2\sum_{{\bf 1}_2}\sum_{\boldsymbol{R}^{(f)}} \,
 T_2(\boldsymbol{R}^{(f)}) 
~=~ 0\mod 3/2\;  ,
\end{eqnarray}
for the $Z_3-G_g-G_g$ anomaly and 
\begin{eqnarray}
\sum_{{\bf 1}_1}\sum_{\boldsymbol{R}^{(f)}}  \dim \boldsymbol{R}^{(f)}
+ 2\sum_{{\bf 1}_2}\sum_{\boldsymbol{R}^{(f)}} \,
 \dim \boldsymbol{R}^{(f)} 
~=~ 0\mod 3/2\;  ,
\end{eqnarray}
for the $Z_3$-gravity-gravity anomaly.

\begin{table}[t]
\begin{center}
\begin{tabular}{|c|c|c|}
\hline
     &${\bf 1}_{k}$ & ${\bf 3}_{[k][\ell]}$ \\ \hline
$\det(b)$&  $\omega^k$      &   $1$    \\ \hline
$\det(a)$&  $1$     &   $1$        \\  \hline
$\det(a')$&  $1$     &   $1$          \\ 
\hline
\end{tabular}
\end{center}
\caption{Determinants on $\Delta(3N^2)$ representations ($N/3\neq$ integer )}
\label{tab:delta-3n-determinants}
\end{table}

\vskip .5cm
{$\bullet$ \bf  ${\bf \Delta (3N^2)}$ (${\bf N/3=}$integer) }

\vskip .2cm
Similarly, we can study anomalies of $\Delta(3N^2)$ with $N/3=$ integer.
As shown in section \ref{sec:Delta-3n}, 
the $\Delta(3N^2)$ group with $N/3 =$ integer has 
nine singlets ${\bf 1}_{k \ell}$ 
and $(N^2 -3)/3$ triplets, ${\bf 3}_{[k][\ell]}$.
All elements $\Delta(3N^2)$  can be written 
by products of $a$, $a'$ and $b$.
On all of triplet representations ${\bf 3}_{[k][\ell]}$, 
their determinants are obtained as 
$\det(a)=\det (a')=\det (b)=1$.
On the other hand, it is found that $\det(a)=\det (a')$ on 
all of nine singlets.
Furthermore, nine singlets are classified by 
values of $\det(a)=\det (a')$ and $\det (b)$.
That is, the determinants of $\det(a)=\det (a')$ and $\det (b)$ 
are obtained as $\det(a)=\det (a') =\omega^\ell$ and 
 $\det (b) = \omega^k$ on  ${\bf 1}_{k \ell}$.
These results are shown in Table \ref{tab:delta-3n-2-determinants}.
This implies that two independent $Z_3$ symmetries can be 
anomalous.
One corresponds to $b$ and the other corresponds to 
$a$.
For the $Z_3$ symmetry corresponding to $b$, 
each representation has the following $Z_3$ charge $q_3$,
\begin{eqnarray}
q_3=0 &:& {\bf 1}_{0 \ell}, \quad {\bf 3}_{[k][\ell]},\nonumber\\                   
q_3=1 &:& {\bf 1}_{1 \ell}, \\
q_3=2 &:& {\bf 1}_{2 \ell} ,\nonumber
\end{eqnarray}
while for $Z'_3$ symmetry corresponding to $a$, 
each representation has the following $Z_3$ charge $q'_3$,
\begin{eqnarray}
q'_3=0 &:& {\bf 1}_{k 0 }, \quad {\bf 3}_{[k][\ell]},\nonumber\\                   
q'_3=1 &:& {\bf 1}_{k 1 }, \\
q'_3=2 &:& {\bf 1}_{k 2 } .\nonumber
\end{eqnarray}
Then, the anomaly-free conditions are written as 
\begin{eqnarray}
\sum_{{\bf 1}_{1 \ell}}\sum_{\boldsymbol{R}^{(f)}}  T_2(\boldsymbol{R}^{(f)})
+ 2\sum_{{\bf 1}_{2 \ell}}\sum_{\boldsymbol{R}^{(f)}} \,
 T_2(\boldsymbol{R}^{(f)}) 
~=~ 0\mod 3/2\;  ,
\end{eqnarray}
for the $Z_3-G_g-G_g$ anomaly and 
\begin{eqnarray}
\sum_{{\bf 1}_{1 \ell}}\sum_{\boldsymbol{R}^{(f)}}  \dim \boldsymbol{R}^{(f)}
+ 2\sum_{{\bf 1}_{2 \ell}}\sum_{\boldsymbol{R}^{(f)}} \,
 \dim \boldsymbol{R}^{(f)} 
~=~ 0\mod 3/2\;  ,
\end{eqnarray}
for the $Z_3$-gravity-gravity anomaly.
Similarly, for the $Z_3'$ symmetry, 
the anomaly-free conditions are written as 
\begin{eqnarray}
\sum_{{\bf 1}_{k 1}}\sum_{\boldsymbol{R}^{(f)}}  T_2(\boldsymbol{R}^{(f)})
+ 2\sum_{{\bf 1}_{k 2 }}\sum_{\boldsymbol{R}^{(f)}} \,
 T_2(\boldsymbol{R}^{(f)}) 
~=~ 0\mod 3/2\;  ,
\end{eqnarray}
for the $Z'_3-G_g-G_g$ anomaly and 
\begin{eqnarray}
\sum_{{\bf 1}_{k 1 }}\sum_{\boldsymbol{R}^{(f)}}  \dim \boldsymbol{R}^{(f)}
+ 2\sum_{{\bf 1}_{k 2 }}\sum_{\boldsymbol{R}^{(f)}} \,
 \dim \boldsymbol{R}^{(f)} 
~=~ 0\mod 3/2\;  ,
\end{eqnarray}
for the $Z'_3$-gravity-gravity anomaly.

\begin{table}[t]
\begin{center}
\begin{tabular}{|c|c|c|}
\hline
     &${\bf 1}_{k \ell}$ & ${\bf 3}_{[k][\ell]}$ \\ \hline
$\det(b)$&  $\omega^k$      &   $1$    \\ \hline
$\det(a)$&  $\omega^\ell$     &   $1$        \\  \hline
$\det(a')$&  $\omega^\ell$     &   $1$          \\ 
\hline
\end{tabular}
\end{center}
\caption{Determinants on $\Delta(3N^2)$ representations ($N/3=$ integer )}
\label{tab:delta-3n-2-determinants}
\end{table}

\vskip .5cm
{$\bullet$ \bf  ${\bf T_7}$ }

\vskip .2cm
We study anomalies of $T_7$.
As shown in section 10, 
the $T_7$ group has 
three singlets, ${\bf 1}_{0,1,2}$, and 
two triplets, ${\bf 3}$ and $\bar{\bf 3}$.
All elements of  $T_7$ can be 
written by products of $a$ and $b$, 
where $a$ and $b$ correspond to the generators of $Z_7$ and $Z_3$, 
respectively.
It is found that $\det(a)=1$ on all of representations.
Thus, these elements are irrelevant to anomalies.
On the other hand, the determinant of $b$ is 
obtained as $\det(b)=1$ for both ${\bf 3}$ and $\bar{\bf 3}$ 
and $\det(b)= \omega^k$ for ${\bf 1}_{k}$ ($k=0,1,2$), as 
shown in Table \ref{tab:T-7-determinants}. 
These results imply that only the $Z_3$  symmetry 
corresponding to $b$ can be anomalous.
Under such a $Z_3$ symmetry, each representation 
has the following $Z_3$ charge $q_3$,
\begin{eqnarray}
q_3=0 &:& {\bf 1}_0, \quad {\bf 3},\quad \bar{\bf 3},\nonumber\\                   
q_3=1 &:& {\bf 1}_1,\\
q_3=2 &:& {\bf 1}_2.\nonumber
\end{eqnarray}
Then, the anomaly-free conditions are written as 
\begin{eqnarray}
\sum_{{\bf 1}_1}\sum_{\boldsymbol{R}^{(f)}}  T_2(\boldsymbol{R}^{(f)})
+ 2\sum_{{\bf 1}_2}\sum_{\boldsymbol{R}^{(f)}} \,
 T_2(\boldsymbol{R}^{(f)}) 
~=~ 0\mod 3/2\;  ,
\end{eqnarray}
for the $Z_3-G_g-G_g$ anomaly and 
\begin{eqnarray}
\sum_{{\bf 1}_1}\sum_{\boldsymbol{R}^{(f)}}  \dim \boldsymbol{R}^{(f)}
+ 2\sum_{{\bf 1}_2}\sum_{\boldsymbol{R}^{(f)}} \,
 \dim \boldsymbol{R}^{(f)} 
~=~ 0\mod 3/2\;  ,
\end{eqnarray}
for the $Z_3$-gravity-gravity anomaly.

\begin{table}[t]
\begin{center}
\begin{tabular}{|c|c|c|c|c|c|c|}
\hline
     &${\bf 1}_{0}$&${\bf 1}_{1}$& ${\bf 1}_2$&${\bf 3}$& $\bar{\bf
       3}$ \\ 
\hline
$\det(a)$&  $1$     &   $1$        &   $1$  &   $1$        &   $1$    \\ \hline
$\det(b)$&  $1$     &   $\omega$        &   $\omega^2$  &   $1$        &   $1$    \\  \hline
\end{tabular}
\end{center}
\caption{Determinants on $T_7$ representations }
\label{tab:T-7-determinants}
\end{table}

\vskip .5cm
{$\bullet$ \bf  ${\bf \Sigma(81)}$ }

\vskip .2cm
We study anomalies of $\Sigma(81)$.
As shown in section 11, 
the $\Sigma(81)$ group has 
nine singlets, ${\bf 1}_{\ell}^{k}$, and 
eight triplets, ${\bf 3}_{A,B,C,D}$ and $\bar{\bf 3}_{A,B,C,D}$.
All elements of  $\Sigma(81)$ can be 
written by products of $a$, $a'$, $a''$, and $b$. 
The determinants of those elements on each representation 
are shown in Table \ref{tab:Sigma-81-determinants}.
These results imply that there are two independent $Z_3$  symmetries, 
which can be anomalous.
One is the $Z_3$ symmetry corresponding to $b$ and the other 
is the $Z_3$ symmetry corresponding to $a$.
For the $Z_3$ symmetry corresponding to $b$, each 
representation has the following $Z_3$ charge $q_3$,
\begin{eqnarray}
q_3=0 &:& {\bf 1}^k_{0}, {\bf 3}_{A,B,C,D}, \bar {\bf
  3}_{A,B,C,D},
\nonumber\\                   
q_3=1 &:& {\bf 1}^k_{1},\\
q_3=2 &:& {\bf 1}^k_{2},\nonumber
\end{eqnarray}
while for the $Z_3'$ symmetry corresponding to $a$ 
each representation has the following $Z'_3$ charge $q'_3$,
\begin{eqnarray}
q'_3=0 &:& {\bf 1}^0_{\ell}, {\bf 3}_{D}, \bar {\bf 3}_{D},
\nonumber\\                   
q'_3=1 &:& {\bf 1}^1_{\ell},  {\bf 3}_{A,B,C}, \\
q'_3=2 &:& {\bf 1}^2_{\ell}, \bar {\bf 3}_{A,B,C}.\nonumber
\end{eqnarray}

Then, the anomaly-free conditions are written as 
\begin{eqnarray}
\sum_{{\bf 1}^k_1}
\sum_{\boldsymbol{R}^{(f)}}  T_2(\boldsymbol{R}^{(f)})
+ 2\sum_{{\bf 1}^k_2}
\sum_{\boldsymbol{R}^{(f)}}  T_2(\boldsymbol{R}^{(f)})
~=~ 0\mod 3/2\;  ,
\end{eqnarray}
for the $Z_3-G_g-G_g$ anomaly and 
\begin{eqnarray}
\sum_{{\bf 1}^k_1}
\sum_{\boldsymbol{R}^{(f)}}  \dim \boldsymbol{R}^{(f)}
+ 2\sum_{{\bf 1}^k_2}
\sum_{\boldsymbol{R}^{(f)}}  \dim \boldsymbol{R}^{(f)}
~=~ 0\mod 3/2\;  ,
\end{eqnarray}
for the $Z_3$-gravity-gravity anomaly.
Similarly, for the $Z_3'$ symmetry corresponding to $a$, 
the anomaly-free conditions are written as 
 \begin{eqnarray}
& & \sum_{{\bf 1}^1_\ell}
\sum_{\boldsymbol{R}^{(f)}}  T_2(\boldsymbol{R}^{(f)})
+ \sum_{{\bf 3}_{A,B,C}}
\sum_{\boldsymbol{R}^{(f)}}  T_2(\boldsymbol{R}^{(f)})
+2\sum_{{\bf 1}^2_\ell}
\sum_{\boldsymbol{R}^{(f)}}  T_2(\boldsymbol{R}^{(f)})  \nonumber \\
 & & ~~~~~~ + 2\sum_{\bar {\bf 3}_{A,B,C}}
\sum_{\boldsymbol{R}^{(f)}}  T_2(\boldsymbol{R}^{(f)})
~=~ 0\mod 3/2\;  ,
\end{eqnarray}
for the $Z_3'-G_g-G_g$ anomaly and 
 \begin{eqnarray}
& & \sum_{{\bf 1}^1_\ell}
\sum_{\boldsymbol{R}^{(f)}}  \dim \boldsymbol{R}^{(f)}
+ \sum_{{\bf 3}_{A,B,C}}
\sum_{\boldsymbol{R}^{(f)}}  \dim \boldsymbol{R}^{(f)}
+2\sum_{{\bf 1}^2_\ell}
\sum_{\boldsymbol{R}^{(f)}}  \dim \boldsymbol{R}^{(f)}  \nonumber \\
 & & ~~~~~~ + 2\sum_{\bar {\bf 3}_{A,B,C}}
\sum_{\boldsymbol{R}^{(f)}}  \dim \boldsymbol{R}^{(f)}
~=~ 0\mod 3/2\;  ,
\end{eqnarray}
for the $Z'_3$-gravity-gravity anomaly.

\begin{table}[t]
\begin{center}
\begin{tabular}{|c|c|c|c|c|c|c|c|c|c|c|}
\hline
     &${\bf 1}_{\ell}^{k}$&${\bf 3}_{A}$&$\bar{\bf 3}_{A}$& ${\bf 3}_B$& $\bar{\bf 3}_B$&${\bf 3}_C$&$\bar{\bf 3}_C$& ${\bf 3}_D$& $\bar{\bf 3}_D$ \\ \hline
$\det(b)$&  $\omega^\ell$     &   $1$        &   $1$  &   $1$        &   $1$ &   $1$        &   $1$  &   $1$        &   $1$    \\ \hline
$\det(a)$&  $\omega^{k}$     &   $\omega$   &   $\omega^2$        &   $\omega$ &   $\omega^2$ &   $\omega$     &   $\omega^2$   &   $1$  &$1$  \\  \hline
$\det(a')$&  $\omega^{k}$     &   $\omega$   &   $\omega^2$        &   $\omega$ &   $\omega^2$ &   $\omega$     &   $\omega^2$   &   $1$  &$1$  \\  \hline
$\det(a'')$&  $\omega^{k}$     &   $\omega$   &   $\omega^2$        &   $\omega$ &   $\omega^2$ &   $\omega$     &   $\omega^2$   &   $1$  &$1$  \\  \hline
\end{tabular}
\end{center}
\caption{Determinants on $\Sigma(81)$ representations }
\label{tab:Sigma-81-determinants}
\end{table}

\vskip .5cm
{$\bullet$ \bf  ${\bf \Delta(54)}$ }

\vskip .2cm
We study anomalies of $\Delta(54)$.
As shown in section 12, 
the $\Delta(54)$ group has 
two singlets, ${\bf 1}_{+,-}$, 
four doublets, ${\bf 2}_{1,2,3,4}$, and 
four triplets, ${\bf 3}_{1(k)}$ and ${\bf 3}_{2(k)}$.
All elements of  $\Delta(54)$ can be 
written by products of $a$, $a'$, $b$ and $c$. 
Determinants of $a$, $a'$ and $b$ on any representation 
are obtained as $\det(a)=\det(a')=\det(b)=1$. 
The determinants of $c$ for ${\bf 1}_+$ and ${\bf 3}_2(k)$ are 
obtained as $\det(c)=1$ while the  
other representations lead to  $\det(c)=-1$.
These results are  
shown in Table \ref{tab:Delta-54-determinants}. 
That implies that only the $Z_2$  symmetry corresponding to 
the generator $c$ can be anomalous.
Under such a $Z_2$ symmetry, each representation 
has the following $Z_2$ charge $q_2$,
\begin{eqnarray}
q_2=0 &:& {\bf 1}_+,{\bf 3}_{2(k)}, \nonumber\\                   
q_2=1 &:& {\bf 1}_-,{\bf 2}_{1,2,3,4},{\bf 3}_{1(k)}.
\end{eqnarray}
Then, the anomaly-free conditions are written as 
\begin{eqnarray}
\sum_{{\bf 1}_-}
\sum_{\boldsymbol{R}^{(f)}}  T_2(\boldsymbol{R}^{(f)})
+\sum_{{\bf 2}_k}
\sum_{\boldsymbol{R}^{(f)}}  T_2(\boldsymbol{R}^{(f)})
+\sum_{{\bf 3}_{1(k)}}
\sum_{\boldsymbol{R}^{(f)}}  T_2(\boldsymbol{R}^{(f)})
~=~ 0\mod 1\;  ,
\end{eqnarray}
for the $Z_2-G_g-G_g$ anomaly.

\begin{table}[t]
\begin{center}
\begin{tabular}{|c|c|c|c|c|c|c|c|c|c|c|c|c|}
\hline
     &${\bf 1}_{+}$&${\bf 1}_{-}$&${\bf 2}_{1}$&${\bf 2}_{2}$&${\bf 2}_{3}$&${\bf 2}_{4}$& ${\bf 3}_{1(k)}$&${\bf 3}_{2(k)}$ \\ \hline
$\det(a)$&  $1$     &   $1$        &   $1$  &  $1$     &   $1$        &   $1$  &   $1$        &   $1$    \\ \hline
$\det(a')$&  $1$     &   $1$        &   $1$  &  $1$     &   $1$        &   $1$  &   $1$        &   $1$    \\ \hline
$\det(b)$&  $1$     &   $1$        &   $1$  &  $1$     &   $1$        &   $1$  &   $1$        &   $1$    \\ \hline
$\det(c)$&  $1$     &   $-1$        &   $-1$  &  $-1$     &   $-1$        &   $-1$  &   $-1$        &   $1$    \\ \hline
\end{tabular}
\end{center}
\caption{Determinants on $\Delta(54)$ representations }
\label{tab:Delta-54-determinants}
\end{table}

\vskip.5cm
Similarly, we can analyze on anomalies for other 
non-Abelian discrete symmetries.

\subsection{Comments on anomalies}

Finally, we comment on the symmetry breaking effects by quantum effect. 
When a discrete (flavor) symmetry is anomalous, 
breaking terms can appear in Lagrangian, e.g. 
by instanton effects, such as 
$\frac{1}{M^n}\Lambda^m \Phi_1 \cdots \Phi_k$, 
where $\Lambda$ is a dynamical scale and $M$ is a 
typical (cut-off) scale.
Within the framework of string theory
discrete anomalies as well as anomalies of continuous 
gauge symmetries can be canceled by the Green-Schwarz (GS) mechanism 
\cite{Green:1984sg}
unless discrete symmetries are accidental.
In the GS mechanism, dilaton and moduli fields, 
i.e. the so-called GS fields $\Phi_{GS}$, transform 
non-linearly under anomalous transformation.
The anomaly cancellation due to the GS mechanism imposes 
certain relations among anomalies.
(See e.g. Ref.~\cite{Araki:2008ek}.)\footnote{
See also Ref.~\cite{Kobayashi:1996pb}.} 
Stringy non-perturbative effects as well as field-theoretical effects 
induce terms in Lagrangian such as 
$\frac{1}{M^n}e^{-a\Phi_{GS}} \Phi_1 \cdots \Phi_k$.
The GS fields $\Phi_{GS}$, i.e. dilaton/moduli fields 
are expected to develop non-vanishing vacuum expectation values 
and the above terms correspond to breaking terms of discrete symmetries.

The above breaking terms may be small.
Such approximate discrete symmetries with small 
breaking terms may be useful in particle physics,\footnote{
See for some applications e.g. \cite{Fukuoka:2009cu}.}
if breaking terms are controllable.
Alternatively, if exact symmetries are necessary, 
one has to arrange matter fields and their 
quantum numbers such that models are free from anomalies.

\clearpage

\section{Flavor Models with  Non-Abelian Discrete Symmetry}
\label{sec:application}

We have shown several group-theoretical aspects 
for various non-Abelian discrete groups.
In this section, we study some phenomenological applications 
of these discrete symmetries.

\subsection{Tri-bimaximal mixing of lepton flavor}

The non-Abelian discrete group has been applied 
 to  the flavor symmetry  of quarks and leptons. 
Especially, the recent experimental data of neutrinos
have encouraged us to  work in  the non-Abelian discrete symmetry of flavors.
 The global fit of the neutrino experimental data in Table \ref{tabledata}
\cite{Schwetz:2008er,Fogli:2008jx,Fogli:2009zza},
 strongly  indicates the tri-bimaximal mixing matrix $U_\text{tribi}$
 for  three lepton flavors
\cite{Harrison:2002er,Harrison:2002kp,Harrison:2003aw,Harrison:2004uh}
as follows:
\begin{equation}
U_\text{tribi} = \begin{pmatrix}
               \frac{2}{\sqrt{6}} &  \frac{1}{\sqrt{3}} & 0 \\
     -\frac{1}{\sqrt{6}} & \frac{1}{\sqrt{3}} &  -\frac{1}{\sqrt{2}} \\
      -\frac{1}{\sqrt{6}} &  \frac{1}{\sqrt{3}} &   \frac{1}{\sqrt{2}}
         \end{pmatrix},
\end{equation}
which favors
the non-Abelian discrete symmetry for the lepton flavor.
Indeed, various types of models leading to the tri-bimaximal mixing 
have been proposed  by assuming 
several types of non-Abelian flavor symmetries
as seen , e.g. in the review by Altarelli and Feruglio \cite{Altarelli:2010gt}.
In this section, we introduce  typical models to reproduce
the lepton flavor mixing.

\begin{table}[tbh]
\begin{center}
\begin{tabular}{|c||c|c|c||c|}
        \hline
        parameter & best fit & 2$\sigma$ & 3$\sigma$ &tri-bimaximal
        \\
        \hline
        &&&&\\[-3mm]
        $\Delta m^2_{21}\: [10^{-5}{\rm eV}^2]$
        & $7.59^{+0.23}_{-0.18}$  & 7.22--8.03 & 7.03--8.27 &*\\[2mm]
        $|\Delta m^2_{31}|\: [10^{-3}{\rm eV}^2]$
        & $2.40^{+0.12}_{-0.11}$  & 2.18--2.64 & 2.07--2.75 &*\\[2mm]
        $\sin^2\theta_{12}$
        & $0.318^{+0.019}_{-0.016}$ & 0.29--0.36 & 0.27--0.38& 1/3\\[2mm]
        $\sin^2\theta_{23}$
        & $0.50^{+0.07}_{-0.06}$ & 0.39--0.63 & 0.36--0.67&1/2\\[2mm]
        $\sin^2\theta_{13}$
        & $0.013^{+0.013}_{-0.009}$  & $\leq$ 0.039 & $\leq$ 0.053 &0\\[2mm]
        \hline
\end{tabular}
\end{center}
\caption{ Best-fit values with 1$\sigma$ errors,
  and 2$\sigma$ and 3$\sigma$ intervals (1 d.o.f) for the
  three--flavor neutrino oscillation parameters from global data
  including solar, atmospheric, reactor (KamLAND and CHOOZ) and
  accelerator (K2K and MINOS) experiments in Ref. \cite{Schwetz:2008er}.}
\label{tabledata}
\end{table}

The neutrino mass matrix with the tri-bimaximal mixing of flavors
 is expressed by the sum of  simple mass matrices
in the flavor diagonal basis of the charged lepton. 
In terms of neutrino mass eigenvalues $m_1$, $m_2$ and $m_3$,
 the neutrino mass matrix is given as  
\begin{eqnarray}
M_\nu &=&U_\text{tribi}^*\begin{pmatrix}
                                         m_1 & 0 & 0 \\ 
                                         0 & m_2 & 0 \\
                                         0 & 0 & m_3
           \end{pmatrix}U_\text{tribi}^\dagger  
\nonumber\\
 &=& \frac{m_1+m_3}{2}\begin{pmatrix}
                                         1 & 0 & 0 \\ 
                                         0 & 1 & 0 \\
                                         0 & 0 & 1
           \end{pmatrix}+\frac{m_2-m_1}{3}\begin{pmatrix}
                           1 & 1 & 1 \\
                           1 & 1 & 1 \\
                           1 & 1 & 1
       \end{pmatrix}+\frac{m_1-m_3}{2}\begin{pmatrix}
                           1 & 0 & 0 \\
                           0 & 0 & 1 \\
                           0 & 1 & 0
                   \end{pmatrix}.
\label{tribimass}
\end{eqnarray}

This neutrino mass matrix can be easily realized in some non-Abelian
discrete symmetry. In the following subsection, we present simple realization
of this neutrino mass matrix,
which arise from the dimension-five non-renormalizable operators
\cite{Weinberg:1979sa}, or the see-saw mechanism 
\cite
{Minkowski:1977sc,Yanagida:1979,Gell-Mann:1979,Glashow:1980,Mohapatra:1979ia}.

\subsection{$A_4$ flavor model}

Natural models realizing the tri-bimaximal mixing  
have  been proposed  based on the non-Abelian finite group $A_4$
\cite{Ma:2001dn}-\cite{Lin:2009bw}.
The $A_4$ flavor model  considered by Alterelli et al 
\cite{Altarelli:2005yp,Altarelli:2005yx}
  realizes  the tri-bimaximal flavor mixing.
The deviation from the tri-bimaximal mixing can also be predicted.
  Actually, one of authors has  investigated
the deviation from the tri-bimaximal mixing including
 higher dimensional operators  in the effective model \cite{Honda:2008rs,Hayakawa:2009va}.

In this subsection, we present the $A_4$ flavor model with the supersymmetry
 including the right-handed neutrinos.
 In the non-Abelian finite group $A_4$, there are twelve group elements 
and four irreducible representations: 
${\bf 1}$, $\bf{1}'$, $\bf{1}''$ and ${\bf 3}$. 
The $A_4$ and $Z_3$ charge assignments of leptons,
Higgs fields and SM-singlets are
listed in Table \ref{A4table}. 
Under the $A_4$ symmetry, the chiral superfields for three families of 
the left-handed lepton doublets
$l=(l_e,l_\mu,l_\tau)$ and right handed neutrinos
 $\nu ^c=(\nu_e^c,\nu_\mu^c,\nu_\tau^c)$ are assumed to transform as
 ${\bf 3}$, 
while the right-handed ones of the charged 
lepton singlets $e^c$, $\mu^c$ and $\tau^c$ are assigned with 
${\bf 1}$, $\bf{1}'$, $\bf{1}''$, respectively. 
The third row of Table \ref{A4table} shows how each chiral multiplet transforms 
under $Z_3$, where $\omega= e^{2\pi i/3}$. 
The flavor symmetry is spontaneously broken by vacuum expectation values (VEVs) of 
two ${\bf 3}$'s, $\phi_T$, $\phi_S$, and by one singlet, $\xi$, 
which are $SU(2)_L\times U(1)_Y$ singlets. 
Their $Z_3$ charges are also shown in Table \ref{A4table}. 
Hereafter, we follow 
the convention that the chiral 
superfield and its lowest component are denoted by the same letter. 

\begin{table}[tbh]
\begin{footnotesize}
\begin{tabular}{|c|ccccc||cc||cccc|}
\hline
              &\!\!$(l_e,l_\mu,l_\tau)$\!\! & \!\!$(\nu_e^c,\nu_\mu^c,\nu_\tau^c)$\!\! & \!\!$e^c$ & $\mu^c$\!\! & \!\!$\tau^c$\!\! & $h_u$ & $h_d$ &\!\!$ \xi $\!\! &  \!\!$(\phi_{T_1},\phi_{T_2},\phi_{T_3})$\!\!  &  \!\!$(\phi_{S_1},\phi_{S_2},\phi_{S_3})$\!\! & \!\!$\Phi$\!\! \\ \hline
\!\!$A_4$ \!\!  & \!\!${\bf 3}$\!\!  & \!\!${\bf 3}$\!\!    &\!\!$\bf1$\!\! &\!\!${\bf 1}''$\!\!&\!\!${\bf 1}'$\!\!  
 &$\bf 1$ & $ \bf 1$ &\!\!$\bf 1$\!\!&\!\! $\bf 3$\!\! &\!\! $\bf 3$\!\! &\!\!$\bf 1$\!\! \\
\!\!$Z_3$  \!\!    &\!\!$\omega$ \!\! &\!\!$\omega^2$ \!\!   &\!\!$\omega^2$ \!\!&\!\!$\omega^2$\!\!&
\!\!$\omega^2$\!\!  & $1$  & $1$ &\!\!$\omega^2$\!\!&\!\! $1$ \!\!&\!\! $\omega^2$\!\!  &\!\!$1$\!\! \\
\!\!$U(1)_{FN}$  \!\!    &\!\!0\!\!  &\!\!0\!\!    &\!\!$2q$\!\! &\!\!$q$\!\! & \!\!0\!\!  & 0 & 0 &\!\!0\!\!&\!\! 0\!\! & \!\!0\!\!
&\!\!$-1$\!\! \\
\hline
\end{tabular}
\end{footnotesize}
\caption{ $A_4$, $Z_3$ and $U(1)_{FN}$ charges}
\label{A4table}
\end{table}


Allowed terms in the superpotential including charged leptons are 
written by
\begin{eqnarray}
w_l
&=&
y_0^e 
  e^c  l\phi_Th_d\frac{\Phi^{2q}}{\Lambda'^{2q}}\frac{1}{\Lambda}
+y_0^\mu
  \mu^c  l\phi_Th_d\frac{\Phi^{q}}{\Lambda'^{q}}\frac{1}{\Lambda}
+y_0^\tau 
  \tau^c  l\phi_Th_d\frac{1}{\Lambda}
\nonumber\\&&
+y_1^e
 e^c l\phi_T\phi_Th_d\frac{\Phi^{2q}}{\Lambda'^{2q}}\frac{1}{\Lambda}
+y_1^\mu
 \mu^c l\phi_T\phi_Th_d\frac{\Phi^{q}}{\Lambda'^{q}}\frac{1}{\Lambda}
 \nonumber\\&&
+y_1^\tau
 \tau^c l\phi_T\phi_Th_d\frac{1}{\Lambda}. 
\label{charged}
\end{eqnarray}
In our notation, all $y$ with some subscripts denote Yukawa couplings 
of order $1$ and $\Lambda$ denotes a cut off scale of the $A_4$ symmetry. 
In order to obtain the natural hierarchy among lepton masses $m_e$, $m_\mu$ 
and $m_\tau$, the Froggatt-Nielsen mechanism  \cite{Froggatt:1978nt}
 is introduced 
as an additional  $U(1)_{FN}$ flavor symmetry 
under which only the right-handed lepton sector is charged. 
$\Lambda'$ is  a cut off scale of the $U(1)_{FN}$ symmetry and
$\Phi$ denotes the Froggatt-Nielsen flavon in Table \ref{A4table}.
The $U(1)_{FN}$ charge values are 
taken as $2q$, $q$ and 0 for $e^c$, $\mu^c$ and $\tau^c$, respectively. 
By assuming that the flavon, 
carrying a negative unit charge of $U(1)_{FN}$, acquires a VEV 
$\left<\Phi\right>/\Lambda'\equiv\lambda \ll 1$, the following 
mass ratio is  realized through the Froggatt-Nielsen charges, 
\begin{eqnarray}
 m_e :  m_\mu:  m_\tau = \lambda^{2q} : \lambda^q :1.
\end{eqnarray}
If we take  $q = 2$,  the value $\lambda\sim 0.2$ is required to be consistent with
the observed charged lepton  mass hierarchy. 
The $U(1)_{FN}$ charges are listed  in the fourth row of Table \ref{A4table}.


The superpotential associated with 
 the Dirac neutrino mass is given as
\begin{eqnarray}
w_D
&=&
y_0^D \nu^c lh_u 
+y_1^D \nu^c lh_u\phi_T\frac{1}{\Lambda},
\label{dirac}
\end{eqnarray}
and for the right-handed Majorana sector, the superpotential  is given as
\begin{equation}
w_N
=
y_0^N \nu^c \nu^c \phi_S
+y_1^N \nu^c \nu^c \xi
+y_2^N \nu^c \nu^c \phi_T\xi\frac{1}{\Lambda}
+y_3^N \nu^c \nu^c \phi_T\phi_S\frac{1}{\Lambda},
\label{majorana}
\end{equation}
where there appear products of $A_4$ triplets 
such as ${\bf 3}\times {\bf 3}\times {\bf 3}$ and 
${\bf 3}\times {\bf 3}\times {\bf 3}\times {\bf 3}$.

The $A_4$ symmetry is spontaneously broken by  VEVs of flavons.
The tri-bimaximal mixing requires 
vacuum alignments of $A_4$ triplets $\phi_T$ and $\phi_S$ as follows:
\begin{eqnarray}
\left<(\phi_{T_1},\phi_{T_2},\phi_{T_3})\right>=v_T(1,0,0), \qquad 
\left<(\phi_{S_1},\phi_{S_2},\phi_{S_3})\right>=v_S(1,1,1).
\end{eqnarray}
These  vacuum alignments are   realized in the scalar potential 
of  the leading order \cite{Altarelli:2005yp,Altarelli:2005yx}.  

 We write  other VEVs as follows:
\begin{eqnarray}
&&\left<h_{u}\right>=v_{u},
\qquad
\left<h_{d}\right>=v_{d},
\qquad
\left<\xi\right>=u .
\end{eqnarray}
By 
inserting these VEVs in  the superpotential of the charged lepton sector 
in Eq.(\ref{charged}), 
we obtain the charged lepton mass matrix $M_E$ as
\begin{eqnarray}
M_E
=\alpha_T v_d
\begin{pmatrix}
      y_0^e\lambda^{2q} &  0   & 0 \\
    0 &   y_0^\mu\lambda^q   &  0\\
    0& 0&   y_0^\tau
\end{pmatrix} \ ,
\label{chargedlepton}
\end{eqnarray}
with 
\begin{equation}
\alpha_T =\frac{v_T}{\Lambda }.
\end{equation}
Since we have 
\begin{eqnarray}
m_e^2={y_0^e}^2\lambda^{4q} 
\alpha_T^2  v_d^2 , \qquad
m_\mu^2={y_0^\mu}^2\lambda^{2q} \alpha_T^2  v_d^2, \qquad
m_\tau^2={y_0^\tau}^2 \alpha_T^2  v_d^2,
\end{eqnarray}
we can determine $\alpha_T$ from the tau lepton mass by fixing $y_0^\tau$: 
\begin{eqnarray}
\alpha_T
=\sqrt{\frac{m_\tau^2}{{y_0^\tau}^2v_d^2}}.
\end{eqnarray}

Now, we present the Dirac neutrino mass matrix in the leading order
 as follows:
\begin{eqnarray}
\label{dirac}
M_D
&=&v_u
\begin{pmatrix}
    y_0^D  & 0 & 0 \\
  0 &   y_0^D & 0\\
  0& 0 &   y_0^D 
\end{pmatrix}.
\end{eqnarray}
On the other hand, the right-handed Majorana mass matrix is given as
\begin{eqnarray}
\label{majorana}
M_N
=2\Lambda 
\begin{pmatrix}
   \frac{2}{3}  y_0^N\alpha_S+ y_1^N\alpha_V & -\frac{1}{3}  y_0^N\alpha_S  & -\frac{1}{3} y_0^N \alpha_S \\
   -\frac{1}{3}  y_0^N\alpha_S & \frac{2}{3} y_0^N \alpha_S
& -\frac{1}{3} y_0^N\alpha_S+y_1^N\alpha_V  \\
   -\frac{1}{3} y_0^N \alpha_S& -\frac{1}{3} y_0^N \alpha_S+y_1^N\alpha_V & \frac{2}{3} y_0^N \alpha_S
\end{pmatrix}   ,
\nonumber\\
\end{eqnarray}
where
\begin{equation}
\alpha _S=\frac{v_S}{\Lambda },\qquad \alpha _V=\frac{u}{\Lambda }.
\end{equation}
By the seesaw mechanism $M_D^TM_R^{-1}M_D$, 
we get  the neutrino mass matrix $M_\nu$, which is rather complicated.
We only display leading matrix elements 
which correspond to the neutrino mass matrix in Ref. \cite{Altarelli:2005yx}:

\begin{eqnarray}
\begin{split}
M_\nu
&=\frac13
 \begin{pmatrix}A+2B   & A-B & A-B \\ 
                   A-B    & A+\frac12B+\frac32 C  &A+\frac12B-\frac32C    \\
                   A-B  &a+\frac12B-\frac32C  & A+\frac12B+\frac32C  \\
 \end{pmatrix}
\\
& = \frac{B+C}{2}\begin{pmatrix}
                                  1 & 0 & 0 \\
                                  0 & 1 & 0 \\
                                  0 & 0 & 1
                             \end{pmatrix} + \frac{A-B}{3}\begin{pmatrix}
                                                   1 & 1 & 1 \\
                                                   1 & 1 & 1 \\
                                                   1 & 1 & 1
                \end{pmatrix} + \frac{B-C}{2}\begin{pmatrix}
                                 1 & 0 & 0 \\
                                 0 & 0 & 1 \\
                                 0 & 1 & 0
                                            \end{pmatrix}\ ,
\end{split}
\end{eqnarray}
where
\begin{eqnarray}
\begin{split}
&&A=k_0({y_0^{N}}^2\alpha_S^2-{y_1^{N}}^2\alpha_V^2),
\quad
B=k_0(y_0^{N}y_1^{N}\alpha_S\alpha_V-{y_1^{N}}^2\alpha_V^2),
\nonumber \\
&&C=k_0(y_0^{N}y_1^{N}\alpha_S\alpha_V+{y_1^{N}}^2\alpha_V^2), \quad
k_0=\frac{{y_0^{D}}^2v_u^2}{({y_0^{N}}^2y_1^N\alpha_V\alpha_S^2-{y_1^{N}}^3\alpha_V^3)\Lambda}\ .
\end{split}
\end{eqnarray}
At the leading order, 
neutrino masses are given as $m_1=B$, $m_2=A$, and $m_3=C$. 
Our neutrino mass  matrix is  diagonalized 
by the tri-bimaximal mixing matrix
$U_{\rm tri}$ in Eq.(\ref{tribimass}).  

Let us estimate magnitudes of $\alpha_S$ and $\alpha_V$
to justify this model.
The mass squared  differences $\Delta m_{\rm atm}^2=\Delta m_{31}^2$ and
 $\Delta m_{\rm sol}^2=\Delta m_{12}^2$ are given as
\begin{eqnarray}
& & \Delta m_{\rm atm}^2\simeq \pm\frac{(y_0^Dv_u)^4}{\Lambda^2}
\frac{y_0^N y_1^N \alpha_S\alpha_V}
{[(y_0^N\alpha_S)^2-(y_1^N\alpha_V)^2]^2}, \nonumber \\
& & \nonumber \\
& & \Delta m_{\rm sol}^2\simeq\frac{(y_0^Nv_u)^4}{4\Lambda^2}
\frac{y_0^N\alpha_S(y_0^N\alpha_S+2y_1^N\alpha_V)}
{(y_1^N\alpha_V)^2(y_0^N\alpha_S+y_1^N\alpha_V)^2}, 
\label{dm2}
\end{eqnarray}
where the sign $+(-)$ in $\Delta m_{\rm atm}^2$ corresponds to
 the normal (inverted)  mass hierarchy.
We can obtain  $\alpha_S$ and $\alpha_V$
from these   equations.
In the case of the normal mass hierarchy, putting 
\begin{equation}
\alpha _S = k\ \alpha _V \quad (k > 0)\ ,
\label{k1}
\end{equation}
 we have 
\begin{align}
\Delta m_\text{atm}^2 \simeq 
\frac{(y_0^Dv_u)^4}{\alpha _V^2\Lambda ^2}\frac{y_0^Ny_1^Nk}{(y_0^Nk+y_1^N)^2(y_0^Nk-y_1^N)^2}, \qquad
\Delta m_\text{sol}^2 \simeq 
\frac{(y_0^Dv_u)^4}{4\alpha _V^2\Lambda ^2}\frac{y_0^Nk(y_0^Nk+2y_1^N)}
{{y_1^N}^2(y_0^Nk+y_1^N)^2}.
\end{align}
The ratio of $\Delta m_\text{atm}^2$ and $\Delta m_\text{sol}^2$ 
is expressed in terms of $k$ and Yukawa couplings as
\begin{equation}
\frac{\Delta m_\text{atm}^2}{\Delta m_\text{sol}^2} 
\simeq \frac{4{(y_1^N)}^3}{(y_0^Nk+2y_1^N)(y_0^Nk-y_1^N)^2} \ .
\label{k}
\end{equation}
Yukawa couplings are expected to be order one since there is no symmetry
to suppress them. Then, by using Eq.(\ref{k}), we get
\begin{eqnarray}
k\simeq 1\pm \frac{2}{\sqrt{3}}
\sqrt{\frac{\Delta m_\text{sol}^2}{\Delta m_\text{atm}^2}}\simeq 1.2, \ \text{or}\ 0.8\ .
\label{k2}
\end{eqnarray}
Thus, $k$ is also expected to be order one,
that is to say, $\alpha_S\sim \alpha_V$, which indicates that 
 symmetry breaking scales of   $\xi$ and  
$\phi_{S}$ are the same order  in the neutrino sector.

 We also obtain a typical value:
\begin{eqnarray}
\alpha_V\sim 5.8 \times10^{-4},
\label{normalalpha}
\end{eqnarray}
 where we put   $\Lambda=2.4\times10^{18}{\rm GeV}$,
$\Delta m^2_{\rm atm}\sim 2.4\times 10^{-3}{\rm eV}^2$, 
$\Delta m^2_{\rm sol}\sim 8.0\times 10^{-5}{\rm eV}^2$ and  
$v_u=165{\rm GeV}$.
In the following  numerical calculations,
 we take  magnitudes of Yukawa couplings
 to be $0.1\sim 1$. It is found that 
$\alpha_V$ is lower than $10^{-3}$, which is much smaller than
$\alpha_T\simeq 0.032$ in the charged lepton sector.

In the case of the inverted mass hierarchy, 
the situation is different from the case of the normal one.
As seen in $\Delta m^2_{\rm atm}$ of Eq.(\ref{dm2}), 
the sign of $y_0^N$ is opposite against $y_1^N$.
Therefore, the value of $(y_0^N\alpha_S+2y_1^N\alpha_V)$ should be suppressed
compared with $(y_1^N\alpha_V)$ in order to be consistent with
the observed ratio $\Delta m^2_{\rm atm}/\Delta m^2_{\rm sol}$.
In terms of the ratio $r$
\begin{equation}
r=\frac{y_1^N\alpha_V}{y_0^N\alpha_S+2y_1^N\alpha_V} \ ,
\end{equation}
we have 
\begin{equation}
\frac{\Delta m_\text{atm}^2}{\Delta m_\text{sol}^2} 
= -r \frac{(y_1^N\alpha_V)^2}{(y_0^N \alpha_S-y_1^N\alpha_V)^2} \ .
\label{r}
\end{equation}
Therefore, we expect $r\sim -100$ for 
$y_0^N \alpha_S \sim -2y_1^N \alpha_V$. 
Then,  we  obtain a typical value:
\begin{eqnarray}
\alpha_V\sim 1.1 \times 10^{-4},
\end{eqnarray}
which is  smaller than the one in the normal hierarchical case 
of Eq.(\ref{normalalpha}).

Thus, the next leading orders ${\cal O}(\alpha_S^2)$ and 
 ${\cal O}(\alpha_V^2)$ are neglected in the lepton mass matrices, and so
the tri-bimaximal flavor mixing is justified in the model.


\subsection{$S_4$ flavor model}

The flavor symmetry is expected to explain  the mass spectrum  and
the mixing matrix of both  quarks and leptons. 
 The tri-bimaximal mixing of leptons has been   understood 
based on the non-Abelian finite group $A_4$ as presented in the previous
subsection.

On the other hand, much  attention has been devoted to the question 
whether these  models  can be  extended to
describe the observed pattern of quark masses and mixing angles,
and whether these can be made compatible with the $SU(5)$ or $SO(10)$ 
grand unified theory (GUT).
The attractive candidate is the $S_4$ symmetry, which
 has been already used for the neutrino masses and mixing 
\cite{Yamanaka:1981pa,Brown:1984dk,Ma:2005pd}.
The exact  tri-bimaximal neutrino mixing is realized
 in the $S_4$ flavor model 
\cite{Lam:2008sh,Bazzocchi:2008ej,Grimus:2009pg,
Bazzocchi:2009pv,Bazzocchi:2009da,Meloni:2009cz}.
The $S_4$ flavor models  have been discussed  for the lepton sector
\cite{Zhang:2006fv}-\cite{Daikoku:2009pi}.
Although an  attempt to unify the quark and lepton sectors was presented
towards a grand unified theory of flavor 
\cite{Hagedorn:2006ug,Cai:2006mf,Caravaglios:2006aq},
  quark mixing angles are not predicted clearly.

We present a  $S_4$ flavor model to unify
the quarks and leptons in the framework of the $SU(5)$ GUT
\cite{Ishimori:2008fi}.
The $S_4$ group has 24 distinct elements and 
has five irreducible representations 
${\bf 1},~{\bf 1}',~{\bf 2},~{\bf 3}$, and ${\bf 3}'$.
 Three generations of $\overline {\bf 5}$-plets in $SU(5)$ are assigned 
to ${\bf 3}$
of ${S_4}$ while the  first and the second generations of 
$\bf 10$-plets  in  $SU(5)$  are assigned to   $\bf 2$ of $S_4$,
and the third generation of $\bf 10$-plet is to  $\bf 1$ of $S_4$.
These  assignments of $S_4$ for $\overline{\bf 5}$ and $\bf 10$ 
lead to the  completely different structure 
of  quark and lepton mass matrices.
Right-handed neutrinos, which are $SU(5)$ gauge singlets, 
are also assigned to $\bf 2$ for the first and second generations,
and ${\bf 1}'$ for  the third generation, respectively.
These  assignments are  essential to realize the tri-bimaximal mixing
of neutrino flavors.
 Assignments of $SU(5)$, $S_4$, $Z_4$ and $U(1)_{FN}$ representations are
 summarized in Table \ref{tables4}, where $\ell$, $m$ and $n$ are positive integers
with the condition $m<n \leq 2m$. 
Taking  vacuum alignments of relevant gauge singlet scalars,
 we  predict the quark  mixing  as well as the tri-bimaximal
mixing of leptons. Especially, the Cabbibo angle is
predicted to be around $15^{\circ }$ under the relevant  vacuum alignments.

We present the  $S_4$ flavor model 
in the framework of $SU(5)$ SUSY GUT. 
The flavor symmetry of quarks and leptons is the discrete group $S_4$
 in our model. The group $S_4$ has irreducible representations 
${\bf 1}$,~${\bf 1}'$,~${\bf 2}$,~${\bf 3}$, and ${\bf 3}'$. 

\begin{table}[h]
\begin{tabular}{|c|ccccc||cccc|}
\hline
&$(T_1,T_2)$ & $T_3$ & $( F_1, F_2, F_3)$ & $(N_e^c,N_\mu ^c)$ & $N_\tau ^c$ & $H_5$ &$H_{\bar 5} $ & $H_{45}$ & $\Theta $ \\ \hline
$SU(5)$ & $10$ & $10$ & $\bar 5$ & $1$ & $1$ & $5$ & $\bar 5$ & $45$ & $1$ \\
$S_4$ & $\bf 2$ & $\bf 1$ & $\bf 3$ & $\bf 2$ & ${\bf 1}'$ & $\bf 1$ & $\bf 1$ & $\bf 1$ & $\bf 1$ \\
$Z_4$ & $-i$ & $-1$ & $i$ & $1$ & $1$ & $1$ & $1$ & $-1$ & $1$ \\
$U(1)_{FN}$ & $\ell $ & 0 & 0 & $m$ & 0 & 0 & 0 & 0 & $-1$ \\
\hline
\end{tabular}
\end{table}
\vspace{-0.5cm}
\begin{table}[h]
\begin{tabular}{|c|cccccc|}
\hline
& $(\chi _1,\chi _2)$ & $(\chi _3,\chi _4)$ & $(\chi _5,\chi _6,\chi _7)$ 
& $(\chi _8,\chi _9,\chi _{10})$ & $(\chi _{11},\chi _{12},\chi _{13})$ & $\chi _{14}$ \\ \hline
$SU(5)$ & $1$ & $1$ & $1$ & $1$ & $1$ & $1$ \\
$S_4$ & $\bf 2$ & $\bf 2$ & ${\bf 3}'$ & $\bf 3$ & $\bf 3$ & $\bf 1$ \\
$Z_4$ & $-i$ & $1$ & $-i$ & $-1$ & $i$ & $i$ \\
$U(1)_{FN}$ & $-\ell $ & $-n$ & 0 & 0 & 0 & $-\ell $ \\
\hline
\end{tabular}
\caption{Assignments of $SU(5)$, $S_4$, $Z_4$, and $U(1)_{FN}$ representations.}
\label{tables4}
\end{table}

Let us present the model of the quark and lepton  flavor 
with the $S_4$ group in $SU(5)$ GUT. 
In $SU(5)$, matter fields are unified into $10(q_1,u^c,e^c)_L$ 
and $\bar 5(d^c,l_e)_L$ dimensional representations. 
Three generations of $\bar 5$, which are denoted by $F_i$,
 are assigned to $\bf 3$ of $S_4$. 
On the other hand, the third generation of the $10$-dimensional 
representation is assigned to $\bf 1$ of $S_4$, 
so that the top quark Yukawa coupling is allowed in tree level. 
While, the first and the second generations are assigned to $\bf 2$ of $S_4$. 
These $10$-dimensional representations are denoted by 
$T_3$ and $(T_1,T_2)$, respectively.
Right-handed neutrinos, which are $SU(5)$ gauge singlets,
are also assigned to ${\bf 1}'$ and  $\bf 2$ for $N^c_\tau$ and  
$(N^c_e,N^c_\mu)$, respectively.

We introduce new scalars 
$\chi_i$ in addition to the $5$-dimensional, 
$\bar 5$-dimensional and $45$-dimensional Higgs of the $SU(5)$, $H_5$, 
$H_{\bar 5} $, and $H_{45} $  which are  assigned to $\bf 1$ of $S_4$. 
These new scalars are supposed to be $SU(5)$ gauge singlets. 
The  $(\chi_1,\chi_2)$ and $(\chi_3,\chi_4)$ scalars 
are assigned to $\bf 2$, $(\chi _5,\chi _6,\chi _7)$ are assigned to ${\bf 3}'$, 
$(\chi _8,\chi _9,\chi _{10})$ and $(\chi _{11},\chi _{12},\chi _{13})$ 
are $\bf 3$, and $\chi_{14}$ is assigned to $\bf 1$ of the $S_4$ representations, respectively. 
 In the leading order, 
 $(\chi _3,\chi _4)$ are  
coupled with the right-handed Majorana neutrino sector, 
$(\chi _5,\chi _6,\chi _7)$ are coupled with the Dirac neutrino sector, 
$(\chi _8,\chi _9,\chi _{10})$ and $(\chi _{11},\chi _{12},\chi _{13})$ 
are coupled with the charged lepton and down-type quark sectors, respectively.
 In the next-leading order, 
$(\chi_1,\chi_2)$ scalars 
are coupled with the  up-type  quark sector, 
and the $S_4$ singlet $\chi_{14}$ contributes 
 to the charged lepton and down-type quark sectors,
 and then the mass ratio of the electron  and the down quark is reproduced
properly.
We also add $Z_4$ symmetry in order to obtain relevant couplings.
In order to obtain the natural hierarchy among quark and lepton masses,
 the Froggatt-Nielsen mechanism \cite{Froggatt:1978nt}
  is introduced as an additional 
$U(1)_{FN}$ flavor symmetry. $\Theta$ denotes the Froggatt-Nielsen flavon.
The particle assignments of $SU(5)$, $S_4$ and $Z_4$ and $U(1)_{FN}$
 are summarized Table {\ref{tables4}}. 
The  $U(1)_{FN}$ charges $\ell$, $m$ and $n$
 will be determined phenomenologically.


 We can now write down  the superpotential 
respecting  $S_4$, $Z_4$  and $U(1)_{FN}$
symmetries
 in terms of the $S_4$ cutoff scale $\Lambda$,  and
the $U(1)_{FN}$ cutoff scale  $\overline \Lambda$.
The $SU(5)$ invariant superpotential 
of the Yukawa  sector up to the linear terms of $\chi_i$ is given as
\begin{align}
w_\text{$SU(5)$} &= y_1^u(T_1,T_2)\otimes T_3\otimes (\chi _1,\chi _2)\otimes H_5/\Lambda + y_2^uT_3\otimes T_3\otimes H_5 \nonumber \\
&\ + y_1^N(N_e^c,N_\mu ^c)\otimes (N_e^c,N_\mu ^c)\otimes \Theta ^{2m}/\bar \Lambda ^{2m-1} \nonumber \\
&\ + y_2^N(N_e^c,N_\mu ^c)\otimes (N_e^c,N_\mu ^c)\otimes (\chi _3,\chi _4)\otimes \Theta ^{2m-n}/\bar \Lambda ^{2m-n} + MN_\tau ^c\otimes N_\tau ^c \nonumber \\
&\ + y_1^D(N_e^c,N_\mu ^c)\otimes (F_1,F_2,F_3)\otimes (\chi _5,\chi _6,\chi _7)\otimes H_5\otimes \Theta ^m/(\Lambda \bar \Lambda ^m) \nonumber \\
&\ + y_2^DN_\tau ^c\otimes (F_1,F_2,F_3)\otimes (\chi _5,\chi _6,\chi _7)\otimes H_5/\Lambda \nonumber \\
&\ + y_1(F_1,F_2,F_3)\otimes (T_1,T_2)\otimes (\chi _8,\chi _9,\chi _{10})\otimes H_{45}\otimes \Theta ^{\ell }/(\Lambda \bar \Lambda ^{\ell }) \nonumber \\
&\ + y_2(F_1,F_2,F_3)\otimes T_3\otimes (\chi _{11},\chi _{12},\chi _{13})\otimes H_{\bar 5}/\Lambda ,
\end{align}
where $y_1^u$, $y_2^u$, $y_1^N$, $y_2^N$, $y_1^D$, $y_2^D$, 
$y_1$, and $y_2$ are Yukawa couplings.
The  $U(1)_{FN}$ charges $\ell$, $m$, and $n$ are integers, and satisfy the conditions 
$m-n<0,\ 2m-n\geq 0$. In our numerical study, we take $\ell =m=1$ and $n=2$. 
Then, some couplings are forbidden in the   superpotential. 
We discuss the feature of the quark and lepton mass matrices and flavor mixing
 based on this superpotential.
However, we will take into account  the  next leading 
 couplings as to  $\chi_i$  in the
 numerical study of the flavor mixing and $CP$ violation.


We begin to discuss the  lepton sector of 
the superpotential $w_{SU(5)}^{(0)}$.
 Denoting Higgs doublets as $h_u$ and $h_d$,
 the superpotential of the Yukawa sector respecting 
the $S_4 \times Z_4 \times U(1)_{FN}$ symmetry  
is given for charged leptons as
\begin{align}
w_l &=\ -3y_1\left [\frac{e^c}{\sqrt2}(l_\mu \chi _9-l_\tau \chi _{10})+\frac{\mu ^c}{\sqrt 6}(-2l_e \chi _8+l_\mu \chi _9+l_\tau \chi _{10})\right ] 
h_{45}\Theta ^{\ell }/(\Lambda \bar \Lambda ^{\ell })\nonumber \\
&\ +y_2\tau ^c (l_e \chi _{11}+l_\mu \chi _{12}+l_\tau\chi _{13})h_d/\Lambda .
\end{align}
For  right-handed Majorana neutrinos,  the superpotential is given as
\begin{align}
w_N &= y_1^N(N_e^cN_e^c+N_\mu ^cN_\mu ^c)\Theta ^{2m}/\bar \Lambda ^{2m-1}
\nonumber \\
&\ +y_2^N\left[(N_e^cN_\mu ^c+N_\mu ^cN_e^c)\chi _3+(N_e^cN_e^c-N_\mu ^cN_\mu ^c)\chi _4\right ]\Theta ^{2m-n}/\bar \Lambda ^{2m-n} + MN_\tau ^cN_\tau ^c ,
\end{align}
and for Dirac neutrino Yukawa couplings,  the superpotential is
\begin{align}
w_D &= y_1^D\left [\frac{N_e^c}{\sqrt 6}(2l_e \chi _5 -l_\mu \chi _6-l_\tau \chi _7) + \frac{N_\mu ^c}{\sqrt2}(l_\mu \chi _6-l_\tau \chi _7)\right ]
 h_u\Theta ^m/(\Lambda \bar \Lambda ^m) \nonumber \\
&\ +y_2^DN_\tau ^c(l_e\chi _5+l_\mu \chi _6+l_\tau \chi _7)h_u/\Lambda .
\end{align}
Higgs doublets $h_u,h_d$ and gauge singlet scalars $\Theta $ and $\chi _i$, 
are assumed to develop their VEVs as follows:
\begin{align}
&\langle h_u\rangle =v_u,
\quad
\langle h_d\rangle =v_d,
\quad
\langle h_{45}\rangle =v_{45},
\quad
\langle \Theta \rangle =\theta , \nonumber \\
&\langle (\chi _3,\chi _4)\rangle =(u_3,u_4), 
\quad 
\langle (\chi _5,\chi _6,\chi _7)\rangle =(u_5,u_6,u_7), \nonumber \\ 
&\langle (\chi _8,\chi _9,\chi _{10})\rangle =(u_8,u_9,u_{10}),
\quad 
\langle (\chi _{11},\chi _{12},\chi _{13})\rangle =(u_{11},u_{12},u_{13}),
\label{alignment1}
\end{align}
which are supposed to be real.
Then, we obtain the mass matrix for charged leptons as
\begin{equation}
M_l = -3y_1\lambda ^\ell v_{45}\begin{pmatrix}
                       0 & \alpha _9/\sqrt 2 & -\alpha _{10}/\sqrt 2 \\
 -2\alpha _8/\sqrt 6 & \alpha _9/\sqrt 6 & \alpha _{10}/\sqrt 6   \\
                                   0 & 0 & 0 
                                \end{pmatrix}
+y_2v_d\begin{pmatrix}
               0 & 0 & 0 \\
               0 & 0 & 0 \\
               \alpha _{11} & \alpha _{12} & \alpha _{13}
          \end{pmatrix}, 
\label{charged}
\end{equation}
while the right-handed Majorana neutrino mass matrix is given as
\begin{equation}
M_N = \begin{pmatrix}
               \lambda ^{2m-n}(y_1^N\lambda ^n\bar \Lambda +y_2^N\alpha _4\Lambda ) & y_2^N\lambda ^{2m-n}\alpha _3\Lambda & 0 \\
               y_2^N\lambda ^{2m-n}\alpha _3\Lambda & \lambda ^{2m-n}(y_1^N\lambda ^n\bar \Lambda -y_2^N\alpha _4\Lambda ) & 0 \\
               0 & 0 & M
         \end{pmatrix}.
\label{majorana}
\end{equation}
Because of the condition $m-n<0$, 
 $(1,3)$, $(2,3)$, $(3,1)$ and $(3,3)$ elements of 
the right-handed Majorana neutrino mass matrix vanish. 
 These are so called SUSY zeros.
The Dirac mass matrix of neutrinos is
\begin{equation}
M_D = y_1^D\lambda ^mv_u\begin{pmatrix}
          2\alpha _5/\sqrt 6 & -\alpha _6/\sqrt 6 & -\alpha _7/\sqrt 6 \\
                            0 & \alpha _6/\sqrt 2 & -\alpha _7/\sqrt 2 \\
                            0 & 0 & 0 \end{pmatrix}+y_2^Dv_u\begin{pmatrix}
                                                               0 & 0 & 0 \\
                                                               0 & 0 & 0 \\
                                  \alpha _5 & \alpha _6 & \alpha _7
                                                             \end{pmatrix},
\label{dirac}
\end{equation}
where we denote 
$\alpha_i \equiv  u_i/\Lambda$ and $\lambda \equiv \theta /\bar \Lambda $.

In order to get the left-handed mixing of  charged leptons,
we investigate $M_l^\dagger M_l$.
If we can take   vacuum alignment 
$(u_8, u_9, u_{10})=(0, u_9, 0)$ and $(u_{11}, u_{12},  u_{13})=(0,0, u_{13})$, that is
 $\alpha _8=\alpha_{10}=\alpha_{11}=\alpha_{12}=0$,
we obtain 
\begin{equation}
M_l = \begin{pmatrix}
                                   0 & -3y_1\lambda ^\ell \alpha _9v_{45}/\sqrt 2 & 0 \\
                                   0 & -3y_1\lambda ^\ell \alpha _9v_{45}/\sqrt 6 & 0 \\
                                   0 & 0 & y_2\alpha _{13}v_d
                                \end{pmatrix},
\end{equation}
then $M_l^\dagger M_l$ is as follows:
\begin{equation}
M_l ^\dagger M_l = v_d^2
\begin{pmatrix}
0 & 0 & 0 \\
0 & 6|\bar y_1\lambda ^\ell \alpha _9|^2 & 0 \\
0 & 0 & |y_2|^2\alpha _{13}^2
\end{pmatrix},
\end{equation}
where we replace $y_1v_{45}$ with $\bar y_1v_d$. We find
 $\theta ^l_{12}=\theta^l_{13}=\theta^l_{23}=0$, where $\theta^l_{ij}$ denote
 left-handed mixing angles to diagonalize the charged lepton mass matrix.
Then, charged lepton masses are 
\begin{align}
m_e^2 = 0 \ ,
\quad
m_\mu ^2 =6|\bar y_1\lambda ^\ell \alpha _9|^2v_d^2\ ,
\quad 
m_\tau ^2=|y_2|^2\alpha _{13}^2v_d^2\ .
\label{chargemass}
\end{align}
It is remarkable that the electron mass vanishes.
We will discuss the electron mass in the next leading order.


 Taking vacuum alignment $(u_3, u_4)=(0, u_4)$
and   $(u_5, u_6, u_7)=(u_5, u_5, u_5)$ in Eq.(\ref{majorana}),
the right-handed Majorana mass matrix of neutrinos turns to 
\begin{equation}
M_N = \begin{pmatrix}
               \lambda ^{2m-n}(y_1^N\lambda ^n\bar \Lambda +y_2^N\alpha _4\Lambda ) & 0 & 0 \\
               0 & \lambda ^{2m-n}(y_1^N\lambda ^n\bar \Lambda -y_2^N\alpha _4\Lambda ) & 0 \\
               0 & 0 & M
         \end{pmatrix},
\end{equation}
and the Dirac mass matrix of neutrinos turns to 
\begin{equation}
M_D = y_1^D\lambda ^mv_u\begin{pmatrix}
        2\alpha _5/\sqrt 6 & -\alpha _5/\sqrt 6 & -\alpha _5/\sqrt 6 \\
           0 & \alpha _5/\sqrt 2 & -\alpha _5/\sqrt 2 \\
                                        0 & 0 & 0 
                                     \end{pmatrix}+y_2^Dv_u\begin{pmatrix}
                                     0 & 0 & 0 \\
                                     0 & 0 & 0 \\
                         \alpha _5 & \alpha _5 & \alpha _5
                                                \end{pmatrix}.
\end{equation}

By using the seesaw mechanism $M_\nu = M_D^TM_N^{-1}M_D$, 
the left-handed Majorana neutrino mass matrix is  written as
\begin{equation}
M_\nu = \begin{pmatrix}
                 a+\frac{2}{3}b & a-\frac{1}{3}b & a-\frac{1}{3}b \\
 a-\frac{1}{3}b& a+\frac{1}{6}b+\frac{1}{2}c & a+\frac{1}{6}b-\frac{1}{2}c \\
 a-\frac{1}{3}b & a+\frac{1}{6}b-\frac{1}{2}c & a+\frac{1}{6}b+\frac{1}{2}c
            \end{pmatrix},
\label{neutrino}
\end{equation}
where
\begin{equation}
a = \frac{(y_2^D\alpha _5v_u)^2}{M},\qquad 
b = \frac{(y_1^D\alpha _5v_u\lambda ^m)^2}{\lambda ^{2m-n}(y_1^N\lambda ^n\bar \Lambda +y_2^N\alpha _4\Lambda )},\qquad 
c = \frac{(y_1^D\alpha _5v_u\lambda ^m)^2}{\lambda ^{2m-n}(y_1^N\lambda ^n\bar \Lambda -y_2^N\alpha _4\Lambda )}.
\label{neutrinomassparameter}
\end{equation}
The neutrino mass matrix is decomposed as
\begin{equation}
M_\nu = \frac{b+c}{2}\begin{pmatrix}
                        1 & 0 & 0 \\
                        0 & 1 & 0 \\
                        0 & 0 & 1
                      \end{pmatrix} + \frac{3a-b}{3}\begin{pmatrix}
                                                      1 & 1 & 1 \\
                                                      1 & 1 & 1 \\
                                                      1 & 1 & 1
             \end{pmatrix} + \frac{b-c}{2}\begin{pmatrix}
     1 & 0 & 0 \\
      0 & 0 & 1 \\
     0 & 1 & 0
     \end{pmatrix},
\end{equation}
which gives the tri-bimaximal mixing matrix 
$U_\text{tri-bi}$ and mass eigenvalues  as follows:
\begin{equation}
U_\text{tri-bi} = \begin{pmatrix}
               \frac{2}{\sqrt{6}} &  \frac{1}{\sqrt{3}} & 0 \\
     -\frac{1}{\sqrt{6}} & \frac{1}{\sqrt{3}} &  -\frac{1}{\sqrt{2}} \\
      -\frac{1}{\sqrt{6}} &  \frac{1}{\sqrt{3}} &   \frac{1}{\sqrt{2}}
         \end{pmatrix},
\qquad m_1 = b\ ,\qquad m_2 = 3a\ ,\qquad m_3 = c\ .
\end{equation}


 The next leading terms of the superpotential are important
to predict the deviation from the tri-bimaximal mixing of leptons,
especially, $U_{e3}$.  
 The relevant superpotential in the charged lepton sector
 is given at the next leading order  as 
\begin{align}
\Delta w&=y_{\Delta _a}(T_1,T_2)\otimes (F_1,F_2,F_3)\otimes (\chi _1,\chi _2)\otimes (\chi _{11},\chi _{12},\chi _{13})\otimes H_{\bar 5}/\Lambda ^2 \nonumber \\
&\ +y_{\Delta _b}(T_1,T_2)\otimes (F_1,F_2,F_3)\otimes (\chi _5,\chi _6,\chi _7)\otimes \chi _{14}\otimes H_{\bar 5}/\Lambda ^2 \nonumber \\
&\ +y_{\Delta _c}(T_1,T_2)\otimes (F_1,F_2,F_3)\otimes (\chi _1,\chi _2)\otimes (\chi _5,\chi _6,\chi _7)\otimes H_{45}/\Lambda ^2 \nonumber \\
&\ +y_{\Delta _d}(T_1,T_2)\otimes (F_1,F_2,F_3)\otimes (\chi _{11},\chi _{12},\chi _{13})\otimes \chi _{14}\otimes H_{45}/\Lambda ^2 \nonumber \\
&\ +y_{\Delta _e}T_3\otimes (F_1,F_2,F_3)\otimes (\chi _5,\chi _6,\chi _7)\otimes (\chi _8,\chi _9,\chi _{10})\otimes H_{\bar 5}\otimes /\Lambda ^2 \nonumber \\
&\ +y_{\Delta _f}T_3\otimes (F_1,F_2,F_3)\otimes (\chi _8,\chi _9,\chi _{10})\otimes (\chi _{11},\chi _{12},\chi _{13})\otimes H_{45}\otimes /\Lambda ^2\ .
\label{nextsusy}
\end{align} 
In order to estimate the effect of this superpotential on the lepton flavor
mixing, we calculate the next leading terms of 
the  charged lepton mass matrix elements $\epsilon_{ij}$,
which   are given as 
\begin{align}
\epsilon_{11}&=y_{\Delta _b}\alpha _5\alpha _{14}v_d
   -3\bar y_{\Delta _{c2}}\alpha _1\alpha _5v_d , \nonumber \\
\epsilon_{12}&=
   -\frac{1}{2}y_{\Delta _b}\alpha _5\alpha _{14} v_d   
   +3\left [  \frac{\sqrt 3}{4}(\sqrt 3-1)\bar y_{\Delta _{c1}}-\frac{1}{4}(\sqrt 3+1)\bar y_{\Delta _{c2}}  
   \right ]\alpha _1\alpha _5v_d , \nonumber \\
\epsilon_{13}&=\left [\left \{ \frac{\sqrt 3}{4}(\sqrt 3-1)y_{\Delta _{a1}}+\frac{1}{4}(\sqrt 3+1)y_{\Delta _{a2}}\right \} \alpha _1\alpha _{13} 
   -\frac{1}{2}y_{\Delta _b}\alpha _5\alpha _{14}\right ]v_d , \nonumber \\
   &\ -3\left [\left \{ -\frac{\sqrt 3}{4}(\sqrt 3+1)\bar y_{\Delta _{c1}}-\frac{1}{4}(\sqrt 3-1)\bar y_{\Delta _{c2}}\right \}\alpha _1\alpha _5
   +\frac{\sqrt 3}{2}\bar y_{\Delta _d}\alpha _{13}\alpha _{14}\right ]v_d , \nonumber \\
\epsilon_{21}&=-3 \bar y_{\Delta _{c1}}\alpha _1\alpha _5v_d , \nonumber \\
\epsilon_{22}&=\frac{\sqrt 3}{2}y_{\Delta _b}\alpha _5\alpha _{14} v_d
   +3\left [ \frac{1}{4}(\sqrt 3-1)\bar y_{\Delta _{c1}}
   +\frac{\sqrt 3}{4}(\sqrt 3+1)\bar y_{\Delta _{c2}}  
   \right ]\alpha _1\alpha _5 v_d , \nonumber \\
\epsilon_{23}&=\left [\left \{ -\frac{1}{4}(\sqrt 3-1)y_{\Delta _{a1}}+\frac{\sqrt 3}{4}(\sqrt 3+1)
    y_{\Delta _{a2}}\right \} \alpha _1\alpha _{13}
   -\frac{\sqrt 3}{2}y_{\Delta _b}\alpha _5\alpha _{14}\right ]v_d , \nonumber \\
   &\ -3\left [\left \{ \frac{1}{4}(\sqrt 3+1)\bar y_{\Delta _{c1}}
   -\frac{\sqrt 3}{4}(\sqrt 3-1)\bar y_{\Delta _{c2}}\right \} \alpha _1\alpha _5
   -\frac{1}{2}\bar y_{\Delta _d}\alpha _{13}\alpha _{14}\right ]v_d , \nonumber \\
\epsilon_{31}&=-y_{\Delta _e}\alpha _5\alpha _9v_d-3\bar y_{\Delta _f}\alpha _9\alpha _{13}v_d , \nonumber \\
\epsilon_{32}&=0, \nonumber \\
\epsilon_{33}&=y_{\Delta _e}\alpha _5\alpha _9v_d .
\label{correction}
\end{align}
Magnitudes of $\epsilon_{ij}$'s are  of ${\cal O}(\tilde\alpha^2)$,
where $\tilde\alpha$ is a linear combination of $\alpha_i$'s.
The charged lepton mass matrix  is written in terms of $\epsilon_{ij}$ as  
\begin{equation}
M_l\simeq
\begin{pmatrix}
\epsilon _{11} & \frac{\sqrt 3m_\mu}{ 2}+\epsilon_{12} & \epsilon _{13} \\
\epsilon _{21} & \frac{m_\mu}{2}+\epsilon_{22} & \epsilon _{23} \\
\epsilon _{31} & \epsilon _{32} & m_\tau+\epsilon_{33}
\end{pmatrix},
\label{nextleading}
\end{equation}
where $m_\mu$ and $m_\tau$ are given in  Eq.(\ref{chargemass}).
Then, $M_l^\dagger M_l$  is not diagonal due to 
next leading terms $\epsilon_{ij}$, which give the non-vanishing
electron mass.
Since we have 
$m_\mu={\cal O}(\lambda \tilde\alpha)$ and  $m_\tau={\cal O}(\tilde\alpha)$
as seen in Eq.(\ref{chargemass}),
and $\epsilon_{ij}={\cal O}(m_e)$,
the left-handed charged lepton mixing matrix is written as
\begin{equation}
U_E=
\begin{pmatrix}
1 & \mathcal{O}\left (\frac{m_e}{m_\mu}\right ) 
& \mathcal{O}\left (\frac{m_e}{m_\tau}\right ) \\
\mathcal{O}\left (\frac{m_e}{m_\mu}\right ) & 1 & 
\mathcal{O}\left (\frac{m_e}{m_\tau} \right ) \\
\mathcal{O}\left (\frac{m_e}{m_\tau} \right ) & 
\mathcal{O}\left (\frac{m_e}{m_\tau} \right ) & 1
\end{pmatrix}.
\end{equation}
Now, the lepton mixing matrix $U_\text{MNS}$ is deviated 
from the tri-bimaximal mixing as follows:
\begin{equation}
U_\text{MNS}=U_E^\dagger U_\text{tri-bi}.
\end{equation}
The lepton mixing matrix elements $U_{e3},\ U_{e2}.\ U_{\mu 3}$ are given as
\begin{equation}
U_{e3}\sim \frac{1}{\sqrt 2}\left (\mathcal{O}\left (\frac{m_e}{m_\mu}\right )\right ),\quad 
U_{e2}\sim \frac{1}{\sqrt 3}\left (1+
\mathcal{O}\left (\frac{m_e}{m_\mu}\right )\right ),\quad U_{\mu 3}\sim -\frac{1}{\sqrt 2}\left (1-\mathcal{O}\left (\frac{m_e}{m_\tau} \right )\right )\ .
\label{deviation}
\end{equation}
Thus, the deviation from the tri-bimaximal mixing is lower than 
 $\mathcal{O}\left (0.01\right )$.

The superpotential of the next leading order for Majorana neutrinos is 
\begin{align}
\Delta w_{SU(5)}^N &= y_{\Delta _1}^N(N_e^c,N_\mu ^c)\otimes (N_e^c,N_\mu ^c)\otimes (\chi _1,\chi _2)\otimes \chi _{14}/\Lambda \nonumber \\
&\ + y_{\Delta _2}^N(N_e^c,N_\mu ^c)\otimes N_\tau ^c\otimes (\chi _5,\chi _6,\chi _7)
\otimes (\chi _{11},\chi _{12},\chi _{13})\otimes \Theta /(\Lambda \bar \Lambda ) \nonumber \\
&\ + y_{\Delta _3}^N(N_e^c,N_\mu ^c)\otimes N_\tau ^c\otimes (\chi _8,\chi _9,\chi _{10})
\otimes (\chi _8,\chi _9,\chi _{10})\otimes \Theta /(\Lambda \bar \Lambda ) \nonumber \\
&\ + y_{\Delta _4}^NN_\tau ^c\otimes N_\tau ^c\otimes (\chi _8,\chi _9,\chi _{10}) \otimes (\chi _8,\chi _9,\chi _{10})/\Lambda .
\end{align}
The dominant matrix elements of the Majorana neutrinos
at the  next leading order are  given as follows:
\begin{align}
&\Delta M_N=\Lambda \times \nonumber \\
&\begin{pmatrix}
y_{\Delta _1}^N\alpha _1\alpha _{14} & y_{\Delta _1}^N\alpha _1\alpha _{14} 
& -\frac{\lambda}{\sqrt 6}y_{\Delta _2}^N \alpha _5\alpha _{13}+\frac{\lambda}{\sqrt 2}y_{\Delta _3}^N\lambda \alpha _9^2 \\
y_{\Delta _1}^N\alpha _1\alpha _{14} & -y_{\Delta _1}^N\alpha _1\alpha _{14} 
& -\frac{\lambda}{\sqrt 2}y_{\Delta _2}^N \alpha _5\alpha _{13}+\frac{\lambda}{\sqrt 6}y_{\Delta _3}^N \alpha _9^2 \\
-\frac{\lambda}{\sqrt 6}y_{\Delta _2}^N \alpha _5\alpha _{13}+\frac{\lambda}{\sqrt 2}y_{\Delta _3}^N \alpha _9^2 
& -\frac{\lambda}{\sqrt 2}y_{\Delta _2}^N \alpha _5\alpha _{13}+\frac{\lambda}{\sqrt 6}y_{\Delta _3}^N \alpha _9^2 & y_{\Delta _4}^N\alpha _9^2
\end{pmatrix}.
\end{align}
Then the $U_{e3}$ is estimated as follows:
\begin{equation}
U_{e3}\sim \frac{y_{\Delta _1}^N\alpha _1\alpha _{14}}{y_2^N\alpha _4}
\sim \mathcal{O}(\tilde \alpha )\ .
\end{equation}

We also consider the Dirac neutrino mass matrix. 
The superpotential at the next leading order for Dirac neutrino is given as
\begin{equation}
\Delta w_{SU(5)}^D =y_\Delta ^D(N_e^c,N_\mu ^c)\otimes (F_1,F_2,F_3)\otimes (\chi _8,\chi _9, \chi _{10})\otimes (\chi _{11},\chi _{12}, \chi _{13})
\otimes H_5\otimes \Theta /(\Lambda ^2\bar \Lambda )\ .
\end{equation}
The dominat matrix elements of the Dirac neutrinos 
at the next leading order are given as follows:
\begin{equation}
\Delta M_D=
\begin{pmatrix}
\ast & \ast & \ast \\
y_{\Delta }^D\lambda \alpha _9\alpha _{13}v_u & \ast & \ast \\
\ast & \ast & \ast .
\end{pmatrix}
\end{equation}
Then,  we can estimate  $U_{e3}$  as follows:
\begin{equation}
U_{e3}\sim -\frac{\sqrt 6y_{\Delta }^D\alpha _9\alpha _{13}}{3y_1^D\alpha _5}\sim \mathcal{O}(\tilde \alpha )\ .
\end{equation}
Thus, the contribution of the next leading terms on $U_{e3}$
is of order  ${\cal O}(\tilde \alpha )$ in the neutrino sector
while that is  ${\cal O}(m_e/m_\mu )$ in the charged lepton  sector.
Therefore, it is concluded that the deviation from the tri-bimaximal mixing 
mainly comes from  the neutrino sector.


Let us discuss the quark sector.
For down-type quarks,  we can write the superpotential as follows:
\begin{align}
w_d &= y_1\left [\frac{ 1}{\sqrt 2}(s^c\chi _9- b^c \chi _{10}) q_1+\frac{1}{\sqrt 6}(-2d^c \chi _8+s^c\chi _9+b^c\chi _{10})q_2\right ]
h_{45}\Theta ^{\ell }/(\Lambda \bar \Lambda ^{\ell }) \nonumber \\
&\ + y_2( d^c\chi _{11}+ s^c \chi _{12}+  b^c\chi _{13})q_3 h_d/\Lambda .
\end{align}
Since the vacuum alignment is fixed in the lepton sector 
as seen in Eq.(\ref{alignment1}), 
the down-type quark mass matrix at the leading order is given as
\begin{equation}
M_d = v_d\begin{pmatrix}
            0 & 0 & 0 \\ 
            \bar y_1\lambda ^\ell \alpha _9/\sqrt 2 & \bar y_1\lambda ^\ell \alpha _9/\sqrt 6 & 0 \\
            0 & 0 & y_2\alpha _{13}
         \end{pmatrix},
\end{equation}
where we denote $\bar y_1v_d=y_1^\prime v_{45}$.
Then, we have
\begin{equation}
M_d^\dagger M_d = v_d^2\begin{pmatrix}
                          \frac{1}{2}|\bar y_1\lambda ^\ell \alpha _9|^2 & \frac{1}{2\sqrt 3}|\bar y_1\lambda ^\ell \alpha _9|^2 & 0 \\
                          \frac{1}{2\sqrt 3}|\bar y_1\lambda ^\ell \alpha _9|^2 & \frac{1}{6}|\bar y_1\lambda ^\ell \alpha _9|^2 & 0 \\
                          0 & 0 & |y_2|^2\alpha _{13}^2
                       \end{pmatrix}.
\end{equation}
This matrix can be diagonalized  by the orthogonal matrix $U_d$ as
\begin{align}
U_d = \begin{pmatrix}
            \cos 60^\circ & \sin 60^\circ & 0 \\
            -\sin 60^\circ & \cos 60^\circ & 0 \\
            0 & 0 & 1
         \end{pmatrix}.
\label{Ud}
\end{align}
The down-type quark masses  are given as 
\begin{align}
&m_d^2=0\ ,
\quad  
m_s^2=\frac{2}{3}|\bar y_1\lambda ^\ell \alpha _9|^2v_d^2\ ,
\quad 
m_b^2\approx
 |y_2|^2\alpha _{13}^2v_d^2\ ,
\label{downmass}
\end{align}
which correspond to  ones of  charged lepton masses in Eq.(\ref{chargemass}). 
The down quark mass vanishes as well as the electron mass, 
however tiny masses appear in  the next leading order. 

The down-type quark mass matrix including the next leading order is
\begin{equation}
M_d\simeq
\begin{pmatrix}
\bar\epsilon _{11} & \bar\epsilon _{21} & \bar\epsilon _{31} \\
\frac{\sqrt 3m_s}{ 2}+\bar\epsilon_{12} & \frac{m_s}{2}+\bar\epsilon_{22} 
& \bar\epsilon _{32} \\
\bar\epsilon _{13} & \bar\epsilon _{23} & m_b+\bar\epsilon_{33}
\end{pmatrix},
\label{nextleading}
\end{equation}
where $\bar\epsilon_{ij}$'s are given 
by replacing  $\bar y_{\Delta_i}$  with  
 $-1/3\bar y_{\Delta_i}\ (i=c1,c2,d,f)$ in Eq.(\ref{correction}),
and  $m_s$ and $m_b$ are given in  Eq.(\ref{downmass}).

By rotating  $M_d^\dagger M_d\ $ with the mixing matrix $U_d$
  in Eq.(\ref{Ud}), we have
\begin{equation}
U_d^\dagger M_d^\dagger M_dU_d\simeq 
\begin{pmatrix}
|m_d|^2 & {\mathcal O}(\tilde \alpha ^2m_s) & \frac{1}{2}(\bar \epsilon _{13}^*-\sqrt 3\bar \epsilon _{23}^*)m_b \\
{\mathcal O}(\tilde \alpha ^2m_s^*) & |m_s|^2 & \frac{1}{2}(\sqrt 3\bar \epsilon _{13}^*+\bar \epsilon _{23}^*)m_b \\
\frac{1}{2}(\bar \epsilon _{13}-\sqrt 3\bar \epsilon _{23})m_b^* & \frac{1}{2}(\sqrt 3\bar \epsilon _{13}+\bar \epsilon _{23})m_b^* & |m_b|^2
\end{pmatrix}.
\label{Md22}
\end{equation}
Then we get mixing angles $\theta _{12}^d,\ \theta _{13}^d,\ \theta _{23}^d$ 
in the mass matrix of  Eq.(\ref{Md22}) as
\begin{eqnarray}
\theta _{12}^d=\mathcal{O}\left (\frac{m_d}{m_s}\right ),\qquad 
\theta _{13}^d=\mathcal{O}\left (\frac{m_d}{m_b}\right ),\qquad 
\theta _{23}^d=\mathcal{O}\left (\frac{m_d}{m_b}\right ),
\end{eqnarray}
where a $CP$ violating phase is neglected.

Let us discuss the up-type quark sector. The superpotential  respecting 
 $S_4 \times Z_4\times U(1)_{FN}$  is given as
\begin{align}
w_u=y_1^u\left [(u^c\chi _1+c^c\chi _2)q_3+t^c(q_1\chi _1+q_2\chi _2)\right ] h_u/\Lambda  +y_2^u t^c q_3h_u \ .
\end{align}
We denote  their VEVs as follows:
\begin{equation}
\langle (\chi _1,\chi _2)\rangle =(u_1,u_2)\ .
\end{equation}
Then, we obtain the mass matrix for up-type quarks as
\begin{equation}
M_u = v_u\begin{pmatrix}
0 & 0 & y_1^u\alpha _1 \\ 
0 & 0 & y_1^u\alpha _2 \\
y_1^u\alpha _1 & y_1^u\alpha _2 & y_2^u
\end{pmatrix}.
\label{Mu}
\end{equation}
 The next leading terms of the superpotential are also important
to predict the $CP$ violation in the quark sector.
The relevant superpotential 
 is given at the next leading order  as 
\begin{align}
\Delta w_u&=y_{\Delta _a}^u(T_1,T_2)\otimes (T_1,T_2)\otimes (\chi _1,\chi _2)\otimes (\chi _1,\chi _2)\otimes H_{5}/\Lambda ^2 \nonumber \\
&\ +y_{\Delta _b}^u(T_1,T_2)\otimes (T_1,T_2)\otimes \chi_{14}\otimes \chi_{14}\otimes H_{5}/\Lambda ^2 \nonumber \\
&\ +y_{\Delta _c}^uT_3 \otimes T_3\otimes (\chi _8,\chi _9,\chi _{10})\otimes (\chi _8,\chi _9,\chi _{10})\otimes H_{5}/\Lambda ^2  .
\end{align}
We have the following mass matrix, in which the next leading terms are added 
to the up-type quark mass matrix of Eq.(\ref{Mu}):
\begin{equation}
M_u=v_u
\begin{pmatrix}
2y_{\Delta _{a1}}^u\alpha _1^2+y_{\Delta _{b}}^u\alpha_{14}^2 & y_{\Delta _{a2}}^u\alpha _1^2 & y_1^u\alpha _1 \\
y_{\Delta _{a2}}^u\alpha _1^2 & 2y_{\Delta _{a1}}^u\alpha _1^2+y_{\Delta _{b}}^u\alpha_{14}^2 & y_1^u\alpha _1 \\
y_1^u\alpha _1 & y_1^u\alpha _1 & y_2^u+y_{\Delta _{c}}^u\alpha _9^2
\end{pmatrix},
\end{equation}
where we take the alignment $\alpha _1=\alpha _2$.
After rotating the mass matrix $M_u$ by  $\theta_{12}=45^\circ$, 
we get
\begin{equation}
\hat M_u \approx 
v_u\begin{pmatrix}
(2y_{\Delta _{a1}}^u-y_{\Delta _{a2}}^u)\alpha _1^2+y_{\Delta _{b}}^u\alpha_{14}^2 & 0 & 0 \\
0 & (2y_{\Delta _{a1}}^u+y_{\Delta _{a2}}^u)\alpha _1^2+y_{\Delta _{b}}^u\alpha_{14}^2 & \sqrt 2y_1^u\alpha _1 \\
0 & \sqrt 2y_1^u\alpha _1 & y_2^u
\end{pmatrix}.
\end{equation}

This  mass matrix is taken to be  real one by  removing phases.
The matrix is diagonalized by the orthogonal transformation
 as   $V_u^T \hat M_u V_\text{F}$, where
\begin{equation}
V_u\simeq 
\begin{pmatrix}
1 & 0  & 0 \\
0 &  r_t & r_c \\
0 & -r_c &  r_t
\end{pmatrix},\quad r_c=\sqrt \frac{m_c}{m_c+m_t}\ , \quad \text{and}\quad r_t=\sqrt \frac{m_t}{m_c+m_t}\ .
\end{equation}

Now we can   discuss  the CKM matrix.
Mixing matrices of up- and down-type quarks are summarized as
\begin{eqnarray}
&&U_u \simeq \begin{pmatrix}
                 \cos 45^\circ & \sin 45^\circ & 0 \\
                 -\sin 45^\circ & \cos 45^\circ & 0 \\
                 0 & 0 & 1
             \end{pmatrix}
\begin{pmatrix}
     1 & 0& 0 \\ 0& e^{-i\rho}& 0\\ 0 & 0 & 1
             \end{pmatrix}
\begin{pmatrix}
1 & 0  & 0 \\
0 &  r_t & r_c \\
0 & -r_c &  r_t
\end{pmatrix},
\nonumber\\
&&U_d \simeq \begin{pmatrix}
                         \cos 60^\circ & \sin 60^\circ & 0 \\
                        -\sin 60^\circ & \cos 60^\circ & 0 \\
                            0 & 0 & 1
                                                 \end{pmatrix}
\begin{pmatrix}
1 & \theta_{12}^d & \theta_{13}^d \\ 
-\theta _{12}^d-\theta _{13}^d\theta _{23}^d  & 1 & \theta _{23}^d \\
-\theta _{13}^d+\theta _{12}^d\theta _{23}^d  & -\theta _{23}^d-\theta _{12}^d\theta _{13}^d & 1 \\
\end{pmatrix}.
\end{eqnarray}
Therefore, the CKM matrix can be written as 
\begin{align}
V^{CKM}= U_u^\dagger U_d
\approx & 
\begin{pmatrix}     
1 & 0 & 0 \\
0 & r_t & -r_c \\
0 & r_c & r_t
\end{pmatrix}
\begin{pmatrix}
1 & 0 & 0 \\
0 & e^{i\rho} & 0 \\
0 & 0 & 1
\end{pmatrix}
\nonumber \\
&\times 
\begin{pmatrix}
\cos 15^\circ & \sin 15^\circ & 0 \\
-\sin 15^\circ & \cos 15^\circ & 0 \\
0 & 0 & 1
\end{pmatrix} 
\begin{pmatrix}
1 & \theta_{12}^d & \theta_{13}^d \\ 
-\theta _{12}^d-\theta _{13}^d\theta _{23}^d  & 1 & \theta _{23}^d \\
-\theta _{13}^d+\theta _{12}^d\theta _{23}^d  & -\theta _{23}^d-\theta _{12}^d\theta _{13}^d & 1 \\
\end{pmatrix}.
\end{align}
The relevant mixing elements are given as
\begin{equation}
\begin{split}
V_{us}
&\approx \theta _{12}^d\cos 15^\circ +\sin 15^\circ ,
\\
V_{ub}
&\approx \theta _{13}^d\cos 15^\circ +\theta _{23}^d\sin 15^\circ ,
\\
V_{cb}
&\approx -r_t\theta _{13}^de^{i\rho }\sin 15^\circ +r_t\theta _{23}^de^{i\rho }\cos 15^\circ -r_c\ ,
\\
V_{td}
&\approx -r_c\sin 15^\circ e^{i\rho }-r_c(\theta _{12}^d+\theta _{13}^d\theta _{23}^d)e^{i\rho }\cos 15^\circ +r_t(-\theta _{13}^d+\theta _{12}^d\theta _{23}^d) \ .
\end{split}
\end{equation}
We can reproduce the experimental values
 with a  parameter set
\begin{eqnarray}
\rho=123^\circ, \quad
\theta _{12}^d= -0.0340,
\quad
\theta _{13}^d=0.00626,
\quad
\theta _{23}^d= -0.00880,
\end{eqnarray}
 by putting typical masses  at the GUT scale 
 $m_u=1.04\times 10^{-3}$GeV, $m_c=302\times 10^{-3}$ GeV, $m_t=129$GeV 
\cite{Fusaoka:1998vc}.

In terms of a phase $\rho$,
we can  also estimate the magnitude of $CP$ violation measure, 
 Jarlskog invariant $J_{CP}$ \cite{Jarlskog:1985ht}, which is given as
\begin{eqnarray}
|J_{CP}|=|\text{Im}\left \{ V_{us} V_{cs}^* V_{ub} V_{cb}^*\right \} | \approx 3.06\times 10^{-5}\ .
\end{eqnarray}
 Our prediction is  consistent with the experimental values
$J_{CP}=3.05^{+0.19}_{-0.20}$.
We also show  $CP$ angles in the unitarity triangle,
  $\phi_1({\rm or}\ \beta)$,  $\phi_2({\rm or}\ \alpha)$  and 
$\phi_3({\rm or}\ \gamma)$,
\begin{eqnarray}
\phi_1=
{\rm arg}\left( -\frac{V_{cd}V_{cb}^*}{V_{td}V_{tb}^*}\right ) ,\qquad
\phi_2=
{\rm arg}\left( -\frac{V_{td}V_{tb}^*}{V_{ud}V_{ub}^*}\right ) ,\qquad
\phi_3=
{\rm arg}\left( -\frac{V_{ud}V_{ub}^*}{V_{cd}V_{ccb}^*}\right ) .
\end{eqnarray}
Putting $\rho=123^\circ$, we get  $\sin2\phi_1=0.693$, 
$\phi_2=89.4^\circ $ and $\phi_3=68.7^\circ $,
which are compared  with experimental values
$\sin 2\phi_1=0.681\pm 0.025$, $\phi_2=(88^{+6}_{-5})^\circ$
and  $\phi_3=(77^{+30}_{-32})^\circ$.


\subsection{$\Delta(54)$ flavor model}
 Certain classes of non-Abelian flavor
symmetries can be derived from superstring theories.
For example, $D_4$ and $\Delta(54)$ flavor symmetries can be obtained
in heterotic orbifold models 
\cite{Kobayashi:2004ya,Kobayashi:2006wq,Ko:2007dz}.
In addition to these flavor symmetries, the $\Delta(27)$ flavor symmetry 
can be derived from magnetized/intersecting D-brane models
\cite{Abe:2009vi,Abe:2009uz,Abe:2010ii}.

Here, we focus on the $\Delta(54)$ discrete symmetry. 
Although it includes several interesting 
aspects, few authors have considered up to now 
\cite{Lam:2008sh,Escobar:2008vc,Ishimori:2008uc,Ishimori:2009ew}. 
The first aspect is that it consists of  
two types of $Z_3$ subgroups and an $S_3$ subgroup.
The $S_3$ group is known as the minimal non-Abelian discrete symmetry, 
and the semi-direct product structure of $\Delta(54)$ 
between $Z_3$ and $S_3$  induces triplet 
irreducible representations.
That suggests that the $\Delta(54)$ symmetry could lead to 
interesting models. 
The $\Delta(54)$ group has irreducible representations 
${\bf 1}_+$, ${\bf 1}_-$, ${\bf 2}_1$, ${\bf 2}_2$, ${\bf 2}_3$, ${\bf 2}_4$, 
${\bf 3}_{1(1)}$, ${\bf 3}_{1{(2)}}$, ${\bf 3}_{2{(1)}}$,
and ${\bf 3}_{2{(2)}}$.
There are four triplets and products of 
${\bf 3}_{1{(1)}}\times {\bf 3}_{1{(2)}}$ 
and ${\bf 3}_{2{(1)}}\times {\bf 3}_{2{(2)}}$ lead to the trivial singlet. 


\begin{table}[h]
\begin{center}
\begin{tabular}{|c|ccc||c||ccc|}
\hline
              &$(l_e,l_\mu,l_\tau)$ & $(e^c,\mu^c,\tau^c)$ & 
$(N_e^c,N_\mu^c,N_\tau^c)$         &$h_{u(d)}$ &$ \chi_1 $&  $(\chi_{2},\chi_3)$&  
$(\chi_{4},\chi_5,\chi_6)$ \\ \hline
$\Delta(54)$ &${\bf 3}_{1{(1)}}$ & ${\bf 3}_{2{(2)}}$ & ${\bf 3}_{1{(2)}}$    
& ${\bf 1}_+$ & ${\bf 1}_-$  & ${\bf 2}_1$ & ${\bf 3}_{1{(2)}}$    \\
\hline
\end{tabular}
\end{center}
\caption{Assignments of $\Delta(54)$ representations}
\label{d54table1}
\end{table}

\subsubsection{Flavor model in lepton sector}
Let us present the model of the lepton flavor with the $\Delta(54)$
group. The triplet representations of the group
correspond to the three generations of leptons.
The left-handed leptons $(l_e,l_\mu,l_\tau)$,  
the right-handed charged leptons  $(e^c,\mu^c,\tau^c)$
and the right-handed neutrinos $(N_e^c,N_\mu^c,N_\tau^c)$ 
are assigned by ${\bf 3}_{1{(1)}}$,
${\bf 3}_{2{(2)}}$, and ${\bf 3}_{1{(2)}}$, respectively. 
Since the product ${\bf 3}_{1{(1)}}\times {\bf 3}_{1{(2)}}$ includes 
the trivial singlet ${\bf 1}_+$, 
only Dirac neutrino Yukawa couplings are allowed
in tree level. On the other hand, charged leptons and 
the right-handed Majorana neutrinos cannot have mass terms 
unless new scalars $\chi_i$ are introduced  in addition to the usual Higgs
doublets, $h_u$ and $h_d$. 
These new scalars are supposed to be $SU(2)$ gauge singlets.
The gauge singlets $\chi_1$, $(\chi_2, \chi_3)$ and
$(\chi_4, \chi_5, \chi_6)$ are assigned to
${\bf 1}_-$, ${\bf 2}_1$, and ${\bf 3}_{1{(2)}}$ of the $\Delta(54)$ 
representations, respectively.
The particle assignments of $\Delta(54)$ are summarized 
in Table \ref{d54table1}.
The usual Higgs doublets $h_u$ and $h_d$ are assigned to  
the trivial singlet ${\bf 1}_+$ of $\Delta(54)$.

In this setup of the particle assignment,
let us consider the superpotential of leptons at the leading order
in terms of the cut-off scale $\Lambda$, which is taken to be
the Planck scale.
For charged leptons, the superpotential of 
the Yukawa sector respecting to $\Delta(54)$ symmetry
is given as 
\begin{eqnarray}
w_l
&=&
y_1^l 
 (  e^c l_e+ \mu^c l_\mu+  \tau^c l_\tau )\chi_1 h_d/\Lambda 
\nonumber\\
&&+y_2^l \ [  
 (\omega  e^c l_e+\omega^2  \mu^c l_\mu+  \tau^c l_\tau )\chi_2 
-( e^c l_e+\omega^2  \mu^c l_\mu+\omega \tau^c l_\tau )\chi_3]
\ h_d/\Lambda. 
\end{eqnarray}
For the right-handed Majorana neutrinos we can write the superpotential
as follows:
\begin{eqnarray}
w_N
&=&y_1 ( N_e^c N_e^c\chi_4+ N_\mu^c N_\mu^c\chi_5+ N_\tau^c N_\tau^c\chi_6)
\nonumber\\&&
+y_2 \ [( N_\mu^c N_\tau^c+ N_\tau^c N_\mu^c)\chi_4
+( N_e^c N_\tau^c+ N_\tau^c N_e^c)\chi_5
+( N_e^c N_\mu^c+ N_\mu^c N_e^c)\chi_6].
\end{eqnarray}
The superpotential for the Dirac neutrinos has tree level contributions 
as
\begin{eqnarray}
w_D
&=&
y_D 
 \ (  N^c_e l_e+N^c_ \mu l_\mu+  N^c_\tau l_\tau )h_u \ .
\end{eqnarray}

We assume that the scalar fields, $h_{u,d}$ and $\chi_i$, develop 
their VEVs as follows:
\begin{eqnarray}
\left<h_u\right>=v_u, \   \left<h_d\right>=v_d,
\
\left<\chi_1\right>=u_1,
\  
\left<(\chi_2,\chi_3)\right>=(u_2,u_3),
\ 
\left<(\chi_4,\chi_5,\chi_6)\right>=(u_4,u_5,u_6).
\end{eqnarray}
Then, we obtain the diagonal mass matrix for   charged leptons
\begin{eqnarray}
M_l
 = 
y_1^lv_d 
\begin{pmatrix}\alpha_1  & 0 & 0 \\ 
           0  & \alpha_1  &  0  \\
                 0  & 0 & \alpha_1   \\
 \end{pmatrix} 
+y_2^l  v_d
\begin{pmatrix} \omega\alpha_2-\alpha_3 & 0 & 0 \\ 
               0    & \omega^2\alpha_2-\omega^2\alpha_3 &   0 \\
                 0 & 0 &  \alpha_2-\omega\alpha_3 \\
 \end{pmatrix}, 
\label{ME}
 \end{eqnarray}
while the right-handed Majorana mass matrix is given as 
\begin{eqnarray}
M_N
&=&
 {y_1  \Lambda}
\begin{pmatrix}\alpha_4  & 0 & 0 \\ 
               0    & \alpha_5  &0    \\
                 0  & 0 &\alpha_6   \\
 \end{pmatrix} 
 + {y_2 \Lambda}
\begin{pmatrix}0   & \alpha_{6} & \alpha_{5} \\ 
                   \alpha_{6}  & 0   &\alpha_{4}    \\
                   \alpha_{5}  & \alpha_{4}  & 0    \\
 \end{pmatrix},
\label{MR}
\end{eqnarray}
and the Dirac mass matrix of neutrinos is obtianed as 
\begin{eqnarray}
M_D = 
y_Dv_u
\begin{pmatrix} 1  & 0 & 0 \\ 
           0  &  1  &  0  \\
                 0  & 0 &  1   \\
 \end{pmatrix},
\label{MD}
 \end{eqnarray}
where we denote $\alpha_i=u_i/\Lambda \ (i=1-6)$.
By using the seesaw mechanism $M_\nu = M_D^T M_N^{-1} M_D$, the neutrino
mass matrix can be written as
\begin{eqnarray}
M_\nu
&=&
 \frac{y_D^2v_u^2}{ \Lambda d}
\begin{pmatrix}y_1^2\alpha_5\alpha_6-y_2^2\alpha_4^2  & -y_1y_2 
\alpha_6^2+y_2^2\alpha_4\alpha_5  & -y_1y_2 \alpha_5^2+y_2^2\alpha_4\alpha_6 \\ 
               -y_1y_2 \alpha_6^2+y_2^2\alpha_4\alpha_5   & 
y_1^2\alpha_4\alpha_6-y_2^2\alpha_5^2 & -y_1y_2 \alpha_4^2+y_2^2 \alpha_5\alpha_6    \\
              -y_1y_2 \alpha_5^2+y_2^2\alpha_4\alpha_6 & -y_1y_2 
\alpha_4^2+y_2^2 \alpha_5\alpha_6 &  y_1^2\alpha_4\alpha_5-y_2^2\alpha_6^2   \\
 \end{pmatrix}, 
\label{neumassmatrix}
 \nonumber\\
 \nonumber\\
 d&=&y_1^3\alpha_4\alpha_5\alpha_6-y_1y_2^2\alpha_4^3-
y_1y_2^2\alpha_5^3-y_1y_2^2\alpha_6^3
+2y_2^3\alpha_4\alpha_5\alpha_6.
\end{eqnarray}

   Since the charged lepton  mass matrix is diagonal one,
we can simply get the mass eigenvalues as
\begin{eqnarray}
\left(  \begin{array}{cc}
m_e  \\ 
m_\mu  \\ 
m_\tau  \\ 
\end{array} \right)
=
 v_d
\left(  \begin{array}{ccc}
1&\omega &-1  \\ 
1&\omega^2 &-\omega^2  \\ 
1&1 &-\omega  \\ 
\end{array} \right)
\left(  \begin{array}{cc}
y^\ell_1\alpha_1  \\ 
y^\ell_2\alpha _2  \\ 
y^\ell_2\alpha _3  \\ 
\end{array} \right) .
\end{eqnarray}
In order to estimate  magnitudes of $\alpha_1$,  $\alpha_2$ and  $\alpha_3$,
we  rewrite  as
\begin{eqnarray}
\left(  \begin{array}{cc}
y^\ell_1\alpha_1  \\ 
y^\ell_2\alpha _2  \\ 
y^\ell_2\alpha _3  \\ 
\end{array} \right)
=
\frac{1 }{3v_d}
\left(  \begin{array}{ccc}
1&1 & 1  \\ 
-\omega-1&\omega & 1   \\ 
-1&-\omega &\omega+1  \\ 
\end{array} \right)
\left(  \begin{array}{cc}
m_e \\ 
m_\mu  \\ 
m_\tau  \\ 
\end{array} \right) ,
\end{eqnarray}
which gives the relation of $|y^\ell_2\alpha_2| = |y^\ell_2\alpha_3|$.
Inserting the experimental values of the charged lepton masses
and $v_d\simeq 55$GeV, which is given by taking $\tan\beta=3$,
we obtain numerical results
\begin{equation}
\begin{pmatrix}
y_1^\ell \alpha _1 \\
y_2^\ell \alpha _2 \\
y_2^\ell \alpha _3 
\end{pmatrix} = \begin{pmatrix}
 1.14\times 10^{-2} \\ 1.05\times 10^{-2} e^{0.016 i\pi } \\      
 1.05\times 10^{-2} e^{0.32 i\pi }   \         
\end{pmatrix}.
\label{alpha123}
\end{equation}
Thus, it is found that  $\alpha_i(i=1,2,3)$ are of  ${\cal O}(10^{-2})$
 if the Yukawa couplings are order one.

In our model,
the lepton mixing comes from the structure of the neutrino mass matrix
of Eq.(\ref{neumassmatrix}).
 In order to reproduce the maximal mixing between
 $\nu_\mu$ and $\nu_\tau$,  we take $\alpha_5=\alpha_6$, and then 
we have
\begin{eqnarray}
\label{mass}
M_\nu 
 &=&
 \frac{y_D^2v_u^2}{ \Lambda d}
\begin{pmatrix}y_1^2\alpha_5^2-y_2^2\alpha_4^2  & -y_1y_2 
\alpha_5^2+y_2^2\alpha_4\alpha_5  & -y_1y_2 \alpha_5^2+y_2^2\alpha_4\alpha_5 \\ 
               -y_1y_2 \alpha_5^2+y_2^2\alpha_4\alpha_5   & 
y_1^2\alpha_4\alpha_5-y_2^2\alpha_5^2 & -y_1y_2 \alpha_4^2+y_2^2 \alpha_5^2    \\
               -y_1y_2 \alpha_5^2+y_2^2\alpha_4\alpha_5  & -y_1y_2 
\alpha_4^2+y_2^2 \alpha_5^2 &  y_1^2\alpha_4\alpha_5-y_2^2\alpha_5^2   \\
 \end{pmatrix}.
\label{neutri}
\end{eqnarray}
 
The tri-bimaximal mixing is realized by the condition of
$M_\nu(1,1)+M_\nu(1,2)=M_\nu(2,2)+M_\nu(2,3)$ in Eq. (\ref{neutri}), 
which turns to
\begin{eqnarray}
(y_1-y_2)(\alpha_4-\alpha_5)(y_1\alpha_5-y_2\alpha_4)=0.
\end{eqnarray}
Therefore, we have  three cases realizing the tri-bimaximal mixing
in Eq.(\ref{neutri}) as
\begin{eqnarray}
y_1=y_2, \qquad \alpha_4=\alpha_5, \qquad y_1\alpha_5=y_2\alpha_4.
\label{limit}
\label{condition}
\end{eqnarray}
Let us investigate the neutrino mass spectrum in these cases.
In general the neutrino mass matrix with the tri-bimaximal mixing is
expressed as the one  in Eq.(\ref{tribimass}).
Actually, the neutrino mass matrix of
Eq.(\ref{neutri}) is decomposed under the condition in Eq.(\ref{condition})
as follows.
In the case of $\alpha_4=\alpha_5$, the neutrino mass matrix is expressed as
\begin{eqnarray}
M_\nu 
 &=&
 \frac{y_D^2v_u ^2\alpha_4^2(y_1-y_2)}{ \Lambda d}
\left[
(y_1+2y_2)
 \begin{pmatrix}1   & 0 & 0 \\ 
                   0    & 1   &0    \\
                   0  & 0 & 1    \\
 \end{pmatrix}
-y_2
 \begin{pmatrix}1   & 1 & 1 \\ 
                   1    & 1   &1    \\
                   1  & 1  & 1    \\
 \end{pmatrix}
\right ].
\label{tri-bi1}
\end{eqnarray}
Therefore, it is found that neutrino masses are given as 
\begin{eqnarray}
\frac{m_1+m_3}{2}
&=&
 \frac{y_D^2v_u^2\alpha_4^2(y_1-y_2)}{ \Lambda d}
(y_1+2y_2),
\nonumber\\
\frac{m_2-m_1}{3}
&=&-\frac{y_D^2v_u^2 \alpha_4^2}{ \Lambda d}
(y_1-y_2) y_2,
\nonumber\\
m_1-m_3&=&0.
\end{eqnarray}
In the case of $y_1=y_2$, the mass matrix is decomposed as
\begin{eqnarray}
M_\nu 
 &=&
 \frac{y_D^2y_1^2v^2(\alpha_4-\alpha_5)}{ \Lambda d}
\left[
\alpha_5
 \begin{pmatrix}1   & 1 & 1 \\ 
                   1    & 1   &1    \\
                   1  & 1  & 1    \\
 \end{pmatrix}
-(\alpha_4+2\alpha_5)
 \begin{pmatrix}1   & 0 & 0 \\ 
                   0    & 0   &1    \\
                   0  & 1 & 0    \\
 \end{pmatrix}\right ], 
\end{eqnarray}
 and we have
\begin{eqnarray}
m_1+m_3&=&0,
\nonumber\\
\frac{m_2-m_1}{3}
&=&\frac{y_D^2 y_1^2 v_u^2(\alpha_4-\alpha_5)}{ \Lambda d}
\alpha_5,
\nonumber\\
\frac{m_1-m_3}{2}
&=&-\frac{y_D^2 y_1^2 v_u^2(\alpha_4-\alpha_5)}{ \Lambda d}
(\alpha_4+2\alpha_5).
\end{eqnarray}
In the last case of
$y_1\alpha_5=y_2\alpha_4$, we have
\begin{eqnarray}
M_\nu 
 &=&
 \frac{y_D^2v_u^2}{ \Lambda d}
y_1^2 \alpha_4\alpha_5 \left (1-\frac{\alpha_5^3}{\alpha_4^3}\right)
\left[
 \begin{pmatrix}1   & 0 & 0 \\ 
                   0    & 1   &0    \\
                   0  & 0 & 1    \\
 \end{pmatrix}
-
 \begin{pmatrix}1   & 0 & 0 \\ 
                   0   & 0   &1    \\
                   0  & 1  & 0    \\
 \end{pmatrix}
\right ]. 
\end{eqnarray}
Then, we obtain
\begin{eqnarray}
&&m_3=
 \frac{2y_D^2v_u^2}{ \Lambda d}
y_1^2 \alpha_4\alpha_5 \left (1-\frac{\alpha_5^3}{\alpha_4^3}\right),
\nonumber\\
&& m_2=m_1=0 \ .
\label{tri-bi2}
\end{eqnarray}

Thus, the tri-bimaximal mixing is not realized for
arbitrary neutrino masses $m_1$, $m_2$ and $m_3$ in our model.
In both conditions of $y_1=y_2$ and $\alpha_4=\alpha_5$,
we have $|m_1|=|m_3|$, which leads to quasi-degenerate neutrino masses
due to the condition of $\Delta m^2_{\rm atm} \gg \Delta m_{\rm sol}^2$.
Therefore, we do not discuss these cases 
because we need fine-tuning of parameters
in order to be consistent with the experimental data of the neutrino
oscillations \cite{Schwetz:2008er,Fogli:2008jx,Fogli:2009zza}.

In the case of $y_1\alpha_5=y_2\alpha_4$, the neutrino mass matrix turns to be
\begin{eqnarray}
M_\nu
=
 \frac{y_D^2y_1^2v_u^2}{ \Lambda d}
\begin{pmatrix}0  & 0  & 0 \\ 
               0   &  \alpha_4\alpha_5- \alpha_5^4/\alpha_4^2 & 
-\alpha_4\alpha_5+ \alpha_5^4/\alpha_4^2    \\
               0  & -\alpha_4\alpha_5+\alpha_5^4/\alpha_4^2 &  
\alpha_4\alpha_5- \alpha_5^4/\alpha_4^2   \\
 \end{pmatrix}.
\label{proto}
\end{eqnarray}
This neutrino matrix is a prototype 
which leads to  the tri-bimaximal mixing  
with the mass hierarchy $m_3\gg m_2\geq m_1$,
then we expect that realistic mass matrix is obtained near the condition
$y_1\alpha_5=y_2\alpha_4$. 

Let us discuss the detail of the mass matrix (\ref{mass}). 
After rotating $\theta_{23}=45^\circ$, we get
\begin{equation}
 \frac{y_D^2v_u^2}{ \Lambda d}
\begin{pmatrix}y_1^2\alpha_5^2-y_2^2\alpha_4^2  & \sqrt2(-y_1y_2 
\alpha_5^2+y_2^2\alpha_4\alpha_5) & 0 \\ 
               \sqrt2(-y_1y_2 \alpha_5^2+y_2^2\alpha_4\alpha_5)   & y_1^2 
\alpha_4\alpha_5-y_1y_2 \alpha_4^2  &  0     \\
                 0   & 0  & y_1^2 \alpha_4\alpha_5+y_1y_2 
\alpha_4^2-2y_2^2\alpha_5^2   \\
 \end{pmatrix}, 
\end{equation}
which  leads $\theta_{13}=0$ and
\begin{eqnarray}
\theta_{12}
=\frac12\arctan\frac{2\sqrt2 y_2\alpha_5}
{y_1\alpha_5+y_2\alpha_4-y_1 \alpha_4} \qquad (y_2\alpha_4\not 
=y_1\alpha_5).
\end{eqnarray}
Neutrino masses are given as
\begin{eqnarray}
m_1&=& \frac{y_D^2v_u^2}{ \Lambda d}
[y_1^2\alpha_5^2-y_2^2\alpha_4^2 -\sqrt2 (-y_1y_2 
\alpha_5^2+y_2^2\alpha_4\alpha_5)\tan\theta_{12}],
\nonumber\\
m_2&=& \frac{y_D^2v_u^2}{ \Lambda d}
[y_1^2 \alpha_4\alpha_5-y_1y_2 \alpha_4^2  + {\sqrt2}  (-y_1y_2 
\alpha_5^2+y_2^2\alpha_4\alpha_5) \tan\theta_{12}],
\nonumber\\
m_3&=&\ \frac{y_D^2v_u^2}{ \Lambda d}
[y_1^2 \alpha_4\alpha_5+y_1y_2 \alpha_4^2-2y_2^2\alpha_5^2],
\label{m3}
\end{eqnarray}
which are reconciled with the normal hierarchy of neutrino masses
in the case of $y_1\alpha_5\simeq y_2\alpha_4$.

Let us estimate magnitudes of $\alpha_i(i=4,5,6)$
by using Eq.(\ref{m3}). Suppose
$\tilde \alpha=\alpha_4\simeq \alpha_5=\alpha_6$.
If we take all Yukawa couplings to be order one,
Eq.(\ref{m3}) turns to be
$v_u^2=\Lambda \tilde\alpha m_3$ because of $d\sim \tilde\alpha^3$.
Putting $v_u\simeq 165$GeV ($\tan\beta=3$),
$m_3\simeq \sqrt{\Delta m_{\rm atm}^2}\simeq 0.05$eV,
and $\Lambda = 2.43\times 10^{18}$GeV, we obtain
$\tilde \alpha={\cal O}(10^{-4}-10^{-3})$.
Thus, values of $\alpha_i (i=4,5,6)$ are enough suppressed
to discuss perturbative series of higher mass operators.

We  show our numerical analysis of neutrino masses and mixing angles
in the normal mass hierarchy.
Neglecting  higher order corrections of mass matrices,
we obtain the allowed region of parameters and predictions of neutrino masses and mixing angles. Here, we neglect the renomarization effect of the 
neutrino mass matrix because we suppose the normal hierarchy of 
neutrino masses and take $\tan\beta = 3$.
\begin{figure}[tbh]
\begin{center}
\includegraphics[width=7 cm]{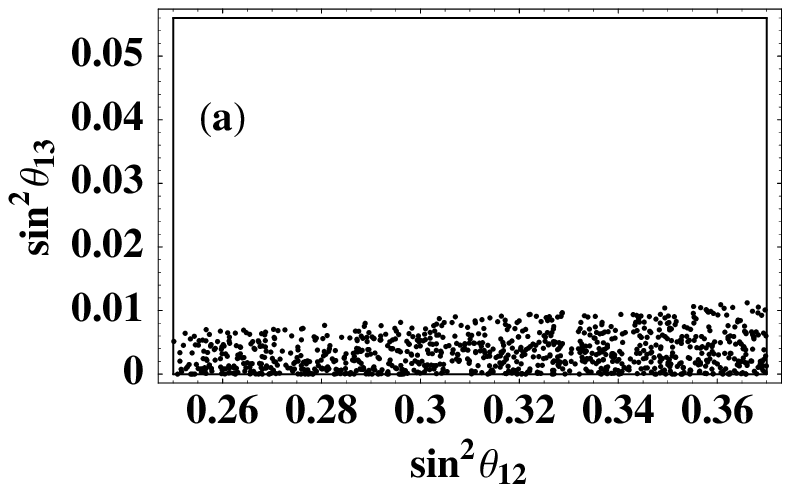}
\quad
\includegraphics[width=7 cm]{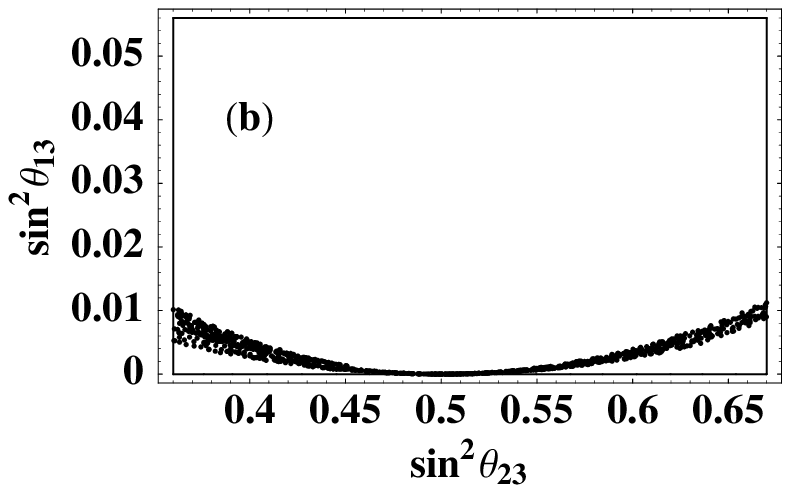}
\quad
\caption{Prediction of the upper bound of  $\sin^2\theta_{13}$
on (a) $\sin^2\theta_{12}$--$\sin^2\theta_{13}$ and
 (b) $\sin^2\theta_{23}$--$\sin^2\theta_{13}$ planes.}
\label{fig:Delta54-model}
\end{center}
\end{figure}

Input data of masses and mixing angles are taken in the  region of 
 3$\sigma$ of the experimental data 
\cite{Schwetz:2008er,Fogli:2008jx,Fogli:2009zza} in Table \ref{tabledata}
and   $\Lambda=2.43 \times 10^{18}$GeV is taken.
We fix $y_D=y_1=1$ as a convention, and vary $y_2/y_1$.
The change of $y_D$ and $y_1$  is absorbed into  the change of 
$\alpha_i(i=4,5,6)$.
If we take a smaller value of $y_1$, values of  $\alpha_i$ scale up.
On the other hand,
if we take a smaller value of $y_D$, the magnitude of $\alpha_i$ scale down.

 We can predict the deviation from the tri-bimaximal mixing.
The remarkable prediction is given in the magnitude of $\sin^2\theta_{13}$.
 In Figures \ref{fig:Delta54-model} (a) and (b), 
we plot the allowed region of mixing angles
 in planes of $\sin^2\theta_{12}$--$\sin^2\theta_{13}$ and
$\sin^2\theta_{23}$--$\sin^2\theta_{13}$, respectively.
It is found that the upper bound of $\sin^2\theta_{13}$ is $0.01$.
It is also found the strong correlation between
$\sin^2\theta_{23}$ and $\sin^2\theta_{13}$.
Unless $\theta_{23}$ is deviated from the maximal mixing considerably,
$\theta_{13}$ remains to be tiny. 
Thus, the model  reproduces the almost tri-bimaximal mixing in 
the parameter region around two vanishing neutrino masses.
Therefore, the  model is testable in the future neutrino experiments.

\subsubsection{Comments on  the $\Delta(54)$ flavor model}

  The $\Delta(54)$ symmetry can appear 
in heterotic string models on factorizable orbifolds 
including the $T^2/Z_3$ orbifold \cite{Kobayashi:2006wq}.
In these string models only singlets and 
triplets appear as fundamental modes, 
but doublets do not appear as fundamental modes.
The doublet plays an role in our model, 
and such doublet could appear, e.g. 
as composite modes of triplets.
On the other hand, doublets could appear as fundamental modes 
within the framework of magnetized/intersecting D-brane models.

As discussed in Eqs.(\ref{tri-bi1})-(\ref{tri-bi2}), the tri-bimaximal mixing is not realized for arbitrary neutrino masses  in our model.
Parameters are adapted to get neutrino masses  consistent 
with  observed values of $\Delta m^2_{\rm atm}$ and $\Delta m_{\rm sol}^2$.
Then, the deviation from  the tri-bimaximal mixing is predicted.

It is also useful to give the following comment on the $\Delta(27)$ 
flavor symmetry. Our mass matrix gives  the same result 
in the   $\Delta(27)$ flavor symmetry  \cite{Ma:2006ip}
where the type II seesaw is used.


 We can present an alternative  $\Delta(54)$ flavor model 
\cite{Ishimori:2009ew}, in which the tri-bimaximal mixing is reproduced
 for arbitrary neutrino masses $m_1$, $m_2$ and $m_3$.
The left-handed leptons $(l_e,l_\mu,l_\tau)$,  
the right-handed charged leptons  $(e^c,\mu^c,\tau^c)$
are assigned to be  ${\bf 3}_{1{(1)}}$ and ${\bf 3}_{2{(2)}}$, respectively. 
On the other hand, 
for right-handed neutrinos, $N_e^c$ is assigned to be ${\bf 1}_+$ and 
$(N_\mu^c,N_\tau^c)$ are assigned to be ${\bf 2}_2$. 
We introduce  new scalars, 
which are supposed to be $SU(2)_L$ gauge singlets 
with vanishing $U(1)_Y$ charge.
Gauge singlets $\chi_1$,  $\chi'_1$$(\chi_2, \chi_3)$, $(\chi_4, \chi_5)$,
$(\chi_6, \chi_7, \chi_8)$, and $(\chi'_6, \chi'_7, \chi'_8)$ are assigned 
to be 
${\bf 1}_-$, ${\bf 1}_1$, ${\bf 2}_1$, ${\bf 2}_2$, ${\bf 3}_{1{(2)}}$,
 and ${\bf 3}_{1{(2)}}$, 
respectively. We also introduce $Z_3$ symmetry and the non-trivial charge 
is assigned. 
The particle assignments of $\Delta(54)$ and $Z_3$ are summarized 
in Table \ref{d54table2}.
\begin{table}[h]
\begin{center}
\begin{tabular}{|c|cccc||c|}
\hline
& $(l_e,l_\mu ,l_\tau )$ & $(e^c,\mu ^c ,\tau ^c )$ & $N_e$ & $(N_\mu ,N_\tau )$ & $h_{u(d)}$ \\
\hline
$\Delta (54)$ & ${\bf 3}_{1{(1)}}$ & ${\bf 3}_{2{(2)}}$ & ${\bf 1}_+$ 
& ${\bf 2}_2$ & ${\bf 1}_+$ \\
$Z_3$ & $1$ & $\omega $ & $\omega $ & $1$ & $1$ \\
\hline
\end{tabular}

\vspace{1mm}
\begin{tabular}{|c|cccccc|}
\hline
& $\chi _1$ & $\chi _1^\prime$ & $(\chi _2,\chi_ 3)$ & $(\chi _4,\chi _5 )$ & $(\chi _6,\chi _7,\chi _8)$ & $(\chi _6^\prime ,\chi _7^\prime ,\chi _8^\prime )$ \\
\hline 
$\Delta (54)$ & ${\bf 1}_-$ & ${\bf 1}_+$ & ${\bf 2}_1$& ${\bf 2}_2$ 
& ${\bf 3}_{1{(2)}}$ & ${\bf 3}_{1{(2)}}$ \\
$Z_3$ & $\omega ^2$ & $\omega $ & $\omega ^2$ & $1$ & $\omega ^2$ & $1$ \\
\hline 
\end{tabular}
\end{center}
\caption{Assignments of $\Delta(54)$ and $Z_3$ representations, 
where $\omega $ is $e^{2\pi i/3}$.}
\label{d54table2}
\end{table}

In this particle assignment,
the charged lepton mass matrix is diagonal while
the right-handed Majorana mass matrix is given as 
\begin{eqnarray}
M_N
&=&
\begin{pmatrix}
 y_1^N\alpha'_1\Lambda  & 0 & 0 \\ 
               0    & y_2^N\alpha_4\Lambda & M  \\
                 0  & M & y_2^N \alpha_5 \Lambda  \\
 \end{pmatrix} ,
\end{eqnarray}
and the Dirac mass matrix of neutrinos is  obtained as
\begin{eqnarray}
M_D
&=&
y_1^Dv_u
\begin{pmatrix}\alpha_6  & \alpha_7 & \alpha_8 \\ 
               0    & 0  &0    \\
                 0  & 0 &0   \\
 \end{pmatrix} 
 +y_2^Dv_u
\begin{pmatrix}0  & 0 & 0 \\ 
               \omega\alpha'_8   & \omega^2\alpha'_6 &\alpha'_7   \\
                \alpha_6  & \omega^2\alpha_7 &\omega\alpha_8   \\
 \end{pmatrix} ,
\end{eqnarray}
where we denote $\alpha_i=u_i/\Lambda$.
The tri-bimaximal mixing is realized by taking following alignments,
\begin{eqnarray}
\alpha_5=\omega\alpha_4\ , \qquad \qquad \alpha_6=\alpha_7=\alpha_8 .
\label{alin}
\end{eqnarray}
By using the seesaw mechanism $M_\nu = M_D^T M_N^{-1} M_D$, the neutrino
mass matrix can be derived as follows:
\begin{eqnarray}
\begin{split}
M_\nu
=&
\begin{pmatrix}1  & 1 & 1 \\ 
               1   & \omega  &\omega^2    \\
                 1  & \omega^2 &\omega   \\
 \end{pmatrix} 
 \begin{pmatrix}y_1^D   & 0 & 0 \\ 
                   0    & y_2^D\omega   &0    \\
                   0  & 0  & y_2^D    \\
 \end{pmatrix}
\begin{pmatrix}\frac{1}{M_1}  & 0 & 0 \\ 
    0 & \frac{y^N\omega \alpha_4\Lambda}{(y^N\alpha_4\Lambda  
)^2\omega-M_2^2} 
  &\frac{-M_2 }{(y^N\alpha_4\Lambda )^2\omega-M_2^2}    \\
                   0  & \frac{-M_2}{(y^N\alpha_4\Lambda
                     )^2\omega-M_2^2}   
& \frac{y^N\alpha_4\Lambda }{(y^N\alpha_4\Lambda )^2\omega-M_2^2}     \\
 \end{pmatrix}
\\&\times
 \begin{pmatrix}y_1^D   & 0 & 0 \\ 
                   0    & y_2^D\omega   &0    \\
                   0  & 0  & y_2^D    \\
 \end{pmatrix} 
\begin{pmatrix}1  & 1 & 1 \\ 
               1   & \omega  &\omega^2    \\
                 1  & \omega^2 &\omega   \\
 \end{pmatrix}\alpha _6^2v_u^2 .
\end{split}
\end{eqnarray}
It can be rewritten as
\begin{eqnarray}
\begin{split}
M_\nu
=3c
\begin{pmatrix}
1 & 0 & 0 \\
0 & 1 & 0 \\
0 & 0 & 1
\end{pmatrix} +(a-b-c)
\begin{pmatrix}
1 & 1 & 1 \\
1 & 1 & 1 \\
1 & 1 & 1
\end{pmatrix} + 3b
\begin{pmatrix}
1 & 0 & 0 \\
0 & 0 & 1 \\
0 & 1 & 0
\end{pmatrix},
\end{split}
\end{eqnarray}
where 
\begin{equation}
\label{abc}
a=\frac{(y_1^D)^2}{M_1}\alpha _6^2v_u^2,\quad 
b=\frac{y^N(y_2^D)^2\alpha _4\Lambda }{(y^N\alpha_4\Lambda
  )^2\omega-M_2^2}
\alpha _6^2v_u^2, \quad 
c=\frac{-(y_2^D)^2\omega M2}{(y^N\alpha_4\Lambda )^2\omega-M_2^2}
\alpha _6^2v_u^2.
\end{equation}
Therefore, our neutrino mass matrix $M_\nu $ gives
 the  tri-bimaximal mixing matrix $U_\text{tribi}$ with
  mass eigenvalues  as follows:
\begin{equation}
m_1=3(b+c),\quad m_2=3a,\quad m_3=3(c-b)\ .
\end{equation}


\subsection{Comment on alternative flavor mixing}

In all the above models, the tri-bimaximal mixing of the lepton flavor
is reproduced.
However, there are other flavor mixing models,
which are  based on the golden ratio or the tri-maximal for the lepton flavor
mixing.

The golden ratio is supposed to  appear in the solar mixing angle
$\theta_{12}$.
One example is proposed as 
$\tan\theta_{12}=1/\phi$  where $\phi=(1 + \sqrt{5})/2\simeq 1.62$
 \cite{Kajiyama:2007gx}.
The rotational icosahedral group, 
which is isomorphic to $A_5$, the alternating
group of five elements, provides a natural context of
the golden ratio $\tan \theta_{12}=1/\phi$  \cite{Everett:2008et}.
The model was
constructed in  a minimal model at tree level in which the solar angle is 
related to the golden ratio,
the atmospheric angle is maximal, and the reactor angle vanishes 
to leading order. 
The approach provides a rich setting in order 
to investigate the flavor puzzle of 
the Standard Model.
Another  context of the golden ratio  $\cos \theta_{12}=\phi/2$
 \cite{Rodejohann:2008ir} is also proposed.
The dihedral group $D_{10}$ 
  derives this ratio  \cite{Adulpravitchai:2009bg}.

One can also  consider the trimaximal lepton mixing \cite{Grimus:2008tt}, 
defined by $|U_{\alpha2}|^2 = 1/3$ for $\alpha =e,  \mu$, 
and so the mixing matrix
 $U$ is given by using an arbitrary angle $\theta $ and a phase $\phi$ as
follows:
\begin{equation}
U_\text{tri} = \begin{pmatrix}
               \frac{2}{\sqrt{6}} &  \frac{1}{\sqrt{3}} & 0 \\
     -\frac{1}{\sqrt{6}} & \frac{1}{\sqrt{3}} &  -\frac{1}{\sqrt{2}} \\
      -\frac{1}{\sqrt{6}} &  \frac{1}{\sqrt{3}} &   \frac{1}{\sqrt{2}}
         \end{pmatrix}
 \begin{pmatrix}
               \cos\theta &  0& \sin\theta e^{-i\phi} \\
     0 & 1 &  0 \\
      -\sin\theta e^{i\phi} &  0  &   \cos\theta
         \end{pmatrix}.
\end{equation}
This corresponds to a two-parameter lepton flavor mixing matrix.
 We present a model
for the lepton sector in which tri-maximal mixing is enforced by softly broken discrete symmetries; one version of the model is based on the group
 $\Delta (27)$. A salient feature
of the  model is that no vacuum alignment is required.

 The bimaximal neutrino mixing \cite{Barger:1998ta}
is also studied in the context of 
the  quark-lepton complementarity of mixing angles  \cite{Minakata:2004xt}
in the  $S_4$ model \cite{Altarelli:2009gn,Merlo:2009zi}.

\subsection{Comments on other applications}

Supersymmetric extension is one of interesting candidates 
for the physics beyond the standard model.
Even if the theory is supersymmetric at high energy, 
supersymmetry must break above the weak scale.
The supersymmetry breaking induces soft supersymmetry 
breaking terms such as gaugino masses, 
sfermion masses and scalar trilinear couplings, 
i.e. the so-called A-terms.
Flavor symmetries control not only quark/lepton mass matrixes 
but also squark/slepton masses and their A-terms.
Suppose that flavor symmetries are exact.
When three families have quantum numbers different from 
each other under flavor symmetries, 
squark/slepton mass-squared matrices are diagonal.
Furthermore, when two (three) of three families correspond to 
doublets (triplets) of flavor symmetries, 
their diagonal squark/slepton masses are degenerate.
That would become an interesting prediction of 
a certain class of flavor models, which could be tested 
if the supersymmetry breaking scale is reachable by 
collider experiments.
Flavor symmetries have similar effects on A-terms.
These results are very important to suppress 
flavor changing neutral currents, which are 
constrained strongly by experiments.
However, the flavor symmetry must break 
to lead to realistic quark/lepton mass matrices.
Such breaking effects deform the above predictions.
How much results are changed depends on breaking patterns.
If masses of superpartners are of ${\cal O}(100)$ GeV, 
some models may be ruled out e.g. by experiments 
on flavor changing neutral currents.
See e.g. Refs.\cite{Babu:2002ki,Kobayashi:2003fh,Ko:2007dz,
Ishimori:2008ns,Ishimori:2008au,Ishimori:2009ew}.

What is the origin of non-Abelian flavor symmetries ?
Some of them are symmetries of geometrical solids.
Thus, its origin may be geometrical aspects of 
extra dimensions.
For example, it is found that the two-dimensional 
orbifold $T^2/Z_2$ with proper values of moduli has 
discrete symmetries such as $A_4$ and $S_4$ 
\cite{Altarelli:2006kg,Adulpravitchai:2009id}.

Superstring theory is a promising candidate 
for unified theory including gravity, 
and predicts extra six dimensions.
Superstring theory on a certain type of six-dimensional 
compact space realizes  a discrete flavor symmetry.
Such a string theory leads to stringy selection rules 
for allowed couplings among matter fields 
in four-dimensional effective field theory.
Such stringy selection rules and geometrical symmetries 
as well as broken (continuous) gauge symmetries
result in discrete flavor symmetries in 
superstring theory. 
For example, 
discrete flavor symmetries in heterotic orbifold models are  
studied in Refs. \cite{Kobayashi:2004ya,Kobayashi:2006wq,Ko:2007dz},
and $D_4$ and $\Delta(54)$ are realized.
Magnetized/intersecting D-brane models also realize 
the same flavor symmetries and other types such as 
$\Delta(27)$ \cite{Abe:2009vi,Abe:2009uz,Abe:2010ii}.
Different types of non-Abelian flavor symmetries 
may be derived in other string models.
Thus, such a study is quite important.

Alternatively, discrete flavor symmetries may 
be originated from continuous (gauge) symmetries 
\cite{deMedeirosVarzielas:2005qg,Adulpravitchai:2009kd,Frampton:2009pr}.

At any rate, the experimental data of quark/lepton 
masses and mixing angles have no symmetry.
Thus, non-Abelian flavor symmetries must be broken.
The breaking direction is important, 
because the forms of mass matrices are determined by 
along which direction the flavor symmetries break.
We need a proper breaking direction to derive 
realistic values of quark/lepton masses and 
mixing angles.

One way to fix the breaking direction is 
to analyze the potential minima of scalar fields with 
non-trivial representations of flavor symmetries.
The number of the potential minima may be finite and 
in one of them the realistic breaking would happen.
That is rather the conventional approach.

Another scenario to fix the breaking direction 
could be realized in theories with extra dimensions.
One can impose the boundary conditions of matter fermions 
\cite{Haba:2006dz} and/or flavon scalars \cite{Kobayashi:2008ih,
Seidl:2008yf,Adulpravitchai:2010na}
in bulk such that zero models for 
some components of irreducible multiplets are projected out, 
that is, the symmetry breaking.
If a proper component of a flavon multiplet remains, 
that can realize a realistic breaking direction.


\clearpage

\section{Discussions and Summary}

 We have reviewed  pedagogically non-Abelian discrete groups,
which  play  an important role in the particle physics.
We have shown  group-theoretical aspects for many concrete groups explicitly, 
such as representations and their tensor products.
We have shown  them
 explicitly for 
non-Abelian discrete groups,
$S_N$, $A_N$, $T'$, $D_N$, $Q_N$, $\Sigma(2N^2)$,  $\Delta(3N^2)$,
$T_7$, $\Sigma(3N^3)$,  and  
$\Delta(6N^2)$.
We have explained  pedagogically how to derive conjugacy classes, 
characters, representations and tensor products for 
these groups (with a finite number).

The origin of non-Abelian flavor symmetries is considered in
the geometrical aspects of extra dimensions such as the two-dimensional 
orbifold $T^2/Z_2$ with proper values of moduli.
Superstring theory on a certain type of six-dimensional 
compact space also  realizes  a discrete flavor symmetry.
On the other hand, 
discrete subgroups of $SU(3)$ would be also  interesting from 
the viewpoint of  phenomenological applications 
for the flavor physics.
Most of them have been shown for subgroups including 
doublets or triplets as the largest dimensional 
irreducible representations.

Here we comment on discrete subgroups of $SU(3)$ with 
larger dimensional irreducible representations, i.e. 
$\Sigma(168)$ and $\Sigma(n\phi)$ with 
$n=36, 72, 216, 360$, where 
$\phi$ is {\it a group homomorphism}. 
The $\Sigma(168)$
\cite{miller, Fairbairn,Luhn:2007yr,King:2009mk,Ludl:2009ft} has one
hundred and sixty eight elements, and has six irreducible
representations; one singlet, two triplets(or one complex
triplet), one sextet, one septet, and one octet. The
$\Sigma(36\phi)$ \cite{miller,Fairbairn,Ludl:2009ft} has one hundred and
eight elements, and has fourteen irreducible representations; four
singlets, eight triplets, and two quartets. The $\Sigma(72\phi)$
\cite{miller,Ludl:2009ft} has two hundreds and sixteen elements, and has
sixteen irreducible representations; four singlets, one doublet,
eight triplets, two sextets(or one complex sextet), and one
octet. The $\Sigma(216\phi)$, which is known as {\it Hessian
group} \cite{miller,Fairbairn,Ludl:2009ft}, has one thousand and eighty
elements, and has sixteen irreducible representations; three
singlets, three doublets, seven triplets, six sextets, three
octets, and two nonets. 
Readers can find  details  in ref. \cite{Ludl:2009ft}.

{}From the viewpoint of model building for 
the flavor physics,
breaking patterns of discrete groups and 
decompositions of multiplets are important.
We  have summarized    these  breaking patterns of 
the non-Abelian discrete groups in section 13. 
 
Symmetries at the tree-level can be 
broken in general  by quantum effects, i.e. anomalies.
Anomalies of continuous symmetries, in particular 
gauge symmetries, have been studied well.
Here we have reviewed  about anomalies of non-Abelian discrete symmetries
by using the path integral approach.
Also we have shown the anomaly-free conditions explicitly 
for several concrete groups.
Similarly, readers could compute anomalies for other 
non-Abelian discrete symmetries.
Those anomalies of non-Abelian discrete flavor symmetries 
would be controlled by string dynamics, when 
such flavor symmetries are originated from superstring theory.
Then, studies on such anomalies would be important 
to provide us with a hint for the question: why 
there appear three families of quarks and leptons, 
because those anomalies are relevant to the generation 
number and the flavor structure.

We hope that this review contributes on the progress for particle physics
with non-Abelian discrete groups.


\vspace{1cm}
\noindent
{\bf Acknowledgement}

The authors would like to thank H.Abe, T.~Araki,  K.S.Choi, Y.~Daikoku, 
J.~Kubo,  H.P.Nilles,  F.Ploger,  S.Raby, S.~Ramos-Sanchez, 
M.~Ratz and P.~K.~S.~Vaudrevange,
for useful discussions.
H.I, T.~K., H.~O.  and M.~T. are supported in part by the
Grant-in-Aid for Scientific Research 
of the Ministry of Education, Science, and Culture of Japan, 
No.21$\cdot$5817, No.~20540266, No.~21$\cdot$897  and  No.~21340055.
T.~K.\/ is also supported in part by
the Grant-in-Aid for the Global COE Program 
"The Next Generation of Physics, Spun from Universality 
and Emergence" from the Ministry of Education, Culture,
Sports, Science and Technology of Japan.
H.~O. is supported by the ICTP grant Project ID 30, the Egyptian Academy for
Scientific Research and Technology, and the Science and Technology Development Fund (STDF) Project ID 437.

\clearpage

\appendix


\section{Useful theorems}

In this appendix, we give simple proofs of useful theorems.
(See also e.g. Refs.~\cite{Hamermesh,Georgi:1982jb,Ludl:2009ft}.)

\vskip .5cm
{$\bullet$ \bf Lagrange's theorem}

The order $N_H$ of a subgroup of a finite group $G$ 
is a divisor of the order $N_G$ of $G$.

\vskip .2cm
Proof)

If $H=G$, the claim is trivial, $N_H=H_G$.
Thus, we consider $H \neq G$.
Let $a_1$ be an element of $G$, but be not contained in $H$.
Here, we denote all of elements in $H$ by 
$\{ e=h_0,h_1,\cdots, h_{N_H -1} \}$.
Then, we consider the products of $a_1$ and elements of $H$,
\begin{eqnarray}
a_1H = \{ a_1, a_1h_1,\cdots, a_1h_{N_h -1} \}.
\end{eqnarray}
All of $a_1 h_i$ are different from each other.
None of $a_1h_i$ are contained in $H$.
If $a_1h_i = h_j$, we could find $a_1=h_jh_i^{-1}$, 
that is, $a_1$ would be an element in $H$.
Thus, the set $a_1H$ includes the $N_H$ elements.
Next, let $a_2$ be an element of $G$, but 
be contained in neither $H$ nor $a_1H$.
If $a_2h_i = a_1h_j$, the element $a_2$ 
would be written as $a_2 =a_1 h_j h_i^{-1}$, that is, 
an element of $a_1H$.
Thus, when $a_2 \notin H$ and $a_2 \notin a_1H$,
the set $a_2H$ yields $N_H$ new elements.
We repeat this process.
Then, we can decompose
\begin{eqnarray}
G = H + a_1H + \cdots + a_{m-1} H.
\end{eqnarray}
That implies $N_G = m N_H$.   \hskip1cm \rule{5pt}{10pt}

\vskip .5cm
{$\bullet$ \bf Theorem}

For a finite group, every representation is equivalent 
to a unitary representation.

\vskip .2cm
Proof)

Every group element $a$ is represented by a matrix $D(a)$, 
which acts on the vector space.
We denote the basis of the representation vector space by 
$\{ \mbox{\boldmath $e$}_1, \cdots, \mbox{\boldmath $e$}_d \} $.
We consider two vectors, $\mbox{\boldmath $v$}$ and $\mbox{\boldmath $w$}$, 
\begin{eqnarray}
\mbox{\boldmath $v$} = \sum^d_{i=1} v_i \mbox{\boldmath $e$}_i, \qquad 
\mbox{\boldmath $w$} = \sum^d_{i=1} w_i \mbox{\boldmath $e$}_i .
\end{eqnarray}
We define the scalar product between ${\bf v}$ and ${\bf w}$ as 
\begin{eqnarray}
(\mbox{\boldmath $v$}, \mbox{\boldmath $w$}) = \sum^d_{i=1} v_i^*w_i.
\end{eqnarray}
Here, we define another scalar product by 
\begin{eqnarray}
\langle \mbox{\boldmath $v$}, \mbox{\boldmath $w$}  \rangle 
= \frac{1}{N_G} \sum_{a\in G}
(D(a)\mbox{\boldmath $v$}, D(a)\mbox{\boldmath $w$}) .
\end{eqnarray}
Then, we find
\begin{eqnarray}
\langle D(b) \mbox{\boldmath $v$}, D(b) \mbox{\boldmath $w$} \rangle &=& 
\frac{1}{N_G} \sum_{a\in G} (D(b)D(a)\mbox{\boldmath $v$}, D(b)D(a)\mbox{\boldmath $w$}) \nonumber\\
&=& \frac{1}{N_G} \sum_{a\in G} (D(ba)\mbox{\boldmath $v$}, D(ba)\mbox{\boldmath $w$}) \nonumber\\
&=& \frac{1}{N_G} \sum_{c\in G} (D(c)\mbox{\boldmath $v$}, D(c)\mbox{\boldmath $w$}) \nonumber\\
&=& \langle \mbox{\boldmath $v$}, \mbox{\boldmath $w$} \rangle.
\end{eqnarray}
That implies that $D(b)$ is unitary with respect to the scalar 
product $\langle \mbox{\boldmath $v$}, \mbox{\boldmath $w$} \rangle$.
The orthogonal bases 
$\{ \mbox{\boldmath $e$}_i\}$ and $\{ \mbox{\boldmath $e$}'_i\}$ for 
the two scalar products 
$(\mbox{\boldmath $v$}, \mbox{\boldmath $w$})$ and $\langle \mbox{\boldmath $v$}, \mbox{\boldmath $w$} \rangle$ 
can be related by the linear transformation $T$ as 
$ \mbox{\boldmath $e$}'_i =T \mbox{\boldmath $e$}_i$, i.e. 
$(\mbox{\boldmath $v$}, \mbox{\boldmath $w$})= \langle T\mbox{\boldmath $v$}, T \mbox{\boldmath $w$} \rangle$.
We define $D'(g) = T^{-1}D(g) T$.
Then, it is found that 
\begin{eqnarray}
(T^{-1}D(a)T \mbox{\boldmath $v$}, T^{-1}D(a)T \mbox{\boldmath $w$}) &=& 
\langle D(a) T \mbox{\boldmath $v$}, D(a) T \mbox{\boldmath $w$} \rangle \nonumber\\
&=& \langle T \mbox{\boldmath $v$},  T \mbox{\boldmath $w$} \rangle \nonumber\\ 
&=& (\mbox{\boldmath $v$}, \mbox{\boldmath $w$}).
\end{eqnarray}
That is, the matrix $D'(g)$ is unitary and is equivalent to 
$D(g)$.   \hskip1cm \rule{5pt}{10pt}

\vskip .5cm
{$\bullet$ \bf Schur's lemma}

(I) Let $D_1(g)$ and $D_2(g)$ be irreducible representations of $G$, 
which are inequivalent to each other.
If 
\begin{eqnarray}\label{eq:schur-lemma-1}
AD_1(g) =  D_2(g)A, \qquad  \forall g \in G,
\end{eqnarray}
the matrix $A$ should vanish, $A=0$.

(II) If 
\begin{eqnarray}\label{eq:schur-lemma-2}
D(g)A = A D(g), \qquad  \forall g \in G,
\end{eqnarray}
the matrix $A$ should be proportional to the 
identity matrix $I$, i.e. $A = \lambda I$.

\vskip .2cm
Proof) (I)
We denote the representation vector spaces for $D_1(g)$ 
and $D_2(g)$ by $V$ and $W$, respectively.
Let the map $A$ be a map $A: V \rightarrow W$ 
such that it satisfies (\ref{eq:schur-lemma-1}).
We consider the kernel of $A$, 
\begin{eqnarray}
Ker (A) = \{ \mbox{\boldmath $v$} \in V | A\mbox{\boldmath $v$} =0 \}.
\end{eqnarray}
Let $\mbox{\boldmath $v$} \in Ker (A)$.
Then, we have 
\begin{eqnarray}
AD_1(g)\mbox{\boldmath $v$} = D_2(g)A \mbox{\boldmath $v$}=0.
\end{eqnarray}
It is found that 
$D_1(g) Ker (A) \subset Ker (A)$, that is, 
$Ker (A)$ is invariant under $D_1(g)$.
Because $D_1(g)$ is irreducible, 
that implies that 
\begin{eqnarray}
 Ker (A) = \{ 0 \}, ~~~~ {\rm or} ~~~~ Ker (A) = V.
\end{eqnarray}
The later, $Ker (A) = V$, can not be realized unless $A= 0$.
Next, we consider the image 
\begin{eqnarray}
Im (A) = \{ A\mbox{\boldmath $v$} | \mbox{\boldmath $v$} \in V \}.
\end{eqnarray}
We find 
\begin{eqnarray}
D_2(g) A \mbox{\boldmath $v$} = A D_1(g) \mbox{\boldmath $v$} \in Im (A).
\end{eqnarray}
That is, $Im (A)$ is invariant under $D_2(g)$.
Because $D_2(g)$ is irreducible, that implies that 
\begin{eqnarray}
 Im (A) = \{ 0 \}, ~~~~ {\rm or} ~~~~ Im (A) = W.
\end{eqnarray}
The former, $ Im (A) = \{ 0 \}$, can not be realized 
unless $A=0$.
As a result, it is found that 
$A$ should satisfy 
\begin{eqnarray}
A=0, ~~~~ {\rm or} ~~~~ AD_1(g)A^{-1}=D_2(g).
\end{eqnarray}
The later means that the representations, 
$D_1(g)$ and $D_2(g)$, are equivalent to each other.
Therefore, $A$ should vanish, $A=0$,
if $D_1(g)$ and $D_2(g)$ are not equivalent.
\hskip1cm \rule{5pt}{10pt}

Proof)(II)
Now, we consider the case with $D(g)=D_1(g)=D_2(g)$ 
and $V=W$.
Here, $A$ is a linear operators on $V$.
The finite dimensional matrix $A$ has at least one eigenvalue, 
because the characteristic equation $det (A - \lambda I)=0$ has 
at lease one root, where $\lambda$ is an eigenvalue.
Then, Eq.~(\ref{eq:schur-lemma-2}) leads to 
\begin{eqnarray}
D(g) \left( A -\lambda I \right) = \left( A -\lambda I \right) D(g), 
\qquad  \forall g \in G.
\end{eqnarray}
Using the above proof of Schur's lemma (I) and 
$Ker (A - \lambda I) \neq \{ 0 \}$, 
we find $Ker (A - \lambda I) = V$, that is, 
$A - \lambda I =0$. \hskip1cm \rule{5pt}{10pt}

\vskip .5cm
{$\bullet$ \bf Theorem}

Let $D_\alpha(g)$ and $D_\beta (g)$ be irreducible 
representations of a group $G$ on the $d_\alpha$ 
and $d_\beta$ dimensional vector spaces.
Then, they satisfy the following orthogonality relation,
\begin{eqnarray}\label{eq:D-D-ortho}
\sum_{a \in G} D_\alpha(a)_{i \ell} D_\beta(a^{-1})_{mj} 
= \frac{N_G}{d_\alpha} \delta_{\alpha \beta} \delta_{i j} 
\delta_{\ell m}.
\end{eqnarray}

\vskip .2cm
Proof)

We define 
\begin{eqnarray}
A= \sum_{a \in G}  D_\alpha(a) B D_\alpha(a^{-1}),
\end{eqnarray}
where $B$ is a $(d_\alpha \times d_\alpha)$ arbitrary matrix.
We find $D(b)A=AD(b)$, since
\begin{eqnarray}
D_\alpha(b)A  &=& \sum_{a \in G} 
D_\alpha(b) D_\alpha(a) B D_\alpha(a^{-1}) \nonumber\\
 &=&  \sum_{a \in G} 
D_\alpha(ba)  B D_\alpha((ba)^{-1}) D_\alpha (b) \nonumber\\
 &=&  \sum_{c \in G} 
D_\alpha(c)  B D_\alpha(c^{-1}) D_\alpha (b).
\end{eqnarray}
That is, by use of Schur's lemma (II) it is found 
that the matrix $A$ should be proportional to 
the $(d_\alpha \times d_\alpha)$ identity matrix.
We choose $B_{ij} = \delta_{i \ell} \delta_{j m}$.
Then, we obtain 
\begin{eqnarray}
A_{ij} = \sum_{a \in G} D_\alpha(a)_{i \ell} 
D_\alpha (a^{-1})_{m j},
\end{eqnarray}
and right hand side (RHS) should be written by 
$\lambda(\ell,m) \delta_{ij}$, that is, 
\begin{eqnarray}
\sum_{a \in G} D_\alpha(a)_{i \ell} 
D_\alpha (a^{-1})_{m j} =\lambda(\ell,m) \delta_{ij}.
\end{eqnarray}
Furthermore, we compute the trace of both sides.
The trace of RHS is computed as
\begin{eqnarray}
\lambda (\ell, m) {\rm tr} \delta_{ij} = d_\alpha \lambda(\ell,m),
\end{eqnarray}
while the trace of left hand side (LHS) is obtained as 
\begin{eqnarray}
\sum_{i=1}^d \sum_{a \in G} D_\alpha(a)_{i \ell} 
D_\alpha (a^{-1})_{m i} &=& \sum_{a \in G} D_\alpha(aa^{-1})_{\ell m }
\nonumber\\
 &=& N_G \delta_{\ell m}.
\end{eqnarray}
By comparing these results, we obtain $\lambda(\ell,m) =
\frac{N_{G}}{d_\alpha} \delta_{\ell m}$.
Then, we find 
\begin{eqnarray}
\sum_{a \in G} D_\alpha(a)_{i \ell} D_\alpha(a^{-1})_{mj} 
= \frac{N_G}{d_\alpha}  \delta_{i j} 
\delta_{\ell m}.
\end{eqnarray}
Similarly, we define 
\begin{eqnarray}
A^{(\alpha \beta)}=\sum_{a \in G}  D_\alpha(a) B D_\beta(a^{-1}),
\end{eqnarray}
where $D_\alpha(a)$ and $D_\beta(a)$ are inequivalent to each other.
Then, we find $D_\alpha(a) A = A D_\beta (a)$.
Similarly to the previous analysis, using Schur's lemma (I), 
we can obtain 
\begin{eqnarray}
\sum_{a \in G} D_\alpha(a)_{i \ell} D_\beta(a^{-1})_{mj} 
= 0.
\end{eqnarray}
Thus, we can obtain Eq.~(\ref{eq:D-D-ortho}).
Furthermore, if the representation is unitary, 
Eq.~(\ref{eq:D-D-ortho}) is written as 
\begin{eqnarray}\label{eq:D-D-ortho*}
\sum_{a \in G} D_\alpha(a)_{i \ell} D^*_\beta(a)_{jm} 
= \frac{N_G}{d_\alpha} \delta_{\alpha \beta} \delta_{i j} 
\delta_{\ell m}. \hskip1cm \rule{5pt}{10pt}
\end{eqnarray}

\vskip .5cm
Because of this orthogonality, we can expand an arbitrary 
function of $a$, $F(a)$, in terms of the matrix elements of 
irreducible representations 
\begin{eqnarray}\label{eq:F-D}
F(a) =  \sum_{\alpha,j,k}c^\alpha_{j,k}D_\alpha(a)_{jk}.
\end{eqnarray}

\vskip .5cm
{$\bullet$ \bf Theorem}

The characters for $D_\alpha(g)$ and $D_\beta(g)$ representations, 
$\chi_\alpha(g)$ and $\chi_\beta(g)$, satisfy the following
orthogonality relation,
\begin{eqnarray}\label{eq:character-1-app}
\sum_{g \in G} \chi_{D_\alpha}(g)^* \chi_{D_\beta}(g) 
= N_G \delta_{\alpha \beta}.
\end{eqnarray}

\vskip .2cm
Proof)

{}From Eq.~(\ref{eq:D-D-ortho*}) we obtain 
\begin{eqnarray}
\sum_{g \in G} D_\alpha(g)_{i i} D^*_\beta(g)_{jj} 
= \frac{N_G}{d_\alpha} \delta_{\alpha \beta} \delta_{i j} .
\end{eqnarray}
Thus, by summing over all $i$ and $j$, we obtain 
Eq.~(\ref{eq:character-1-app}). \hskip1cm \rule{5pt}{10pt}

\vskip .5cm
The {\bf class function} is defined as a function of $a$, 
$F(a)$, which satisfies 
\begin{eqnarray}
F(g^{-1}ag) =F(a), \qquad \forall g \in G.
\end{eqnarray}

\vskip .5cm
{$\bullet$ \bf Theorem}

The number of irreducible representations 
is equal to the number of conjugacy classes.

\vskip .2cm
Proof)
The class function can also be expanded in terms of the matrix elements of 
the irreducible representations as (\ref{eq:F-D}).
Then, it is found that 
\begin{eqnarray}
F(a) &=& \frac{1}{N_G}\sum_{g \in G} F(g^{-1}ag) \nonumber\\
     &=& \frac{1}{N_G}\sum_{g \in G}
\sum_{\alpha,j,k}c^\alpha_{j,k}D_\alpha(g^{-1}ag)_{jk} \nonumber\\
 &=& \frac{1}{N_G}\sum_{g \in G}
\sum_{\alpha,j,k} c^\alpha_{j,k} \left (D_\alpha(g^{-1}) D_\alpha(a)
D_\alpha(g) \right)_{jk} .
\end{eqnarray}
By using the orthogonality relation (\ref{eq:D-D-ortho*}),
we obtain 
\begin{eqnarray}
F(a) &=& \sum_{\alpha,j,\ell} \frac{1}{d_\alpha} c^\alpha_{j,j}
D_\alpha(a)_{\ell \ell} \nonumber\\
 &=&   \sum_{\alpha,j} \frac{1}{d_\alpha} c^\alpha_{j,j}
\chi_\alpha (a).
\end{eqnarray}
That is, any class function, $F(a)$, which 
is constant on conjugacy classes, can be 
expanded by the characters $\chi_\alpha(a)$.
That implies that the number of irreducible representations 
is equal to the number of conjugacy classes. \hskip1cm \rule{5pt}{10pt}

\vskip .5cm
{$\bullet$ \bf Theorem}

The characters  satisfy the following orthogonality relation,
\begin{eqnarray}\label{eq:character-2-app}
\sum_{\alpha} \chi_{D_\alpha}(C_i)^* \chi_{D_\alpha}(C_j) 
= \frac{N_G}{n_i} \delta_{C_i C_j},
\end{eqnarray}
 where $C_i$ and $C_j$ denote the conjugacy classes 
and $n_i$ is the number of elements in the conjugacy class 
$C_i$.

\vskip .2cm
Proof)

We define the following matrix $V_{i \alpha}$,
\begin{eqnarray}
V_{i \alpha} = \sqrt{\frac{n_i}{N_G}} \chi_\alpha(C_i),
\end{eqnarray}
where $n_i$ is the number of elements in the conjugacy class $C_i$.
Note that $i$ and $\alpha$  label the conjugacy class $C_i$ 
and the irreducible representation, respectively. 
The matrix $V_{i \alpha}$ is a square matrix because 
the number of irreducible representations 
is equal to the number of conjugacy classes.
By use of $V_{i \alpha}$, the orthogonality relation 
(\ref{eq:character-1-app}) can be rewritten as 
$V^\dag V=1$, that is, $V$ is unitary.
Thus, we also obtain $V V^\dag =1$.
That means Eq.~(\ref{eq:character-2-app}).
\hskip1cm \rule{5pt}{10pt}

\clearpage


\section{Representations of $S_4$ in several bases}

For the $S_4$ group, several bases of representations have been used 
in the literature.
Most of group-theoretical aspects such as  
conjugacy classes and characters are independent of 
the basis of representations.
Tensor products are also independent of 
the basis.
For example, we always have 
\begin{eqnarray}
{\bf 2} \otimes {\bf 2} = {\bf 1}_1 \oplus {\bf 1}_2 \oplus {\bf 2},
\end{eqnarray}
in any basis.
However, it depends on the basis of representation how this equation 
is written by components. 
For example, the singlets ${\bf 1}_1$  and $ {\bf 1}_2$ in RHS  
are represented by components of ${\bf 2}$ in LHS, 
but their forms depend on the basis of representations 
as we will see below.
For applications, it is useful to show explicitly 
the transformation of bases and tensor products 
for several bases.
That is shown below.

First, we show the basis in section \ref{subsec:S4}.
All of the $S_4$ elements are written by products of 
the generators $b_1$ and $d_4$, 
which  satisfy
\begin{equation}
(b_1)^3=(d_4)^4=e,\quad d_4(b_1)^2d_4=b_1,\quad d_4b_1d_4=b_1(d_4)^2b_1\ .
\end{equation}
These generators are represented on 
${\bf 2}$, ${\bf 3}$ and ${\bf 3}'$ as follows,
\begin{equation}
b_1=\mat2{\omega}{0}{0}{\omega^2},\quad
d_4=\mat2{0}{1}{1}{0}, \qquad {\rm~~on~~{\bf 2}},
\end{equation}
\begin{equation} 
b_1=\Mat3{0}{0}{1} {1}{0}{0} {0}{1}{0},\quad d_4=\Mat3{-1}{0}{0} 
{0}{0}{-1} {0}{1}{0}, \qquad {\rm~~on~~{\bf 3}},
\end{equation}
\begin{equation} 
b_1=\Mat3{0}{0}{1} {1}{0}{0} 
{0}{1}{0},\quad d_4=\Mat3{1}{0}{0} {0}{0}{1} {0}{-1}{0}, 
\qquad {\rm~~on~~{\bf 3}'}.
\end{equation}


Next, we consider another basis, which is used e.g. in 
Ref.~\cite{Hagedorn:2006ug}. 
Following Ref.~\cite{Hagedorn:2006ug}, we denote 
the generators $b_1$ and $d_4$ by $b=b_1$ and $a=d_4$.
In this basis, the generators, $a$ and $b$, are 
represented as 
\begin{equation}
a=\mat2{-1}{0}{0}{1},\quad b=-\frac12\mat2{1}{\sqrt3}{-\sqrt3}{1}, 
\qquad {\rm~~on~~{\bf 2}},
\end{equation}
\begin{equation}
a=\Mat3{-1}{0}{0} {0}{0}{-1} {0}{1}{0},\quad b=\Mat3{0}{0}{1} {1}{0}{0} 
{0}{1}{0}, \qquad {\rm~~on~~{\bf 3}_1},
\end{equation}
\begin{equation}
a=\Mat3{1}{0}{0} {0}{0}{1} {0}{-1}{0},\quad b=\Mat3{0}{0}{1} {1}{0}{0} 
{0}{1}{0},  \qquad {\rm~~on~~{\bf 3}_2},
\end{equation}
where we define as ${\bf 3}_1\equiv {\bf 3}$ and ${\bf 3}_2\equiv {\bf
  3'}$ hereafter.
These generators, $a$ and $b$,  are represented in the real basis. 
On the other hand, the above generators, $b_1$ and $d_4$, 
 are represented in  
the complex basis.
These bases for ${\bf 2}$ are transformed 
by the unitary transformation, $U^\dagger gU$, 
where
\begin{equation}
U=\frac1{\sqrt2}\mat2{1}{i}{-1}{i}.
\end{equation}
That is, the elements $a$ and $b$ 
are written by $b_1$ and $d_4$ as 
\begin{equation}
b=U^{\dagger}b_1U=-\frac12\mat2{1}{\sqrt3}{-\sqrt3}{1},\quad
a=U^{\dagger}d_4U=\mat2{-1}{0}{0}{1},
\end{equation}
in the real basis.
For the triplets, the $(b_1,d_4)$ basis is the same as 
the $(b,a)$ basis.

Therefore, the multiplication rules are obtained as follows:
\begin{align}
\begin{pmatrix}
a_1 \\
a_2
\end{pmatrix}_{\bf 2} \otimes  \begin{pmatrix}
                                      b_1 \\
                                      b_2
                                  \end{pmatrix}_{\bf 2}
 &= (a_1b_1+a_2b_2)_{{\bf 1}_1}  \oplus (-a_1b_2+a_2b_1)_{{\bf 1}_2} 
  \oplus  \begin{pmatrix}
             a_1b_2+a_2b_1 \\
             a_1b_1-a_2b_2
         \end{pmatrix}_{{\bf 2}\ ,} \\
\begin{pmatrix}
a_1 \\
a_2
\end{pmatrix}_{\bf 2} \otimes  \begin{pmatrix}
                                      b_1 \\
                                      b_2 \\
                                      b_3
                                  \end{pmatrix}_{{\bf 3}_1}
 &= \begin{pmatrix}
          a_2b_1 \\
          -\frac{1}{2}(\sqrt 3a_1b_2+a_2b_2) \\
          \frac{1}{2}(\sqrt 3a_1b_3-a_2b_3)
      \end{pmatrix}_{{\bf 3}_1} \oplus \begin{pmatrix}
                                        a_1b_1 \\
                                        \frac{1}{2}(\sqrt 3a_2b_2-a_1b_2) 
\\
                                        -\frac{1}{2}(\sqrt 3a_2b_3+a_1b_3)
                                   \end{pmatrix}_{{\bf 3}_2\ ,} \\
\begin{pmatrix}
a_1 \\
a_2
\end{pmatrix}_{\bf 2} \otimes  \begin{pmatrix}
                                      b_1 \\
                                      b_2 \\
                                      b_3
                                  \end{pmatrix}_{{\bf 3}_2}
&= \begin{pmatrix}
         a_1b_1 \\
         \frac{1}{2}(\sqrt 3a_2b_2-a_1b_2) \\
         -\frac{1}{2}(\sqrt 3a_2b_3+a_1b_3)
     \end{pmatrix}_{{\bf 3}_1} \oplus
      \begin{pmatrix}
                                      a_2b_1 \\
                                      -\frac{1}{2}(\sqrt 3a_1b_2+a_2b_2) \\
                                      \frac{1}{2}(\sqrt 3a_1b_3-a_2b_3)
                                  \end{pmatrix}_{{\bf 3}_2\ ,} \\
\begin{pmatrix}
a_1 \\
a_2 \\
a_3
\end{pmatrix}_{{\bf 3}_1} \otimes  \begin{pmatrix}
                                      b_1 \\
                                      b_2 \\
                                      b_3
                                  \end{pmatrix}_{{\bf 3}_1}
 &= (a_1b_1+a_2b_2+a_3b_3)_{{\bf 1}_1} 
  \oplus \begin{pmatrix}
             \frac{1}{\sqrt 2}(a_2b_2-a_3b_3) \\                                            
             \frac{1}{\sqrt 6}(-2a_1b_1+a_2b_2+a_3b_3)
         \end{pmatrix}_{\bf 2} \nonumber \\
 &\ \oplus \begin{pmatrix}
            a_2b_3+a_3b_2 \\
            a_1b_3+a_3b_1 \\
            a_1b_2+a_2b_1
         \end{pmatrix}_{{\bf 3}_1} \oplus \begin{pmatrix}
                                          a_3b_2-a_2b_3 \\
                                          a_1b_3-a_3b_1 \\
                                          a_2b_1-a_1b_2
                                       \end{pmatrix}_{{\bf 3}_2\ ,} \\
\begin{pmatrix}
a_1 \\
a_2 \\
a_3
\end{pmatrix}_{{\bf 3}_2} \otimes  \begin{pmatrix}
                                      b_1 \\
                                      b_2 \\
                                      b_3
                                  \end{pmatrix}_{{\bf 3}_2}
 &= (a_1b_1+a_2b_2+a_3b_3)_{{\bf 1}_1}
  \oplus \begin{pmatrix}
             \frac{1}{\sqrt 2}(a_2b_2-a_3b_3) \\                                            
             \frac{1}{\sqrt 6}(-2a_1b_1+a_2b_2+a_3b_3)
         \end{pmatrix}_{\bf 2} \nonumber \\
 &\ \oplus \begin{pmatrix}
            a_2b_3+a_3b_2 \\
            a_1b_3+a_3b_1 \\
            a_1b_2+a_2b_1
         \end{pmatrix}_{{\bf 3}_1} \oplus \begin{pmatrix}
                                          a_3b_2-a_2b_3 \\
                                          a_1b_3-a_3b_1 \\
                                          a_2b_1-a_1b_2
                                       \end{pmatrix}_{{\bf 3}_2\ ,} \\
\begin{pmatrix}
a_1 \\
a_2 \\
a_3
\end{pmatrix}_{{\bf 3}_1} \otimes  \begin{pmatrix}
                                      b_1 \\
                                      b_2 \\
                                      b_3
                                  \end{pmatrix}_{{\bf 3}_2}
 &= (a_1b_1+a_2b_2+a_3b_3)_{{\bf 1}_2}  
 \oplus \begin{pmatrix}
             \frac{1}{\sqrt 6}(2a_1b_1-a_2b_2-a_3b_3) \\
             \frac{1}{\sqrt 2}(a_2b_2-a_3b_3)
         \end{pmatrix}_{\bf 2} \nonumber \\
 &\ \oplus \begin{pmatrix}
            a_3b_2-a_2b_3 \\
            a_1b_3-a_3b_1 \\
            a_2b_1-a_1b_2
         \end{pmatrix}_{{\bf 3}_1} \oplus \begin{pmatrix}
                                          a_2b_3+a_3b_2 \\
                                          a_1b_3+a_3b_1 \\
                                          a_1b_2+a_2b_1
                                       \end{pmatrix}_{{\bf 3}_2\ .}
\end{align}

\newpage 

Next, we consider a different basis, which is used, e.g.  
in Ref.~\cite{Bazzocchi:2009pv}, 
with the generator $s$ and $t$ corresponding to $d_4$ and $b_1$, 
respectively. 
These generators are represented 
\begin{equation}
s=\mat2{0}{1}{1}{0},\quad t=\mat2{\omega }{0}{0}{\omega ^2},
\qquad {\rm~~on~~{\bf 2}},
\end{equation}
\begin{equation}
s=\frac13\Mat3{-1}{2\omega }{2\omega ^2} {2\omega }{2\omega ^2}{-1} 
{2\omega ^2}{-1}{2\omega },
\quad t=\Mat3{1}{0}{0} {0}{\omega ^2}{0} {0}{0}{\omega },
\qquad {\rm~~on~~{\bf 3}_1},
\end{equation}
\begin{equation}
s=\frac13\Mat3{1}{-2\omega }{-2\omega ^2} {-2\omega }{-2\omega ^2}{1} 
{-2\omega ^2}{1}{-2\omega },
\quad t=\Mat3{1}{0}{0} {0}{\omega ^2}{0} {0}{0}{\omega },
\qquad {\rm~~on~~{\bf 3}_2}.
\end{equation}
The doublet of this basis \cite{Bazzocchi:2009pv} is the same as 
the $(d_4,b_1) $ basis.
In the representations ${\bf 3}_1$ and ${\bf 3}_2$, 
the $(s,t)$ basis and $(d_4,b_1) $ basis are transformed by 
the following unitary matrix $U_\omega $:
\begin{eqnarray}
U_\omega =\frac1{\sqrt3}\Mat3{1}{1}{1} {1}{\omega }{\omega ^2} {1}{\omega 
^2}{\omega },
\end{eqnarray}
which is the so-called magic matrix.
That is, the elements $s$ and $t$ are written 
by $d_4$ and $b_1$ as 
\begin{equation}
s=U_\omega ^{\dagger}d_4U_\omega =\frac13\Mat3{-1}{2\omega }{2\omega ^2} 
{2\omega }{2\omega ^2}{-1} {2\omega ^2}{-1}{2\omega ^2},\quad
t=U_\omega ^{\dagger}b_1U_\omega =\Mat3{1}{0}{0} {0}{\omega ^2}{0} 
{0}{0}{\omega }.
\end{equation}
For ${\bf 3}_2$, 
we also find $s$ and $t$ in the same way.

Therefore, the multiplication rules are obtained as follows:
\begin{align}
\begin{pmatrix}
a_1 \\
a_2
\end{pmatrix}_{\bf 2} \otimes  \begin{pmatrix}
                                      b_1 \\
                                      b_2
                                  \end{pmatrix}_{\bf 2}
 &= (a_1b_2+a_2b_1)_{{\bf 1}_1}  \oplus (a_1b_2-a_2b_1)_{{\bf 1}_2}  
 \oplus \begin{pmatrix}
             a_2b_2 \\
             a_1b_1
         \end{pmatrix}_{2\ ,} \\
\begin{pmatrix}
a_1 \\
a_2
\end{pmatrix}_{\bf 2} \otimes  \begin{pmatrix}
                                      b_1 \\
                                      b_2 \\
                                      b_3
                                  \end{pmatrix}_{{\bf 3}_1}
 &= \begin{pmatrix}
        a_1b_2+a_2b_3 \\
        a_1b_3+a_2b_1 \\
        a_1b_1+a_2b_2
      \end{pmatrix}_{{\bf 3}_1} \oplus \begin{pmatrix}
                                       a_1b_2-a_2b_3 \\
                                       a_1b_3-a_2b_1 \\
                                       a_1b_1-a_2b_2
                                    \end{pmatrix}_{{\bf 3}_2\ ,} \\
\begin{pmatrix}
a_1 \\
a_2
\end{pmatrix}_{\bf 2} \otimes  \begin{pmatrix}
                                      b_1 \\
                                      b_2 \\
                                      b_3
                                  \end{pmatrix}_{{\bf 3}_2}
&= \begin{pmatrix}
        a_1b_2-a_2b_3 \\
        a_1b_3-a_2b_1 \\
        a_1b_1-a_2b_2
     \end{pmatrix}_{{\bf 3}_1} \oplus \begin{pmatrix}
                                      a_1b_2+a_2b_3 \\
                                      a_1b_3+a_2b_1 \\
                                      a_1b_1+a_2b_2
                                   \end{pmatrix}_{{\bf 3}_2\ ,} \\
\begin{pmatrix}
a_1 \\
a_2 \\
a_3
\end{pmatrix}_{{\bf 3}_1} \otimes  \begin{pmatrix}
                                      b_1 \\
                                      b_2 \\
                                      b_3
                                  \end{pmatrix}_{{\bf 3}_1}
 &= (a_1b_1+a_2b_3+a_3b_2)_{{\bf 1}_1}
\oplus \begin{pmatrix}
             a_2b_2+a_1b_3+a_3b_1 \\
             a_3b_3+a_1b_2+a_2b_1
         \end{pmatrix}_{\bf 2} \nonumber \\
 &\ \oplus \begin{pmatrix}
            2a_1b_1-a_2b_3-a_3b_2 \\
            2a_3b_3-a_1b_2-a_2b_1 \\
            2a_2b_2-a_1b_3-a_3b_1
         \end{pmatrix}_{{\bf 3}_1} \oplus \begin{pmatrix}
                                          a_2b_3-a_3b_2 \\
                                          a_1b_2-a_2b_1 \\
                                          a_3b_1-a_1b_3
                                       \end{pmatrix}_{{\bf 3}_2\ ,} \\
\begin{pmatrix}
a_1 \\
a_2 \\
a_3
\end{pmatrix}_{{\bf 3}_2} \otimes  \begin{pmatrix}
                                      b_1 \\
                                      b_2 \\
                                      b_3
                                  \end{pmatrix}_{{\bf 3}_2}
 &= (a_1b_1+a_2b_3+a_3b_2)_{{\bf 1}_1}
\oplus \begin{pmatrix}
             a_2b_2+a_1b_3+a_3b_1 \\
             a_3b_3+a_1b_2+a_2b_1
         \end{pmatrix}_{\bf 2} \nonumber \\
 &\ \oplus \begin{pmatrix}
            2a_1b_1-a_2b_3-a_3b_2 \\
            2a_3b_3-a_1b_2-a_2b_1 \\
            2a_2b_2-a_1b_3-a_3b_1
         \end{pmatrix}_{{\bf 3}_1} \oplus \begin{pmatrix}
                                          a_2b_3-a_3b_2 \\
                                          a_1b_2-a_2b_1 \\
                                          a_3b_1-a_1b_3
                                       \end{pmatrix}_{{\bf 3}_2\ ,} \\
\begin{pmatrix}
a_1 \\
a_2 \\
a_3
\end{pmatrix}_{{\bf 3}_1} \otimes  \begin{pmatrix}
                                      b_1 \\
                                      b_2 \\
                                      b_3
                                  \end{pmatrix}_{{\bf 3}_2}
 &= (a_1b_1+a_2b_3+a_3b_2)_{{\bf 1}_2}
\oplus \begin{pmatrix}
             a_2b_2+a_1b_3+a_3b_1 \\
             -a_3b_3-a_1b_2-a_2b_1
         \end{pmatrix}_{\bf 2} \nonumber \\
 &\ \oplus \begin{pmatrix}
            a_2b_3-a_3b_2 \\
            a_1b_2-a_2b_1 \\
            a_3b_1-a_1b_3
         \end{pmatrix}_{{\bf 3}_1} \oplus \begin{pmatrix}
                                          2a_1b_1-a_2b_3-a_3b_2 \\
                                          2a_3b_3-a_1b_2-a_2b_1 \\
                                          2a_2b_2-a_1b_3-a_3b_1
         \end{pmatrix}_{{\bf 3}_2\ .}
\end{align}


Here, we consider another basis, which is used, e.g.  
in Ref.~
\cite{Altarelli:2009gn}, 
with the generator $\tilde t$ and $\tilde s$ satisfying 
\begin{equation}
\tilde t^4=\tilde s^2=(\tilde s \tilde t)^3=(\tilde t \tilde s)^3=e.
\end{equation}
These generators are represented as 
\begin{equation}
\tilde t=\mat2{1}{0}{0}{-1},\quad \tilde 
s=\frac12\mat2{-1}{\sqrt3}{\sqrt3}{1},\quad 
\tilde s \tilde t=\frac12\mat2{-1}{-\sqrt3}{\sqrt3}{-1},
\quad {\rm~~on~~{\bf 2}},
\end{equation}
\begin{eqnarray}
& & \tilde t=\Mat3{-1}{0}{0} {0}{-i}{0} {0}{0}{i},
\quad \tilde s=\Mat3{0}{-\frac{1}{\sqrt2}}{-\frac{1}{\sqrt{2}}} 
{-\frac{1}{\sqrt{2}}}{\frac12}{-\frac12} 
{-\frac{1}{\sqrt{2}}}{-\frac12}{\frac12},  \nonumber \\
& & \qquad \tilde s\tilde t=\Mat3{0}{\frac{i}{\sqrt2}}{-\frac{i}{\sqrt2}} 
{\frac{1}{\sqrt2}}{-\frac{i}{2}}{-\frac{i}{2}}
{\frac{1}{\sqrt2}}{\frac{i}{2}}{\frac{i}{2}},
\qquad \qquad {\rm on~~{\bf 3}_1},
\end{eqnarray}
\begin{eqnarray}
& & \tilde t=\Mat3{1}{0}{0} {0}{i}{0} {0}{0}{-i},
\quad \tilde s=\Mat3{0}{\frac{1}{\sqrt2}}{\frac{1}{\sqrt{2}}} 
{\frac{1}{\sqrt{2}}}{-\frac12}{\frac12} 
{\frac{1}{\sqrt{2}}}{\frac12}{-\frac12},  \nonumber \\
& & \qquad \tilde s\tilde t=\Mat3{0}{\frac{i}{\sqrt2}}{-\frac{i}{\sqrt2}} 
{\frac{1}{\sqrt2}}{-\frac{i}{2}}{-\frac{i}{2}}
{\frac{1}{\sqrt2}}{\frac{i}{2}}{\frac{i}{2}}, 
\qquad \qquad {\rm~~on~~{\bf 3}_2}.
\end{eqnarray}
For the representation ${\bf 2}$, the 
following unitary transformation
matrix $U_\text{doublet}$: 
\begin{equation}
U_\text{doublet}=\frac{1}{\sqrt2}\mat2{1}{i}{1}{-i},
\end{equation}
is used and 
the elements $\tilde t$ and $\tilde s \tilde t $ are written 
by $d_1$ and $b_1$ 
as 
\begin{equation}
\tilde t=U_\text{doublet}^\dagger 
d_4U_\text{doublet}=\mat2{1}{0}{0}{-1},\quad 
\tilde s\tilde t=U_\text{doublet}^\dagger 
b_1U_\text{doublet}=\frac12\mat2{-1}{-\sqrt3}{\sqrt3}{-1}.
\end{equation}
On the other hand, for the representation ${\bf 3}_1$ and ${\bf 3}_2$, 
the following unitary transformation 
matrix $U_\text{triplet}$:  
\begin{equation}
U_\text{triplet}=\Mat3{1}{0}{0} {0}{\frac{1}{\sqrt2}}{\frac{1}{\sqrt2}} 
{0}{\frac{i}{\sqrt2}}{-\frac{i}{\sqrt2}},
\end{equation}
is used. 
For ${\bf 3}_1$, the elements $\tilde t$ and $\tilde s\tilde t$ are written 
by $d_4$ and $b_1$ 
as 
\begin{eqnarray}
& & \tilde t=U_\text{triplet}^\dagger d_4U_\text{triplet}=\Mat3{-1}{0}{0} 
{0}{-i}{0} 
{0}{0}{i},  \nonumber \\
& &  \tilde s\tilde t=U_\text{triplet}^\dagger 
b_1U_\text{triplet}=\Mat3{0}{\frac{i}{\sqrt2}}{-\frac{i}{\sqrt2}} 
{\frac{1}{\sqrt2}}{-\frac{i}{2}}{-\frac{i}{2}} 
{\frac{1}{\sqrt2}}{\frac{i}{2}}{\frac{i}{2}}.
\end{eqnarray}
For ${\bf 3}_2$, we also find the same transformations.

Therefore, the multiplication rules are as follows:
\begin{align}
\begin{pmatrix}
a_1 \\
a_2
\end{pmatrix}_{\bf 2} \otimes  \begin{pmatrix}
                                      b_1 \\
                                      b_2
                                  \end{pmatrix}_{\bf 2}
 &= (a_1b_1+a_2b_2)_{{\bf 1}_1}  \oplus (a_1b_2-a_2b_1)_{{\bf 1}_2}
\oplus \begin{pmatrix}
             a_2b_2-a_1b_1 \\
             a_1b_2+a_2b_1
         \end{pmatrix}_{{\bf 2}\ ,} \\
\begin{pmatrix}
a_1 \\
a_2
\end{pmatrix}_{\bf 2} \otimes  \begin{pmatrix}
                                      b_1 \\
                                      b_2 \\
                                      b_3
                                  \end{pmatrix}_{{\bf 3}_1}
 &= \begin{pmatrix}
       a_1b_1 \\
       \frac{\sqrt 3}{2}a_2b_3-\frac{1}{2}a_1b_2 \\
       \frac{\sqrt 3}{2}a_2b_2-a_1b_3
      \end{pmatrix}_{{\bf 3}_1} \oplus \begin{pmatrix}
                                       -a_2b_1 \\
                                       \frac{\sqrt 
3}{2}a_1b_3+\frac{1}{2}a_2b_2 \\
                                       \frac{\sqrt 
3}{2}a_1b_2+\frac{1}{2}a_2b_3
                                    \end{pmatrix}_{{\bf 3}_2\ ,} \\
\begin{pmatrix}
a_1 \\
a_2
\end{pmatrix}_{\bf 2} \otimes  \begin{pmatrix}
                                      b_1 \\
                                      b_2 \\
                                      b_3
                                  \end{pmatrix}_{{\bf 3}_2}
&= \begin{pmatrix}
        -a_2b_1 \\
        \frac{\sqrt 3}{2}a_1b_3+\frac{1}{2}a_2b_2 \\
        \frac{\sqrt 3}{2}a_1b_2+\frac{1}{2}a_2b_3
     \end{pmatrix}_{{\bf 3}_1} \oplus \begin{pmatrix}
                                      a_1b_1 \\
                                      \frac{\sqrt 
3}{2}a_2b_3-\frac{1}{2}a_1b_2 \\
                                      \frac{\sqrt 3}{2}a_2b_2-a_1b_3
                                   \end{pmatrix}_{{\bf 3}_2\ ,} \\
\begin{pmatrix}
a_1 \\
a_2 \\
a_3
\end{pmatrix}_{{\bf 3}_1} \otimes  \begin{pmatrix}
                                      b_1 \\
                                      b_2 \\
                                      b_3
                                  \end{pmatrix}_{{\bf 3}_1}
 &= (a_1b_1+a_2b_3+a_3b_2)_{1_1}  
 \oplus \begin{pmatrix}
            a_1b_1-\frac{1}{2}(a_2b_3+a_3b_2) \\
            \frac{\sqrt 3}{2}(a_2b_2+a_3b_3)
         \end{pmatrix}_{\bf 2} \nonumber \\
 &\ \oplus \begin{pmatrix}
            a_3b_3-a_2b_2 \\
            a_1b_3+a_3b_1 \\
            -a_1b_2-a_2b_1
         \end{pmatrix}_{{\bf 3}_1} \oplus \begin{pmatrix}
                                          a_3b_2-a_2b_3 \\
                                          a_2b_1-a_1b_2 \\
                                          a_1b_3-a_3b_1
                                       \end{pmatrix}_{{\bf 3}_2\ ,} \\
\begin{pmatrix}
a_1 \\
a_2 \\
a_3
\end{pmatrix}_{{\bf 3}_2} \otimes  \begin{pmatrix}
                                      b_1 \\
                                      b_2 \\
                                      b_3
                                  \end{pmatrix}_{{\bf 3}_2}
 &= (a_1b_1+a_2b_3+a_3b_2)_{{\bf 1}_1}
\oplus \begin{pmatrix}
            a_1b_1-\frac{1}{2}(a_2b_3+a_3b_2) \\
            \frac{\sqrt 3}{2}(a_2b_2+a_3b_3)
         \end{pmatrix}_{\bf 2} \nonumber \\
 &\ \oplus \begin{pmatrix}
            a_3b_3-a_2b_2 \\
            a_1b_3+a_3b_1 \\
            -a_1b_2-a_2b_1
         \end{pmatrix}_{{\bf 3}_1} \oplus \begin{pmatrix}
                                          a_3b_2-a_2b_3 \\
                                          a_2b_1-a_1b_2 \\
                                          a_1b_3-a_3b_1
                                       \end{pmatrix}_{{\bf 3}_2\ ,} \\
\begin{pmatrix}
a_1 \\
a_2 \\
a_3
\end{pmatrix}_{{\bf 3}_1} \otimes  \begin{pmatrix}
                                      b_1 \\
                                      b_2 \\
                                      b_3
                                  \end{pmatrix}_{{\bf 3}_2}
 &= (a_1b_1+a_2b_3+a_3b_2)_{{\bf 1}_2} \nonumber \\
& \oplus \begin{pmatrix}
            \frac{\sqrt 3}{2}(a_2b_2+a_3b_3)
            -a_1b_1+\frac{1}{2}(a_2b_3+a_3b_2) \\
         \end{pmatrix}_{\bf 2} \nonumber \\
 &\ \oplus \begin{pmatrix}
            a_3b_2-a_2b_3 \\
            a_2b_1-a_1b_2 \\
            a_1b_3-a_3b_1
         \end{pmatrix}_{{\bf 3}_1} \oplus \begin{pmatrix}
                                          a_3b_3-a_2b_2 \\
                                          a_1b_3+a_3b_1 \\
                                          -a_1b_2-a_2b_1
                                       \end{pmatrix}_{{\bf 3}_2\ .}
\end{align}

\clearpage


\section{Representations of $A_4$ in different basis}

Here, we show another basis for representations of the $A_4$ group.
First, we show the basis in section \ref{sec:A4}.
All of the $A_4$ elements are written by products of the generators, 
$s$ and $t$, which  satisfy
\begin{equation}
s^2=t^3=(st)^3=e\ .
\end{equation}
On the representation ${\bf 3}$, 
these generators are represented as 
\begin{equation} 
s=a_2=\Mat3{1}{0}{0} {0}{-1}{0} {0}{0}{-1},\quad t=b_1=\Mat3{0}{0}{1} 
{1}{0}{0} 
{0}{1}{0}.
\end{equation}


Next, we consider another basis, which is used, 
e.g. in Ref.~\cite{Altarelli:2005yx}. 
In this basis, we denote the generators $a$ and $b$, 
which correspond to $s$ and $t$, respectively,
and these generators are represented as 
\begin{equation} 
a=\frac13\Mat3{-1}{2}{2} {2}{-1}{2} {2}{2}{-1},\quad b=\Mat3{1}{0}{0} 
{0}{\omega 
^2}{0} {0}{0}{\omega },
\end{equation}
on the representation ${\bf 3}$.
These bases are transformed by the following unitary transformation matrix 
$U_\omega $ as
\begin{eqnarray}
U_\omega =\frac1{\sqrt3}\Mat3{1}{1}{1} {1}{\omega }{\omega ^2} {1}{\omega 
^2}{\omega },
\end{eqnarray}
and the elements $a$ and $b$ are written as 
\begin{eqnarray}
&&a=U_\omega ^{\dagger}sU_\omega =\frac13\Mat3{-1}{2}{2} {2}{-1}{2} 
{2}{2}{-1},\quad
b=U_\omega ^{\dagger}tU_\omega =\Mat3{1}{0}{0} {0}{\omega ^2}{0} 
{0}{0}{\omega }.
\end{eqnarray}

Therefore, the multiplication rule of the triplet is 
obtained as follows,
\begin{align}
\begin{pmatrix}
a_1\\
a_2\\
a_3
\end{pmatrix}_{\bf 3}
\otimes 
\begin{pmatrix}
b_1\\
b_2\\
b_3
\end{pmatrix}_{\bf 3}
&=\left (a_1b_1+a_2b_3+a_3b_2\right )_{\bf 1} 
\oplus \left (a_3b_3+a_1b_2+a_2b_1\right )_{{\bf 1}'} \nonumber \\
& \oplus \left (a_2b_2+a_1b_3+a_3b_1\right )_{{\bf 1}''} \nonumber \\
&\oplus \frac13
\begin{pmatrix}
2a_1b_1-a_2b_3-a_3b_2 \\
2a_3b_3-a_1b_2-a_2b_1 \\
2a_2b_2-a_1b_3-a_3b_1
\end{pmatrix}_{{\bf 3}}
\oplus \frac12
\begin{pmatrix}
a_2b_3-a_3b_2 \\
a_1b_2-a_2b_1 \\
a_1b_3-a_3b_1
\end{pmatrix}_{{\bf 3}\ .}
\end{align}

\end{document}